\documentclass[12pt]{article}

\usepackage{palatino}
\usepackage{amssymb,amsmath,amsfonts,eurosym,geometry,ulem,graphicx,caption,color,setspace,sectsty,comment,footmisc,caption,pdflscape,subfigure,array, dsfont}
\usepackage{bbm}
\usepackage{tabularray}
\definecolor{Black}{RGB}{0,0,0}
\usepackage{rotating}
\usepackage{graphicx}
\usepackage{afterpage}
\usepackage[colorlinks,linkcolor=blue, citecolor=blue, anchorcolor=blue,urlcolor=blue]{hyperref}
\DeclareMathOperator*{\argmax}{arg\,max}
\DeclareMathOperator*{\argmin}{arg\,min}
\usepackage{tablefootnote}
\usepackage{float}
\usepackage{threeparttable}
\usepackage{threeparttablex}
\usepackage{multicol}
\usepackage{xcolor}
\usepackage{amsmath}
\usepackage{etoolbox}
\usepackage{natbib}
\usepackage{pdflscape} 
\usepackage{graphicx}  
\usepackage{adjustbox} 
\usepackage{booktabs}         
\usepackage{longtable}        
\usepackage{geometry} 
\usepackage{authblk}
\usepackage{algorithm,algorithmic}

\usepackage{graphicx}
\usepackage{subcaption}
\usepackage{hyperref}
\newcommand{\subsubsubsection}[1]{\paragraph{#1}\mbox{}\\}
\newcolumntype{L}[1]{>{\raggedright\let\newline\\arraybackslash\hspace{0pt}}m{#1}}
\newcolumntype{C}[1]{>{\centering\let\newline\\arraybackslash\hspace{0pt}}m{#1}}
\newcolumntype{R}[1]{>{\raggedleft\let\newline\\arraybackslash\hspace{0pt}}m{#1}}

\begin{document}

\begin{titlepage}
    \title{Electricity Market Predictability: \\\ Virtues of Machine Learning and Links to the Macroeconomy}
    \author{
        Jinbo Cai\thanks{School of Banking \& Finance, UNSW Business School, University of New South Wales, Sydney, Australia, 2033. Email: \texttt{jinbo.cai@student.unsw.edu.au}} \quad
        Wenze Li\thanks{Division of Economics, School of Social Sciences, Nanyang Technological University, Singapore, 639818. Email: \texttt{wenze001@e.ntu.edu.sg}} \quad
        Wenjie Wang\thanks{Division of Economics, School of Social Sciences, Nanyang Technological University, Singapore, 639818. Email: \texttt{wang.wj@ntu.edu.sg}}
    }
    \date{\textit{Version:} \today}
    \maketitle
    
    \begin{abstract}
    \noindent With stakeholder-level in-market data, we conduct a comparative analysis of machine learning (ML) for forecasting electricity prices in Singapore, spanning 15 individual models and 4 ensemble approaches. Our empirical findings justify the three virtues of ML models: (1) the virtue of capturing non-linearity, (2) the complexity \citep{kelly2024virtue} and (3) the $l2$-norm and bagging techniques in a weak factor environment \citep{shen2024can}. Simulation also supports the first virtue. Penalizing prediction correlation improves ensemble performance when individual models are highly correlated. The predictability can be translated into sizable economic gains under the mean-variance framework. We also reveal significant patterns of time-series heterogeneous predictability across macro regimes: predictability is clustered in expansion, volatile market and extreme geopolitical risk periods. Our feature importance results agree with the complex dynamics of Singapore’s electricity market after de-regulation, yet highlight its relatively supply-driven nature with the continued presence of strong regulatory influences.
    \vspace{1cm}
    
    \noindent \textit{JEL} classifications: C52, C53, C63, L94, Q40, Q41, Q47\\

    \end{abstract}
    \setcounter{page}{0}
    \thispagestyle{empty}
\end{titlepage}

\pagebreak \newpage
\doublespacing

\section{Introduction}
Forecasting electricity prices is a fascinating endeavor that attracts both academics and industry practitioners. Accurate prediction is crucial for stabilizing the power system, mitigating price fluctuations, and enabling more efficient power generation planning. It also empowers consumers to adjust their usage through demand response programs, ultimately enhancing market resilience \citep{hu2022impact, olabi2021critical, van2018techno, sai2023event}. Crucially, electricity markets do not operate in a vacuum: price swings cascade through production costs, corporate investment decisions, and household budgets, linking electricity prices to broader macroeconomic cycles, energy market volatility, and even geopolitical risks. Robust forecasting, therefore, must account for these interconnected dynamics to inform both policy and practice. Yet the market’s complexity—especially in a small, open economy like Singapore—intensifies the challenge of predictive accuracy. Although Singapore’s power sector has undergone substantial deregulation since the mid-1990s, “market-driven” dynamics have not always yielded stable price signals. Adding to the volatility is Singapore’s heavy reliance on imported natural gas, which exposes the local market to global shocks and necessitates a balancing act among stakeholders such as suppliers, consumers, and regulators. Understanding these interactions and the broader links between electricity prices and the macroeconomy remains a core policy concern.

To pursue predictive accuracy, previous research has applied machine learning (ML) to electricity price forecasting through linear and tree-based models \citep{kath2018value, sai2023event}, neural network-based approaches \citep{panapakidis2016day, dudek2016multilayer, wang2017multi, bento2018bat}, and ensemble methods \citep{mirakyan2017composite, bordignon2013combining, ghasemi2016novel}. These studies concentrate on various but small sets of predictors \footnote{Such as previous electricity price \citep{kath2018value, sai2023event, panapakidis2016day, wang2017multi, bento2018bat, mirakyan2017composite, bordignon2013combining, ghasemi2016novel}, load data \citep{kath2018value, panapakidis2016day, dudek2016multilayer, ghasemi2016novel}, time indicators \citep{kath2018value, panapakidis2016day, bento2018bat}, gas price \citep{panapakidis2016day, bordignon2013combining}, renewable energy sources (RES) generation \citep{kath2018value, panapakidis2016day}, demand forecast \citep{bordignon2013combining}, available capacity margin \citep{bordignon2013combining}, and volatility \citep{bordignon2013combining}}, and the lack of in-market stakeholder-level data makes it difficult to understand the interplay between different parties within the electricity market, let alone the interactions and linkages between local market players and the global macroeconomy. 

Against this backdrop, this paper conducts a comprehensive comparative analysis of ML methods for forecasting electricity prices (in log returns), using a rich set of 619 features that include stakeholder-level market data from Singapore’s Energy Market Company (EMC), domestic macroeconomic indicators, and international macro factors. We evaluate 15 individual models, spanning a broad range of approaches: Ordinary Least Squares (OLS); penalized linear regressions—including LASSO, Ridge, and Elastic Net (ENet); dimension reduction techniques such as Principal Component Regression (PCR) and Partial Least Squares (PLS); Generalized Linear Models (GLM); tree-based methods—including Random Forest (RF) and two types of Gradient Boosted Decision Trees; and Feedforward Neural Networks (NN) with one to five hidden layers. In addition, we consider four ensemble approaches, including equal weighting, optimized weighting with and without correlation penalties, and the economic-driven combination from \citet{rapach2010out}. The study aims to: (1) verify three core virtues of ML in forecasting—its ability to capture non-linear relationships, the virtue of complexity in high-dimensional settings \citep{kelly2024virtue}, and the virtue of $l2$-norm and bagging in weak-factor environments \citep{shen2024can}; and (2) explore the macroeconomic underpinnings of electricity price predictability, disentangling the roles of global shocks, regulatory design, and in-market participants in shaping Singapore’s electricity prices. The forecasting spans the period from Jan 2003 to Dec 2023, employing an expanding window approach to conduct daily-level predictions.

The out-of-sample (OOS) findings from the time-series analysis indicate that all $R^2_{OOS}$ values are positive and statistically significant at the 1\% level according to the one-sided Diebold-Mariano (DM) test. Thus, at the broadest level, our main empirical finding is that ML as a whole has the potential to improve our understanding of electricity returns. It digests our predictor data set, which is massive from the perspective of the existing literature, into a return forecasting model that dominates traditional approaches. The immediate implication is that ML aids in solving practical problems, such as production plans for suppliers, energy futures/options positions for suppliers, market interventions for regulators, hedging positions in electricity futures for the demand side, etc. Next, we compare individual models to verify the virtue of sophisticated ML models in processing the high-dimensional feature set. Across all $R^2_{OOS}$ metrics and relative root mean squared error ($rRMSE$), we identify GLM,  Gradient Boosting with Huber (XGB(+H)), and Light Gradient Boosting Machine with Huber (LGBM(+H)) as the top three models, with 36.85\% (27.66\%, 43.70\%, 43.70\%, 0.9204), 37.87\% (28.83\%, 43.66\%, 43.66\%, 0.9207) and 37.40\% (28.29\%, 43.26\%, 43.26\%, 0.9239) in $R^2_{OOS,lagprice}$ ($R^2_{OOS,AR(1)}$, $R^2_{OOS,mean}$, $R^2_{OOS,zero}$, and $rRMSE$) metric respectively. When model performance is averaged by category, the ranking is as follows: tree-based models $\sim$ GLM \(>\) NN5 \(>\) dimension reduction models \(>\) penalized linear models \(>\) OLS. This is not surprising, as these best-performing models excel in accommodating the nonlinear predictive relationships, and allowing for nonlinearities substantially improves predictions. This finding is robust across various statistical tests, the average weights in the ensemble model, and a framework in which we evaluate the degree of conditional mean misspecification across various models by comparing the predictive conditional variance losses under the same variance model specification.

Regarding the ensemble models, we find that when individual models' predictions are highly correlated, introducing a prediction correlation penalty when optimizing the ensemble weights effectively enhances performance to beat the best individual models. Moreover, our results also reveal the virtue of employing $l2-$ norm and bagging techniques when factor environment is relatively weak to some extent. Given that from $R^2_{OOS,mean}$ (also, $R^2_{OOS,zero}$) to $R^2_{OOS,lagprice}$ then $R^2_{OOS, AR(1)}$, the factor environment is from strong to less strong, we find that Ridge performs worse than LASSO in $R^2_{OOS, mean}$ (also, $R^2_{OOS, zero}$), but beats the LASSO in $R^2_{OOS, lagprice}$ and more economically significantly in $R^2_{OOS, AR(1)}$, while RF unexpectedly underperforms penalized linear and dimension-reduction models in \( R^2_{OOS, mean} \) and \( R^2_{OOS, zero} \) but reverses in \( R^2_{OOS, lagprice} \) and \( R^2_{OOS, AR(1)} \). Although RF still performs worse than the boosted trees (XGB(+H) and LGBM(+H)) in $R^2_{OOS,lagprice}$ and $R^2_{OOS,AR(1)}$, the magnitude of the difference between them narrows compared to that in $R^2_{OOS,mean}$ (and $R^2_{OOS,zero}$). Finally, with model-specific time-varying $R^2_{OOS}$ metrics and model complexity proxy, we conduct a panel regression with both model and OOS month fixed effects to justify the virtue of complexity. We find that a one-standard-deviation increase in model complexity corresponds to a 1.35\% increase in $R^2_{OOS,lagprice}$ and a 1.05\% increase in $R^2_{OOS,AR(1)}$ (though these are not statistically significant), but a 1.48\% statistically significant increase in the mean and zero bench-marked $R^2_{OOS}$ metrics. 

Simulation provides further insight into our ML findings. In Appendix \ref{simulation}, we perform Monte Carlo simulations that support the above interpretations of our analysis in terms of the virtue of capturing non-linearity. We apply ML to simulated data from two different data generating processes. Both produce data from a high dimensional predictor set. In one, individual predictors enter only linearly and additively, while in the other, predictors can enter through nonlinear transformations and pairwise interactions. When we apply our ML repertoire to the simulated datasets, we find that linear models dominate in the linear and non-interactive setting, whereas tree-based methods and neural networks significantly outperform in the nonlinear and interactive setting.

We also show that predictability can be translated into sizable economic gains. Employing mean-variance portfolio optimization framework, we find that top-performing models—XGB(+H), LGBM(+H), and GLM—consistently deliver the highest cumulative, average, and total utility, underscoring their economic robustness and practical relevance for market participants. We interpret such utility gains in terms of certainty equivalent return (CER), for example, a 77.15\% risk-free rate that a mean-variance-preference investor would be willing to accept in lieu of holding a XGB(+H)-risky portfolio. 

We then investigate the underlying sources of performance gain for our ML models both statistically and economically. Statistically, we design a trend-based probability decomposition for models' OOS performance. Using the lag model as a benchmark, we find: (1) Top individual models (GLM, XGB(+H), LGBM(+H)) show significantly fewer false and excessive trend predictions, with greater accuracy in identifying true trends; (2) Ensemble models outperform all individuals, with significantly fewer performance losses and notably more correct weak or strong trend predictions.

We further investigate the economic drivers of predictability by linking the OOS model performance to the real economy. Firstly, we scrutinize the return predictability across different macro states or regimes. We find that OOS gains for the forecasts are often clustered in extreme periods, especially a bullish electricity market and high-growth night light intensity periods, with 5\% and 1\% significant-level \textit{t}-test results respectively. Since night light intensity is a proxy for GDP, the expansion-concentrated pattern of electricity return predictability is quite similar to that found in the U.S. government bond market, as documented by \citet{bianchi2021bond}. Moreover, we observe strong predictability when the electricity market is volatile, which is similar to findings in several U.S. equity market studies (\citet{nagel2012evaporating}, \citet{stambaugh2012short} and \citet{avramov2023machine}). Also, predictability is clustered in regimes with high-growth or low-growth geopolitical risk periods, but there is little difference between these two regimes. Secondly, we plot the predictability heterogeneity across calendar months to see whether this helps identify structural breaks in OOS performance. We find that certain events likely trigger regime shifts to different levels of predictability, including the US-China Trade War, Organization of the Petroleum Exporting Countries (OPEC) Oil Production Cut, the Energy Crisis (wave 1), and the Russia-Ukraine War (also, wave 2 of the Energy Crisis).

Last but not least, we interpret the ML models in terms of feature importance and marginal associations, with the aim of disentangling different parties’ pricing power in the Singapore electricity market and the linkages between local market players and the global macroeconomy. We employ two model-agnostic measures to compute the contributions of each feature, namely $R^2$ reduction and the Shapley value. Statically, both measures agree that the interaction terms between macroeconomic and in-market features are more important than individual features, revealing that $Macro \rightarrow In-market \rightarrow price$ channel is potentially more important than $Macro \rightarrow price$ and $In-market \rightarrow price$. Furthermore, based on both measures, the composition of the top 20 important features identified across models incorporates many categories of predictors, showing no consensus on which particular player in the Singapore electricity market dominates, which is consistent with the complex nature of the market. When examining individual features based on both measures, the lagged Uniform Singapore Energy Price (USEP) stands out the most. It is followed by the supply cushion (from the supply category) and the contingency reserve price (from the regulation category). However, the conventional single-feature $\Delta R^2$ implicitly subsumes all interaction effects between that feature and the remainders, thereby somehow overestimating its isolated contribution per se. Likewise, although Shapley value decomposition formally allocates joint effects, it offers limited insight into our three principal policy questions given that the interaction terms dominate the top 20 predictors in our setting: (1) Has Singapore’s electricity market been fully liberalized, or does significant administrative regulation remain in place? (2) Does the market more closely resemble a domestic or an international trading environment? (3) Which group—supply-side firms, demand-side consumers, or the regulatory authority—exerts the greatest influence on price formation? To address these challenges, we adopt a best-subset selection framework and develop a group-level importance measure. Drawing on the spirit of Shapley value allocation, our metric decomposes each group's overall $\Delta R^2$ into two orthogonal components-its own (self-attribution) effect and the aggregate interaction effects with all other groups-thus enabling precise, policy-relevant attribution of explanatory power at both the group and system levels. Eventually, we can gain some insights from the sub-group analysis that the Singapore electricity market is relatively supply-side driven and heavily affected by regulatory factors. Dynamically, we link the dynamics of feature importance to the macroeconomy, especially the structural breaks. We conduct Bai-Perron Multiple Breakpoints Test and Pettitt test to detect the breakpoints and evaluate the presence and statistical significance of structural changes. We identify the OOS month 45 (Sep 2021) as the most frequently and significantly detected breakpoint across all features, coinciding with the onset of the first wave of energy crisis. Additionally, we find that the relationship between certain features and electricity return is quite consistent with our economic intuitions and basic theory.

Our paper contributes to the literature on electricity price forecasting using ML methods. Electricity price prediction is inherently complex, requiring the integration of physical conditions such as temperature, wind speed, and generation dynamics \citep{ghoddusi2019machine}. A pioneering study is \citet{conejo2005forecasting}, who compare time series models, neural networks, and wavelet transforms for Pennsylvania-New Jersey-Maryland (PJM)’s day-ahead market, finding time series methods most effective. Since then, a broad range of ML methods have emerged \footnote{ENet and XGBoost have shown strong forecasting performance in Germany \citep{kath2018value} and Singapore \citep{sai2023event}, respectively. Neural networks and hybrid models are widely used: see \citet{panapakidis2016day}, \citet{dudek2016multilayer}, and \citet{mirakyan2017composite} for Artificial Neural Network (ANN)-based ensembles, and \citet{wang2017multi}, \citet{bento2018bat}, and \citet{ghasemi2016novel} for decomposition-based hybrids with advanced optimization.}. Collectively, these studies highlight the growing sophistication and predictive power of ML-based approaches in electricity markets globally. However, for one thing, these studies mostly pursue improvements of forecast accuracy through novel model design, rather than interpret the ML black box. Also mentioned previously, these studies concentrate on various but limited sets of predictors, and the lack of in-market stakeholder-level data makes it hard to disentangle the complexities of the electricity market. Our paper differs from these studies in two aspects. Firstly, on top of verifying the three virtues of ML in terms of predictive accuracy justified in recent asset pricing literature (\citet{gu2020empirical}, \citet{kelly2024virtue} and \citet{shen2024can}), methodologically speaking, we propose a novel ensemble methodology \footnote{Ensemble weights can be defined via methods like minimum variance (\citet{dickinson1973some}, \citet{dickinson1975some}) or time-varying squared forecasting errors (MSE) (\citet{mirakyan2017composite}, \citet{bordignon2013combining}, \citet{ghasemi2016novel}). Yet, simple equal-weight combinations often outperform sophisticated schemes in practice (\citet{palm1992combine}, \citet{smith2009simple}, \citet{stock2006forecasting}, \citet{rapach2010out}, \citet{yuan2024naive}). In electricity price forecasting, simple averaging and inverse-RMSE perform well (\citet{nowotarski2014empirical}, \citet{mirakyan2017composite}), while in equity premium forecasting, economically motivated weights based on historical errors can help (\citet{rapach2010out}). Most studies do not explicitly model prediction correlations, relying instead on model diversity.}, that is, adding penalty to the prediction correlations among individual models, which, on the one hand, improves the ensemble performance dramatically when individual models are highly correlated, and, on the other hand, achieves robust and strong predictions before and after 2021 energy crisis \footnote{Previous studies, such as \citet{weron2006modeling} and \citet{knittel2005empirical}, have found that their forecasting performance deteriorates significantly during and after periods of crisis.}. Secondly, we emphasize linking our model predictions to the macroeconomy and, by leveraging in-market stakeholder-level data, disentangling the interplay among different parties in the Singapore electricity market.

Our paper also contributes to the literature on heterogeneous market return predictability, more specifically, time-varying heterogeneity in predictability. While there are many studies, such as \citet{fama1977asset}, \citet{campbell1988dividend} and \citet{lo1988stock}, that support predictability of equity return premium, \citet{welch2008comprehensive} argue for the lack of such predictability on an OOS basis. Subsequent studies, such as \citet{campbell2008predicting}, \citet{rapach2010out}, \citet{henkel2011time}, \citet{pettenuzzo2014forecasting}, \citet{bianchi2021bond}, \citet{cong2024mosaics}, \citet{chen2023chatgpt} and \citet{chen2025chatgpt} find that financial markets (e.g., U.S. equities, corporate bonds, treasury bills, etc.) exhibit predictability even OOS. \footnote{Specifically, research on time-varying return predictability shows that stock market returns are more predictable during recessions (\citet{henkel2011time}, \citet{dangl2012predictive}, \citet{rapach2010out}, \citet{cong2024mosaics}, etc.), periods of high market volatility (\citet{avramov2023machine}), and after certain year, for instance, \citet{green2017characteristics} document a decline in the significance of characteristics since 2003, signifying a sharp decrease in the return predictability. Recently, the pockets (short periods with significant predictability) of predictability have been revisited in \citet{farmer2023pockets} and \citet{cakici2024pockets}, providing evidence in time-variation in market return predictability.}. \citet{goyal2024comprehensive} recently raise doubts about market predictability by pointing out declining OOS performance of predictors over time. Our paper first differs from these studies in the market it focuses on; specifically, our target market is the electricity market, which is more volatile and exhibits stronger seasonal predictability than traditional financial markets. Second, although electricity differs from financial assets, we reaffirm market return predictability in this context—a foundation of almost all modern asset pricing models with time-varying risk premia.

Our paper makes a novel contribution to the large literature on understanding the interplay of different players in the electricity market. Existing studies typically adopt multi-agent models or supply-demand based structural models to capture market interactions. For example, \citet{ruibal2008forecasting} propose a fundamental bid-based stochastic model for forecasting hourly and average electricity prices, showing that a decrease in the number of market participants significantly raises expected prices. \citet{batlle2005strategic} extend traditional production-cost models by incorporating strategic bidding through conjectural variations, showing that prices increase under oligopolistic behavior. \citet{ziel2016electricity} introduce the X-Model by using auction-based supply and demand curves, integrating market structure insights with modern econometric techniques such as LASSO estimation and dimension reduction. \citet{johnsen2001demand} develop and estimate a weekly supply–demand model for the deregulated Norwegian electricity market, demonstrating that stochastic inflows, snow, and temperature significantly explain variations in generation, demand, and prices. Unlike these studies, our access to stakeholder-level in-market data allows us to use statistically driven $R^2$ reduction and economically driven Shapley value to study individual feature importance, along with our proposed group-level $R^2$ reduction decomposition, to better understand the interplay among market participants. This approach provides a novel perspective on market interactions.

The remainder of this paper is structured as follows.  Section \ref{emc:backgroundNdata} outlines the institutional background of the Singapore electricity market and the data, including the target variable and features, while Section \ref{emc:methodology} explains the methodology employed in the study. Section \ref{emc:result} presents the results, covering model performance, complexity, feature importance, marginal associations, and utility analysis. Section \ref{emc:conclusion} concludes.

\section{Institutional Background and Data}
\label{emc:backgroundNdata}
\subsection{Institutional Background}
The USEP serves as the central settlement price in Singapore’s wholesale electricity market. Accurate forecasting of USEP is therefore operationally valuable for market participants and strategically important for policy evaluation. Yet, USEP is not solely a reflection of real-time supply and demand; it is also influenced by institutional features such as reserve and vesting contracts. These policy-linked mechanisms create a hybrid structure in which regulatory interventions coexist with competitive market bidding, challenging purely market-based interpretations of price dynamics.

This duality makes Singapore’s electricity market an analytically compelling and policy-relevant environment for electricity price forecasting. Its compact, self-contained design, characterized by centralized clearing and strong government oversight, offers a uniquely clean setting for disentangling the effects of operational fundamentals and institutional rules. Moreover, the market's heavy reliance on imported natural gas and oil results in strong price transmission from global fuel markets, facilitating the integration of international variables. Unlike many large-scale systems with diverse fuels, locational pricing, or regional interconnectivity, Singapore provides a data-rich and relatively controlled empirical platform. As such, forecasting USEP in Singapore not only supports local market operations and policy evaluation but also generates insights that are generalizable to other deregulated electricity markets navigating the balance between efficiency, volatility, and regulatory stability. To contextualize our analysis, we provide an overview of the Singapore electricity market below, along with detailed market reform milestones in Appendix \ref{Institutional Background Supplements}.

\subsubsection{The Wholesale Market of NEMS}
The wholesale electricity market in Singapore is structured into two main segments: the “real-time market” and the “procurement market.” The real-time market, managed by EMC, serves as a dynamic platform for trading energy, reserve, and regulation services between market participants. Functioning on an auction-based pricing mechanism, this market clears every half-hour, setting the optimal production quantities for each generation facility, alongside reserve and regulation requirements and corresponding market prices.
	
Market quantities and prices in the real-time segment are determined based on price-quantity offers from generators and load forecasts developed by EMC, which relies on demand projections provided by the Power System Operator (PSO). This auction process is centralized through a computational model, the Market Clearing Engine (MCE), which recalculates the least-cost dispatch schedule and corresponding prices every half-hour. The MCE optimizes resource allocation by considering a broad range of system constraints, including reserve and regulation requirements. This model produces location-specific energy prices, known as nodal prices, which may vary across the network. Variations in these nodal prices capture the effects of transmission losses and physical constraints within the transmission system, accurately reflecting localized costs and enhancing overall grid efficiency. Generators receive payments based on their respective nodal prices, while buyers pay the USEP, a weighted average of nodal prices at all off-take nodes.
	
Figure \ref{fig:1} provides an example of the market-clearing process. In this example, three generators (A, B, and C) submit offers in four price-quantity tranches each. The market-clearing price is determined where the cumulative consumer demand intersects with the offer tranches. Offers below the market-clearing price are accepted, while offers above it are excluded. This structure incentivizes generators to operate more efficiently, as their profitability increases with the difference between their offer price and the market-clearing price. This process illustrates that the USEP is market-driven, providing a reliable basis for forecasting.
	
\subsubsection{The Retail Market of NEMS}
The liberalization of Singapore's retail electricity market has been introduced progressively. Starting in July 2001, consumers with a maximum power requirement of 2MW and above were granted the option to select their preferred electricity provider. This contestability threshold was subsequently lowered to 20,000 kWh and above in June 2003, further reduced to 10,000 kWh in December 2003, and eventually lowered to 2,000 kWh in July 2015. Non-contestable consumers, however, continue to be supplied by the Market Support Services Licensee (MSSL), with SP Services acting as the sole MSSL.
	
In late 2018, the Open Electricity Market (OEM) was launched nationwide \citep{emcreport2018market}. Under the OEM, customers have three purchasing options. The first option is the regulated tariff, allowing consumers to purchase electricity from SP Services at a rate reviewed quarterly. The second option lets consumers buy electricity from a licensed retailer, with prices determined by the retailer’s package offerings. Customers with an average monthly consumption of at least 2,000 kWh can choose from 14 licensed retailers, while those with consumption below 2,000 kWh can select from 10 participating retailers in the OEM. The third option allows consumers to buy electricity directly from the wholesale market, where they pay half-hourly wholesale prices to SP Services. This choice carries greater risk due to price fluctuations, as wholesale prices vary every half-hour based on market demand and supply \citep{OEMWebsite}. Therefore, if consumers can predict the USEP, they may strategically benefit by selecting plans based on expected price trends. For example, choosing a wholesale plan when market surpluses suggest lower prices.

\subsection{Data}
We compile a comprehensive dataset for electricity price returns forecasting, incorporating 35 in-market variables, 7 domestic macroeconomic indicators, and 9 international macroeconomic factors. To explore the interplay between market dynamics and macroeconomic conditions, we include interaction terms between each in-market variable and each macroeconomic factor, resulting in 560 ($35\times(7+9)$) interaction terms. Along with lag price and 7 weekday indicators, bringing the total number of features to 619 ($35+7+9+1+7+560$). The dataset spans a 21-year period from 2003 to 2023. Details on data processing procedures—including handling missing values, feature standardization, and target stationarity—as well as full descriptions of each variable with summary statistics are provided in Appendix \ref{Variable Definitions and Summary Statistics} and Appendix \ref{Data Processing}.
	
\subsubsection{In-market Features}
We obtain the in-market feature data from EMC. The dataset consists of 368,160 observations collected at a half-hourly frequency. We utilize daily averages instead of high-frequency half-hourly data. This choice is informed by two primary considerations. First, the factor model does not account for historical sequences, making it less effective for handling high-frequency data compared to sequential models. Second, a structural mismatch in data frequency exists, as macroeconomic features are reported either daily or monthly, while in-market features are recorded at 30-minute intervals. Therefore, we choose the daily level as a compromise to address this structural mismatch. This approach aligns with existing literature, which predominantly uses daily forecasts, as in studies such as \citet{mirakyan2017composite}. 

The 35 in-market features can be further classified into three categories, based on its corresponding stakeholder, consisting of supply, demand and regulation. We have 6 features for demand category (offer-ratio metrics and estimated demand metrics), 16 for supply (generator-level metrics) and 13 for regulation group (regulated price metrics, vesting contract and reserve-related metrics).

\subsubsection{Domestic Macroeconomics Features}
In addition, we compile domestic macroeconomic variables, covering both meteorological and economic indices for Singapore. Daily meteorological data, obtained from the Meteorological Service Singapore, includes temperature, total rainfall, and wind speed. These variables represent nonstrategic uncertainties that can influence electricity prices \citep{aggarwal2009electricity}. Monthly economic indices are sourced from \citet{foo2023forecasting} and include the Manufactured Producer Price Index (MPPI), Purchasing Managers' Index (PMI), Consumer Price Index (CPI), and Industrial Production Index (IPI). This dataset provides insight into Singapore’s economic conditions, where higher levels of economic prosperity generally correlate with increased electricity demand.
	
\subsubsection{International Macroeconomics Features}
We collect international macroeconomic variables, including geopolitical risk (GPR) metrics, raw material prices, and the exchange rate between the US dollar and Singapore dollar.
	
Given that geopolitical tensions contribute to heightened volatility in global energy markets, which may, in turn, drive up wholesale electricity prices \citep{ema_electricity_prices}, we collect geopolitical risk indicators from \citet{caldara2022measuring}. The GPR variables consist of geopolitical risk daily (GPRD), geopolitical risk daily acts (GPRD\_ACT), and geopolitical risk daily threats (GPRD\_THREAT). GPRD serves as a news-based measure that quantifies adverse geopolitical events and their associated risks, encompassing both potential threats and realized events.
	
Raw material prices in our dataset include the fuel oil price, oil futures prices, natural gas spot and futures prices, and the Energy Consumer Price Index (ECPI). Energy prices are considered nonstrategic uncertainties impacting electricity pricing \citep{aggarwal2009electricity}. Monthly fuel oil price data is sourced from EMC, while daily natural gas spot and futures prices are obtained from the U.S. Energy Information Administration (\href{https://view.officeapps.live.com/op/view.aspx?src=https%3A%2F%2Fwww.eia.gov%2Fdnav%2Fng%2Fhist_xls%2FRNGWHHDd.xls&wdOrigin=BROWSELINK}{EIA}), with the average of Natural Gas Futures Contracts 1 through 4 representing the futures price. Daily West Texas Intermediate (WTI) Crude Oil Futures data is sourced from \href{https://www.investing.com/commodities/crude-oil-historical-data}{Investing.com}, and monthly ECPI data is obtained from the World Bank. For the daily exchange rate between the US dollar and Singapore dollar, we source data from \href{https://sg.finance.yahoo.com/quote/SGD%3DX/history?period1=1070236800&period2=1711324800&interval=1d&filter=history&frequency=1d&includeAdjustedClose=true}{Yahoo Finance} and \href{https://www.macrotrends.net/2561/us-dollar-singapore-exchange-rate-historical-chart}{Macrotrends}.

\section{Methodology}
\label{emc:methodology}
\subsection{Sample Splitting Scheme}
Our analysis covers January 2003 to December 2023. The in-sample period (training + validation) spans 2003–2017 (180 months), while the OOS test period spans 2018–2023 (72 months), capturing both pre- and post-global energy crisis regimes.

To respect the temporal structure critical for time-series forecasting, we avoid cross-validation and instead adopt a recursive (expanding window) training scheme, following \citet{gu2020empirical} and \citet{kelly2019characteristics}. This method incrementally expands the training set while holding the validation set fixed, ensuring models are always trained on past data and evaluated on future outcomes.\footnote{This approach latently assumes temporal stability in the feature-target relationship, so that learning from distant history remains predictive.} Formally, let $\mathcal{T}_1$, $\mathcal{T}_2$, and $\mathcal{T}_3$ denote the training, validation, and test sets. We train the models in $\mathcal{T}_1$, select the optimal hyperparameters via grid search by minimizing the validation loss $\mathcal{T}_2$, and do forecasting in $\mathcal{T}_3$. Note that, due to the recursive scheme, for each day $t \in \mathcal{T}_3$, there is a unique corresponding in-sample training set $\mathcal{T}_{t,1}$, pseudo-OOS validation set $\mathcal{T}_{t,2}$, and OOS test set $\mathcal{T}_{t,3}$. An illustrative example is provided in the Appendix \ref{An illustration of Sample Splitting Scheme}.

\subsection{Models}
\label{emc:Models}
This subsection provides an overview of the ML methods employed in our study. Ideally, we can model the realized USEP returns as:
\renewcommand{\theequation}{\arabic{equation}}
\setcounter{equation}{0}
\begin{equation}
\begin{aligned}
\label{equ:1}
& r_{t+1} = \mu_{t+1} + \epsilon_{t+1}, \quad \epsilon_{t+1} \sim WN(0),
\end{aligned}
\end{equation}
where
\begin{equation}
\begin{aligned}
\label{equ:2}
& \mu_{t+1} = E[r_{t+1} | \mathcal{F}_t] = E_t[r_{t+1}] = f_t(\mathcal{F}_t; \theta^m).
\end{aligned}
\end{equation}
Days are indexed by \( t = 1, \dots, T \). Eq \ref{equ:1} is based on the rationale that any time series can be decomposed into a conditional mean, \( \mu_{t+1} \), and a zero-mean white noise process, \( \epsilon_{t+1} \), which captures the forecasting errors \citep{f8815cbf-36f9-35f3-9962-055fa42e3de8}. The assumption $E_t\left[\epsilon_{t+1}\right]=0$ ensures that the prediction of expected return is unbiased. Our objective is to forecast the conditional mean by approximating \( E[r_{t+1} | \mathcal{F}_t] \) as a function of features that maximizes the OOS explanatory power for the realized return \( r_{t+1} \). 
	
The function \( f_t(\cdot) \) is a flexible mapping of these predictors, parameterized by model-specific parameters \( \theta^m \), updated by day $t$, namely, $f_t(\cdot, \theta^m) \equiv f(\cdot, \theta^m_t)$. In our study, the candidate models, $m \in \mathcal{M} :=$ \{OLS, LASSO, Ridge, ENet, PCR, PLS, GLM, XGB(+H), LGBM(+H), RF, NN1, NN2, NN3, NN4, NN5\}, are briefly described in the following and detailed in the Appendix \ref{appendix models}. The computation algorithms determine the optimal model-specific parameters \( \theta^{m*} \) by minimizing the loss function. This yields the optimal specification \( f^*_t(\cdot) \) and the corresponding "optimal" point estimates \( \mu^*_{t+1} \), i.e., $\mu^*_{t+1} = f_t^*(\mathcal{F}_t; \theta^m) = f(\mathcal{F}_t; \theta_t^{m*})$.

\subsubsection{Individual Models}
\textbf{Simple Linear Model: OLS.} The most simple model, which follows:
\begin{equation}
\begin{aligned}
\label{equ:3}
& f_t(\mathcal{F}_t; \theta) = \mathcal{F}^{\prime}_t \theta,
\end{aligned}
\end{equation}
where the $\theta$ can be estimated by
\begin{equation}
\begin{aligned}
\label{equ:4}
\theta^* = \argmin_{\theta} \frac{1}{T} \sum^T_{t=1} \left(r_{t+1} - f_t(\mathcal{F}_t; \theta)\right)^2.
\end{aligned}
\end{equation}

\noindent \textbf{Penalized Linear Models: LASSO, Ridge, ENet.} Penalized linear models mirror the OLS (Eq \ref{equ:3}) but extend the framework by incorporating regularization terms into Eq \ref{equ:4}, for instance, an ENet penalty is formulated as:
\begin{equation}
\begin{aligned}
\label{equ:5}
\theta^*  = \argmin_{\theta} \frac{1}{T} \sum_{t=1}^{T} \left(r_{t+1} - f_t(\mathcal{F}_t; \theta) \right)^2 + \lambda \left[ \rho \sum_{j=1}^{P} |\theta_j| + (1-\rho) \sum_{j=1}^{P} \theta_j^2 \right],
\end{aligned}
\end{equation}
where \(\lambda > 0\) controls the overall regularization strength,  \(\rho \in [0, 1]\) determines the balance between \(l_1\)- and \(l_2\)-regularization, and \( P \) is the number of predictors. LASSO ($\rho=1$) employs $l_1$-regularization, while Ridge ($\rho=0$) applies $l_2$-regularization. ENet ($0<\rho<1$) combines the strengths of both by balancing variable selection and coefficient shrinkage.

\noindent \textbf{Dimension Reduction Models: PCR, PLS.} Both models can be expressed as reducing the dimensionality of predictors from \(P\) to \(K\) through a linear transformation:
\begin{equation}
\begin{aligned}
\label{equ:6}
R = (\mathcal{F}\Omega_K)\theta_K + \tilde{E},
\end{aligned}
\end{equation}
where \(R\) is the $T \times 1$ vector of $r_{t+1}$, \(\mathcal{F}\) is the $T \times P$ matrix of stacked predictors $\mathcal{F}_t$, \(\Omega_K\) is the transformation matrix that condenses predictors into \(K\) components, and \(\theta_K\) is the regression coefficient vector for the reduced components. In PCR, \(\Omega_K\) is derived through singular value decomposition to maximize predictor variance, while in PLS, \(\Omega_K\) is optimized to maximize the covariance between predictors and the response. Finally, given estimated $\Omega_K$, $\theta_K$ is estimated via OLS regression of $R$ on $Z \Omega_K$. 

\noindent \textbf{Generalized Linear Model: GLM.} GLM extends Eq \ref{equ:3} by using a \( K \)-term spline series expansion for each predictor:
\begin{equation}
\label{equ:7}
f_t = \sum_{j=1}^P p(\mathcal{F}_{t,j})^\prime \theta_j,
\end{equation}
where \( p(\cdot) \in \mathbb{R}^K \) is a vector of basis functions and \( \theta_j \in \mathbb{R}^K \). We adopt quadratic spline bases:
\begin{equation}
\label{equ:8}
p(z) = \left(1, z, (z - c_1)^2, \dots, (z - c_{K-2})^2 \right),
\end{equation}
with nodes \( c_1, \dots, c_{K-2} \) defining the spline regions. To enforce sparsity and interpretability, we apply group LASSO:
\begin{equation}
\label{equ:9}
\phi(\theta) = \lambda \sum_{j=1}^P \left( \sum_{k=1}^K \theta_{j,k}^2 \right)^{1/2}.
\end{equation}
The final estimator minimizes a penalized loss:
\begin{equation}
\label{equ:10}
\theta^* = \argmin_{\theta} \frac{1}{T} \sum_{t=1}^T (r_{t+1} - f_t)^2 + \phi(\theta),
\end{equation}
tuned over \( \lambda \) and \( K \).

\noindent \textbf{Tree-based Ensemble Models: RF, XGB, LGBM.} The RF, XGB and LGBM are based on the regression trees method. To mitigate the overfitting problem and unstable predictions in a single regression tree, RF, XGB and LGBM employ three distinct ensemble techniques, i.e., combining multiple regression trees. 

Specifically, \textit{RF} employs bagging by building \( B \) trees on bootstrap samples and random feature subsets \( \mathcal{F}_{b,t} \). The final prediction is the average:
\begin{equation}
\label{equ:11}
f_t = \frac{1}{B} \sum_{b=1}^{B} f_{b,t}(\theta; \mathcal{F}_{b,t}, L, K),
\end{equation}
where \( L \) is tree depth and \( K \) the number of randomly selected features.

\textit{XGB} uses boosting, sequentially adding trees to correct prior errors by minimizing an objective function with learning rate \( \eta \) and regularization parameter \( \lambda \):
\begin{equation}
\label{equ:12}
f_t = \sum_{b=1}^{B} \eta f_{t}(\theta; \mathcal{F}_t, \lambda).
\end{equation}

\textit{LGBM} also uses boosting but grows trees leaf-wise instead of level-wise. It focuses on the leaf with the highest gain at each step, enabling deeper, more efficient trees. Complexity is controlled by the maximum number of leaves \( H \):
\begin{equation}
\label{equ:13}
f_t = \sum_{b=1}^{B} \eta f_{t}(\theta; \mathcal{F}_t, \lambda, H).
\end{equation}

\noindent \textbf{Feed-forward Neural Network Models: NN$\#$.} Unlike tree ensembles, Feed-forward Neural Networks (NN$\#$) use multi-layer perceptrons (MLPs) to model complex non-linear relationships. A general form is:
\begin{equation}
\label{equ:14}
f_t = A_D(W_D \cdots A_2(W_2 A_1(W_1 \mathcal{F}_t + b_1) + b_2) \cdots + b_D),
\end{equation}
where \( D \) is the number of hidden layers, \( W_d, b_d \) are weights and biases, and \( A_d \) denotes activation functions (e.g., ReLU, Tanh). NN$\#$ learns parameters by minimizing a loss function via gradient-based methods like SGD or Adam:
\begin{equation}
\label{equ:15}
\min_{W, b} \sum_{t=1}^{T} \mathcal{L}(f_t, r_{t+1}),
\end{equation}
where \( \mathcal{L}(\cdot) \) is the loss function. Regularization (e.g., dropout, \( \ell_1/\ell_2 \) penalties, early stopping) helps prevent overfitting. Details on the NN architecture, regularization and optimization algorithms are provided in the Appendix \ref{appendix models nn}.

In practice, we further impose two restrictions to $f$'s flexibility. Firstly, time-serial heterogeneity restriction due to the high computational costs of updating parameters daily. That is, we restrict the \( f_t(\cdot) \) as \( f_{\mathcal{T}_{t,3}}(\cdot) \) so that the $f$ (equivalently, parameters $\theta^m$) is updated by clustered periods (monthly, since the length $|\mathcal{T}_{t,3}|$ for any day $t$ is month). $t \in \mathcal{T}_{t,3}$ shares same model specification \( f_{\mathcal{T}_{t,3}}(\cdot) \). Secondly, information set restriction. Following \citet{gu2020empirical}, we further impose that \( E_t[r_{t+1}] = f_{\mathcal{T}_{t,3}}(X_t; \theta^m) =  f(X_t; \theta^m_{\mathcal{T}_{t,3}})\), implying that our prediction does not utilize information prior to day $t$. By maintaining the same form over small cluster/period, the model lends stability to point estimates of conditional mean \footnote{Meanwhile, the byproducts are twofold: (1) this framework also imposes one latent restriction to $f(\cdot)$'s flexibility, that is, $f(\cdot)$ is time-serial \textit{iid}, as \( f_{\mathcal{T}_{t,3}}(\cdot) \) combines all $t \in \mathcal{T}_{t,3}$ under one homogeneous model, neglecting the potential heterogeneity of return predictability in more granular daily setting; (2) potential false cluster: periodical cluster $\mathcal{T}_{t,3}$ follows calendar sequential orders due to our spanning window scheme, failing to capture macro-driven regime shifts.}.

\subsubsection{Ensemble Models}
Ensemble methods are widely recognized for their ability to match or often surpass the performance of the best individual models, addressing two key forms of risk. The first, selection risk, arises from the possibility of choosing a suboptimal individual model or combination of models. The second, prediction risk, refers to the likelihood of incurring large forecasting errors \citep{bordignon2013combining}. Formally, given Eq \ref{equ:1} and \ref{equ:2}, after imposing two restrictions, the ensemble model, denoted as \( g \), is specified as follows:
\begin{equation}
\label{equ:16}
g_{\mathcal{T}_{t,3}}(X_{t}; \theta^m, \omega_m) = \sum_{m \in \mathcal{M}} \omega_m \cdot f_{\mathcal{T}_{t,3}}(X_{t}; \theta^m),
\end{equation}
where $g_{\mathcal{T}_{t,3}}(X_{t}; \theta^m, \omega_m)$ is the conditional mean forecasts generated by the ensemble model, $f_{\mathcal{T}_{t,3}}(X_{t}; \theta^m)$ is the predictions from individual model $m$, and $\omega_m$ is the corresponding weight. In our study, we consider four types of weights:
	
\noindent \textbf{Time-invariant weights: Equal weights.} $\omega_m = \frac{1}{|\mathcal{M}|}$, where $|\mathcal{M}|$ is the number of model. It is well known in the forecasting literature \citep{rapach2010out, timmermann2006forecast} that the simple equal weighting scheme performs reasonably well in practice.
	
\noindent \textbf{Time-variant and data-driven weights: a hyperparameter tuned in $\mathcal{T}_{t,2}$.} Namely, $\omega_m = \omega^*_{m,t}$, which is subjected to $\sum_{m \in \mathcal{M}} \omega_{m,t}^* = 1$ and tuned via 10-fold cross-validation within $\mathcal{T}_{t,2}$.
	
\noindent \textbf{Time-variant and data-driven weights: a hyperparameter tuned in $\mathcal{T}_{t,2}$ with correlation penalty.} From Figure \ref{fig:4} and \ref{fig:6}, we observe that models share similar patterns in OOS performance dynamics, revealing high correlations between models' predictions, which may hinder the ensemble method's performance. Therefore, we further incorporate a penalty for model prediction's correlation into the MSE loss function:
\begin{equation}
\label{equ:17}
\argmin_{w_{m,t}} \sum_{t \in \mathcal{T}_{t,2}}\left(r_t-\sum_{m \in \mathcal{M}} w_{m,t} \cdot \hat{r}_{m, t}\right)^2+\lambda \sum_{m, m^{\prime} \in \mathcal{M}} w_{m,t} w_{m^{\prime}, t} \rho_{m, m^{\prime}} \quad \text{s.t.} \quad \sum_{m \in \mathcal{M}} \omega_{m,t} = 1,
\end{equation}
where the first term minimizes the error of the ensemble, the second term penalizes the correlation between models, weighted by their contributions $w_{m,t}$ and $w_{m^{\prime}, t}$, and $\lambda$ is a hyperparameter that controls the trade-off between accuracy and diversity. The correlation $\rho_{m, m^{\prime}}$ is calculated as the Pearson correlation between the predictions of models $m$ and $m^{\prime}$.
	
\noindent \textbf{Time-variant and economic-driven weights: \citet{stock2004combination} and \citet{rapach2010out}.} The combining weights $\omega_{m,t}$ are functions of the historical forecasting performance of the individual models over the validation period $\mathcal{T}_{t,2}$. Their discount mean square prediction error combining method employs the following weights:
\begin{equation}
\label{equ:18}
\omega_{m,t} = \frac{\phi_{m,t}^{-1}}{\sum_{m \in \mathcal{M}} \phi_{m,t}^{-1}},
\end{equation} 
where 
\begin{equation}
\label{equ:19}
\phi_{m,t} = \sum_{t^{\prime} \in \mathcal{T}_{t,2}} \theta^{|\mathcal{T}_{t,2}|-t^{\prime}} \left(r_{t^{\prime}+1} - \hat{r}_{m, t^{\prime}+1} \right)^2,
\end{equation}
and $\theta$ is a discount factor. This method thus assigns greater weights to individual model forecasts that have lower MSE values (better forecasting performance) over the validation periods. When $\theta=1$, there is no discounting, and it produces the optimal forecast derived by \citet{bates1969combination} for the case where the individual forecasts are uncorrelated. When $\theta<1$, greater weight is attached to the recent forecast accuracy of the individual models. We consider the two values of 1.0 and 0.9 for $\theta$. \footnote{Note that, we can make this approach more data-driven: we tune the $\omega_{m,t}$ by setting a grid for the discount factor $\theta$, and then choose the best $\theta^*$ that minimizes the loss function MSE. In such setting, $g_{\mathcal{T}_{t,3}}(X_{t}; \theta^*, \omega^*_{m,t}) = \sum_{m \in \mathcal{M}} \omega^*_{m,t} \cdot f^{m, *}_{\mathcal{T}_{t,3}} + \epsilon_{t+1}$. Besides, if $\theta$ is flexible, we can also incorporate the correlation penalty into the MSE loss to tune the $\theta$ and $\lambda$ simultaneously. But for simplicity, we leave it for future study.}

\subsection{Out-of-Sample Performance Evaluation}
\subsubsection{Out-of-Sample $R^2$}
\label{emc: oosR2}
We utilize three types of benchmarks: lag, historical mean, and zero, among which the "lag benchmark" is our proposed new benchmark that is suitable for our data setting, the remaining two have been well-studied in financial forecasting literature \footnote{Specifically, zero for individual stocks (like \citet{kelly2019characteristics}, \citet{gu2020empirical}, etc.), the historical mean for the market index (like \citet{campbell2008predicting, rapach2010out}, etc.).}.

The "lag benchmark" consists of both the lag price ($p_t$) and AR(1) (namely, $\widehat{\beta}_t p_t$) benchmarks, defined as follows:
\begin{equation}
\begin{aligned}
\label{equ:20}
R_{OOS,lagprice, m}^2 = 1 - \frac{\sum_{t \in \mathcal{T}_3}\left(p_{t+1} - \widehat{p}_{m,t+1}\right)^2}{\sum_{t \in \mathcal{T}_3} (p_{t+1} - p_{t})^2},
\end{aligned}
\end{equation}

\begin{equation}
\begin{aligned}
\label{equ:21}
R_{OOS,AR(1),m}^2 = 1 - \frac{\sum_{t \in \mathcal{T}_3}\left(p_{t+1} - \widehat{p}_{m,t+1}\right)^2}{\sum_{t \in \mathcal{T}_3} (p_{t+1} - \widehat{\beta}_t p_t)^2},
\end{aligned}
\end{equation}
where $\widehat{p}_{m,t+1} = p_t \times \exp(\widehat{r}_{m,t+1})$. $\widehat{\beta}_t p_t$ is the prediction from pseudo AR(1) regression: $r_{t+1} = \alpha + \beta p_t + \epsilon_{t+1}$, the subscript $t$ signifies that we follow the same recursive scheme to update the persistence coefficient $\beta$ monthly \footnote{Please note that, a key difference when fitting our pseudo AR(1) and OLS, compared to other model candidates, is that pseudo AR(1) and OLS do not involve any hyperparameter to be tuned. Consequently, a validation set is not required. However, in order to keep fairness and consistence of input data across models, we fit AR(1) and OLS with learned coefficients from $\mathcal{T}_1$ to the $\mathcal{T}_3$, ignoring the data in $\mathcal{T}_2$.}.

The reasons for employing these two "lag benchmarks" are twofold: (1) given inclusion of lag price $p_t$ in feature set $X_t$, it is important to quantify the explanatory power of features on top of this pseudo auto-correlated term \footnote{Readers may ask why we do not use AR term $r_t$ instead. By definition, \(r_t = \ln(p_t/p_{t-1}) = \ln(p_t) - \ln(p_{t-1})\). Including \(r_t\) in the feature set would implicitly incorporate information from \(t-1\) into predictions for \(t+1\), which violates our information set restriction that base predictions only on data available at time \(t\). Therefore, we use the lagged price at time \(t\) (rather than the return) to ensure the model adheres to this restriction.}; (2) the lag price tends to be a strong factor; we can test the model's performance in a relatively weak factors environment \citep{shen2024can} by partialling it out. The AR(1) benchmark is more robust than the lag price one when testing the above two motivations, since it provides time-variant predictive variance as our model candidates \footnote{We can understand this better by referring to the well-known bias-variance decomposition of expected OOS MSE:
\begin{equation}
\label{equ:22}
\begin{aligned}
E_t\left[(r_{t+1} - \hat{r}_{t+1})^2\right] & = \left[Bias(\hat{r}_{t+1})\right]^2 + Var(\hat{r}_{t+1}) + \sigma^2_{\epsilon_{t+1}} \\
& = \left[E(\hat{r}_{t+1}) - r_{t+1}\right]^2 + E_t\left[(r_{t+1}-E(\hat{r}_{t+1}))^2\right] + \sigma^2_{\epsilon_{t+1}}.
\end{aligned}
\end{equation} When predictive benchmark is $p_t$, $Var_t(\hat{r}_{t+1})=Var_{t+1}(\hat{r}_{t+2})$. But for AR(1) benchmark and our model candidates, the updated $\theta^m_t$ (or $\beta_t$ in AR(1)) makes sure that $Var_t(\hat{r}_{t+1}) \ne Var_{t+1}(\hat{r}_{t+2})$, so that the difference between model candidates and AR(1) benchmark is the effects of remaining features.}.

The robust predictive benchmarks in the literature: zero and historical mean, are defined as follows:
\begin{equation}
\begin{aligned}
\label{equ:23}
R_{OOS, mean,m}^2 = 1 - \frac{\sum_{t \in \mathcal{T}_3}\left(r_{t+1} - \widehat{r}_{m,t+1}\right)^2}{\sum_{t \in \mathcal{T}_3} (r_{t+1} - \overline{r}_{\mathcal{T}_{1,2}})^2},
\end{aligned}
\end{equation}
where \(\mathcal{T}_{1,2}\) represents the union of training set and validation set, the $\overline{r}_{\mathcal{T}_{1,2}}$ is the historical mean calculated using in-sample data.
\begin{equation}
\begin{aligned}
\label{equ:24}
R_{OOS, zero, m}^2 = 1 - \frac{\sum_{t \in \mathcal{T}_3}\left(r_{t+1} - \widehat{r}_{m, t+1}\right)^2}{\sum_{t \in \mathcal{T}_3} (r_{t+1})^2}.
\end{aligned}
\end{equation}
When historical mean is quite near to zero, we expect that $R_{OOS, mean,m}^2$ is similar to $R_{OOS, zero, m}^2$.

\subsubsection{Relative Root Mean Square Errors}
$R^2_{OOS, m}$ compares the model's prediction relative to the three selected benchmarks: lag, historical mean and zero. Comparisons across models tell us the relative performance of models with respect to the benchmarks. However, for one thing, historical mean and zero benchmarks provide zero predictive variance (namely, $Var(\hat{r}_{t+1})=0$ in Eq \ref{equ:22}), for the other thing, $R^2_{OOS, m}$ does not provide a direct measure of the relative performance comparisons of sophisticated models relative to the most simple model, namely, OLS. To address this, following \citet{hong2024forecasting}, we set the OLS as benchmark, and calculate the relative rRMSE as
\begin{equation}
\label{equ:25}
rRMSE_{m} = \frac{1}{|\mathcal{T}_3|}\sum_{t \in \mathcal{T}_3} \frac{RMSE_{m,t}}{RMSE_{\text{OLS},t}},
\end{equation}
where
\begin{equation}
\label{equ:26}
RMSE_{m,t} = \sqrt{\frac{1}{|\mathcal{T}_{3}|}\sum_{t \in \mathcal{T}_{3}}(\widehat{r}_{m,t} - r_{t})^2}.
\end{equation}

When $m$ is OLS, $rRMSE_{OLS} = 1$; when model $m$ outperforms OLS, the $rRMSE_{m} < 1$; if not, $rRMSE_{m} > 1$.\footnote{Also, due to recursive scheme, we can also obtain OOS-month-specific $rRMSE_{m,\mathcal{T}_{t,3}}$, i.e., 
\begin{equation}
\label{equ:27}
rRMSE_{m,\mathcal{T}_{t,3}} = \frac{RMSE_{m,\mathcal{T}_{t,3}}}{RMSE_{\text{OLS},\mathcal{T}_{t,3}}},
\end{equation}
where
\begin{equation}
\label{equ:28}
RMSE_{m,\mathcal{T}_{t,3}} = \sqrt{\frac{1}{|\mathcal{T}_{3,t}|}\sum_{t \in \mathcal{T}_{3,t}}(\widehat{r}_{m,t} - r_{t})^2}.
\end{equation}}

\subsubsection{Statistical Tests}
The calculated $R^2_{OOS,m}$ and $rRMSE_{m}$ are both model-specific, to obtain statistical inference, we can (1) conduct two-sided DM test \citep{francis1995comparing} of model's relative performance with respect to each other; (2) conduct one-sided DM and Equal Predictive Ability (EPA) tests \citep{bianchi2021bond, goulet2022machine} of model's relative performance with respect to selected benchmarks (lag, mean and zero for $R^2_{OOS,m}$ and OLS for $rRMSE_{m}$), such as the MSFE-adjusted Clark-West (CW) \citep{clark2007approximately} test, Model Confidence Sets (MCS) \citep{hansen2011model} test, and Giacomini and White (GW) \citep{giacomini2006tests} test. We leave the details of these tests in the Appendix \ref{appendix DM} and Appendix \ref{Equal Predictive Ability Test} for readers interested in rigorous treatments.

\subsubsection{Decomposing  Out-of-Sample Performance: A Trend-based Probability Framework}
In this framework, we first conduct trend‐based classification (False Trend (FT), Right Weak Trend (RWT), and Right Strong Trend (RST)), then compare the model performance relative to the benchmark (measured by whether the model prediction falls inside or outside the interval). Pseudocode describing our trend-based probability framework for decomposing the OOS performance of each model is provided in Algorithm \ref{alg:1}. Please note that our proposed lag benchmarks reduce the possible scenarios to four, since the lag benchmark itself serves as the basis for calculating the trend. In Appendix \ref{Trend-based Probability Framework}, we illustrate how to decompose OOS model performance relative to the lag benchmark using our proposed framework.

\subsection{A Tale of Two Stories: Out-of-Sample Performance \& Out-of-Sample Predictability}
With the time-variant OOS performance matrices, it is natural to link the dynamic performance of models to macroeconomic conditions (i.e., the real economy) in order to identify when or under what macroeconomic environments the models perform better or worse. Equivalently, from a data perspective, we aim to determine when or under what macroeconomic conditions the return is more predictable. \citet{cong2024mosaics} propose that in-sample $R^2$ of a predictive model serves as a natural measure of data predictability. Instead, in our data setting, we propose that OOS $R^2$, specifically, $R^2_{OOS, mean, m, \mathcal{T}_{t,3}}$ and $R^2_{OOS, zero, m, \mathcal{T}_{t,3}}$, is a tale of two stories: (1) from the perspective of the model, we can interpret it as a property of the model, on top of the relative performance with respect to the benchmark, representing the proportion of return variation explained by the model. We will justify this rationale afterwards. (2) from a data perspective, we view it as a measure of how much variation in a specific dataset can be explained by the best-fitted ML model. 

We can revisit the calculation of in-sample $R^2$ and compare it with our OOS one (Eq \ref{equ:23} and \ref{equ:24}) to justify the story (1) and (2):
\begin{equation}
\begin{aligned}
\label{equ:29}
R^2_{IS, m, t} & = 1 - \frac{\sum_{t \in \mathcal{T}_{IS}}(r_{t} - \hat{r}_t)^2}{\sum_{t \in \mathcal{T}_{IS}}(r_{t} - \bar{r}_{\mathcal{T}_{IS}})^2} = 1 - \frac{Var(\epsilon_{t})}{Var(r_t)} := 1 - \frac{\sigma^2_{\epsilon_{t}}}{\sigma^2_{r_t}} \quad \text{if Eq \ref{equ:1}, story (2)} \\
& = 1 - \frac{\sum_{t \in \mathcal{T}_{t,3}}(r_{t} - \hat{r}_t)^2}{\sum_{t \in \mathcal{T}_{t,3}}(r_{t} - \bar{r}_{\mathcal{T}_{t,3}})^2} \quad \text{if $IS$ $\perp$ $OOS$, story (1)} \\
& = 1 - \frac{\sum_{t \in \mathcal{T}_{t,3}}(r_{t} - \hat{r}_t)^2}{\sum_{t \in \mathcal{T}_{t,3}}(r_{t} - \bar{r}_{\mathcal{T}_{1,2}})^2} \equiv R^2_{OOS, mean, m, \mathcal{T}_{t,3}} \quad \text{if $\bar{r}_{\mathcal{T}_{1,2}}^2 = \bar{r}_{\mathcal{T}_{3}}^2$, story (1)} \\
& = 1 - \frac{\sum_{t \in \mathcal{T}_{t,3}}(r_{t} - \hat{r}_t)^2}{\sum_{t \in \mathcal{T}_{t,3}}(r_{t})^2} \equiv R^2_{OOS, zero, m, \mathcal{T}_{t,3}} \quad \text{if $\bar{r}_{\mathcal{T}_{1,2}}^2 = \bar{r}_{\mathcal{T}_{3}}^2 = 0$, story (1)}
\end{aligned}
\end{equation}

From line 1, we can see that in-sample $R^2$ can be interpreted as the signal-to-noise ratio \citep{cong2024mosaics}. Subscript $t$ emphasizes its time-series variability. Conceptually, when $R^2_t$ is high for a specific period $t$, it is relatively easier for a predictive model to capture the conditional mean. Conversely, if the noise is large, even with the true knowledge of the conditional mean, the resulting $R^2_t$ would still be low, let alone when learning the conditional mean from noisy data. Thus, with an appropriate model approximating the conditional mean (usually using ML), the in-sample $R^2$ provides a reasonable measure of the signal-to-noise ratio, reflecting the return predictability across different periods.

In our data setting, under the recursive scheme, the second line condition (if $IS$ $\perp$ $OOS$) naturally holds, and our sample USEP return is near zero, thus the fourth line condition approximately holds. In \citet{cong2024mosaics}'s setting, the primary reason for not using OOS $R^2$ is that the second line condition does not hold, and when predicting stock returns, the OOS $R^2$ generally remains below 1\%, making it hard to produce robust results based on OOS $R^2$ in the environments with low signal-to-noise ratios. However, based on our OOS performance results, we are unlikely to be in an environment with a low signal-to-noise ratio, since the average OOS $R^2$ exceeds 30\%. Therefore, we propose our $R^2_{OOS, mean, m, \mathcal{T}_{t,3}}$ and $R^2_{OOS, zero, m, \mathcal{T}_{t,3}}$ are one tale of two stories.

To tell the story (1), following \citet{rapach2010out}, \citet{henkel2011time} and \citet{huang2015investor}, we compute the $R^2_{OOS,m}$ in different states separately,
\begin{equation}
\begin{aligned}
\label{equ:30}
R_{OOS, benchmark, m, s}^2 = 1-\frac{\sum_{t \in \mathcal{T}_3}\left(r_{t+1, m}-\widehat{r}_{t+1, m}\right)^2 \cdot 1(t \in \mathcal{T}_{s})}{\sum_{t \in \mathcal{T}_3} (r_{t+1, m} - benchmark)^2 \cdot 1(t \in \mathcal{T}_{s})},
\end{aligned}
\end{equation}
and to tell the story (2), we further define the predictability of state $s$ as:
\begin{equation}
\begin{aligned}
\label{equ:31}
Predictability_{s} = \argmax_{m \in \mathcal{M}} R_{OOS, benchmark, m, s}^2,
\end{aligned}
\end{equation}
where \( m \) represents the model, \( s \) denotes the state and $benchmark \in \{zero, mean\}$. $1(t \in \mathcal{T}_{s})$ is the indicator that takes a value of one when day $t$ is in the state $s$ period set $\mathcal{T}_{s}$. \footnote{Given the Eq \ref{equ:31} can be time-varying, namely, $R_{OOS, benchmark, m, s, \mathcal{T}_{t,3}}^2$, we also get the predictability of state $s$ in period $\mathcal{T}_{t,3}$:
\begin{equation}
\begin{aligned}
Predictability_{s, \mathcal{T}_{t,3}} = \argmax_{m \in \mathcal{M}} R_{OOS, benchmark, m, s, \mathcal{T}_{t,3}}^2 \nonumber.
\end{aligned}
\end{equation}}

\subsection{Model Interpretations}
Recall that the underlying assumption of the recursive scheme is that the relationships between features and the response are stable over long time periods, both in sign and in magnitude. 
Additionally, non-linear ML models are often criticized for being black boxes. To interpret the prediction model, we first focus on the importance of each feature to provide evidence on their magnitudes, and then explore the marginal relationships between returns and characteristics to provide evidence on their signs.

\subsubsection{Individual Feature Importance}
\subsubsubsection{Reduction in \(R^2\)}
The reduction in \(R^2\) \citep{kelly2019characteristics, gu2020empirical, hong2024forecasting} is a model-agnostic method that is applicable to all types of predictive models. This approach evaluates the importance of a given predictor by setting its values to zero, recalculating the \(R^2\) within each training sample, and measuring the resulting reduction. The reductions in \(R^2\) are then averaged across all training samples to derive a single importance score for each predictor.
	
We define the reduction in \(R^2\) of feature $j$ in model $m$ as a metric denoted \(FI_{m,j}\), which is formalized as follows:
\begin{equation}
\begin{aligned}
\label{equ:32}
FI_{m,j} = \frac{1}{K} \sum_{k=1}^{K} \left( R^2_{m,\mathcal{T}_{1,k}} - R^2_{m,\mathcal{T}_{1,k}(-j)} \right),
\end{aligned}
\end{equation}
where \(R^2_{m,\mathcal{T}_{1, k}}\) denotes the \(R^2\) value of the \(m\)-th model in the \(k\)-th training sample \footnote{Also, due to recursive scheme, we can also obtain training-set-specific $FI_{m,j,\mathcal{T}_{1, k}}$, i.e.,
\begin{equation}
\begin{aligned}
\label{equ:33}
FI_{m,j, \mathcal{T}_{1, k}} = R^2_{m,\mathcal{T}_{1,k}} - R^2_{m,\mathcal{T}_{1,k}(-j)}.
\end{aligned}
\end{equation}} (with a total of $K$ samples), while \(R^2_{m,\mathcal{T}_{1, k}(-j)}\) is the \(R^2\) value of the same model and training sample after setting the \(j\)-th predictor to zero. The \(FI_{m,j}\) values are computed for each predictor and subsequently normalized to sum to 1 within each model $m$, ensuring that the metric reflects the relative importance of each predictor. This normalization captures the proportional contribution of each predictor to the model's predictive performance and facilitates direct comparison across predictors.
	
\subsubsubsection{Shapley Value}
The Shapley value, introduced by \citet{shapley1953value}, is a method from game theory to allocate rewards fairly among players, having been widely adopted in ML to evaluate the importance of individual predictors \citep{chen2022expected, filippou2024short}. In this context, features act as players in a cooperative game, and the model's performance (such as $R^2$, MSE, etc.) serves as the total payoff. For a feature \(j\), its Shapley value represents the average marginal contribution it makes to all possible subsets of the other features. This method is model-agnostic and gradient-free, making it suitable for any predictive model.
	
Shapley value satisfies four properties: (i) Efficiency (Pareto optimality); (ii) Equal Treatment (given equal marginal contribution); (iii) Additivity (sum of multiple games, thus dynamic) and (iv) Null Player (no contribution, no payoff), which make Shapley value a robust and principled method for assigning credit to individual features. If a feature adds no information to any subset of features, its Shapley value is zero. If two features contribute identically to every subset, their Shapley values are equal. Additionally, the sum of all Shapley value equals the total prediction of the model, ensuring consistency in credit allocation.
	
Formally, for each model $m$, the Shapley value for a feature \(j\) is defined as:
\begin{equation}
\begin{aligned}
\label{equ:34}
SV_{m,j} := \frac{1}{|\mathcal{F}|} \sum_{C \subseteq \mathcal{F} \setminus \{j\}} \binom{|\mathcal{F}| - 1}{|C|}^{-1} \left( \mathcal{V}(C \cup \{j\}) - \mathcal{V}(C) \right),
\end{aligned}
\end{equation}
where \(\mathcal{F}\) is the set of all features, \(|\mathcal{F}|\) represents the total number of features, \(C\) is a subset (coalition) of features, and \(\mathcal{V}(C)\) is the value function that assigns a score to the coalition \(C\), such as the \(R^2\) of a model trained only on features in \(C\). The term \(\mathcal{V}(C \cup \{j\}) - \mathcal{V}(C)\) represents the marginal contribution of feature \(j\) to the coalition \(C\). Thus, $SV_{m,j}$ is essentially the average marginal contribution of $j$ when added to a coalition.

Both reduction in $R^2$ and Shapley value are model-agnostic measures. Additionally, we provide model-specific measures for robustness, including Sum of Squared Deviation (SSD; introduced by \citet{dimopoulos1995use} and widely applied in stock return \citep{gu2020empirical} and bond return \citep{bianchi2021bond} predictions), and Mean Decrease Gini (MDG, introduced by \citet{breiman2002manual}), as detailed in Appendix \ref{appendix Feature Importance}.

\subsubsection{Sub-group Feature Importance and $R^2_{OOS}$ Decomposition}
We investigate the relative importance of each individual feature for the performance of each ML model over time, using the well-studied feature importance measures discussed in previous sub-sections. There are two drawbacks of the previous individual feature importance analysis. Firstly, given the inclusion of the interaction between macro and in-market predictors, the effect of each single feature is never truly removed from the model \footnote{The $R^2$ reduction of single feature within group actually captures the interaction effect. For instance, the $R^2$ reduction for contingency reserve price, i.e., $R^2_{OOS}(\cdot + \text{contingency reserve price}) - R^2_{OOS}(\cdot)$, is not an ideal method to depict the pricing power of contingency reserve price, since (1) $R^2_{OOS}(\cdot + \text{contingency reserve price})$ fully consider the interactions of $\cdot$ and contingency reserve price; and (2) in our setting, $\cdot$ contains the interaction terms of macro features and contingency reserve price.}. Secondly, although individual feature importance analysis helps reveal the pricing power of each predictor and pricing channel due to the inclusion of interaction terms, if the interactions dominate the individual feature importance, it hinders us from answering the policy-concerned questions:  (1) Is the Singapore electricity market fully liberalized or does it remain heavily administratively regulated even after the transformation? (2) Is the Singapore electricity market more similar to a domestic or an international market? (3) Which players dominate the market: supply side, demand side or regulator? Alternatively, we consider the best subset selection approach, in which only a subset of features is fitted at a time, and compare the OOS performance across different subsets.

Our analysis involves 31 subsets of five feature groups in total, comprising one five-group subset, five four-group subsets, ten three-group subsets, ten two-group subsets, and five one-group subsets. By utilizing these subsets, we retrain our ML models and calculate the $R^2_{OOS}$ for each subset to evaluate their contributions. To save computational resources and keep the presentation manageable, we focus on the LGBM(+H) model, one of the top three individual models identified in our data setting. Also note that, given the exclusion of auto-correlated lag price term during training, we employ the mean and zero benchmarks for $R^2_{OOS}$.

As explained in the footnote, $R^2_{OOS}(X+Y)$ actually captures both the individual and interactive importance of the groups $X$ and $Y$. We adopt Shapley value, i.e., Eq \ref{equ:34}, to decompose the total $R^2_{OOS}$ into group-specific $R^2_{OOS}$ using OOS subset predictions within a coalitional game framework. Formally, the feature importance for group $g$ is defined as:
\begin{equation}
    \phi(g) = \sum_{H \subseteq \{ F / g\}} \frac{|H|!(|F| - |H| - 1)!}{|F|!} \left[R^2_{OOS}(H \cup g) - R^2_{OOS}(H)\right],
\end{equation}
where $H$ represents a subset of the feature groups used in the model, $F$ denotes the set of all feature groups, $g$ is the target feature group. $|\cdot|$ indicates the number of groups in the feature set. Thus, the importance of each feature group is a weighted average of all differences in $R^2_{OOS}$'s observed when the LGBM(+H) model is trained with and without feature group $g$\footnote{To illustrate this calculation, we consider the Supply group as an example. When $g = S$, $F = \{S, D, R, DM, IM\}$ and $F \setminus \{S\} = \{D, R, DM, IM\}$, we have:
\begin{equation}
   \begin{aligned}
        \phi(S)
        =
        \underbrace{
        \sum_{\substack{H \subseteq F \setminus \{S\}\\H \neq \varnothing}}
        \frac{\lvert H\rvert!\,(\lvert F\rvert - \lvert H\rvert - 1)!}{\lvert F\rvert!}
        \bigl[R^2_{OOS}(H \cup \{S\}) - R^2_{OOS}(H)\bigr]
        }_{\text{Interaction Effect}}
        \;+\;
        \underbrace{
        \frac{0!\,(\lvert F\rvert - 1)!}{\lvert F\rvert!}
        \bigl[R^2_{OOS}(\{S\}) - R^2_{OOS}(\varnothing)\bigr]
        }_{\text{Self Effect}}.
   \end{aligned}
\end{equation}}. This group importance measure has three desirable properties: (1) it includes both individual and interactive contributions; (2) it is grounded in OOS performance and (3) the sum of group importance corresponds to the total $R^2_{OOS}$. Different from the $R^2$ reduction and Shapley value discussed in the previous sub-sections, which primarily focus on the in-sample marginal dependence of models on individual features, this group-based measure assesses the relative contributions of feature groups to OOS predictive accuracy.

\subsubsection{Marginal Association}
We follow the approach of \citet{gu2020empirical}. In this approach, the targeted characteristic \(j\) is normalized to the \((-1, 1)\) interval, while all other features are fixed at their median value, which is zero. The characteristic \(j\) is then systematically varied from \(-1\) to \(1\), enabling the observation of its marginal effect with respect to the target variable.
	
\subsection{Economic Value: Utility Analysis}
In this section, we evaluate the economic value of ML forecasts for daily electricity returns within an investment framework that maximizes the utility of a mean-variance-preference investor.

The mean-variance investor allocates between a "risky asset," represented by the USEP, and a "risk-free asset," such as risk-free bills. At the end of day \(t\), the investor optimizes his/her portfolio allocation to maximize expected utility. This optimization problem is formalized as follows \citep{jagannathan2003risk, filippou2024cryptocurrency, rapach2010out}:
\begin{equation}
\begin{aligned}
\label{equ:35}
\arg \max_{w_{t+1}} w_{t+1} \mu_{t+1} - \frac{1}{2} \gamma w^2_{t+1} \hat{\sigma}^2_{t+1}
\end{aligned}
\end{equation}
where \(\gamma\) denotes the investor's relative risk aversion, \(w_{t+1}\) represents the proportion allocated to the risky asset, and \(1-w_{t+1}\) corresponds to the allocation to the risk-free asset. The conditional mean return \(\mu_{t+1}\), estimated as \(\hat{r}_{t+1}\), is predicted using our ML models, while the conditional variance \(\hat{\sigma}^2_{t+1}\) is modeled through a GJR-GARCH(\(p, q\)) framework. The lag parameters \(p\) and \(q\) are selected based on the Bayesian Information Criterion (BIC).
	
The solution to this optimization problem is:
\begin{equation}
\begin{aligned}
\label{equ:36}
w_{t+1}^* = \frac{1}{\gamma} \frac{\hat{r}_{t+1}}{\hat{\sigma}^2_{t+1}}
\end{aligned}
\end{equation}
This optimal allocation indicates that the weight assigned to the USEP is inversely proportional to the investor's risk aversion (\(\gamma\)) and directly proportional to the ratio of the expected return to its conditional variance. Consequently, the realized portfolio return at time $t+1$, denoted as $r^p_{t+1}$, is expressed by the following equation:
\begin{equation}
\begin{aligned}
\label{equ:37}
r^p_{t+1} = w_{t} r_{t+1} + (1-w_t) r^f_{t+1}.
\end{aligned}
\end{equation}
The CER of the portfolio is computed as:
\begin{equation}
\begin{aligned}
\label{equ:38}
CER_p = \hat{\mu}_p - 0.5 \gamma \hat{\sigma}^2_p,
\end{aligned}
\end{equation}
where $\hat{\mu}_p$ and $\hat{\sigma}^2_p$ are the sample mean and variance of the investor’s portfolio over the forecast evaluation period, respectively. The CER can be interpreted as the risk-free return that an investor would be willing to accept in lieu of holding a risky portfolio. This approach allows us to directly measure the economic value derived from return predictability.

\section{Results}
\label{emc:result}
In this section, we present our findings, covering model performance, model complexity, OOS predictability and the links to macroeconomy, feature importance, marginal associations, and utility analysis.

\subsection{Out-of-Sample Model Performance}
\subsubsection{Model Performance Based on \(R_{OOS}^2\)}
Table \ref{tab:1} summarizes the evaluation of model performance based on \(R_{OOS}^2\). \(R_{OOS}^2\) values are reported for the entire OOS period, with columns (1) to (4) calculated using Eqs. \ref{equ:20}, \ref{equ:21}, \ref{equ:23}, and \ref{equ:24}. All \( R^2_{OOS} \) values are positive and statistically significant at the 1\% level according to the DM test, indicating that all models consistently outperform lag price, AR (1), historical mean, and zero benchmarks. Next, we dive into each $R^2_{OOS}$ metric. $R^2_{OOS,mean}$ and $R^2_{OOS,zero}$ share a similar scale, as in our data setting, the historical mean of USEP return is close to zero. The diminishing magnitude from $R^2_{OOS,mean}$ (also, $R^2_{OOS,zero}$) to $R^2_{OOS,lagprice}$ and then $R^2_{OOS, AR(1)}$ aligns with the factor environment from strong to less strong. Considering the lag price is the most important feature based on our results in Section \ref{emc: feature_importance}, $R^2_{OOS,mean}$ (also, $R^2_{OOS,zero}$) does not exclude the impact of lag price, while $R^2_{OOS,lagprice}$ naively partials it out and $R^2_{OOS, AR(1)}$ excludes it with learned persistence. The positive $R^2_{OOS,lagprice}$ and $R^2_{OOS, AR(1)}$ also indicate that on top of the lag price, remaining features still contribute sizable explanatory power as on average for individual models, $R^2_{OOS,lagprice}$ reaches about 28\% and $R^2_{OOS, AR(1)}$ 18\%. We also conclude that in the most robust setting ($R^2_{OOS, AR(1)}$), we are still far away from the weak factor environment ($R^2$ below 5\%) in econometrics or finance literature, and our setting is more like a semi-strong factor environment (in a strong environment, $R^2$ is generally above 50\%). For robustness, we also report one-sided CW test results in the parentheses. We can observe that all models' $R^2_{OOS, AR(1)}$ and some models in $R^2_{OOS, lagprice}$ metrics may lose statistical significance, while $R^2_{OOS,mean}$ (also, $R^2_{OOS,zero}$) remains statistically significant at the 1\% level as one-sided DM.

Then, we compare the individual models to verify the virtue of sophisticated ML models. Please note that, our rigorous experimental design, i.e., with the same data input, train-validation-test scheme and benchmarks (also, factor environment), makes it possible to fairly compare individual models within the same $R^2_{OOS}$ metrics. The only difference between models lies in how they process the features. Across \(R_{OOS, lagprice}^2\), \(R_{OOS, AR(1)}^2\), $R^{2}_{OOS, mean}$, and \(R_{OOS, zero}^2\), we identify GLM, XGB(+H), and LGBM(+H) as the top three models. Moreover, when we identify the best-performing model for each of the 72 OOS months—based on $R^2_{OOS, mean}$ and $R^2_{OOS, zero}$—LGBM(+H) is most frequently selected as the top model (13 times). For the $R^2_{OOS, lagprice}$ and $R^2_{OOS, AR(1)}$ metrics, GLM achieves the highest frequency as the top performer (12 times). When model performance is averaged by category, the ranking is as follows: tree-based models $\sim$ GLM \(>\) NN5 \(>\) dimension reduction models \(>\) penalized linear models \(>\) OLS. 

For robustness, we conduct the two-sided DM test between each pair of models (see Table \ref{tab:2}), where a positive statistic indicates that the column model outperforms the row model. We observe that the top three individual models are still GLM, XGB(+H), and LGBM(+H), underscoring their strong and reliable performance across various evaluation metrics. Furthermore, we report the two-sided CW test results for the nested models: (1) ENet is nested with LASSO and Ridge; (2) NN5 is nested with NN1 to NN4 (each deeper network being nested within its shallower counterparts), see Table \ref{tab:A5} in Appendix \ref{Additional Tables}.

Also, given that misspecification of the conditional mean model typically induces misspecification of the conditional variance model, and with access to high-frequency intraday electricity price series, we further evaluate the robustness of the identified best-performing conditional mean models from the perspective of conditional variance. For detailed methodology, results, and analysis, refer to Appendix \ref{Evaluating Conditional Mean Misspecification via Conditional Variance}, Table \ref{tab:A12}, Table \ref{tab:A13}, and Figure \ref{fig:a4} in Appendix \ref{Result Supplements}.

Additionally, combination weights among the ensemble models provide statistical and economic perspectives for comparing individual models. The ensemble model can be interpreted as a representative investor who seeks to minimize forecast errors (or maximize forecast accuracy), with each individual model treated as a forecast analyst. The ensemble weights represent the proportions that the representative investor assigns to each forecast analyst. Accordingly, we expect that the best individual models should receive higher weights. The results are presented in Table \ref{tab:3}. The reported weight of each model is averaged over 72 OOS months. We can observe that the best three individual models identified by $R^2_{OOS}$ receive the highest weights (NN5 as well).

The virtue of ML models in processing high-dimensional features is generally twofold: (1) their ability to capture highly non-linear interrelations among features; and (2) their capacity to handle complex models, especially when the number of parameters exceeds the number of observations. Both aspects are well-studied and documented in the literature (see \citet{gu2020empirical} for the first rationale and \citet{kelly2024virtue} for the second). To further justify these rationales, we conduct a Monte Carlo simulation exercise; see Appendix \ref{simulation}.

Moreover, our results also reveal the virtue of employing $l2-$ norm and bagging techniques when factor environment is relatively weak (\citet{shen2024can}) to some extent. However, as previously emphasized, our data setting is more like a semi-strong factor environment. As a result, we do not expect to observe the perfect virtue of Ridge (with $l2-$ norm) and RF (bagging) over their counterparts, LASSO (with $l1-$ norm) and XGB(+H) (or LGBM(+H), boosting) respectively. Recall that, from $R^2_{OOS,mean}$ (also, $R^2_{OOS,zero}$) to $R^2_{OOS,lagprice}$ then to $R^2_{OOS, AR(1)}$, the factor environment is from strong to less strong. In Table \ref{tab:1}, Ridge performs worse than LASSO in $R^2_{OOS, mean}$ (also, $R^2_{OOS, zero}$), but beats the LASSO in $R^2_{OOS, lagprice}$ and more economically significantly in $R^2_{OOS, AR(1)}$, verifying the virtue of $l2-$ norm. In \( R^2_{OOS, mean} \) and \( R^2_{OOS, zero} \), RF unexpectedly underperforms penalized linear and dimension-reduction models. However, in \( R^2_{OOS, lagprice} \) and \( R^2_{OOS, AR(1)} \), this performance gap reverses, with RF outperforming both. Additionally, although RF continues to underperform boosted trees (XGB(+H) and LGBM(+H)) in terms of $R^2_{\text{OOS, lagprice}}$ and $R^2_{\text{OOS, AR(1)}}$, the performance gap narrows compared to that observed under $R^2_{\text{OOS, mean}}$ and $R^2_{\text{OOS, zero}}$. These two observations may reveal the virtue of bagging in some extent. \footnote{\citet{shen2024can} also verifies the virtue of $l2-$ norm via comparing the NNs only with $l2-$ norm (alternatively, NNs only with early-stopping) with NNs only with $l1-$ norm. In our setting, it is difficult to compare NNs' performance from $R^2_{OOS,mean}$ (also, $R^2_{OOS,zero}$) to $R^2_{OOS,lagprice}$ then $R^2_{OOS, AR(1)}$, since our NNs are designed with $l1-$ norm and early-stopping but without $l2-$ norm. This is a possible reason why our NNs exhibit unstable performance in $R^2_{OOS,lagprice}$ and $R^2_{OOS, AR(1)}$ metrics.}

Finally, we look at the ensemble models. In Table \ref{tab:1}, the performance ranks as \(Ensemble_{wp} > Ensemble_{op} > Ensemble_{avg} > Ensemble_{\theta}\) across all \(R_{OOS}^2\) metrics, namely, data-driven (\(Ensemble_{wp}, \\ Ensemble_{op}\)) $>$ naive $>$ economic-driven combination (\(Ensemble_{\theta}\)). Data-driven approaches are more prominent than economic-driven approach and the hard-to-be-beaten naive average. Moreover, our novel combination weight strategy, i.e., when models' predictions are highly correlated, introducing a prediction correlation penalty when optimizing the ensemble weights, effectively enhances performance to beat the best individual models. Furthermore, simpler approaches, such as naive time-invariant equal-weight ensemble method, outperform more sophisticated methods like the \(Ensemble_{\theta}\), but still fail to outperform some sophisticated individual models, aligning with the findings of \citet{palm1992combine}, \citet{smith2009simple}, \citet{stock2006forecasting}, \citet{rapach2010out}, and \citet{yuan2024naive}. Additionally, Figure \ref{fig:2} shows that the weight dynamics of $Ensemble_{wp}$ (Panel B) and $Ensemble_{op}$ (Panel C) differ dramatically from the one of $Ensemble_{\theta}$ (Panel A). We observe that $Ensemble_{wp}$ and $Ensemble_{op}$ tend to lean on single well-performing individual models rather than smoothing the weights across all individual models. Relative to $Ensemble_{op}$, our proposed $Ensemble_{wp}$ assigns more concentrated weights to well-performing individual models after accounting for prediction correlations.

Before moving on to set OLS as the benchmark, we can better understand the sources of performance gain through our proposed trend-based probability decomposition. We leave this part in the Table \ref{tab:A7} in Appendix \ref {Additional Tables} for interested readers to explore further. If we take the lag as the benchmark, the major conclusions are twofold: (1) relative to other individual models, the best three (GLM, XGB(+H), LGBM(+H)) make significantly fewer false trend predictions and excessive trend predictions (performance loss), and insignificantly more appropriate weak trend predictions and strong trend predictions (performance gain); (2) relative to individual models, the ensemble models incur less performance loss, and achieve bottom-line significantly more performance gain.

\subsubsection{Model Performance Based on $rRMSE$}
To further justify the virtue of sophisticated ML models, we present the results for $rRMSE$. As shown in column (2) of Table \ref{tab:4}, all sophisticated models, except Ridge, outperform the simplest model, OLS. The relatively poor performance of Ridge is due to its underperformance in the presence of strong factors, as noted in the \(R_{OOS}^2\) section. These results are consistent with those based on \(R_{OOS}^2\), including the identified top three individual models, the ranking of model categories, and the comparison among ensemble methods.

As a robustness check, we further report the \textit{p-values} of EPA tests, comparing the performance between sophisticated ML models and simple OLS. Please refer to Table \ref{tab:A6} in Appendix \ref{Additional Tables}.

\subsubsection{Model Sophisticate and Performance: the Virtue of Complexity}
\citet{kelly2024virtue} provides theoretical proof of the virtue of complexity when employing ML models to predict returns. In this part, we investigate this motivation: holding the model type fixed, does its increased sophistication (complexity) contribute to better OOS performance? Following the methodology of \citet{gu2020empirical}, for penalized linear models, we track the number of non-zero coefficients during the training process for each OOS month prediction. For dimension reduction models, we record the number of selected components, while for tree-based ensembles, we document the number of distinct features included in the trees. These measures provide insights into the complexity (\# of features) of each model. 

Figure \ref{fig:3} illustrates the complexity levels across different models. We observe that the three best-performing models demonstrate higher complexity compared to others, as measured by the number of selected features. Specifically, XGB(+H) and LGBM(+H) select around 350 features. GLM initially selects approximately 2400 features before 2022, stabilizing at around 1600. In contrast, LASSO selects about 100 features and ENet fewer than 160 before 2023, eventually increasing to approximately 220. PCR initially selects around 80 components before 2023, later increasing to around 100, while PLS consistently selects fewer than 14 components throughout the evaluation period. The number of features selected by RF fluctuates upward, increasing from 104 to approximately 111. However, these observed patterns should be interpreted with caution, as the number of selected features for each model is constrained by a fixed hyperparameter grid that varies across models. To further justify the relationship between model complexity and performance, we need to control for such model-specific and time-invariant factors, thus we conduct a panel regression:
\begin{equation}
\label{equ:39}
\begin{aligned}
R^{2}_{OOS, m, \mathcal{T}_{t,3}} = \alpha_0 + \alpha_1 \cdot model\_complexity_{m,\mathcal{T}_{t,3}} + \lambda_m + \Gamma_{\mathcal{T}_{t,3}} + \varepsilon_{m, \mathcal{T}_{t,3}},
\end{aligned}
\end{equation}
where $\lambda_m$ is the model fixed effect, $\Gamma_{\mathcal{T}_{t,3}}$ is the OOS month fixed effect.
		
To ensure comparability, we standardize the model complexity within each model. The results, presented in Table \ref{tab:5}, show that \(\alpha_1\) in all \( R^2_{OOS} \) metrics is positive, indicating that higher model complexity is associated with better performance. Specifically, the economic significance is sizable, as a one-standard-deviation increase in model complexity corresponds to an increase of 1.35\% in \(R^2_{OOS, lagprice}\) and 1.05\% in \(R^2_{OOS, AR(1)}\), although these increases are not statistically significant. However, we observe a statistically significant 1.48\% increase in the mean and zero benchmark-based \(R^2_{OOS}\) metrics \footnote{To further disentangle the explanatory power of $Model\_complexity$, model fixed effects and OOS month fixed effects on the panel $R^2_{OOS}$s' variations, we conduct variance decomposition. We find that OOS month fixed effects almost dominate, while model fixed effects and $Model\_complexity$'s explanatory power is small}.

To further disentangle the model fixed effects, we also plot the OLS-fitted line with 95\% CI of model complexity against four $R^2_{OOS}$s for each model in the Panel A of Figure \ref{fig:a5} in \hyperref[Online Appendix]{Appendix}. For the three best-performing individual models (LGBM(+H), XGB(+H), GLM) and RF, we observe an upward trend; for the dimensional reduction models (PLS and PCR), the fitted lines tend to be flat; while for the penalized linear models (LASSO and ENet), we observe a downward trend. \footnote{Additionally, our sub-group analysis provides an ideal setting to see the relationship between \# of features fitted in the model and OOS performance. The result is shown in the Panel B of Figure \ref{fig:a5} in \hyperref[Online Appendix]{Appendix}. We observe an apparent upward line with 95\% CI above the zero.}

\subsection{Out-of-Sample Predictability and Links to Real Economy}
As explained in the methodology section, in our data setting, the $R^2_{OOS, mean}$ and $R^2_{OOS,zero}$ tell a tale of two stories. By dividing the sample into different macro regimes $\mathcal{T}_s$ based on macro variables, a natural question arises: does OOS return predictability vary across macro states or regimes? For robustness, we also directly use calendar month as a splitting variable. This approach helps to answer whether the observed heterogeneity in predictability across calendar months can help to identify or predict structural breaks in OOS \footnote{Notably, the time-series heterogeneity analysis in this section first involves splitting the full sample period from 2003 to 2023 based on macro features, with the goal of detecting regimes over the entire sample. Then, for analyzing OOS return predictability across different macro states and for predicting OOS structural breaks, we focus exclusively on the OOS periods.}.

\subsubsection{Heterogeneous Predictability of Macro States}
We begin by using aggregate macro predictors to segment the time periods and identify regimes of predictability. This approach divides the entire sample based on macro-level variables, providing clear and intuitive interpretations. Notably, because the regimes are defined by macro values, a single regime may not correspond to a continuous time horizon.

In particular, we focus on two sets of macro predictors: (1) predictors specific to Singapore electricity market, including market conditions (the sample's unconditional mean and variance); (2) macroeconomic predictors, including domestic macro predictors (such as night light intensity) and international macro predictors (such as global geopolitical risk). Following \citet{rapach2010out}, the method of state classification ($\mathcal{T}_s$) is relatively straightforward \footnote{For instance, we first rank the daily unconditional return in the full sample, then categorize the days into bearish (bottom third), normal (middle third), and bullish (top third) market states. For volatile/tranquil market division, we follow the same simple criteria, but ranking the monthly unconditional variance of returns in the full sample, then categorize the 72 OOS months into tranquil (bottom third), normal (middle third), and volatile (top third) market states. As for features, we first rank the daily growth rate of target feature in the full sample, then categorize the days into low (bottom third), normal (middle third), and high (top third) market states of targeted feature.}. After $1(t \in \mathcal{T}_s)$ classification, we compute the state-specific $R^2_{OOS, m, s}$ for each model based on Eq \ref{equ:30} and the predictability of state $s$ via Eq \ref{equ:31}. The results are presented in Table \ref{tab:6}.

From Panels A and C, predictability are often concentrated in extreme periods (significant one-sided DM and CW tests results), especially bullish and high-growth night light intensity periods, with \textit{t}-test results significant at the 5\% and 1\% levels, respectively. Notably, in comparisons between bearish and bullish electricity market conditions, our three best-performing individual models (GLM, XGB(+H), and LGBM(+H)), as well as RF, perform better during bearish conditions, in contrast to the patterns observed for the other models. This may be a potential economic source of their performance gains in our data setting. The $R^2_{OOS}$ statistics are always higher during high-growth night light intensity compared to low-growth periods, with an economically significant average difference of about 5\%. Given that night light intensity is a good proxy for GDP, we find stronger predictability of Singapore electricity returns during periods of economic expansion. Such a finding is in contrast to the long-established studies investigating time-varying return predictability in the U.S. stock market, which highlight stronger predictability during recessions (e.g., \citet{rapach2010out}, \citet{henkel2011time},  and \citet{cong2024mosaics}). However, the expansion-concentrated pattern of predictability is quite similar to the U.S. government bonds market as \citet{bianchi2021bond}. Both the contradiction and the alignment are rational, as the underlying dynamics of electricity markets differ markedly from those of equity markets but quite similar to government bond market. For one thing, equities are subject to a mix of investor sentiment, liquidity issues, and behavioral biases. These factors might not be as prevalent in the electricity market. For another, government bond and Singapore's electricity market seem to align more closely with real economic activity.

From Panel B, we observe strong predictability when the electricity market is volatile, since $R^2_{OOS}$ of volatile regime shows significant one-sided DM and CW test results across all models while such statistical significance patterns disappear mostly in tranquil state, and the magnitude of the mean difference reaches 11\%. Such a finding coincides with the U.S. equity return forecasting literature, such as \citet{nagel2012evaporating}, \citet{stambaugh2012short} and \citet{avramov2023machine}. This pattern is also evident in the dynamics of $R^2_{OOS}$ (Figure \ref{fig:4}), where several sharp dips in $R^2_{OOS}$ occur during months of low volatility, as measured by candle length. We can also observe that the ensemble models systematically perform well (although the one-sided CW test is not significant) in the tranquil state compared to the volatile one, while individual models systematically perform well in the volatile market. Such distinct learning ability of ensemble models during stable markets may be the economic source of their performance gains relative to the individual ones. From Panel D, OOS gains for the forecasts are often concentrated in extreme periods, and regimes with high-growth geopolitical risk are both statistically and economically insignificantly predictable than the low-growth counterpart.

We report the predictability of state $s$, measured with and without ensemble models, in the last two rows of each panel. $Predictability_{oE}$ selects the $R^2_{OOS,s}$ of the best three individual models (GLM, XGB(+H) and LGBM(+H)), as well as NN5, as the predictability of state $s$. $Predictability_{wE}$ consistently selects the value from $Ensemble_{wp}$, which enhances the robustness of our identification of well-performing models.

\subsubsection{Structural Breaks by Calendar Months}
Splitting by aggregate predictors generates non-continuous regimes on the time horizon. For robustness, we also examine results from splitting by calendar months. One advantage of this approach is that calendar months progress sequentially over time, leading to simple, continuous regimes which resemble structural breaks \citep{cong2024mosaics}. In the following, we plot the time-varying predictability by month (regime) across our 72 OOS months.

Figure \ref{fig:5} illustrates that electricity return predictability remains relatively stable across different time horizons and does not show a declining trend over time. A similar pattern is also evident in Figure \ref{fig:4}, alleviating concerns raised in previous studies about potential attenuation in model performance over time \citep{harvey2016and, cong2021alphaportfolio}, or a drop in forecasting accuracy to disappointing levels following major shocks \citep{weron2006modeling, knittel2005empirical, weron2014electricity}. Certain events, like US-China Trade War (Sep 2018 - Oct 2018), OPEC Oil Production Cut (Feb 2019 - Mar 2019), the beginning of the Energy Crisis (Sep 2021 - Dec 2021) and Russia-Ukraine war (Feb 2022 - Dec 2022), likely triggered regime shifts to different levels of predictability. These structural breaks are also identified by the model-specific dynamics of cumulative sum of squared prediction errors (CUMSFE) (Figure \ref{fig:6}), defined as

\begin{equation}
CUMSFE_{m} = \sum_{t \in \mathcal{T}_3} (r_t - \hat{r}_{m,t})^2 \nonumber.
\end{equation}

Additionally, since the 2021 energy crisis represents the longest macroeconomic event within the observation period, another calendar-based split is performed by analyzing two periods separately: before and after 2021. Specifically, we define state $s=Crisis$ and $\mathcal{T}_{Crisis} := \{\text{$t$ after 2021}\}$ as in Eq \ref{equ:30}. Table \ref{tab:7} indicates that, on average, the period before the energy crisis is more predictable.

\subsection{Model Interpretations}
\subsubsection{Individual Feature Importance}
\label{emc: feature_importance}
\subsubsubsection{\(R^2\) Reduction}
In this subsection, we present the feature importance results using the \(R^2\) reduction method. For each model, the importance of a given predictor is determined by calculating the reduction in \(R^2\) when all values of that predictor are set to zero within each training set. These reductions are then averaged across all training samples to produce a single importance measure for each predictor. Figure \ref{fig:7} highlights the resultant importance of the top 20 features for each model. To ensure comparability, variable importance within each model is normalized to sum to one, enabling interpretation of relative importance within each model.

We find that the majority of the top 20 important features are interaction terms rather than single features—even in linear models.  This finding provides a rationale for our motivation to incorporate macro and in-market interactions. It is also consistent with the logic borrowed from the Euler equation of factor pricing in the asset pricing literature: $E_t(r_{i, t+1})=\beta_{i,t}^{\prime} \lambda_t$. In our setting, when $\lambda_t$ is international macro factor, the cross-section index $i$ can be interpreted as country $i$, $\beta_{i,t}$ is the factor loadings specific to Singapore market and $\lambda_t$ is the common factor affecting all electricity markets around the world. When $\lambda_t$ is domestic macro factor, the cross-sectional index $i$ can be interpreted as players in the Singapore electricity market (demand side, supplier or regulator), $\beta_{i,t}$ is the factor loadings specific to certain player within the Singapore electricity market and $\lambda_t$ is the common factor affecting all participants in the market. The interactions beat individual predictors, signifying that the determinants of Singapore energy price are potentially driven by complex interactions of macro features and in-market traits simultaneously. In other words, the $Macro \rightarrow In-market \rightarrow price$ channel may be more important than $Macro \rightarrow  price$ and $In-market \rightarrow price$ channels.

Furthermore, the heterogeneity of feature importance magnitudes across models (sparsity patterns) is evident. We find that the magnitudes for penalized linear (except for Ridge regression with the $l_2$ norm, but including GLM) and dimension reduction models are highly skewed toward USEP\_lag. In contrast, tree-based models and neural networks tend to utilize a broader set of features, drawing predictive information from a wider range of variables, which is consistent with the findings of \citet{gu2020empirical}.

Next, we look at the composition of the top 20 important features identified across models. We find that the composition includes many categories of factors, and there is no consensus on which category of features matters most. Such a finding aligns well with the multi-faceted nature of Singapore energy market, as gas is the main source of Singapore electricity generation, which is internationally driven and represents the supply side, whereas electricity consumption is domestically driven and represents the demand side. On top of this, the Singapore energy market is reforming from highly administrative regulation to a more liberalized regime, the regulator (government) is still a vital player in this market. Thus, if we look at the pricing power of individual features separately, it is quite hard to answer such policy-concerned questions: (1) Is the Singapore electricity market fully free or still heavily administratively regulated even after the transformation? (2) Is the Singapore electricity market more like a domestic or an international market? (3) Which players dominate the market, supply side, demand side, or the regulator? Additionally, models also disagree on which specific features are most vital. For instance, the auto-correlated term, USEP\_lag, provides the strongest explanatory power to the linear models, such as LASSO, ENet, RCR, PLS, and GLM, but contributes little to the non-linear models, except for XGB(+H).

Finally, we analyze each feature. Though models disagree on which categories of features matter, or more specifically, which factors matter most, we can gain consistent insights from the relative frequency. We observe that USEP\_lag emerges as the most frequently ranked top feature across models, reflecting its critical role in capturing past price behavior, either momentum or mean reversion effects. Contingency reserve price follows as the second most common top feature, emphasizing its importance in influencing USEP returns by accounting for the cost of maintaining reserve capacity to address potential supply shocks. The remaining top five include raw material prices such as gas price and oil price, supply cushion (a measure of spare generation capacity relative to demand; it is critical for understanding market elasticity and predicting price volatility) and vesting/vested quantity (a regulation that stabilizes market revenues and limits market power through predefined contractual agreements). 

\subsubsubsection{Shapley Value}
For robustness, in this subsection, we present the results of feature importance analysis using Shapley value. However, it should be noted that $R^2$ reduction is statistically driven, while Shapley value is economically driven. Therefore, we do not expect perfect alignment between the results, but rather aim to highlight the robustness of common patterns and the differences between these two methods.

Figure \ref{fig:8} illustrates the feature importance results aggregated across all models. Consistent with the individual model analysis, USEP\_lag remains the top-ranked feature due to the high model consensus. Also, compared with the color distributions and sparsity based on $R^2$ reduction in Figure \ref{fig:a6} and Figure \ref{fig:7} respectively, the distribution based on Shapley value does not display model similarity and feature reliance in penalized linear models, as Shapley value cannot reflect the statistical properties of the models. For further robustness, we also calculate the SSD and impurity reduction, as shown in Figures \ref{fig:a8}, \ref{fig:a9}, \ref{fig:a10} and \ref{fig:a11} in Appendix \ref{Additional Figures}.

\subsubsubsection{Feature Importance Dynamics and Links to Macroeconomy}
In this subsection, we link the dynamics of feature importance to the macroeconomy, especially the structural breaks. To identify potential structural changes, we apply the Bai-Perron Multiple Breakpoints Test to detect breakpoints in the \(R^2\) reduction-based feature importance data and use the Pettitt Test \footnote{We leave the basic methodology of these two tests in Appendix \ref{appendix Statistical Tests} and Appendix \ref{Trend-based Probability Framework}.} to independently evaluate the presence and significance of structural changes. We conduct the analysis both at the individual model level and at the aggregated level. For individual models, the Bai-Perron Multiple Breakpoints Test identifies OOS month 45 (September 2021) as the most frequently detected breakpoint across features (excluding boundary points), coinciding with the onset of the energy crisis. Among the cases where OOS month 45 is detected as a breakpoint by the Bai-Perron test, the Pettitt Test independently confirms significant structural changes in 76.06\% of these cases. Notably, the breakpoints identified by the Pettitt Test do not necessarily coincide with those detected by the Bai-Perron test. At the aggregated level, which combines feature importance across all models, OOS month 45 is also identified as the most significant breakpoint. This consistency underscores the widespread impact of the energy crisis on signal importance ranks thus model behavior.

\subsubsection{Sub-group Feature Importance and $R^2_{OOS}$ Decomposition}
Given that the most important features identified are predominantly interaction terms, it is challenging to directly address the policy-relevant questions discussed earlier. Therefore, we proceed by grouping features into five categories: supply, demand, regulation, domestic macro, and international macro. The specific feature composition for each group is detailed in the Table \ref{tab:A2} in \hyperref[Online Appendix]{Appendix}.

For expositional convenience, we pick one of the best three models, LGBM(+H), as an example to calculate the sub-group results by decomposing the $R^2_{OOS}$. Figure~\ref{fig:9} presents the results of the subgroup feature importance analysis. Panel B displays the static $R^2_{OOS}$ decomposition. The results show that regulatory factors contribute the most to predictive performance, followed by supply-side features. This suggests that the Singapore electricity market is primarily driven by supply and remains highly sensitive to regulatory interventions.

From a policy perspective, our findings indicate the following: (1) the Singapore electricity market continues to be heavily regulated, even after market reforms; (2) domestic macroeconomic features play a larger role than international ones, suggesting that the market behaves more like a domestic market; and (3) regulatory factors are the dominant drivers in the market.

Panel C shows the dynamic $R^2_{OOS}$ decomposition over time. We can observe that (1) albeit volatile, the magnitude of the regulation group is apparently more sizable than the other groups; (2) in April 2019, the demand group dominates, while the regulation and supply groups move into negative territory; (3) the disparity in importance across feature groups has narrowed since September 2021.

\subsubsection{Marginal Associations}
In this subsection, we present the results of the marginal association. We choose the important features in Section \ref{emc: feature_importance}, including USEP\_lag, contingency reserve price, supply cushion, GPRD\_ATC, gas spot price, fuel oil price, and exchange rate.
		
Figure \ref{fig:10} illustrates the results of the marginal association, revealing patterns aligned with economic and energy market theories. For example, USEP\_lag demonstrates a negative effect on USEP returns, indicating a reversion effect where elevated past prices dampen future returns. An increase in the contingency reserve price is associated with higher USEP returns, reflecting the increased costs of maintaining reserve capacity during periods of heightened demand or supply uncertainty. Conversely, a higher supply cushion, representing a substantial surplus of supply relative to demand, leads to lower USEP returns, driven by reduced scarcity pressures. Additionally, GPRD\_ATC raises electricity prices by disrupting gas and fuel oil supply chains, amplifying price volatility, and increasing production costs. Interestingly, the signs of raw material prices (e.g., gas spot price and fuel oil price) and the exchange rate are indeterminate.

\subsection{Economic Value: Utility Analysis}
We further assess model performance using the mean-variance optimal portfolio approach. In this framework, a mean-variance investor allocates between a "risky asset," represented by the USEP, and a "risk-free asset," such as risk-free bills. Models with superior OOS performance are expected to deliver the highest cumulative and average utility. At the portfolio level, i.e., after combining the individual models' forecast with the risk-free asset (we employ 10-year treasury bills in the SGX as a proxy), we expect that the best individual models have the highest CER, as CER can be interpreted as the equivalent risk-free return that an investor would be willing to accept when holding such a risky asset. Table \ref{tab:8} and Figure \ref{fig:11} demonstrate that, if assuming the investor's relative risk aversion as \(\gamma = 3\), the top-performing models, XGB(+H), LGBM(+H), and GLM, consistently achieve the highest cumulative, average utility, and CER across all models, highlighting their effectiveness and robustness. Specifically, a 77.15\% risk-free rate that a mean-variance-preference investor would be willing to accept in lieu of holding a XGB(+H)-risky portfolio. Results remain robust for alternative plausible values of \(\gamma\), as demonstrated in Tables \ref{tab:A9}, \ref{tab:A10}, \ref{tab:A11} and Figures \ref{fig:a12}, \ref{fig:a13}, \ref{fig:a14} in Appendix \ref{Result Supplements}.

\section{Conclusions}
\label{emc:conclusion}
This study demonstrates the power of ML in forecasting Singapore’s electricity prices, emphasizing three core virtues: (1) capturing nonlinear relationships through tree-based models (XGBoost, LightGBM), as well as GLM and NN5, which outperform linear methods; (2) leveraging complexity in high-dimensional settings to process 619 features spanning stakeholder-level dynamics, macroeconomic indicators, and global shocks; and (3) adapting to weak-factor environments via $l2$-regularization and bagging, crucial for Singapore’s volatile, still highly regularized and import-dependent electricity market. Such predictability can be translated into sizable economic gains under the mean-variance framework.

Our novel ensemble methodology, i.e., by adding a penalty to the prediction correlations among individual models, has two key advantages. First, it dramatically improves ensemble performance when individual models are highly correlated. Second, it achieves robust and strong predictions both before and after the 2021 energy crisis, alleviating concerns about deteriorating performance during and after periods of energy crisis (\citet{weron2006modeling} and \citet{knittel2005empirical}).

Our findings reveal heterogeneous predictability: ML models achieve significant OOS accuracy, particularly during expansion and volatile periods, mirroring the patterns observed in broader financial markets. Moreover, predictability peaks during extreme geopolitical risk regimes. Structural breaks also affect predictability, underscoring the interplay between global shocks and local market design.

Feature importance analysis highlights the complex nature of the Singapore electricity market. On the one hand, the interaction terms between macro and in-market features are more important than individual features, revealing that $Macro \rightarrow In-market \rightarrow price$ channel is potentially more important than $Macro \rightarrow price$ and $In-market \rightarrow price$. On the other hand, the composition of the top 20 important features identified across models spans many categories of predictors. However, when looking at sub-group $R^2_{OOS}$ decomposition, we gain some insights: the Singapore electricity market is relatively supply-side driven and heavily affected by regulatory factors.

These insights advance electricity price forecasting by bridging ML's predictive power with economic interpretability, offering stakeholders actionable strategies for risk management and policy. Future work could extend adaptive frameworks for real-time forecasting by loosening the two main model restrictions mentioned previously, and could also conduct causal analysis of market design on price stability.

\section*{Acknowledgments}
This paper is based on our \textit{Econometric Model Design, Approach and Methodology Report} for EMC, Singapore. Confidential data is provided by EMC. The non-disclosure agreement applies.

\newpage
\section*{Tables and Figures}
\label{Tables and Figures}

\setcounter{figure}{0}
\begin{figure}[H]
\caption{Example of the Market Clearing Process}
\caption*{\fontsize{10pt}{0.35cm}\selectfont From \citet{authority2009introduction}, electricity prices in this figure can be negative due to oversupply. Negative prices mean that power companies need to "pay" users or grid operators to consume the excess electricity. This phenomenon reflects the unique characteristics of the electricity market, as electricity is difficult to store, and excess supply can drive prices into negative territory.}
\centering
\includegraphics[width=0.6\textwidth]{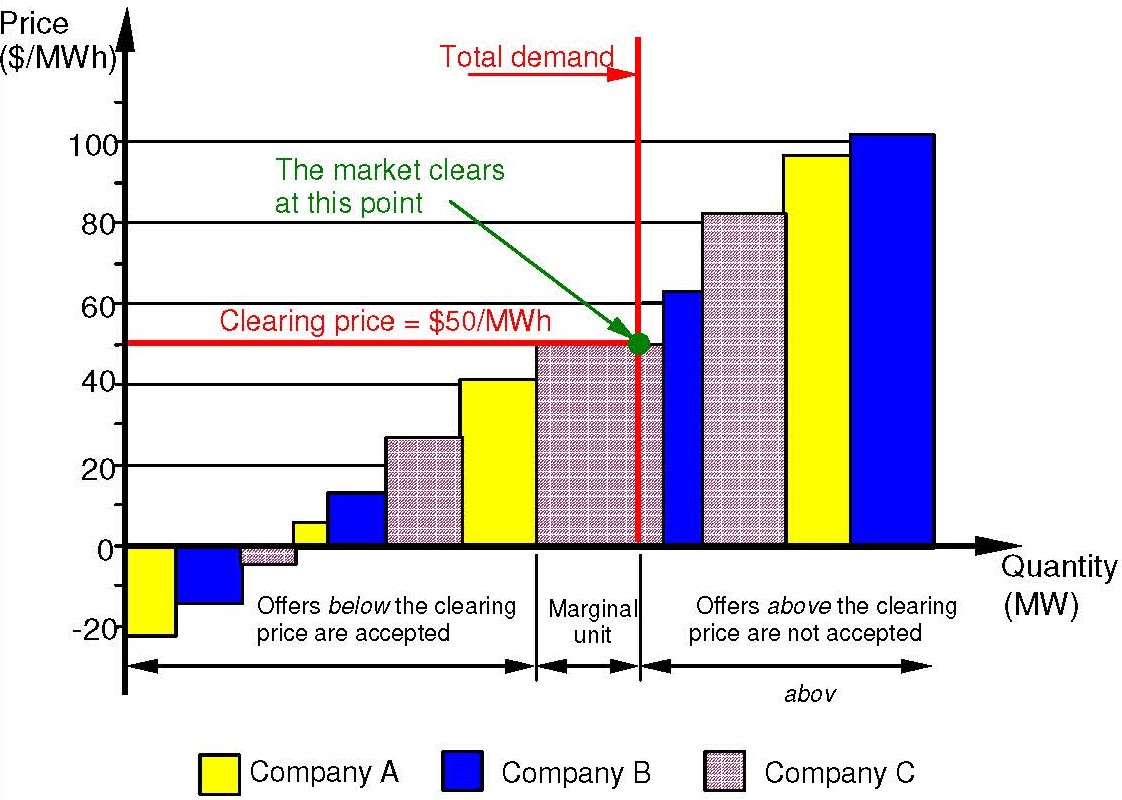}
\label{fig:1}
\end{figure}

\begin{figure}[H]
\centering
\caption{Weight Dynamics of Ensemble Models}
Panel A: $Ensemble_{\theta=1}$\\
\includegraphics[width=1\textwidth]{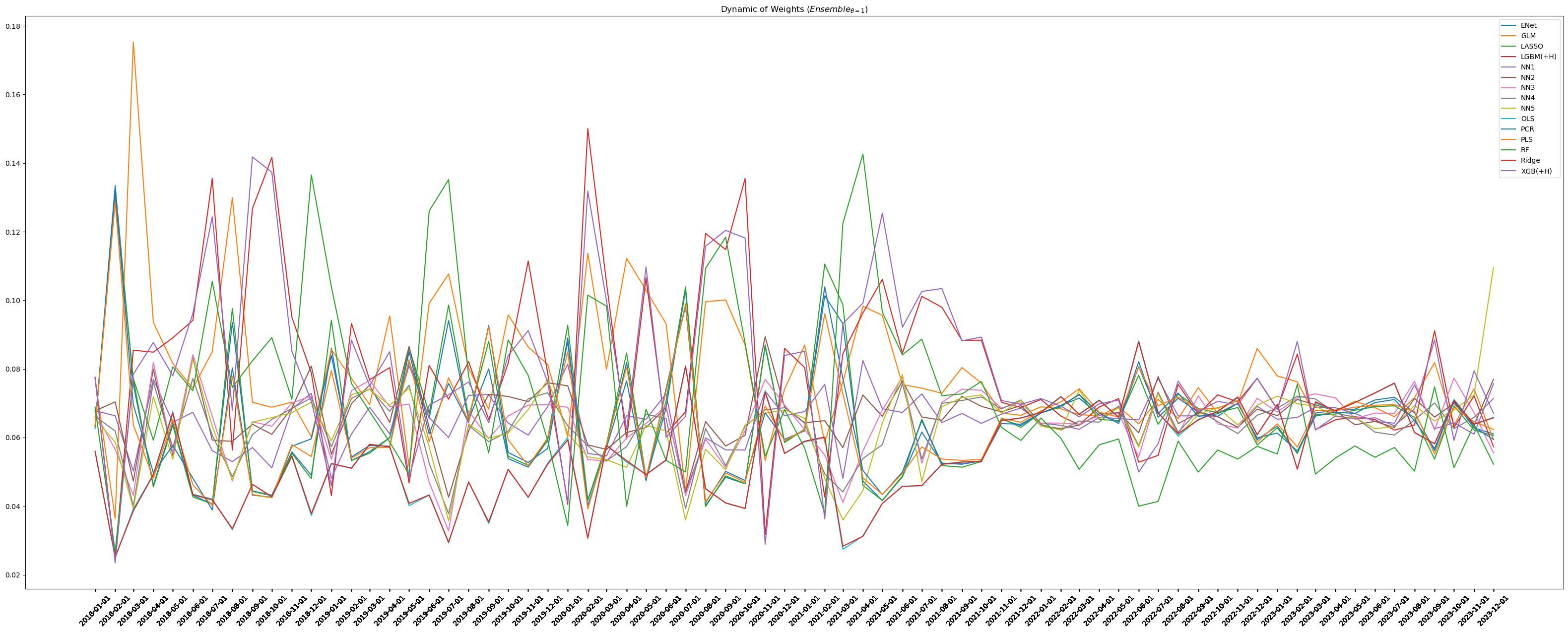}
\vspace{0.05cm}
Panel B: $Ensemble_{wp}$\\
\includegraphics[width=1\textwidth]{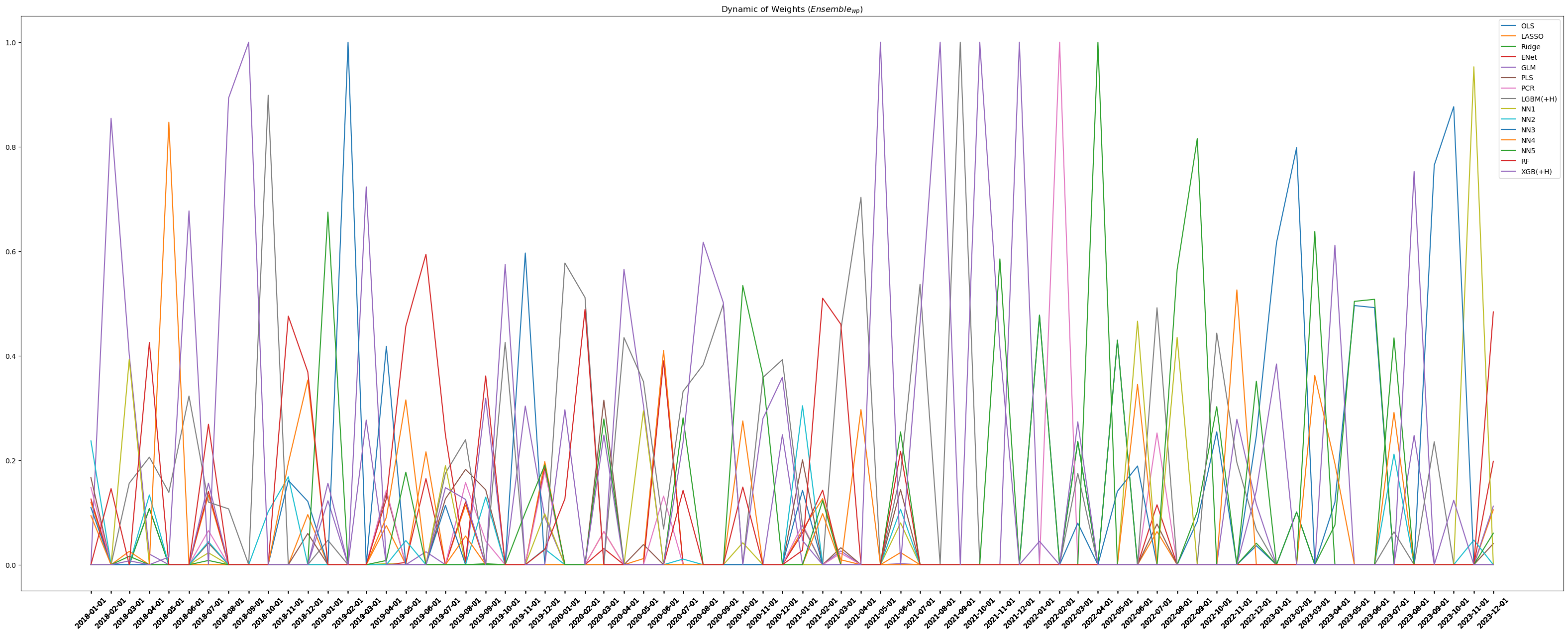}
\vspace{0.05cm}
Panel C: $Ensemble_{op}$\\
\includegraphics[width=1\textwidth]{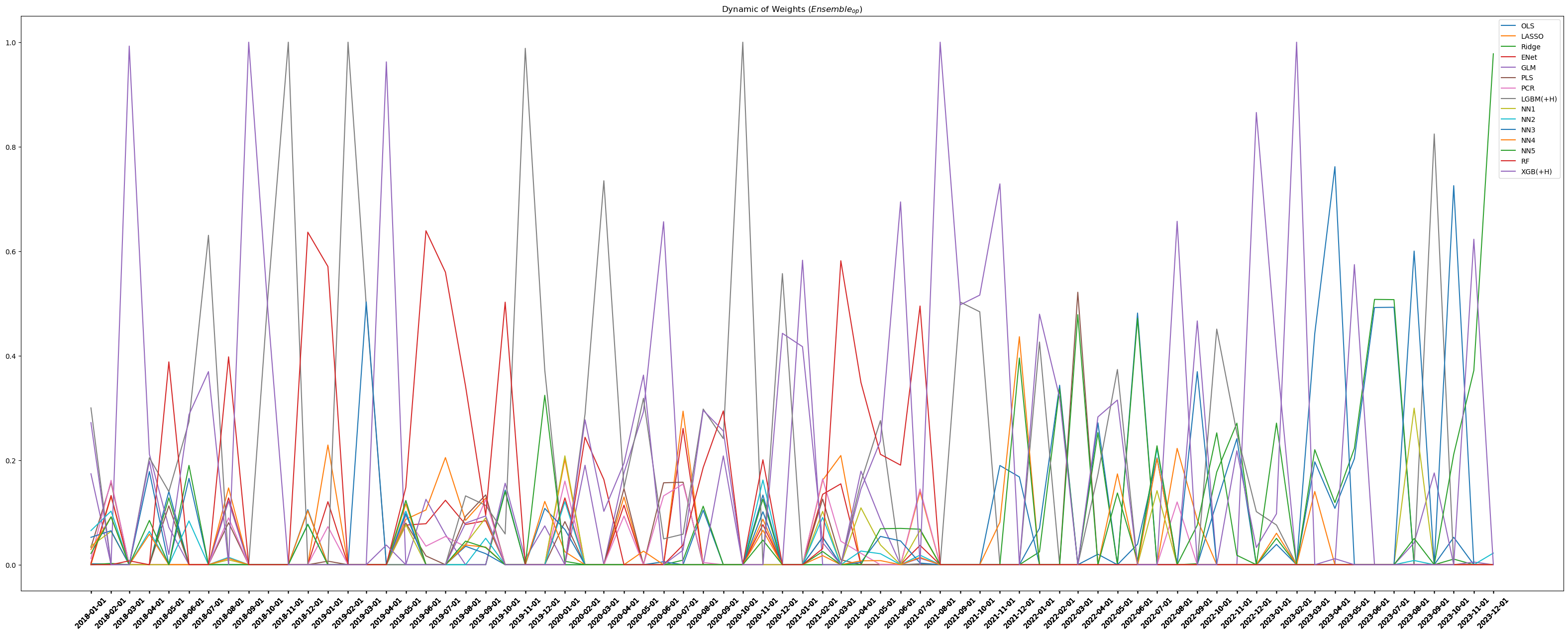}
\label{fig:2}
\end{figure}

\begin{figure}[H]
\caption{Model Complexity}
\caption*{\fontsize{10pt}{0.35cm}\selectfont This figure demonstrates the model's complexity for LASSO, ENet, PCR, PLS, GLM, RF, XGB(+H) and LGBM(+H) in each training sample of our 72-months recursive OOS analysis. For LASSO, ENet and GLM, we report the number of features selected to have nonzero coefficients; for PCR and PLS, we report the number of selected components; for RF, we report the average number of leaf nodes; and, for XGB(+H) and LGBM(+H), we report the number of distinct characteristics entering into the trees.}
\resizebox{\textwidth}{!}{
  \begin{minipage}{\textwidth}
    \begin{multicols}{2}
      \includegraphics[width=\linewidth]{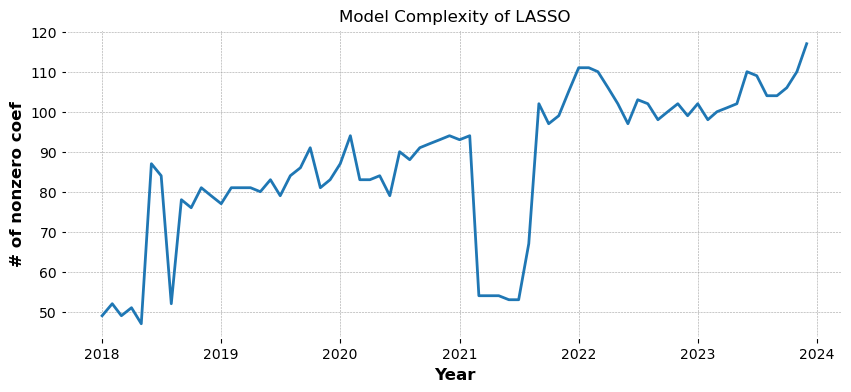}
      \includegraphics[width=\linewidth]{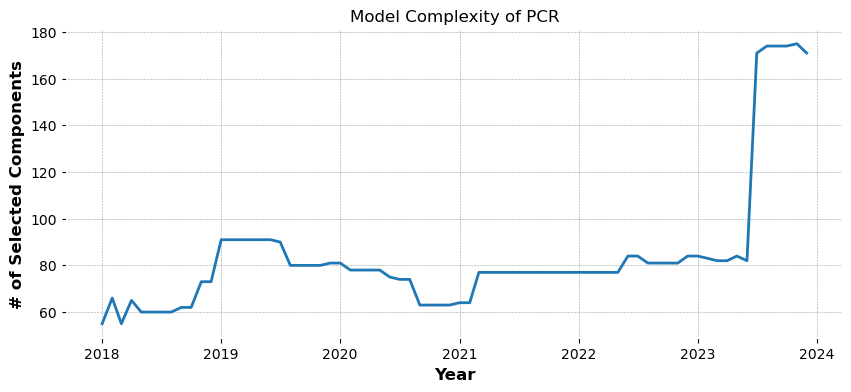}
      \includegraphics[width=\linewidth]{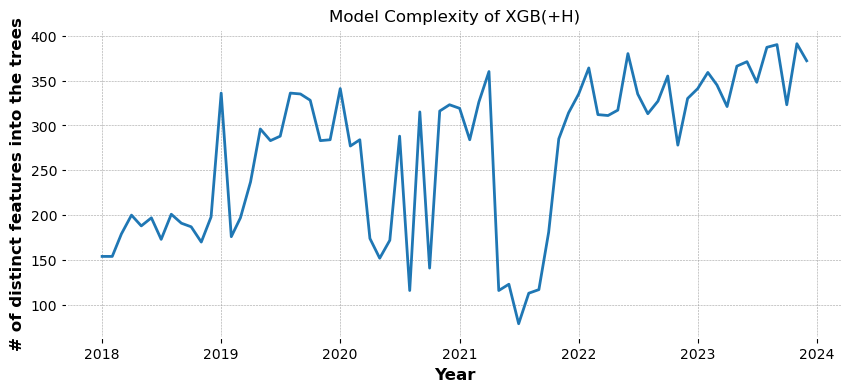}
      \includegraphics[width=\linewidth]{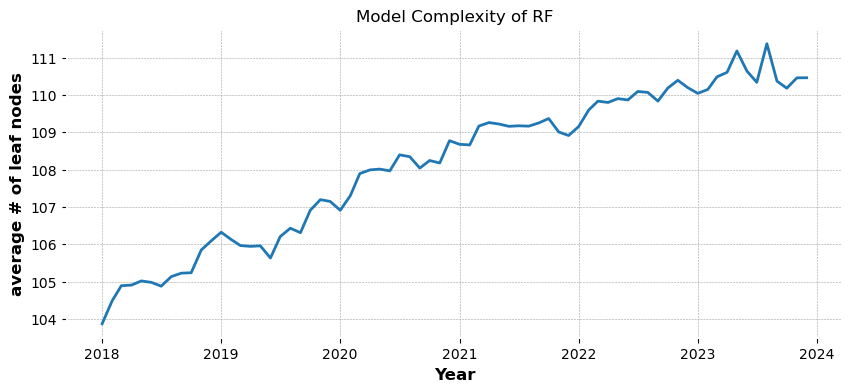}

      \includegraphics[width=\linewidth]{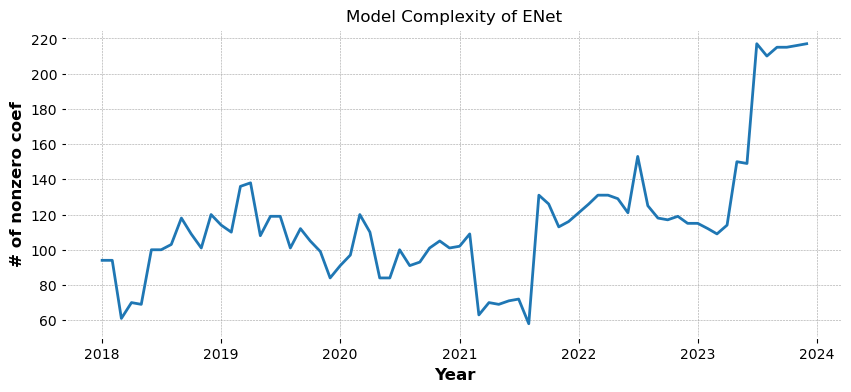}
      \includegraphics[width=\linewidth]{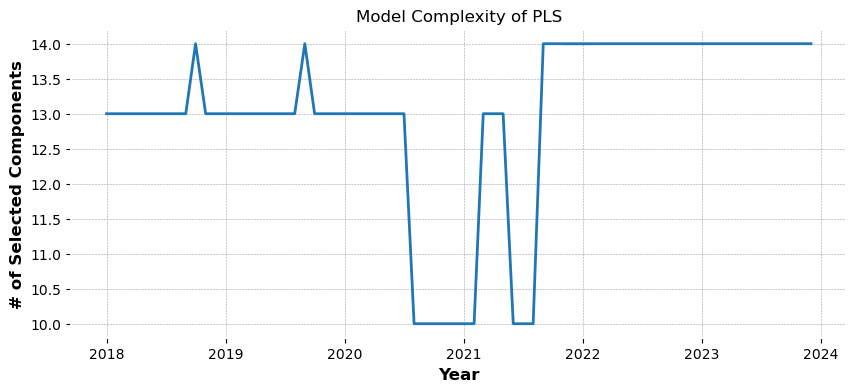}
      \includegraphics[width=\linewidth]{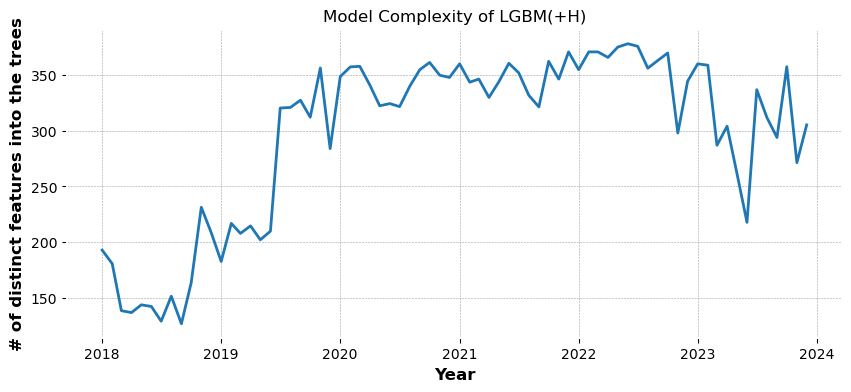}
      \includegraphics[width=\linewidth]{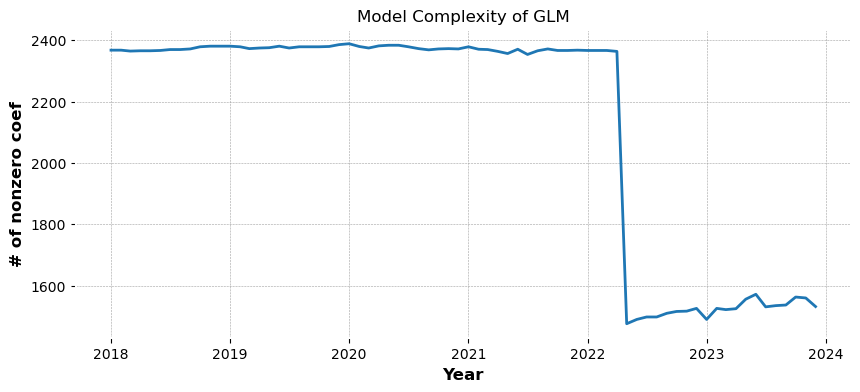}
    \end{multicols}
  \end{minipage}
}
\label{fig:3}
\end{figure}

\begin{figure}[H]
\caption{\centering Dynamics of Out-of-Sample Electricity Returns \& $R^2_{OOS,mean}$}
\caption*{\fontsize{10pt}{0.35cm}\selectfont The upper sub-figure plots the candlestick of USEP returns from Jan 2018 to Dec 2023 (72 OOS months). The candlestick is constructed monthly, based on the monthly highest, lowest, begin and close prices. Red stands for bearish, while green represents bullish. We can observe that (1) the obvious \textit{doji} patterns (short range of body) and (2) long range of box, which is consistent with two facts of electricity returns: (1) historical mean is near to zero and (2) volatile. The lower sub-figure plots the time-series dynamic of $R^2_{OOS,mean}$ from Jan 2018 to Dec 2023 (72 OOS months) across all model candidates.}
\includegraphics[width=1\linewidth]{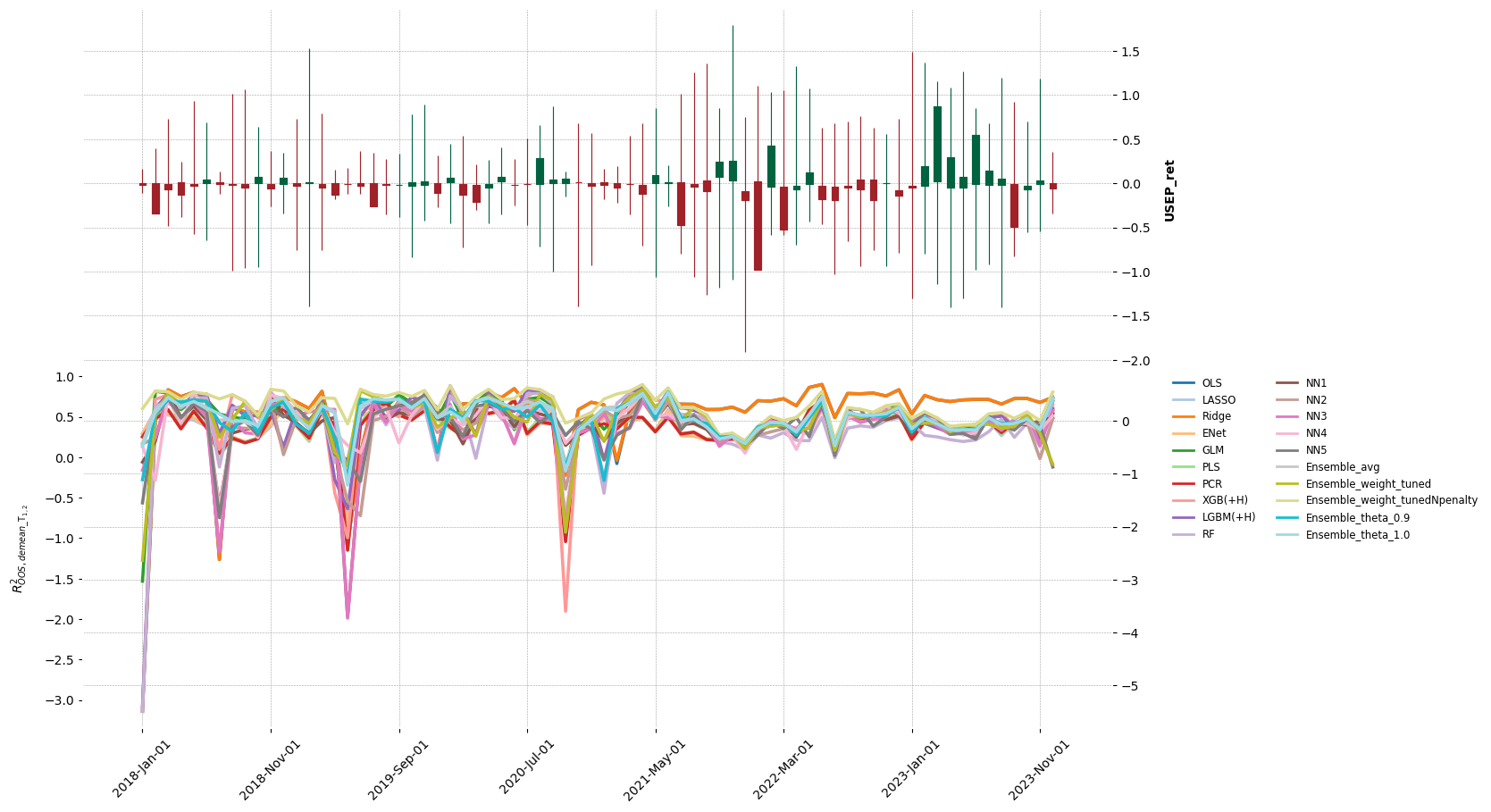}
\label{fig:4}
\end{figure}

\begin{figure}[H]
\caption{\centering Dynamics of Out-of-Sample Regime Predictability}
\caption*{\fontsize{10pt}{0.35cm}\selectfont This figure plots the OOS regime predictability based on $R^2_{OOS,mean}$ from Jan 2018 to Dec 2023 (72 OOS months). Regime here is defined as each OOS month. Within each month, the regime (monthly) predictability is calculated as \ref{equ:31} by setting $s=\mathcal{T}_{3,t}$ and $benchmark = mean$. We consider the ensemble models when calculating the state predictability, thus the blue line is essentially the OOS dynamics of $R^2_{OOS,mean}$ for $Ensemble_{wp}$. We select the 25\%, 50\%, 75\% and 100\% percentile as the cutoffs to classify each month into four types of regime predictability. The predictability increases as the color turns from red, blue, green to yellow. Shaded areas denote some vital macro events related to energy market.}
\includegraphics[width=1\linewidth]{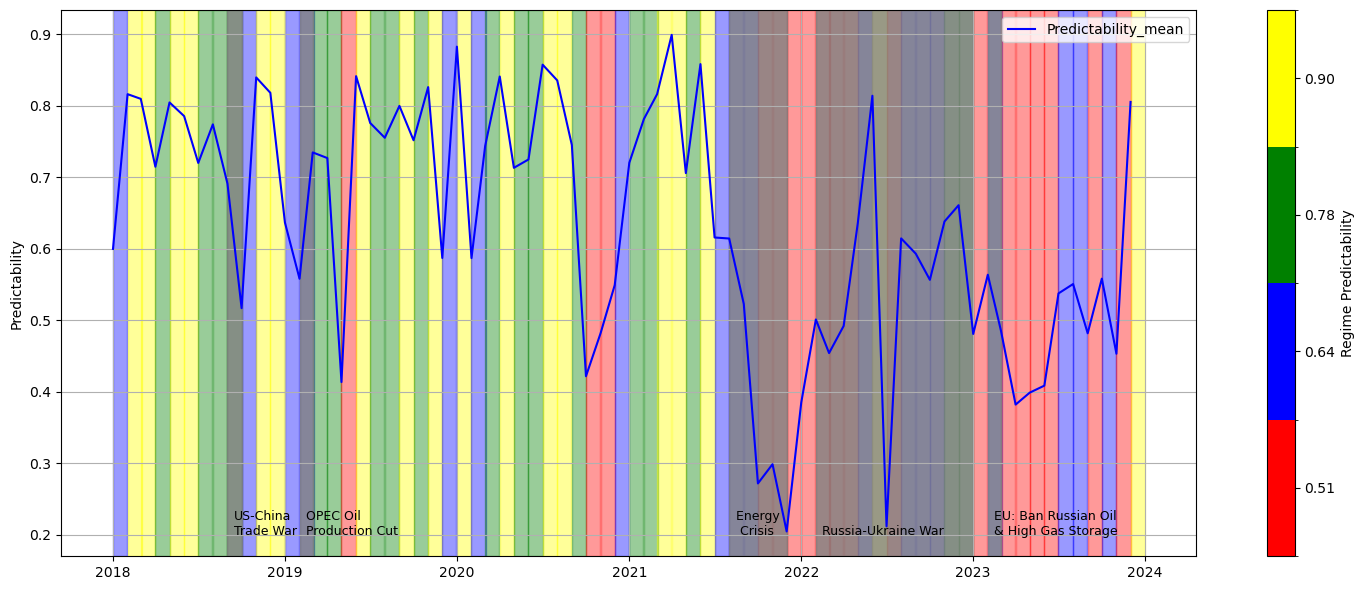}
\label{fig:5}
\end{figure}

\begin{figure}[H]
\caption{\centering Dynamics of Out-of-Sample CUMSFE}
\caption*{\fontsize{10pt}{0.35cm}\selectfont This figure plots the OOS cumulative mean sum squared of forecasting errors (CUMSFE) from Jan 2018 to Dec 2023 (72 OOS months) across all models. Shaded areas denote some vital macro events related to energy market.}
\includegraphics[width=1\linewidth]{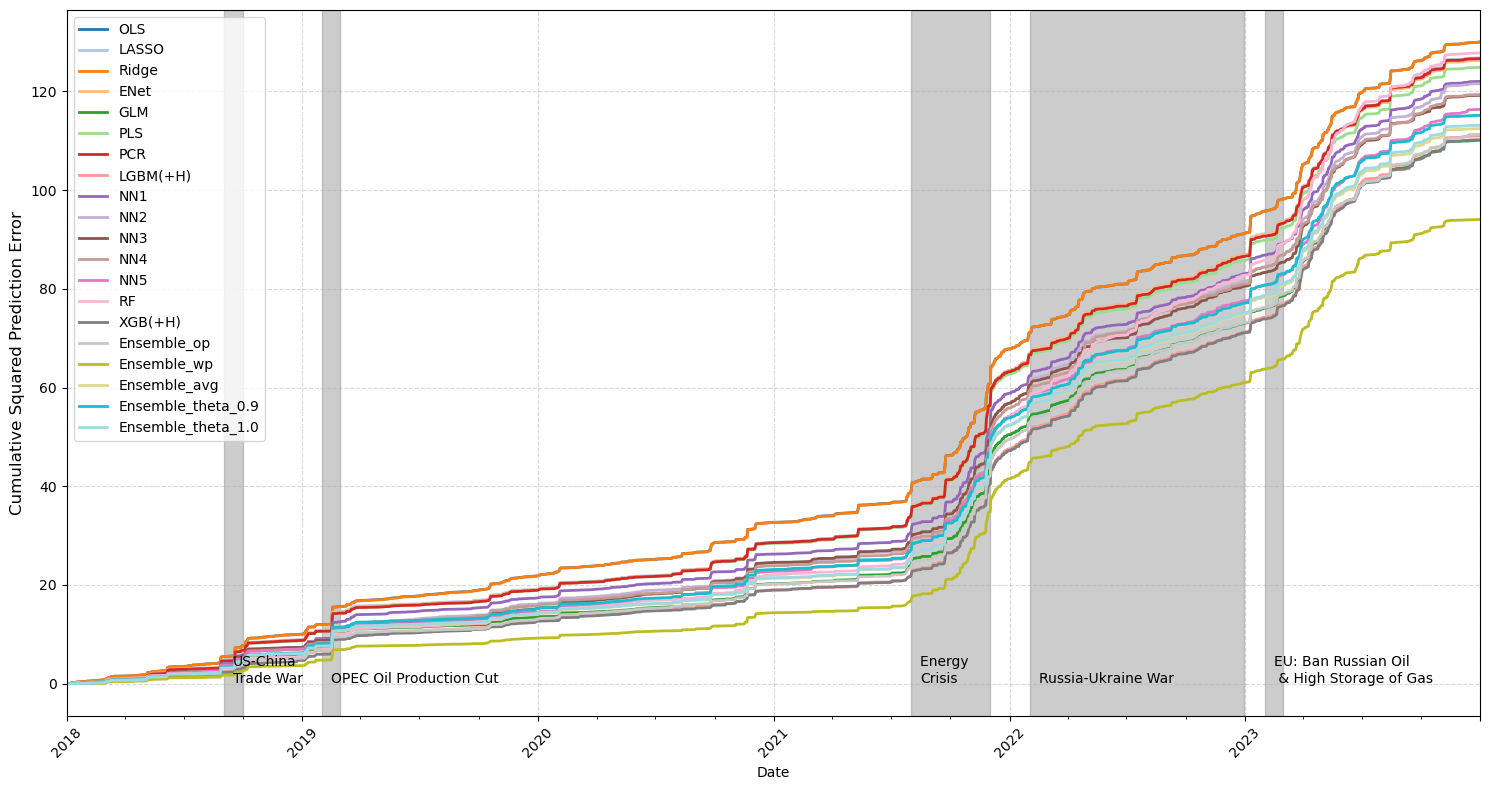}
\label{fig:6}
\end{figure}

\begin{figure}[H]
\caption{\centering Feature Importance Based on \(R^2\) Reduction Across Each Model}
\caption*{\fontsize{10pt}{0.35cm}\selectfont Variable importance for the top-20 most influential variables in each model. Variable importance is an average over all training samples. Variable importance within each model is normalized to sum to one.}
\begin{multicols}{2}
\begin{figure}[H]
\includegraphics[width=1\linewidth]{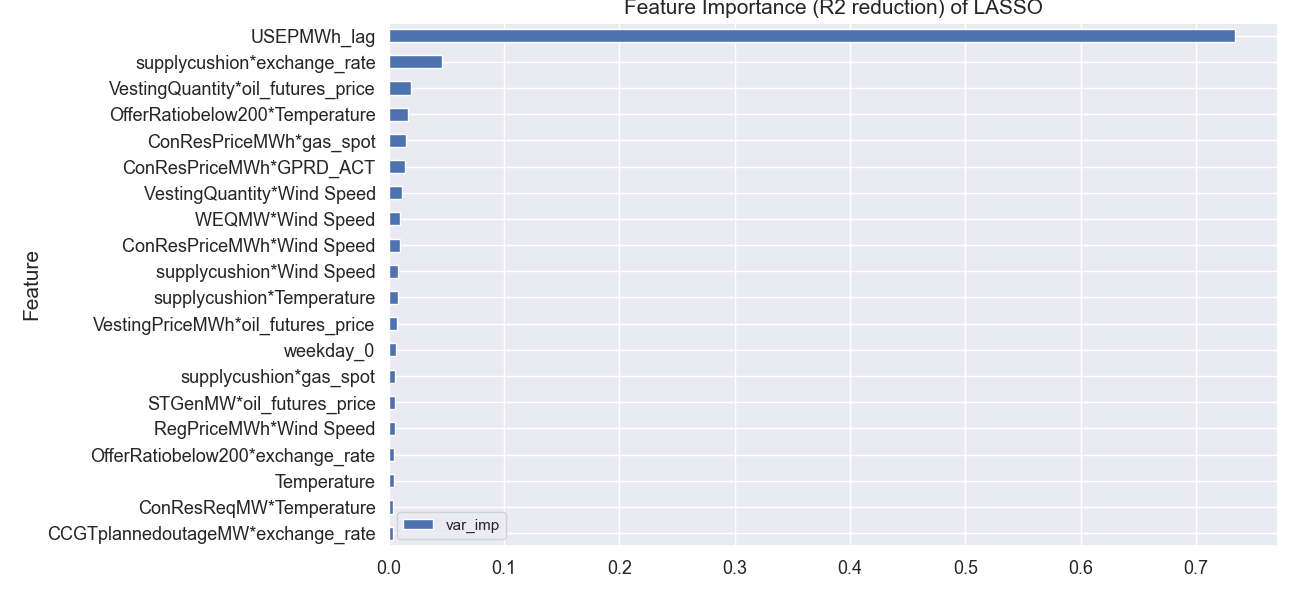}
\includegraphics[width=1\linewidth]{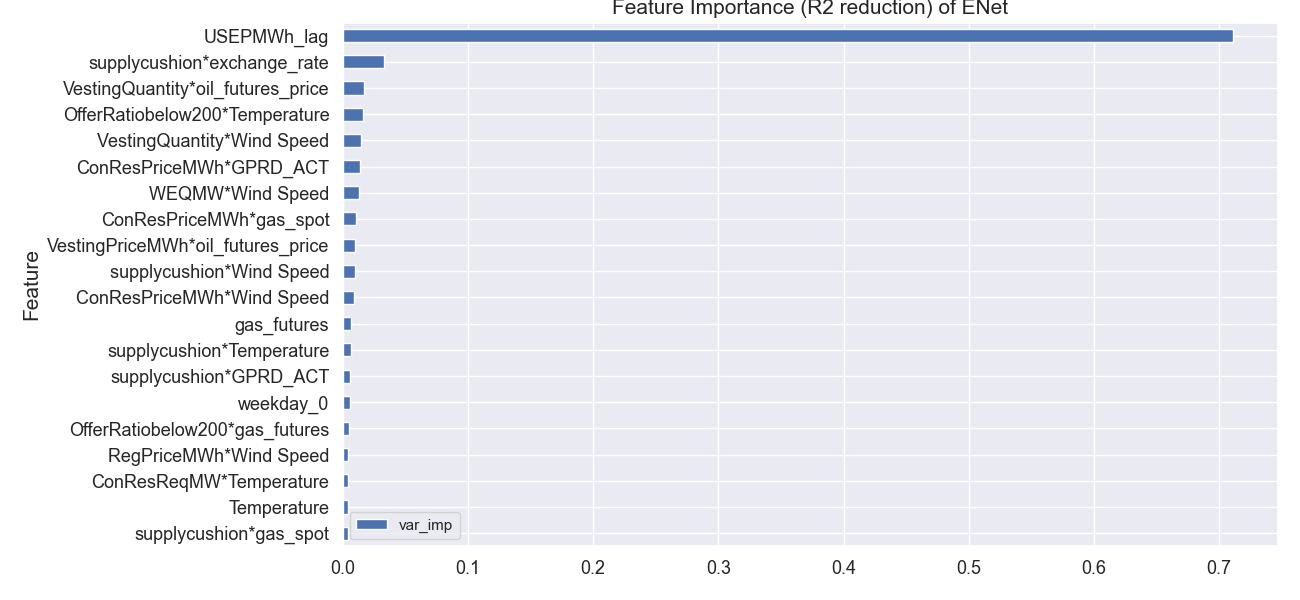}
\includegraphics[width=1\linewidth]{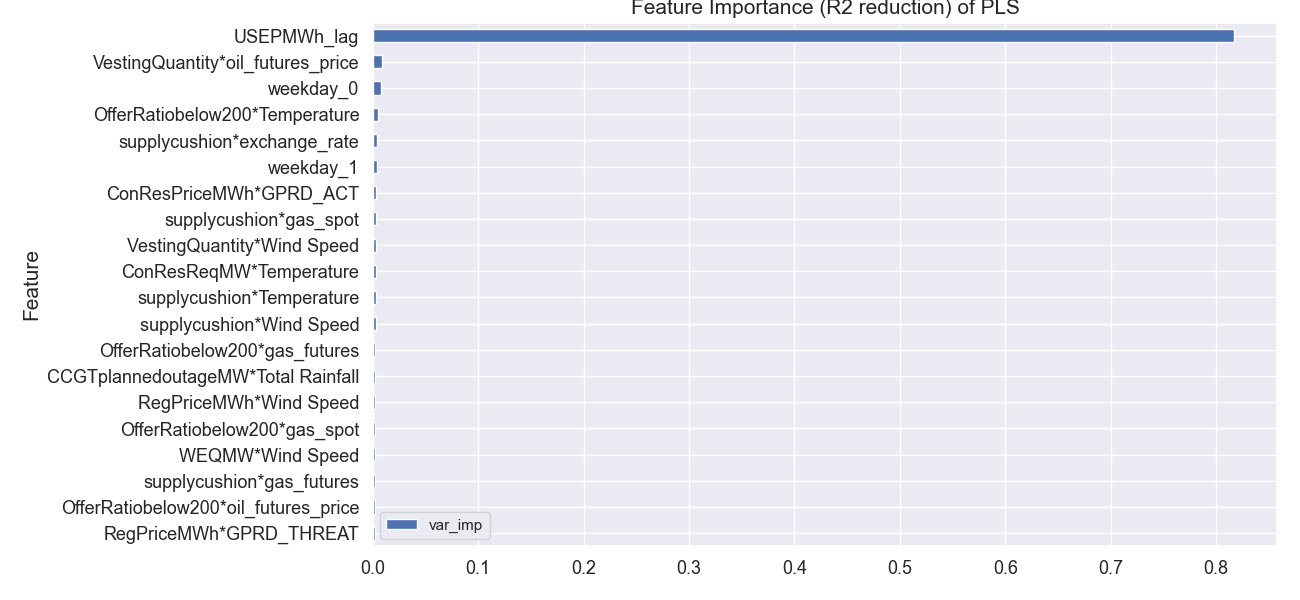}
\includegraphics[width=1\linewidth]{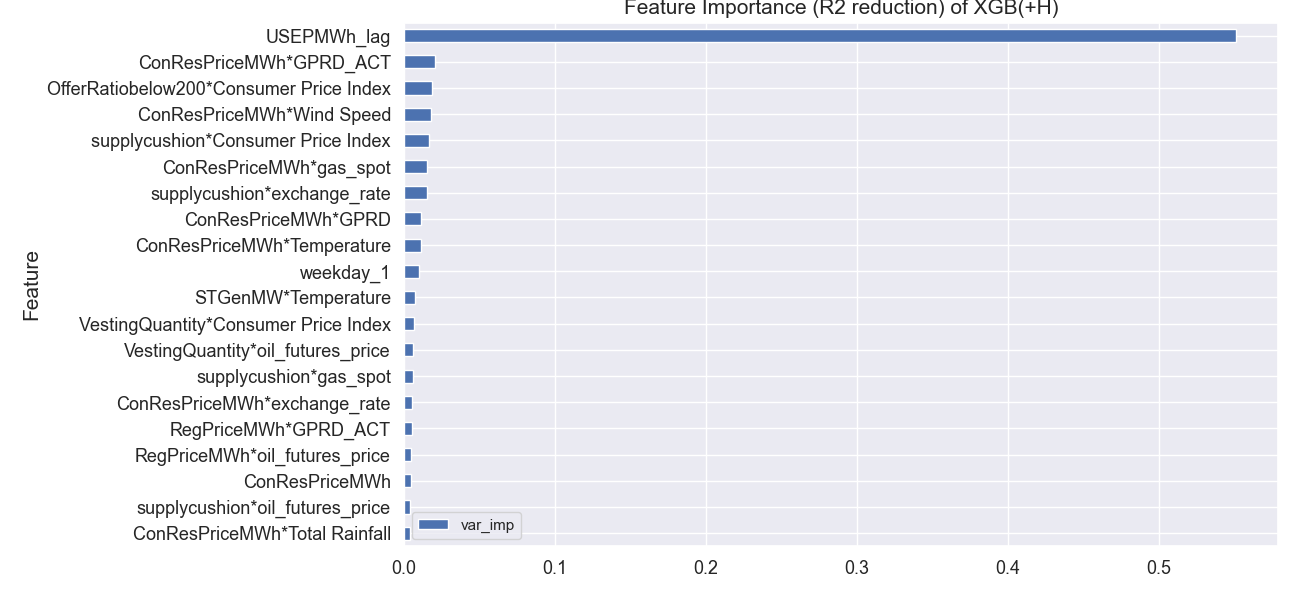}
\includegraphics[width=1\linewidth]{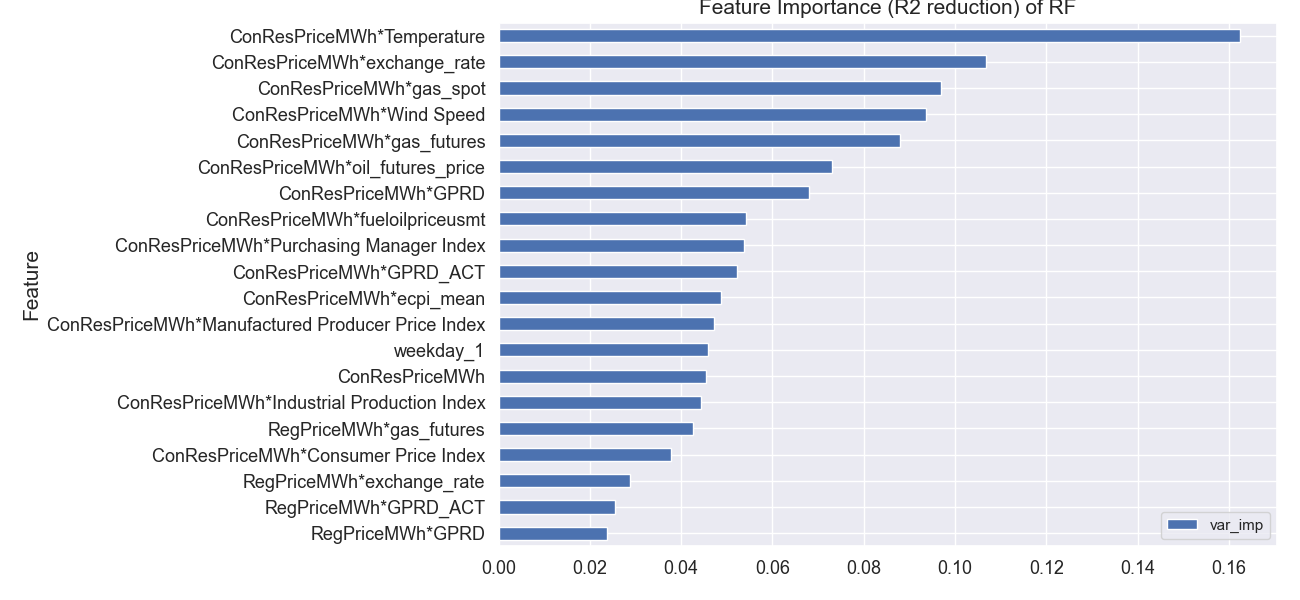}
\end{figure}
\begin{figure}[H]
\includegraphics[width=1\linewidth]{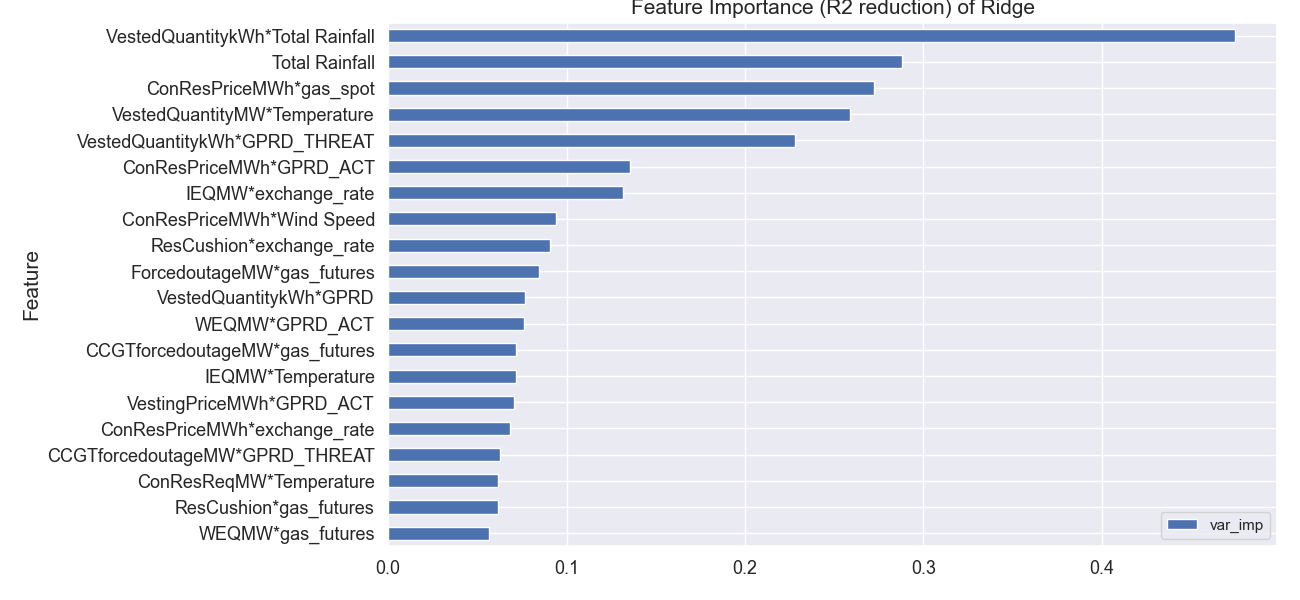}
\includegraphics[width=1\linewidth]{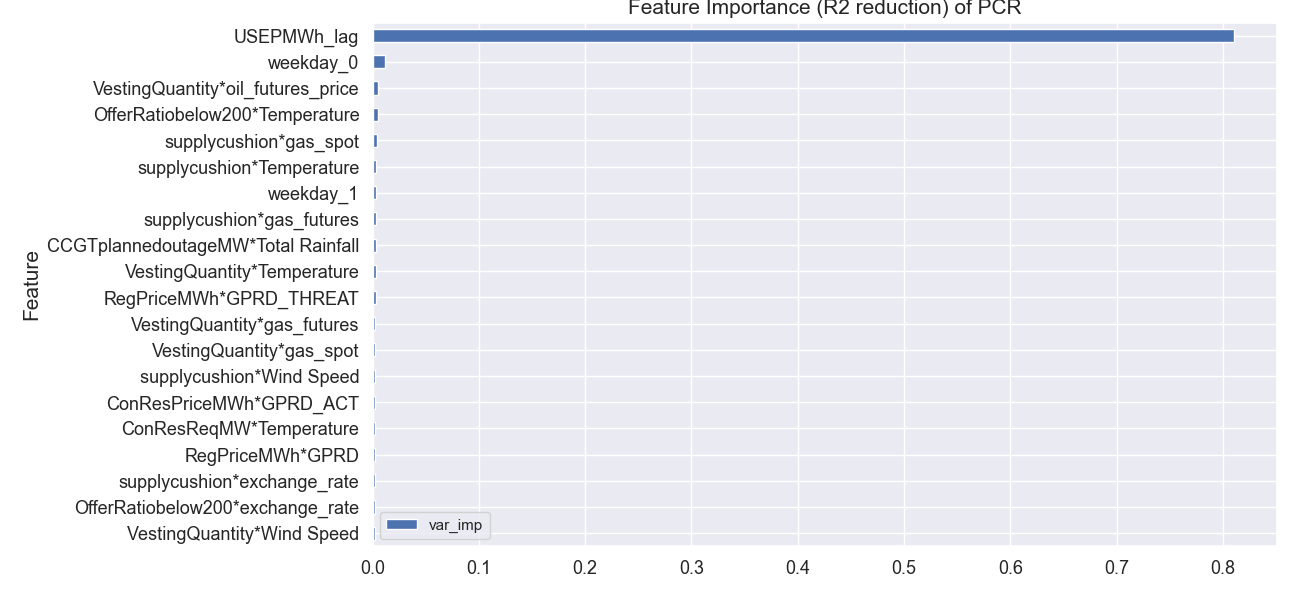}
\includegraphics[width=1\linewidth]{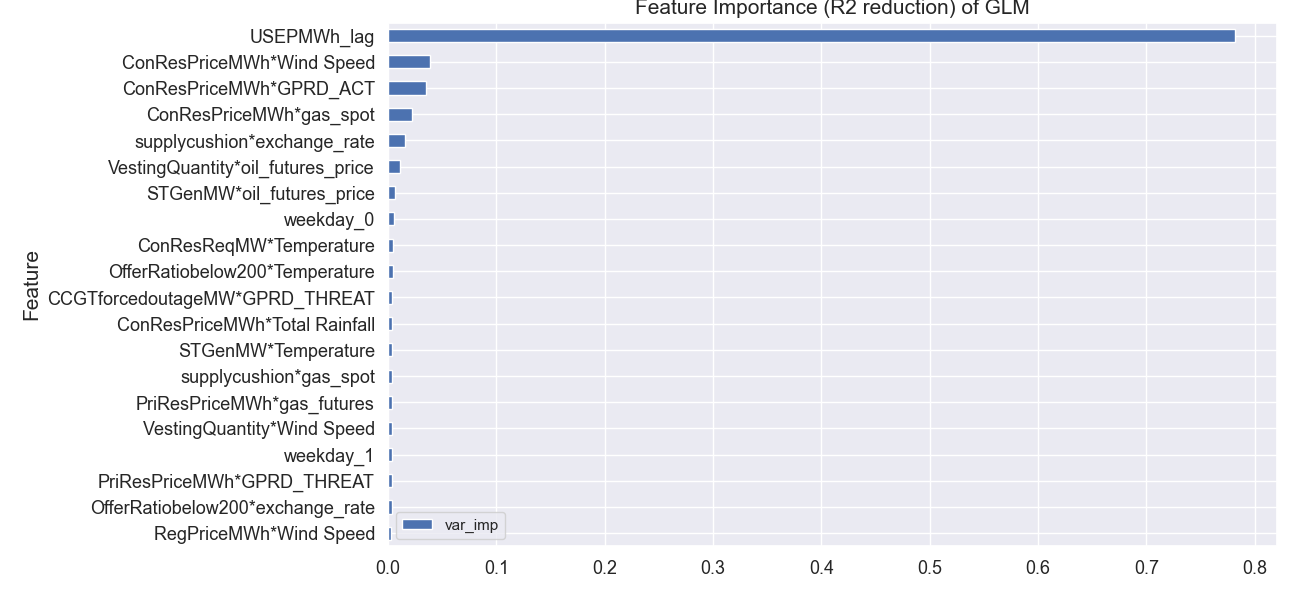}
\includegraphics[width=1\linewidth]{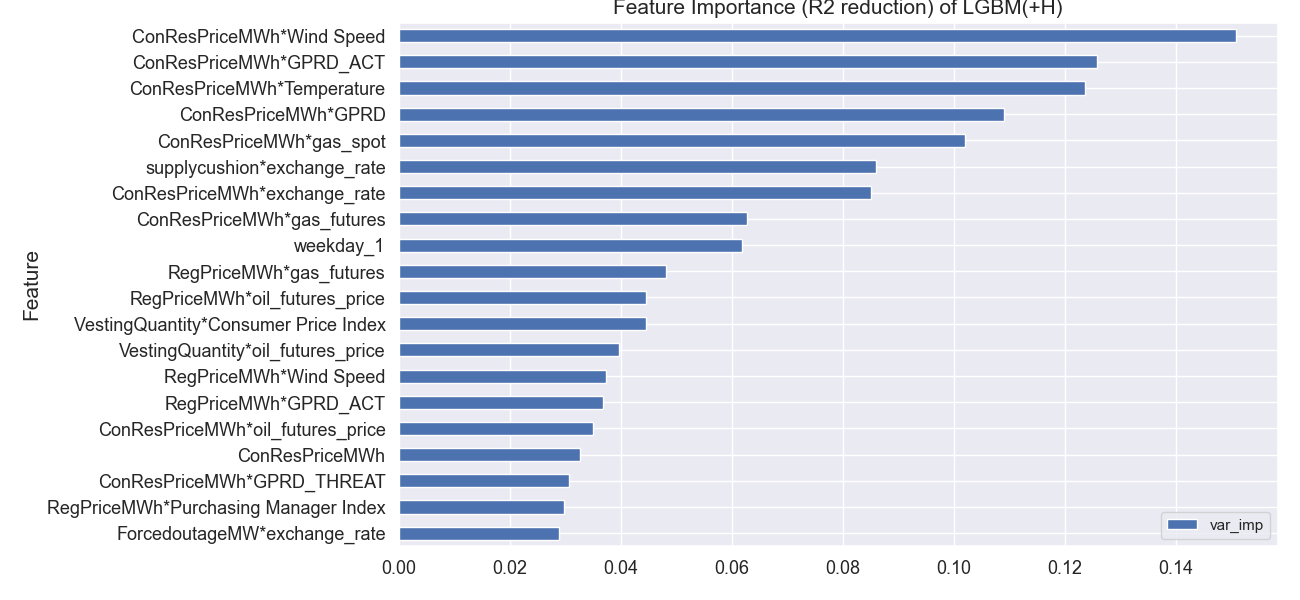}
\includegraphics[width=1\linewidth]{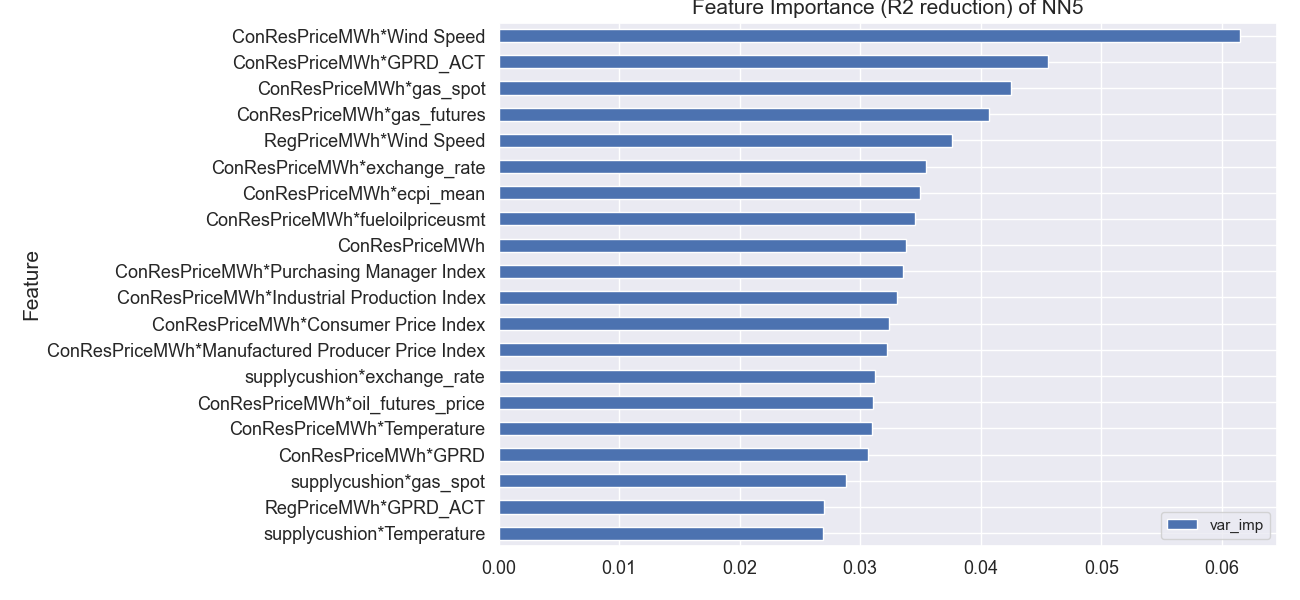}
\end{figure}
\end{multicols}
\label{fig:7}
\end{figure}

\begin{figure}[H]
\caption{\centering Feature Importance Based on Shapley Value Across All Models}
\caption*{\fontsize{10pt}{0.35cm}\selectfont Rankings of top-50 most influential variables in terms of overall model contribution. Characteristics are ordered based on the sum of their ranks over all models, with the most influential characteristics on the top and the least influential on the bottom. Columns correspond to the individual models,
and the color gradients within each column indicate the most influential (dark blue) to the least influential (white) variables.}
\includegraphics[width=1\linewidth]{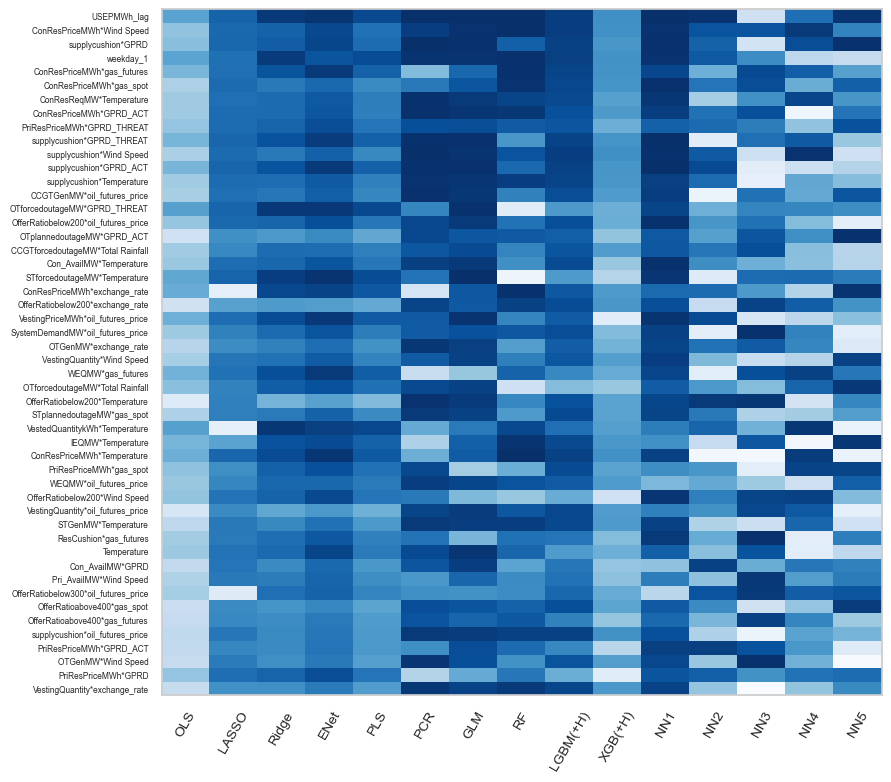}
\label{fig:8}
\end{figure}

\begin{figure}[H]
\centering
\caption{\centering Sub-group Feature Importance and $R^2_{OOS}$ Decomposition}
\caption*{\fontsize{10pt}{0.35cm}\selectfont Group-level feature importance for each scenario and each group. In the figure, $S$ stands for Supply group, $D$ for Demand group, $R$ for Regulation group, $DM$ for Domestic Macro group, $IM$ for International Macro group. The feature composition of each group can be referred to the \hyperref[Online Appendix]{Appendix}}
Panel A: Sub-group Feature Importance\\
\begin{minipage}{\linewidth}
    \centering
    \includegraphics[width=0.6\linewidth]{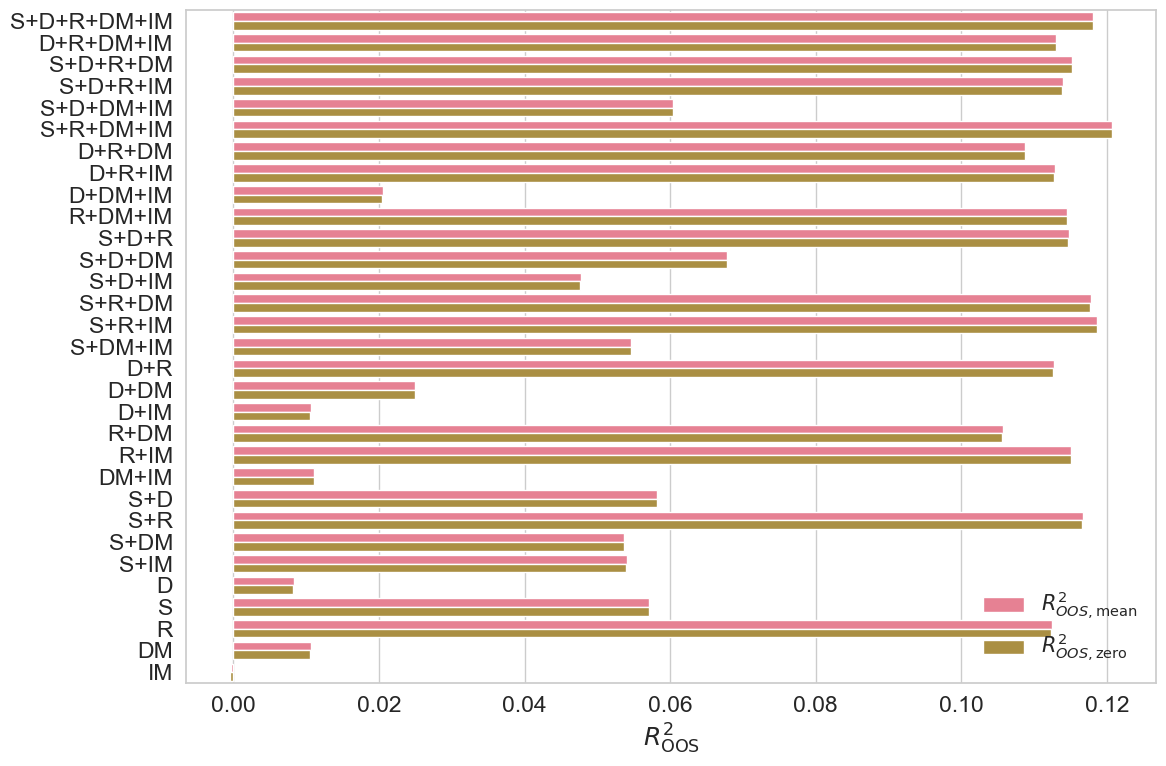}
\end{minipage}
\vspace{0.5em}
Panel B: Static $R^2_{OOS}$ Decomposition\\
\begin{minipage}{0.48\linewidth}
    \centering
    \includegraphics[width=\linewidth]{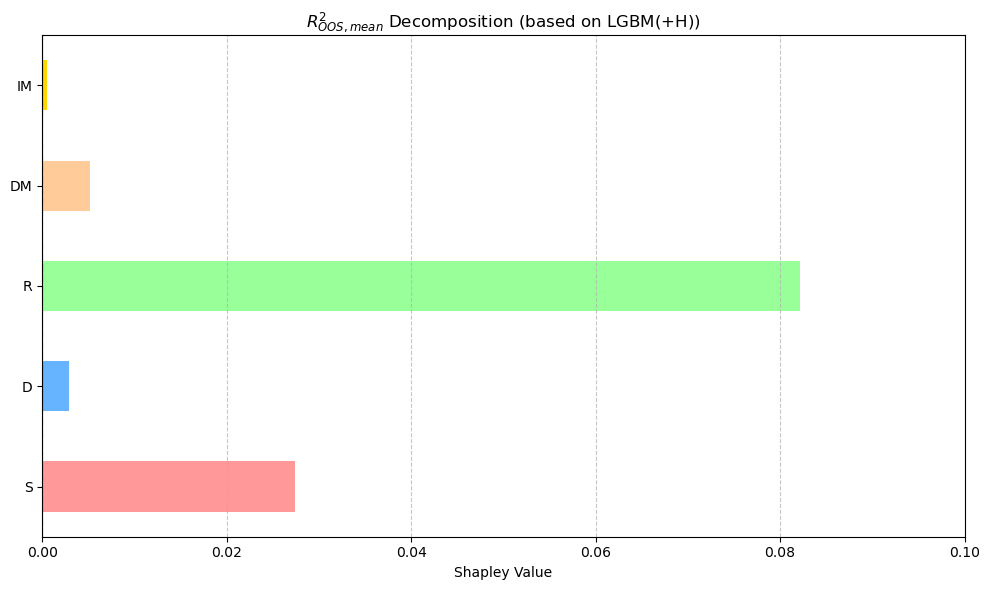}
\end{minipage}
\hfill
\begin{minipage}{0.48\linewidth}
    \centering
    \includegraphics[width=\linewidth]{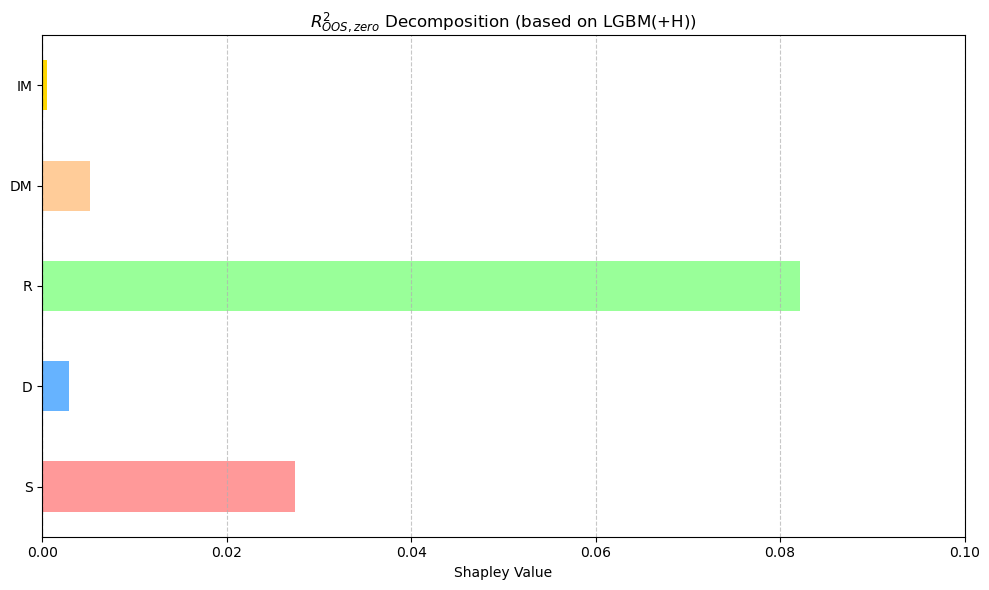}
\end{minipage}
\vspace{0.5em}
Panel C: Dynamic $R^2_{OOS}$ Decomposition\\
\begin{minipage}{\linewidth}
    \centering
    \includegraphics[width=0.6\linewidth]{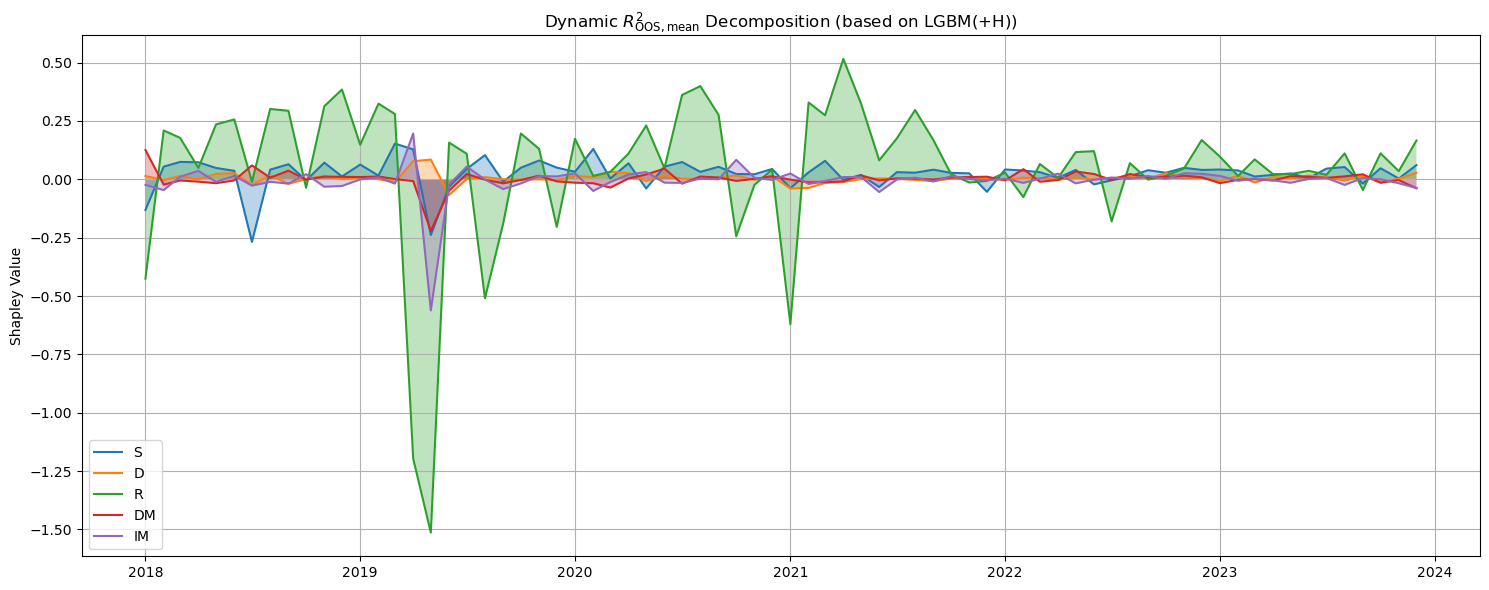}
    \includegraphics[width=0.6\linewidth]{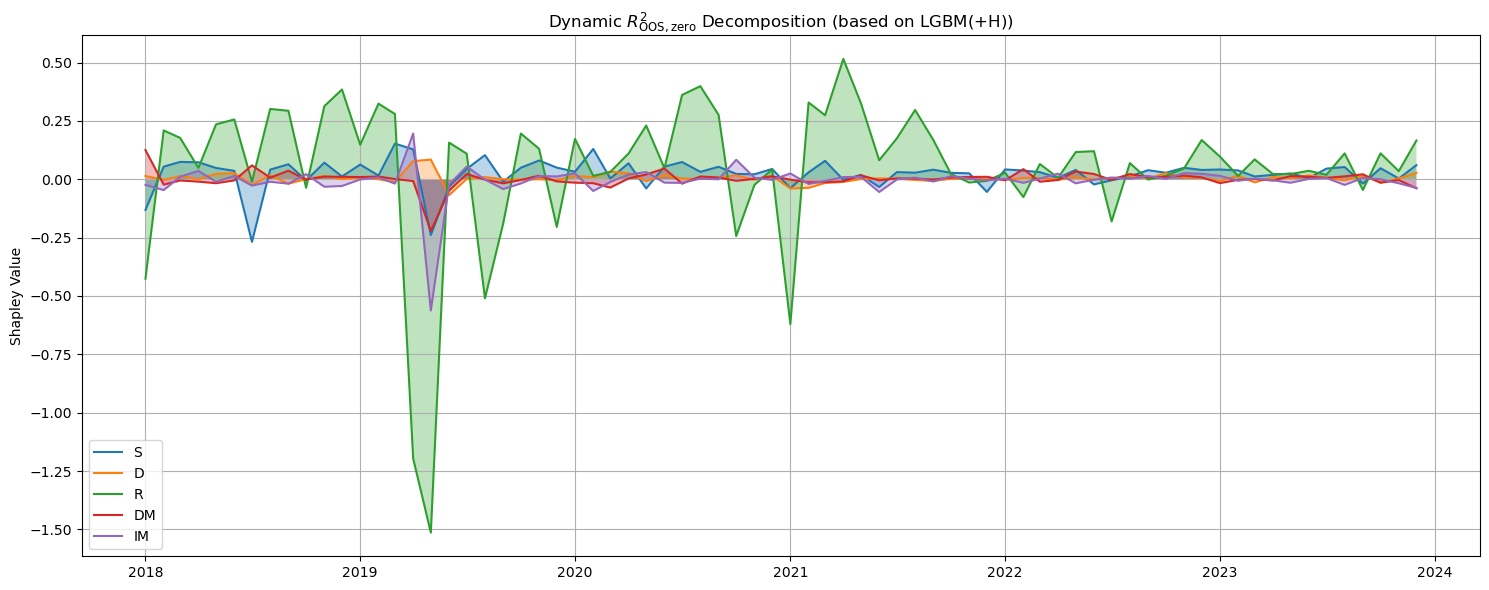}
\end{minipage}
\label{fig:9}
\end{figure}

\begin{figure}[H]
\caption{Marginal Associations}
\caption*{\fontsize{10pt}{0.35cm}\selectfont The panels show the sensitivity of expected daily electricity returns (vertical axis) to the individual characteristics (holding all other covariates fixed at their median values) across all models.}
\begin{multicols}{2}
\begin{figure}[H]
\includegraphics[width=1\linewidth]{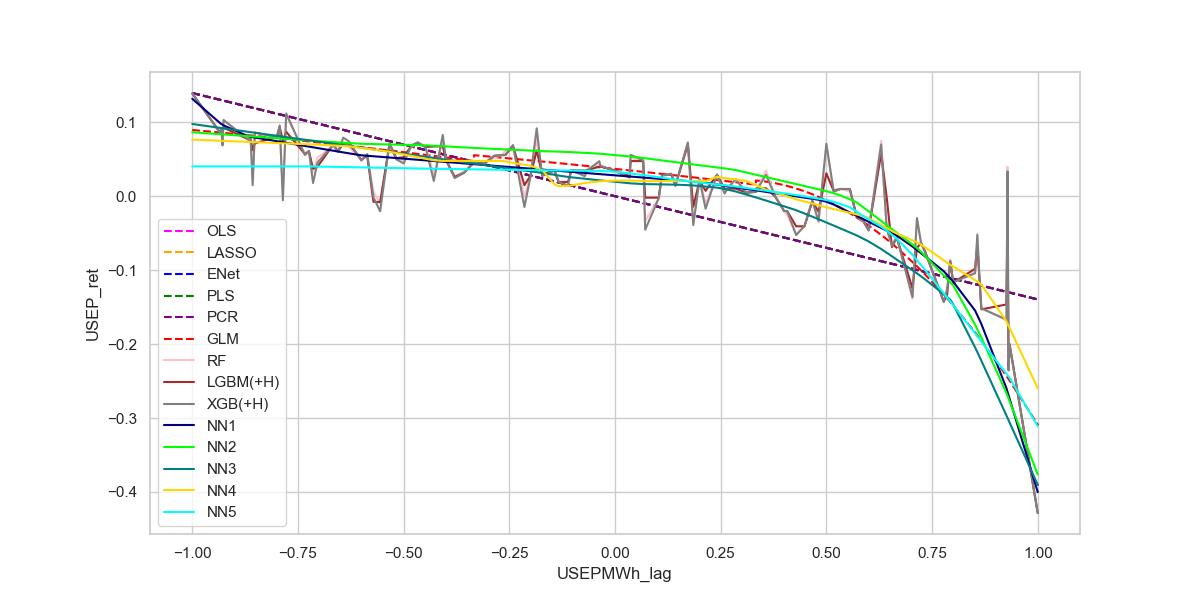}
\includegraphics[width=1\linewidth]{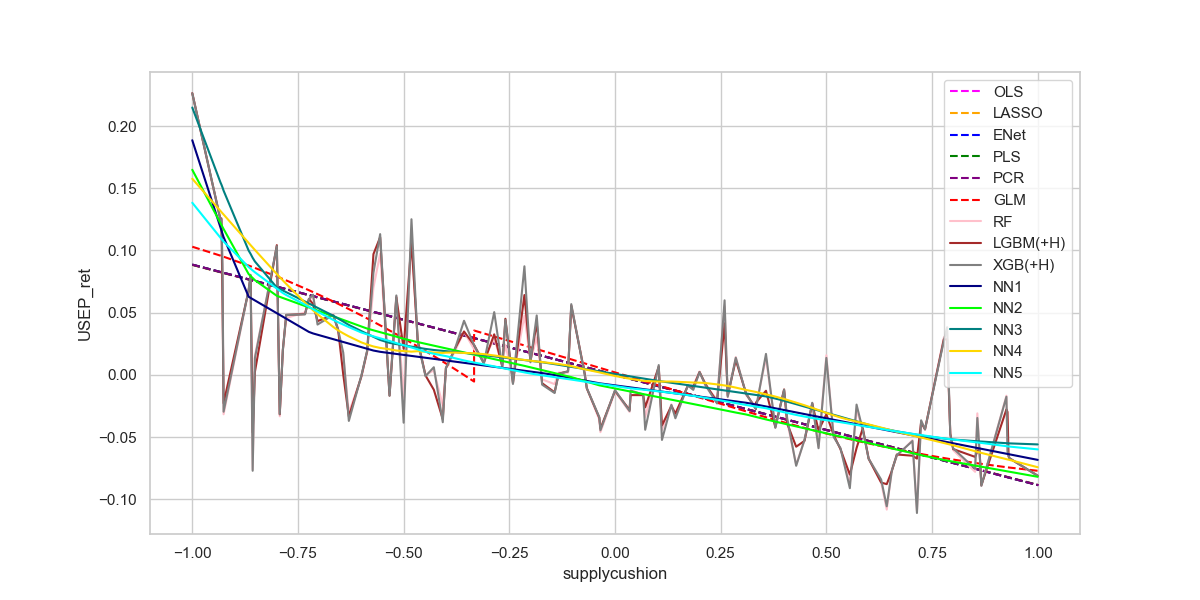}
\includegraphics[width=1\linewidth]{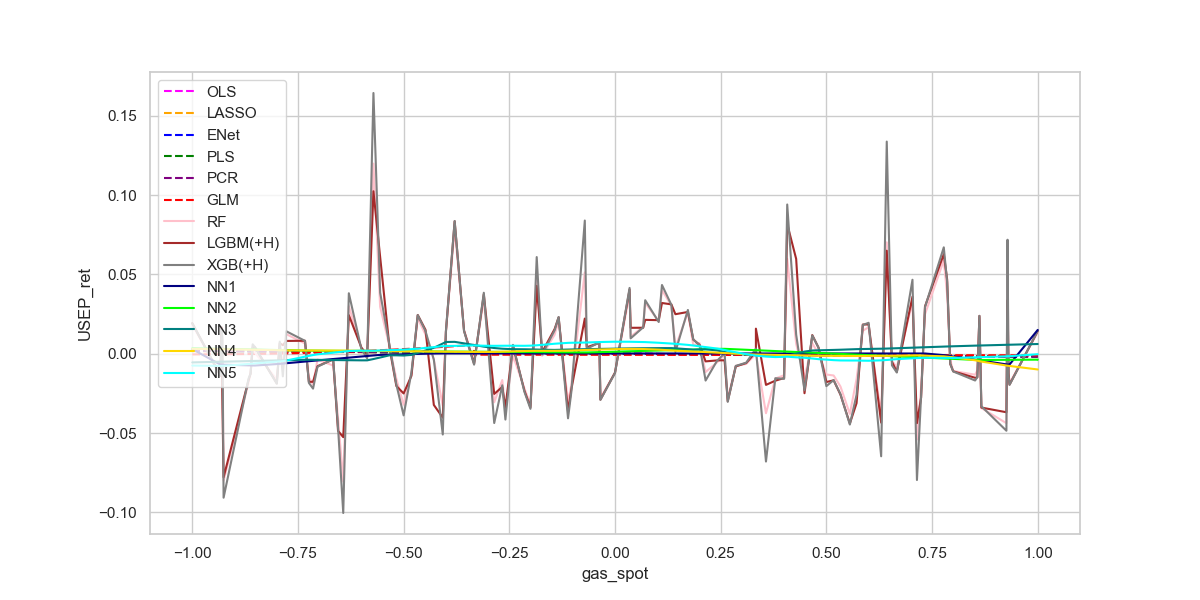}
\includegraphics[width=1\linewidth]{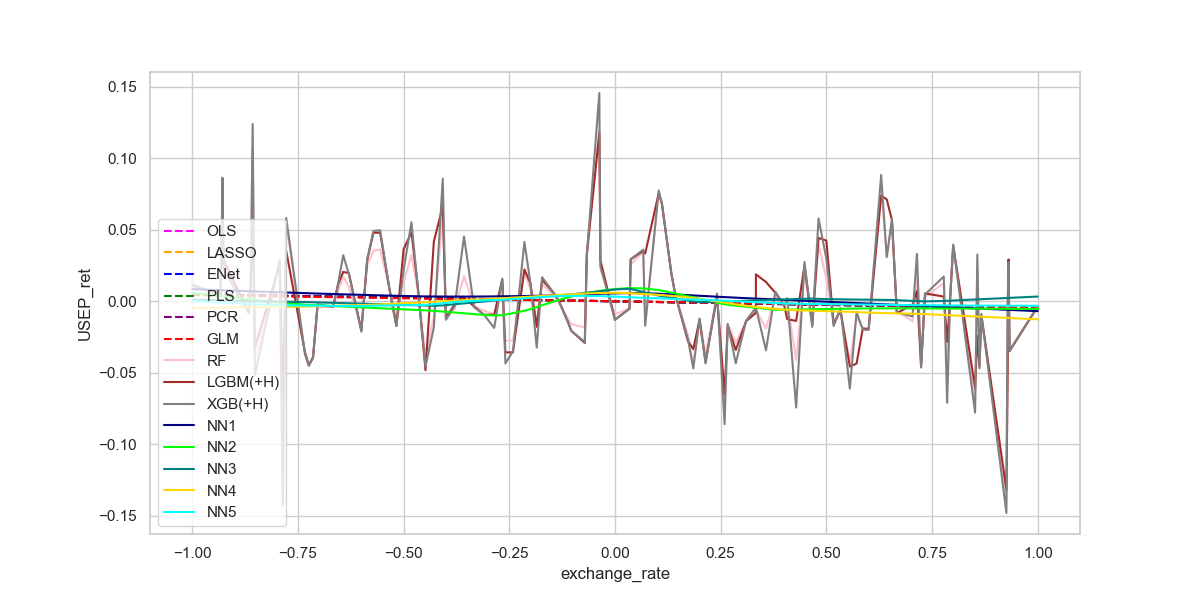}
\end{figure}

\begin{figure}[H]
\includegraphics[width=1\linewidth]{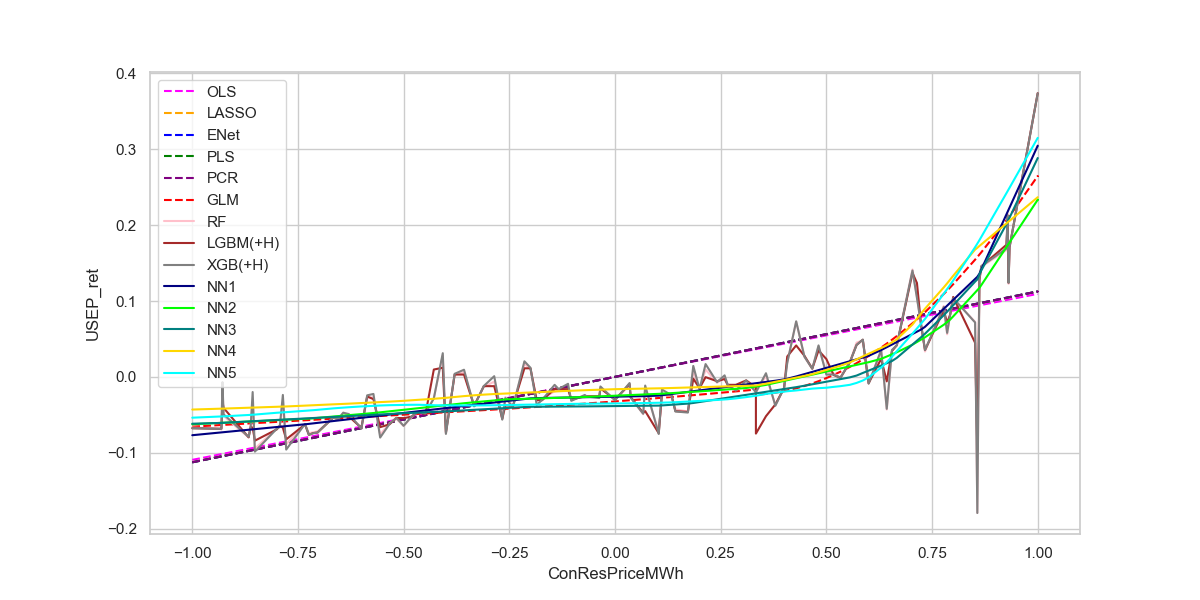}
\includegraphics[width=1\linewidth]{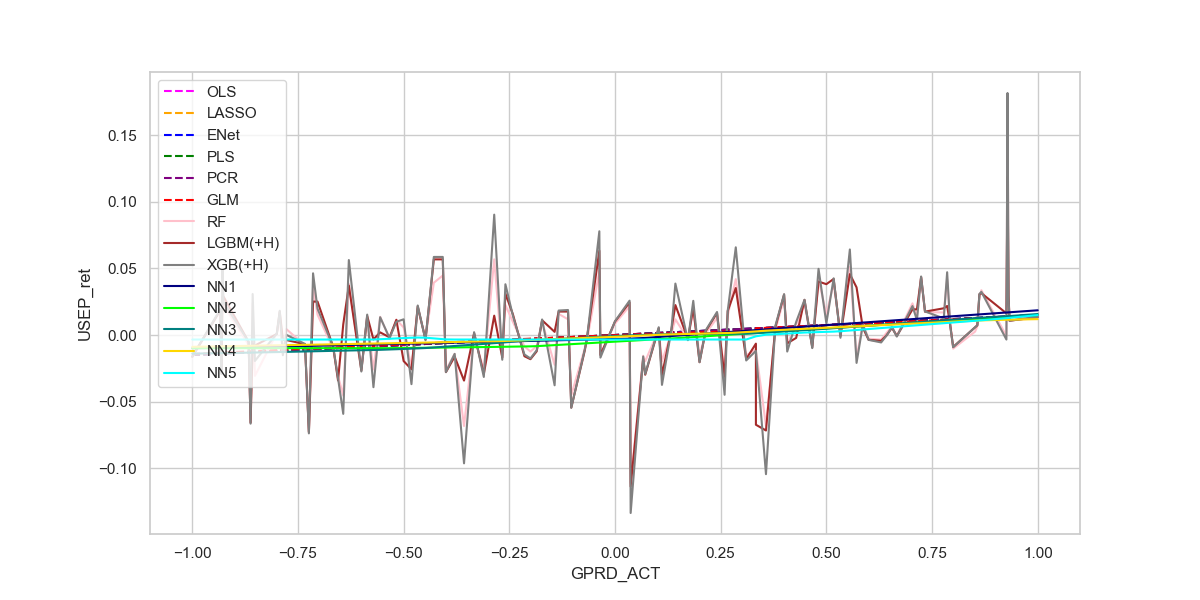}
\includegraphics[width=1\linewidth]{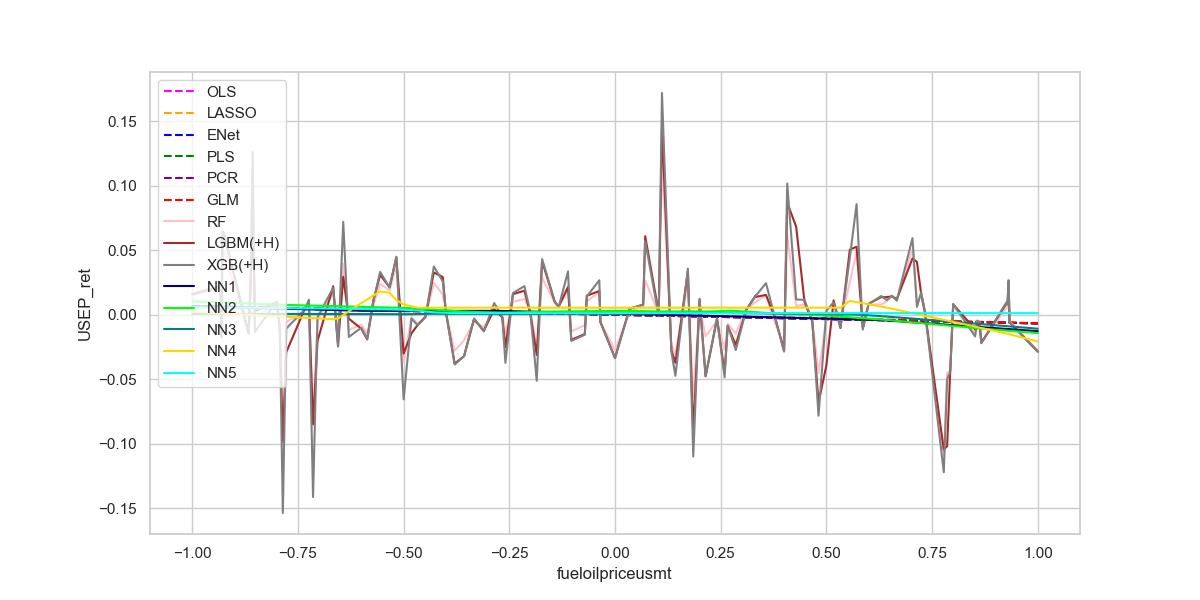}			
\end{figure}
\end{multicols}
\label{fig:10}
\end{figure}

\begin{figure}[H]
\caption{Utility under Mean-Variance Framework ($\gamma$=3)}
\caption*{\fontsize{10pt}{0.35cm}\selectfont The upper panel plots the cumulative utility of each individual model from Jan 2018 to Dec 2023 (72 OOS months). We highlight our identified best three individual models (GLM, XGB(+H) and LGBM(+H)) by soliding the lines while keeping other models' lines dash. The lower panel consists of two plots, the left one records the average utility while the right one depicts the total utility of each individual model from Jan 2018 to Dec 2023 (72 OOS months). For comparison convenience, we report the values on the top of each bar.}
\includegraphics[width=1\linewidth]{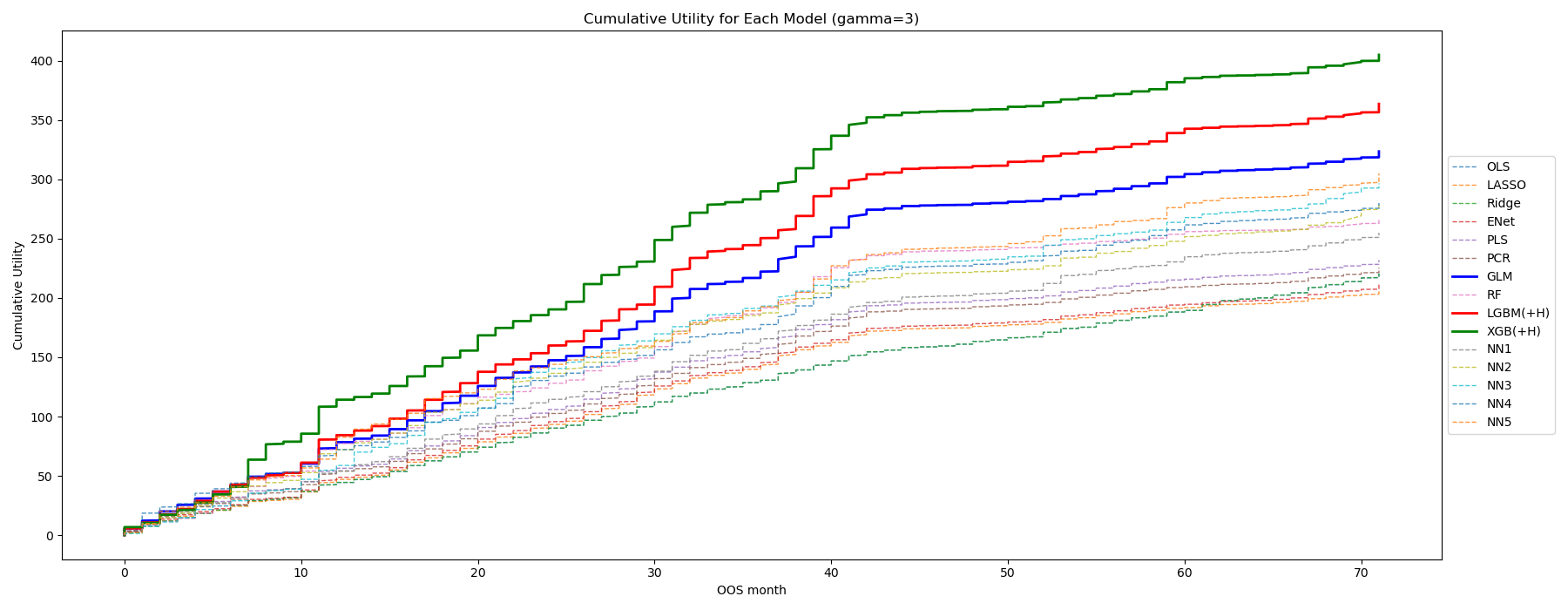}
\includegraphics[width=1\linewidth]{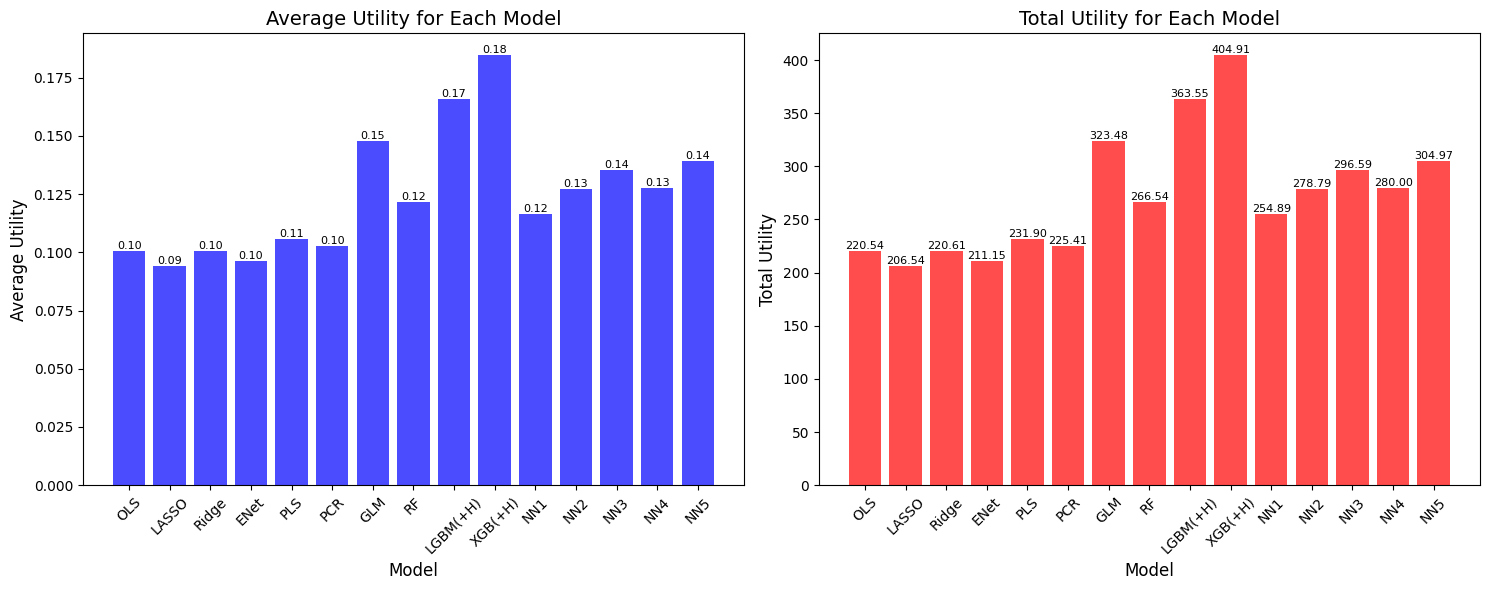}
\label{fig:11}
\end{figure}

\renewcommand{\thetable}{\arabic{table}}
\setcounter{table}{0}
\begin{table}[H]
\fontsize{9pt}{0.38cm}\selectfont
\centering
\caption{\centering Performance Evaluation Using \(R_{OOS}^2\) for All Out-of-Sample Months}
\label{tab:1}
\setlength{\tabcolsep}{7pt}
\renewcommand{\arraystretch}{1.1}
\begin{threeparttable}
\begin{tabular}{p{2.5cm} p{2.1cm} p{2.2cm} p{2.7cm} p{2.1cm}}
\hline	
&(1)&(2)&(3)&(4)\\	
Model &$R_{OOS, lagprice}^2$ &$R_{OOS, AR(1)}^2$& $R^{2}_{OOS, mean}$ & $R^2_{OOS, zero}$ \\
\hline
OLS&25.06***(***)&14.16***()&33.54***(***)&33.53***(***)\\
LASSO&23.63***(***)&12.52***()&35.17***(***)&35.17***(***)\\
Ridge&24.98***(***)&14.06***()&33.53***(***)&33.53***(***)\\
ENet&23.80***(***)&12.72***()&35.44***(***)&35.44***(***)\\
PCR&23.35***(***)&12.20***()&35.25***(***)&35.25***(***)\\
PLS&24.68***(***)&13.72***()&36.16***(***)&36.15***(***)\\
GLM&36.85***(***)&27.66***()&43.70***(***)&43.70***(***)\\
XGB(+H)&37.87***(***)&28.83***()&43.66***(***)&43.66***(***)\\
LGBM(+H)&37.40***(***)&28.29***()&43.26***(***)&43.26***(***)\\
RF&32.67***(***)&22.88***()&34.67***(***)&34.66***(***)\\
NN1&26.31***(***)&15.59***()&37.61***(***)&37.61***(***)\\
NN2&24.71***()&13.75***()&37.85***(***)&37.84***(***)\\
NN3&26.68***()&16.00***()&39.06***(***)&39.05***(***)\\
NN4&21.72***()&10.33***()&38.96***(***)&38.95***(***)\\
NN5&30.18***(***)&20.02***()&40.51***(***)&40.51***(***)\\
$Ensemble_{avg}$&32.14***(***)&22.27***()&42.50***(***)&42.50***(***)\\
$Ensemble_{op}$&33.08***(***)&23.35***()&43.10***(***)&43.09***(***)\\
$Ensemble_{wp}$&41.32***(***)&32.78***()&51.92***(***)&51.91***(***)\\
$Ensemble_{\theta=0.9}$&31.82***(***)&21.90***()&41.13***(***)&41.13***(***)\\
$Ensemble_{\theta=1}$&31.48***(***)&21.51***()&42.16***(***)&42.16***(***)\\
\hline
\end{tabular}
\begin{tablenotes}
\fontsize{8pt}{0.35cm}\selectfont
\item \textit{Notes}: All values are in \%. $Ensemble_{op}$ means ensemble model is without prediction correlation penalty to tune the weight, while $Ensemble_{wp}$ is the counterpart. Statistical significance for the \(R_{OOS}^2\) statistic is assessed using $p$-values from the CW OOS MSPE-adjusted statistic and t-statistics from the DM test, with Newey–West robust standard errors. These tests correspond to a one-sided hypothesis test, where the null hypothesis states that the forecast from our model has an expected square prediction error equal to that of the benchmark forecast, and the alternative hypothesis states that our forecast has a lower expected square prediction error than the benchmark. \(^{*}, ^{**},\) and \(^{***}\) indicate significance at the 10\%, 5\%, and 1\% levels for both the CW and DM tests, with CW significance levels presented outside the parentheses and DM within the parentheses.
\end{tablenotes}
\end{threeparttable}
\end{table}

\begin{landscape} 
\begin{table}[H]
\fontsize{8pt}{0.32cm}\selectfont
\centering
\caption{Model Performance Based on DM Test}
\label{tab:2}
\setlength{\tabcolsep}{3pt} 
\renewcommand{\arraystretch}{1} 
\begin{threeparttable}
\begin{tabular}{lcccccccccccccccccccc}
\hline
&(1)&(2)&(3)&(4)&(5)&(6)&(7)&(8)&(9)&(10)&(11)&(12)&(13)&(14)&(15)&(16)&(17)&(18)&(19)&(20)\\
& OLS & LASSO & Ridge & ENet & PCR & PLS & GLM 
& XGB(+H) & LGBM(+H) & RF & NN1 & NN2 & NN3 & NN4 & NN5
& $Ens_{avg}$ & $Ens_{op}$ & $Ens_{wp}$ & $Ens_{\theta=0.9}$ & $Ens_{\theta=1}$ \\ 
\hline
OLS  & 0 & 2.31** & -0.04 & 2.86*** & 2.44** & 4.24*** & 6.82*** & 4.67** & 4.69*** & 0.52 & 3.70*** & 2.85*** & 3.30*** & 3.29*** & 4.12*** & 7.39*** & 5.40*** & 8.58*** & 6.17*** & 7.18*** \\  
LASSO & -2.31** & 0 & -2.32** & 2.56** & 0.29 & 4.07*** & 5.47*** & 3.82*** & 3.82*** & -0.24 & 2.13** & 1.71* & 2.20** & 2.30** & 2.99*** & 5.72*** & 4.20*** & 7.25*** & 4.49*** & 5.60*** \\  
Ridge & 0.04 & 2.32** & 0 & 2.87*** & 2.45** & 4.26*** & 6.82*** & 4.67*** & 4.70*** & 0.52 & 3.71*** & 2.85*** & 3.30*** & 3.29*** & 4.12*** & 7.41*** & 5.41*** & 8.59*** & 6.19*** & 7.20*** \\  
ENet  & -2.86*** & -2.56** & -2.87*** & 0 & -0.79 & 3.20*** & 5.40*** & 3.73*** & 3.72*** & -0.37 & 1.94* & 1.57 & 2.08** & 2.17** & 2.87*** & 5.60*** & 4.09*** & 7.24*** & 4.32*** & 5.46*** \\  
PCR   & -2.44** & -0.29 & -2.45** & 0.79 & 0 & 4.26*** & 5.67*** & 3.92*** & 3.93*** & -0.29 & 2.23** & 1.76* & 2.29** & 2.37** & 3.07*** & 6.18*** & 4.31*** & 7.49*** & 4.72*** & 6.00*** \\  
PLS   & -4.24*** & -4.07*** & -4.26*** & -3.20*** & -4.26*** & 0 & 5.13*** & 3.50*** & 3.48*** & -0.72 & 1.41 & 1.15 & 1.74* & 1.81* & 2.60*** & 5.22*** & 3.85*** & 7.18*** & 3.86*** & 5.09*** \\  
GLM   & -6.82*** & -5.47*** & -6.82*** & -5.40*** & -5.67*** & -5.13*** & 0 & -0.04 & -0.44 & -7.15*** & -5.60*** & -5.57*** & -4.02*** & -3.56*** & -2.70*** & -3.99*** & -0.92 & 6.87*** & -4.42*** & -3.67*** \\  
XGB(+H) & -4.67*** & -3.82*** & -4.67*** & -3.73*** & -3.92***  & -3.50*** & 0.04 & 0  & -0.70 & -8.46*** & -3.66*** & -3.90*** & -2.97*** & -2.79*** & -2.14** & -2.11**  & -0.83   &	7.37***	& -2.71***&	-2.02**  \\
LGBM(+H) & -4.69*** & -3.82*** & -4.70*** & -3.72*** & -3.93***  & -3.48*** & 0.44 & 0.70 & 0 & -8.66*** & -3.57*** & -3.90*** & -2.91*** & -2.60*** & -1.90*  & -1.96** & -0.47  & 7.83*** & -2.60***& -1.86*   \\
RF & -0.52 & 0.24 & -0.52 & 0.37 & 0.29  & 0.72 & 7.15***  & 8.46*** & 8.66*** & 0 & 1.73* & 2.12** & 2.70*** & 2.51**  & 3.33*** & 4.49*** & 6.20***& 10.10***& 3.62*** & 4.79*** \\
NN1 &  -3.70*** &	-2.13**	& -3.71*** & -1.94*	& -2.23**  & -1.41 & 5.60*** & 3.66*** & 3.57*** & -1.73* & 0 & 0.22 & 1.22 & 1.20 & 2.58*** & 4.97*** & 4.15***& 8.57*** & 2.83*** &	5.00***  \\
NN2 & -2.85*** & -1.71* & -2.85*** & -1.57 & -1.76*  & -1.15 & 5.57*** & 3.90*** 	& 3.90*** & -2.12** & -0.22 & 0 & 1.27 & 0.84 & 2.28**  & 3.30*** & 4.17***&	8.96*** & 1.77*	  & 3.56***  \\
NN3	& -3.30*** & -2.20** & -3.30*** & -2.08** & -2.29** & -1.74* & 4.02*** & 2.97*** & 2.91*** & -2.70*** & -1.22 & -1.27 & 0 & -0.07 & 1.26    & 1.63    & 2.83***&	8.70*** & 0.56    &	1.79*    \\
NN4	& -3.29*** & -2.30** & -3.29*** & -2.17** & -2.37**  & -1.81* & 3.56*** &	2.79*** & 2.60*** & -2.51** & -1.20 & -0.84 & 0.07 & 0  & 1.13    & 1.64    & 2.76***& 7.86*** & 0.67    & 1.83*    \\
NN5	& -4.12*** & -2.99*** & -4.12*** & -2.87***	& -3.07*** & -2.60*** & 2.70*** & 2.14** & 1.90* & -3.33*** & -2.58*** & -2.28** & -1.26 & -1.13  & 0       & 0.26    & 1.86*  &	8.84*** & -0.60   &	0.42     \\
$Ens_{avg}$ & -7.39*** & -5.72*** & -7.41*** & -5.60*** & -6.18*** & -5.22*** & 3.99*** & -2.11** & -1.96** & -4.49*** & -4.97*** & -3.30*** & -1.63 & -1.64 & 0.26 & 0 & 1.88* & 7.58*** & -2.32** & 0.94 \\  
$Ens_{op}$ & -5.40*** & -4.20*** & -5.41*** & -4.09*** & -4.31*** & -3.85*** & 0.92 & -0.83 & -0.47 & -6.20*** & -4.15*** & -4.17*** & -2.83*** & -2.76*** & 1.86* & 1.88* & 0 & 7.79*** & -2.87*** & -1.92* \\  
$Ens_{wp}$ & -8.58*** & -7.25*** & -8.59*** & -7.24*** & -7.49*** & -7.18*** & -6.87*** & -7.37*** & -7.83*** & -10.10*** & -8.57*** & -8.96*** & -8.70*** & -7.86*** & -8.84*** & -7.58*** & -7.79*** & 0 & -7.72*** & -7.68*** \\  
$Ens_{\theta=0.9}$ & -6.17*** & -4.49*** & -6.19*** & -4.32*** & -4.72*** & -3.86*** & 4.42*** & 2.71*** & 2.60*** & -3.62*** & -2.83*** & -1.77* & -0.56 & -0.67 & 0.60 & 2.32** & 2.87*** & 7.72*** & 0 & 2.85*** \\  
$Ens_{\theta=1}$ & -7.18*** & -5.60*** & -7.20*** & -5.46*** & -6.00*** & -5.09*** &3.67*** & 2.02** & 1.86* & -4.79*** &	-5.00*** &	-3.56*** & -1.79* &	-1.83* & -0.42 & -0.94  & 1.92*   &	7.68*** & -2.85*** & 0 \\
\hline
\end{tabular}
\begin{tablenotes}
\fontsize{8pt}{0.3cm}\selectfont
\item \textit{Notes}: This table reports pairwise DM test statistics (with Newey–West robust standard errors) comparing the OOS performance among thirteen models. The numbers represent t-statistics. Positive numbers indicate the column model outperforms the row model. Significance levels: * 10\%, ** 5\%, *** 1\%.
\end{tablenotes}
\end{threeparttable}
\end{table}
\end{landscape} 

\begin{table}[H]
\fontsize{9pt}{0.38cm}\selectfont
\centering
\caption{\centering Weights of Individual Model in Ensemble Models}
\label{tab:3}
\setlength{\tabcolsep}{7pt}
\renewcommand{\arraystretch}{1.1}
\begin{threeparttable}
\begin{tabular}{p{2cm} p{2cm} p{2cm} p{2cm}}
\hline
&(1)&(2)&(3)\\
& $Ensemble_{\theta=1}$ & $Ensemble_{op}$ & $Ensemble_{wp}$ \\
\hline
OLS & 0.054922 & 0.050483 & 0.039347 \\
LASSO & 0.065417 & 0.036767 & 0.025442 \\
Ridge  & 0.054950 & 0.051258 & 0.040083 \\
ENet & 0.064924 & 0.024015 & 0.023320 \\
PCR & 0.064199 & 0.023866 & 0.031719 \\
PLS & 0.064069 & 0.026571 & 0.023683 \\
GLM & 0.078487 & 0.109413 & 0.090686 \\
XGB(+H) & 0.079120 & 0.182078 & 0.167393 \\
LGBM(+H) & 0.079459 & 0.195527 & 0.157618 \\
RF & 0.071153 & 0.100975 & 0.084873 \\
NN1 & 0.065270 & 0.017775 & 0.041319 \\
NN2 & 0.066505 & 0.017392 & 0.021808 \\
NN3 & 0.064354 & 0.072036 & 0.088524 \\
NN4 & 0.063408 & 0.025868 & 0.060606 \\
NN5 & 0.063762 & 0.065976 & 0.103580 \\
\hline
\end{tabular}
\begin{tablenotes}
\fontsize{8pt}{0.35cm}\selectfont
\item \textit{Notes}: $Ensemble_{op}$ means ensemble model is without prediction correlation penalty to tune the weight, while $Ensemble_{wp}$ is the counterpart. The weights of individual model in ensemble models are averaged across 72 OOS months.
\end{tablenotes}
\end{threeparttable}
\end{table}

\begin{table}[H]
\fontsize{9pt}{0.38cm}\selectfont
\centering
\caption{\centering Performance Evaluation Using $RMSE$ and $rRMSE$}
\label{tab:4}
\setlength{\tabcolsep}{7pt}
\renewcommand{\arraystretch}{1.1}
\begin{threeparttable}
\begin{tabular}{p{6cm} p{4cm} p{4cm}}
\hline
&(1)&(2)\\
Model & $RMSE$ & $rRMSE$ \\
\hline
OLS & 0.2436 & 1.0000 \\
LASSO & 0.2406 & 0.9876(**) \\
Ridge & 0.2436 & 1.0000 \\
ENet & 0.2401 & 0.9855(***) \\
PCR & 0.2404 & 0.9870(***) \\
PLS & 0.2387 & 0.9801(***) \\
GLM & 0.2242 & 0.9204(***) \\
RF & 0.2415 & 0.9914 \\
XGB(+H) & 0.2243 & 0.9207(***) \\
LGBM(+H) & 0.2250 & 0.9239(***) \\
NN1 & 0.2360 & 0.9689(***) \\
NN2 & 0.2355 & 0.9670(***) \\
NN3 & 0.2332 & 0.9576(***) \\
NN4 & 0.2334 & 0.9583(***) \\
NN5 & 0.2304 & 0.9461(***) \\
$Ensemble_{avg}$ & 0.2299 & 0.9439***(***) \\
$Ensemble_{op}$ & 0.2260 & 0.9280(***) \\
$Ensemble_{wp}$ & 0.2070 & 0.8500***(***) \\
$Ensemble_{\theta=0.9}$ & 0.2319 & 0.9521(***) \\
$Ensemble_{\theta=1}$ & 0.2296 & 0.9424***(***) \\
\hline
\end{tabular}
\begin{tablenotes}
\fontsize{8pt}{0.35cm}\selectfont
\item \textit{Notes}: $Ensemble_{op}$ means ensemble model is without prediction correlation penalty to tune the weight, while $Ensemble_{wp}$ is the counterpart. Statistical significance for the \(R_{OOS}^2\) statistic is assessed using $p$-values from the CW OOS MSPE-adjusted statistic and t-statistics from the DM test, with Newey–West robust standard errors. These tests correspond to a one-sided hypothesis test, where the null hypothesis states that the forecast from our model has an expected square prediction error equal to that of the benchmark forecast, and the alternative hypothesis states that our forecast has a lower expected square prediction error than the benchmark. \(^{*}, ^{**},\) and \(^{***}\) indicate significance at the 10\%, 5\%, and 1\% levels for both the CW and DM tests, with CW significance levels presented outside the parentheses and DM within the parentheses.
\end{tablenotes}
\end{threeparttable}
\end{table}

\begin{table}[H]
\fontsize{9pt}{0.38cm}\selectfont
\centering
\caption{\centering Model Complexity and Out-of-Sample Performance}
\label{tab:5}
\setlength{\tabcolsep}{7pt}
\renewcommand{\arraystretch}{1.1}
\begin{threeparttable}
\begin{tabular}{p{5cm} p{2.3cm} p{2.3cm} p{2.3cm} p{2.3cm}}
\hline
& $R^2_{OOS, lagprice}$ & $R^2_{OOS, AR(1)}$ & $R_{OOS, mean}^2$ & $R^2_{OOS, zero}$ \\
\hline
$Model\_Complexity$ & 0.0134919 & 0.0104735 & 0.0147623** & 0.0147617**  \\
& (0.0115421) & (0.0079751) & (0.0051509) & (0.005153)\\
\hline
Model FEs & Yes & Yes & Yes & Yes \\
OOS month FEs & Yes & Yes & Yes & Yes \\
N & 576 & 576 & 576 & 576 \\
Adjusted $R^2$ & 0.4984 & 0.4446 & 0.6037 & 0.6034 \\
\hline
\end{tabular}
\begin{tablenotes}
\fontsize{8pt}{0.35cm}\selectfont
\item \textit{Notes}: Standard errors are clustered at the model level. Statistical significance levels are denoted as follows: \(^{***} p < 0.01\), \(^{**} p < 0.05\), \(^{*} p < 0.1\).
\end{tablenotes}
\end{threeparttable}
\end{table}

\begin{ThreePartTable} 
\begin{TableNotes} 
\item \textit{Notes}: $Ensemble_{op}$ means ensemble model is without prediction correlation penalty to tune the weight, while $Ensemble_{wp}$ is the counterpart. $Predictability_{wE}$ is the predictability of state $s$ considering ensemble models, and $Predictability_{oE}$ opts out the ensemble models. Statistical significance for the \(R_{OOS}^2\) statistic is assessed using p-values from the CW OOS MSPE-adjusted statistic and t-statistics from the DM test, with Newey–West robust standard errors. These tests correspond to a one-sided hypothesis test, where the null hypothesis states that the forecast from our model has an expected square prediction error equal to that of the benchmark forecast, and the alternative hypothesis states that our forecast has a lower expected square prediction error than the benchmark. \(^{*}, ^{**},\) and \(^{***}\) indicate significance at the 10\%, 5\%, and 1\% levels for both the DM, CW tests and t-test, with CW significance levels presented outside the parentheses, DM within the parentheses and t-test presented in the last row of each panel.
\end{TableNotes}

\fontsize{8pt}{0.38cm}\selectfont
\centering
\begin{longtable}[H]{p{5.5cm} p{0.02cm} p{1.5cm} p{0.01cm} p{1.5cm} p{0.02cm} p{1.5cm} p{0.01cm} p{1.5cm}}
\label{tab:6} \\
\caption{\centering Heterogeneous Predictability of Macro States} \\
\\
\toprule
&&\multicolumn{3}{c}{$R^2_{OOS,mean}$}&&\multicolumn{3}{c}{$R^2_{OOS,zero}$} \\
\cmidrule{3-5} \cmidrule{7-9}
\endhead

\midrule
\multicolumn{9}{r}{\textit{Continued on next page}}
\endfoot

\bottomrule 
\insertTableNotes \\ 
\endlastfoot

\textit{Panel A: Bearish V.S. Bullish} &&  Bearish &&  Bullish &&  Bearish && Bullish \\
\cmidrule{3-5} \cmidrule{7-9}
    OLS   &       & 0.32077***(***) &       & 0.40442***(***) &       & 0.31908***(***) &       & 0.40577***(***) \\
    LASSO &       & 0.35281***(***) &       & 0.37944***(***) &       & 0.3512***(***) &       & 0.38084***(***) \\
    Ridge &       & 0.32055***(***) &       & 0.4047***(***) &       & 0.31886***(***) &       & 0.40605***(***) \\
    ENet  &       & 0.35096***(***) &       & 0.38754***(***) &       & 0.34935***(***) &       & 0.38893***(***) \\
    GLM   &       & \textbf{0.4721***(***)} &       & \textbf{0.43316***(***)} &       & \textbf{0.47078***(***)} &       & \textbf{0.43445***(***)} \\
    PLS   &       & 0.35435***(***) &       & 0.40041***(***) &       & 0.35275***(***) &       & 0.40177***(***) \\
    PCR   &       & 0.34485***(***) &       & 0.39207***(***) &       & 0.34323***(***) &       & 0.39345***(***) \\
    XGB(+H) &       & \textbf{0.47618***(***)} &       & \textbf{0.43565***(***)} &       & \textbf{0.47488***(***)} &       & \textbf{0.43692***(***)} \\
    LGBM(+H) &       & \textbf{0.47315***(***)} &       & \textbf{0.42624***(***)} &       & \textbf{0.47184***(***)} &       & \textbf{0.42754***(***)} \\
    RF    &       & 0.37839***(***) &       & 0.3433***(***) &       & 0.37684***(***) &       & 0.34479***(***) \\
    NN1   &       & 0.34411***(***) &       & 0.43707***(***) &       & 0.34248***(***) &       & 0.43834***(***) \\
    NN2   &       & 0.33454***(***) &       & 0.45807***(***) &       & 0.33289***(***) &       & 0.45929***(***) \\
    NN3   &       & 0.34112***(***) &       & 0.46972***(***) &       & 0.33948***(***) &       & 0.47092***(***) \\
    NN4   &       & 0.35374***(***) &       & 0.45243***(***) &       & 0.35214***(***) &       & 0.45367***(***) \\
    NN5   &       & 0.36008***(***) &       & 0.47338***(***) &       & 0.35849***(***) &       & 0.47457***(***) \\
    $Ensemble_{avg}$ &       & 0.39509***(***) &       & 0.4413***(***) &       & 0.39358***(***) &       & 0.44256***(***) \\
    $Ensemble_{\theta=0.9}$ &       & 0.38692***(***) &       & 0.42843***(***) &       & 0.3854***(***) &       & 0.42973***(***) \\
    $Ensemble_{\theta=1}$ &       & 0.39524***(***) &       & 0.44414***(***) &       & 0.39374***(***) &       & 0.44539***(***) \\
    $Ensemble_{op}$ &       & 0.43527***(***) &       & 0.448***(***) &       & 0.43387***(***) &       & 0.44925***(***) \\
    $Ensemble_{wp}$ &       & 0.50117***(***) &       & 0.55633***(***) &       & 0.49993***(***) &       & 0.55734***(***) \\
\toprule
Mean(Bearish)-Mean(Bullish)& & & &-0.04622**& & & &-0.04904** \\
    $Predictability_{wE}$&       & 0.50117         &       & 0.55633         &       & 0.49993          &       & 0.55734 \\
    $Predictability_{oE}$&       & 0.47618         &       & 0.47338         &       & 0.47488          &       & 0.47457 \\
\toprule

\textit{Panel B: Volatile V.S. Tranquil}&&Volatile&&Tranquil&&Volatile&&Tranquil\\
\cmidrule{3-5} \cmidrule{7-9}
    OLS   &       & 0.34083***(***) &       & -0.49984() &       & 0.3408***(***) &       & -0.49973() \\
    LASSO &       & 0.3375***(***) &       & 0.37084**() &       & 0.33746***(***) &       & 0.37089**() \\
    Ridge &       & 0.34076***(***) &       & -0.49163() &       & 0.34073***(***) &       & -0.49152() \\
    ENet  &       & 0.34051***(***) &       & 0.36587**() &       & 0.34047***(***) &       & 0.36592**() \\
    GLM   &       & 0.41267***(***) &       & 0.38412***() &       & 0.41264***(***) &       & 0.38417***() \\
    PLS   &       & 0.34798***(***) &       & 0.33699**() &       & 0.34794***(***) &       & 0.33704**() \\
    PCR   &       & 0.33898***(***) &       & 0.30069**() &       & 0.33895***(***) &       & 0.30074**() \\
    XGB(+H) &       & 0.41945***(***) &       & 0.01814() &       & 0.41941***(***) &       & 0.01821() \\
    LGBM(+H) &       & 0.41129***(***) &       & 0.15483() &       & 0.41126***(***) &       & 0.15489() \\
    RF    &       & 0.32098***(***) &       & 0.22807*() &       & 0.32094***(***) &       & 0.22812*() \\
    NN1   &       & 0.35993***(***) &       & 0.21887() &       & 0.3599***(***) &       & 0.21892() \\
    NN2   &       & 0.36369***(***) &       & 0.13160() &       & 0.36365***(***) &       & 0.13167() \\
    NN3   &       & 0.37999***(***) &       & 0.14738() &       & 0.37996***(***) &       & 0.14744() \\
    NN4   &       & 0.37316***(***) &       & 0.35160**() &       & 0.37312***(***) &       & 0.35165**() \\
    NN5   &       & 0.39348***(***) &       & 0.25848**() &       & 0.39345***(***) &       & 0.25853**() \\
    $Ensemble_{avg}$ &       & 0.3819***(***) &       & 0.58791***() &       & 0.38186***(***) &       & 0.58794***() \\
    $Ensemble_{\theta=0.9}$ &       & 0.37308***(***) &       & 0.53187***() &       & 0.37305***(***) &       & 0.53190***() \\
    $Ensemble_{\theta=1}$ &       & 0.38287***(***) &       & 0.60327***() &       & 0.38284***(***) &       & 0.60330***() \\
    $Ensemble_{op}$ &       & 0.40789***(***) &       & 0.47804***() &       & 0.40785***(***) &       & 0.47807***() \\
    $Ensemble_{wp}$ &       & 0.48937***(***) &       & 0.78660***() &       & 0.48934***(***) &       & 0.78661***() \\
\toprule
Mean(Volatile)-Mean(Tranquil)& && &0.11263*& && &0.11254* \\
    $Predictability_{wE}$&       & 0.48937         &       & 0.78660         &       & 0.48934          &       & 0.78661 \\
    $Predictability_{oE}$&       & 0.41267         &       & 0.38412         &       & 0.41941          &       & 0.38417 \\
\toprule

\textit{Panel C: High V.S. Low Night Light Intensity}&&High&&Low&&High&&Low\\
\cmidrule{3-5} \cmidrule{7-9}
    OLS   &       & 0.36188***(***) &       & 0.3044***(**) &       & 0.36191***(***) &       & 0.30424***(**) \\
    LASSO &       & 0.36783***(***) &       & 0.3234***(***) &       & 0.36786***(***) &       & 0.32324***(***) \\
    Ridge &       & 0.36196***(***) &       & 0.30429***(**) &       & 0.36198***(***) &       & 0.30412***(**) \\
    ENet  &       & 0.37145***(***) &       & 0.3262***(***) &       & 0.37148***(***) &       & 0.32604***(***) \\
    GLM   &       & 0.4638***(***) &       & 0.401***(***) &       & 0.46382***(***) &       & 0.40086***(***) \\
    PLS   &       & 0.37953***(***) &       & 0.33217***(***) &       & 0.37955***(***) &       & 0.33201***(***) \\
    PCR   &       & 0.36697***(***) &       & 0.32363***(***) &       & 0.367***(***) &       & 0.32347***(***) \\
    XGB(+H) &       & 0.47399***(***) &       & 0.39434***(***) &       & 0.47401***(***) &       & 0.39419***(***) \\
    LGBM(+H) &       & 0.46654***(***) &       & 0.3926***(***) &       & 0.46656***(***) &       & 0.39246***(***) \\
    RF    &       & 0.37785***(***) &       & 0.30818***(***) &       & 0.37788***(***) &       & 0.30801***(***) \\
    NN1   &       & 0.38932***(***) &       & 0.35495***(***) &       & 0.38934***(***) &       & 0.3548***(***) \\
    NN2   &       & 0.39382***(***) &       & 0.37884***(***) &       & 0.39385***(***) &       & 0.37869***(***) \\
    NN3   &       & 0.4057***(***) &       & 0.38847***(***) &       & 0.40572***(***) &       & 0.38833***(***) \\
    NN4   &       & 0.41399***(***) &       & 0.36671***(***) &       & 0.41402***(***) &       & 0.36656***(***) \\
    NN5   &       & 0.42026***(***) &       & 0.40547***(***) &       & 0.42028***(***) &       & 0.40533***(***) \\
    $Ensemble_{avg}$ &       & 0.43144***(***) &       & 0.38184***(***) &       & 0.43146***(***) &       & 0.38169***(***) \\
    $Ensemble_{\theta=0.9}$ &       & 0.4227***(***) &       & 0.36473***(***) &       & 0.42273***(***) &       & 0.36458***(***) \\
    $Ensemble_{\theta=1}$ &       & 0.43486***(***) &       & 0.37982***(***) &       & 0.43488***(***) &       & 0.37967***(***) \\
    $Ensemble_{op}$ &       & 0.45526***(***) &       & 0.39264***(***) &       & 0.45528***(***) &       & 0.39249***(***) \\
    $Ensemble_{wp}$ &       & 0.54284***(***) &       & 0.49492***(***) &       & 0.54286***(***) &       & 0.4948***(***) \\
\toprule
Mean(High)-Mean(Low)& && &0.04917***& && &0.04934*** \\
    $Predictability_{wE}$&       & 0.54284         &       & 0.49492         &       & 0.54286         &       & 0.4948 \\
    $Predictability_{oE}$&       & 0.47399         &       & 0.40547         &       & 0.47401          &       & 0.40533 \\
\toprule

\textit{Panel D: High V.S. Low Geopolitical Risk}&&High&&Low&&High&&Low\\
\cmidrule{3-5} \cmidrule{7-9}
    OLS   &       & 0.33442***(***) &       & 0.31906***(**) &       & 0.33447***(***) &       & 0.31892***(**) \\
    LASSO &       & 0.34964***(***) &       & 0.3384***(***) &       & 0.34968***(***) &       & 0.33827***(***) \\
    Ridge &       & 0.3351***(***) &       & 0.31838***(**) &       & 0.33515***(***) &       & 0.31825***(**) \\
    ENet  &       & 0.35199***(***) &       & 0.34123***(***) &       & 0.35203***(***) &       & 0.3411***(***) \\
    GLM   &       & 0.44494***(***) &       & 0.42075***(***) &       & 0.44498***(***) &       & 0.42064***(***) \\
    PLS   &       & 0.3619***(***) &       & 0.34646***(***) &       & 0.36195***(***) &       & 0.34633***(***) \\
    PCR   &       & 0.35562***(***) &       & 0.33846***(***) &       & 0.35567***(***) &       & 0.33833***(***) \\
    XGB(+H) &       & 0.4611***(***) &       & 0.42977***(***) &       & 0.46113***(***) &       & 0.42966***(***) \\
    LGBM(+H) &       & 0.44413***(***) &       & 0.42813***(***) &       & 0.44416***(***) &       & 0.42801***(***) \\
    RF    &       & 0.36629***(***) &       & 0.34645***(***) &       & 0.36634***(***) &       & 0.34632***(***) \\
    NN1   &       & 0.3813***(***) &       & 0.35699***(***) &       & 0.38135***(***) &       & 0.35686***(***) \\
    NN2   &       & 0.37298***(***) &       & 0.35116***(***) &       & 0.37302***(***) &       & 0.35103***(***) \\
    NN3   &       & 0.38422***(***) &       & 0.36876***(***) &       & 0.38426***(***) &       & 0.36863***(***) \\
    NN4   &       & 0.3972***(***) &       & 0.39166***(***) &       & 0.39724***(***) &       & 0.39154***(***) \\
    NN5   &       & 0.41677***(***) &       & 0.3816***(***) &       & 0.41681***(***) &       & 0.38148***(***) \\
    $Ensemble_{avg}$ &       & 0.4133***(***) &       & 0.39118***(***) &       & 0.41335***(***) &       & 0.39106***(***) \\
    $Ensemble_{\theta=0.9}$ &       & 0.41015***(***) &       & 0.3791***(***) &       & 0.41019***(***) &       & 0.37898***(***) \\
    $Ensemble_{\theta=1}$ &       & 0.41731***(***) &       & 0.39378***(***) &       & 0.41735***(***) &       & 0.39366***(***) \\
    $Ensemble_{op}$ &       & 0.4398***(***) &       & 0.41015***(***) &       & 0.43984***(***) &       & 0.41004***(***) \\
    $Ensemble_{wp}$ &       & 0.53134***(***) &       & 0.51254***(***) &       & 0.53137***(***) &       & 0.51245***(***) \\
\toprule
Mean(High)-Mean(Low)& && &0.0409& && &0.00515 \\
    $Predictability_{wE}$&       & 0.53134         &       & 0.51254         &       & 0.53137         &       & 0.51245 \\
    $Predictability_{oE}$&       & 0.44494         &       & 0.42977         &       & 0.46113          &       & 0.42966 \\
\end{longtable} 
\end{ThreePartTable}

\begin{ThreePartTable} 
\begin{TableNotes} 
\item \textit{Notes}: All values are in \%. $Ensemble_{op}$ means ensemble model is without prediction correlation penalty to tune the weight, while $Ensemble_{wp}$ is the counterpart. $Predictability_{wE}$ is the predictability of state $s$ considering ensemble models, and $Predictability_{oE}$ opts out the ensemble models. Statistical significance for the \(R_{OOS}^2\) statistic is assessed using $p$-values from the CW OOS MSPE-adjusted statistic and t-statistics from the DM test, with Newey–West robust standard errors. These tests correspond to a one-sided hypothesis test, where the null hypothesis states that the forecast from our model has an expected square prediction error equal to that of the benchmark forecast, and the alternative hypothesis states that our forecast has a lower expected square prediction error than the benchmark. \(^{*}, ^{**},\) and \(^{***}\) indicate significance at the 10\%, 5\%, and 1\% levels for both the DM, CW tests and t-test, with CW significance levels presented outside the parentheses, DM within the parentheses and t-test presented in the last row of each panel.
\end{TableNotes}

\fontsize{8pt}{0.38cm}\selectfont
\centering
\begin{longtable}[H]{p{5.5cm} p{0.02cm} p{1.5cm} p{0.01cm} p{1.5cm} p{0.02cm} p{1.5cm} p{0.01cm} p{1.5cm}}
\label{tab:7} \\
\caption{\centering Heterogeneous Predictability before/after Energy Crisis} \\
\\
\toprule
&&\multicolumn{3}{c}{$R^2_{OOS,mean}$}&&\multicolumn{3}{c}{$R^2_{OOS,zero}$} \\
\cmidrule{3-5} \cmidrule{7-9}
\endhead

\midrule
\multicolumn{9}{r}{\textit{Continued on next page}}
\endfoot

\bottomrule 
\insertTableNotes \\ 
\endlastfoot

\textit{Panel A: before/after Energy Crisis}&&before&&after&&before&&after\\
\cmidrule{3-5} \cmidrule{7-9}
    OLS   &       & 29.85(***) &       & 34.69***(***) &       & 29.84(***) &       & 34.69***(***) \\
    LASSO &       & 38.44(***) &       & 34.15***(***) &       & 38.44(***) &       & 34.15***(***) \\
    Ridge &       & 29.90(***) &       & 34.67***(***) &       & 29.90(***) &       & 34.67***(***) \\
    ENet  &       & 38.62(***) &       & 34.45***(***) &       & 38.62(***) &       & 34.44***(***) \\
    GLM   &       & 56.55***(***) &       & 39.68***(***) &       & 56.54***(***) &       & 39.68***(***) \\
    PLS   &       & 38.99(***) &       & 35.27***(***) &       & 38.99(***) &       & 35.27***(***) \\
    PCR   &       & 38.68(***) &       & 34.18***(***) &       & 38.67(***) &       & 34.18***(***) \\
    XGB(+H) &       & 59.35(***) &       & 38.75***(***) &       & 59.35(***) &       & 38.75***(***) \\
    LGBM(+H) &       & 59.27*(***) &       & 38.25***(***) &       & 59.27*(***) &       & 38.25***(***) \\
    RF    &       & 52.61***(***) &       & 29.05***(***) &       & 52.61***(***) &       & 29.05***(***) \\
    NN1   &       &43.66(***) &       & 35.72***(***) &       & 43.66(***) &       & 35.72***(***) \\
    NN2   &       & 47.55(***) &       & 34.81***(***) &       & 47.55(***) &       & 34.80***(***) \\
    NN3   &       & 51.19**(***) &       & 36.48***(***) &       & 51.19**(***) &       & 36.47***(***) \\
    NN4   &       & 48.75**(***) &       & 35.90***(***) &       & 48.75**(***) &       & 35.89***(***) \\
    NN5   &       & 51.19**(***) &       & 37.17***(***) &       & 51.19**(***) &       & 37.16***(***) \\
    $Ensemble_{avg}$ &       & 54.19***(***) &       & 38.85***(***) &       & 54.19***(***) &       & 38.84***(***) \\
    $Ensemble_{\theta=0.9}$ &       &50.38***(***) &       & 38.24***(***) &       & 50.38***(***) &       & 38.23***(***) \\
    $Ensemble_{\theta=1}$ &       & 53.91***(***) &       & 38.48***(***) &       & 53.91***(***) &       & 38.48***(***) \\
    $Ensemble_{op}$ &       & 56.67***(***) &       & 38.85***(***) &       & 56.67***(***) &       & 38.85***(***) \\
    $Ensemble_{wp}$ &       & 69.15***(***) &       & 46.53***(***) &       & 69.15***(***) &       & 46.53***(***) \\
\end{longtable} 
\end{ThreePartTable}

\begin{table}[H]
\fontsize{9pt}{0.38cm}\selectfont
\centering
\caption{\centering Mean-Variance Portfolio Results ($\gamma$=3)}
\label{tab:8}
\setlength{\tabcolsep}{7pt}
\renewcommand{\arraystretch}{1.1}
\begin{threeparttable}
\begin{tabular}{p{1.5cm} p{2.5cm} p{2.5cm} p{2cm} p{2.5cm} p{2cm}}
\hline
&Average $w_{m,t}$&Average $w_{r_f,t}$&$\sigma_{w_{m,t}}$&Average Utility&CER\\
\hline
OLS&0.07092&0.92908&1.57235&0.10066&31.757284\\
LASSO&0.06308&0.93692&1.95284&0.09427&35.401559\\
Ridge&0.07082&0.92918&1.57339&0.10069&31.817969\\
ENet&0.05760&0.94240&1.97606&0.09637&36.999837\\
PCR&0.05609&0.94391&2.03234&0.10288&38.612024\\
PLS&0.05615&0.94385&2.07878&0.10584&40.008014\\
GLM&0.10460&0.89540&2.80676&0.14764&66.962009\\
XGB+H&0.15399&0.84601&3.12115&0.18480&77.153650\\
LGBM+H&0.16988&0.83012&2.85457&0.16593&66.555418\\
RF&0.13977&0.86023&2.63051&0.12165&53.453153\\
NN1&0.20019&0.79981&2.16978&0.11634&45.145728\\
NN2&0.24084&0.75916&2.29679&0.12724&51.996737\\
NN3&0.26925&0.73075&2.34244&0.13537&51.361202\\
NN4&0.24619&0.75381&2.3760&0.12779&49.663786\\
NN5&0.27412&0.72588&2.39578&0.13919&51.603603\\
\hline
\end{tabular}
\begin{tablenotes}
\fontsize{8pt}{0.35cm}\selectfont
\item \textit{Notes}: $w_{m,t}$ is the weight of model $m$ at day $t$, average $w_{m,t}$ is the average of the weight of model $m$ across all days in $\mathcal{T}_3$. $w_{r_f,t}$ is the weight of risk-free asset $r_f$ at day $t$. $\sigma_{w_{m,t}}$ is the sample standard deviation of the weight of model $m$ across all days in $\mathcal{T}_3$.
\end{tablenotes}
\end{threeparttable}
\end{table}

\begin{algorithm}[H]
\caption{Decomposing OOS Performance: A Trend-based Probability Framework}
\label{alg:1}
\textbf{INPUT:} $r_t$, $r_{t+1}$, $\hat{r}_{t+1}$, $\hat{r}_{\text{benchmark}, t+1}$ \\
\textbf{OUTPUT:} $\text{TrendClass}$, $\text{PerfClass}$ \\[0.5em]
\textbf{STEP 1: Compute Trend Classification}
{\footnotesize
\begin{algorithmic}[1]
    \IF{$r_{t+1} \ge r_t$} 
        \IF{$\hat{r}_{t+1} < r_t$}
            \STATE $\text{TrendClass} \gets \text{False Trend (FT)}$
        \ELSIF{$r_t \le \hat{r}_{t+1} < r_{t+1}$}
            \STATE $\text{TrendClass} \gets \text{Right Weak Trend (RWT)}$
        \ELSE
            \STATE $\text{TrendClass} \gets \text{Right Strong Trend (RST)}$
        \ENDIF
    \ELSE 
        \IF{$\hat{r}_{t+1} > r_t$}
            \STATE $\text{TrendClass} \gets \text{False Trend (FT)}$
        \ELSIF{$r_{t+1} < \hat{r}_{t+1} \le r_t$}
            \STATE $\text{TrendClass} \gets \text{Right Weak Trend (RWT)}$
        \ELSE
            \STATE $\text{TrendClass} \gets \text{Right Strong Trend (RST)}$
        \ENDIF
    \ENDIF
\end{algorithmic}
} 

\vspace{0.5em}
\textbf{STEP 2: Compute Benchmark-Based Performance Classification}

{\footnotesize
\begin{algorithmic}[1]
    \STATE $L \gets r_{t+1} - |\hat{r}_{\text{benchmark}, t+1} - r_{t+1}|$
    \STATE $U \gets r_{t+1} + |\hat{r}_{\text{benchmark}, t+1} - r_{t+1}|$
    
    \IF{$\hat{r}_{t+1} < L$}
        \STATE $\text{PerfClass} \gets \text{Downward Loss (DL)}$
    \ELSIF{$L \le \hat{r}_{t+1} < r_{t+1}$}
        \STATE $\text{PerfClass} \gets \text{Downward Gain (DG)}$
    \ELSIF{$r_{t+1} < \hat{r}_{t+1} \le U$}
        \STATE $\text{PerfClass} \gets \text{Upward Gain (UG)}$
    \ELSE
        \STATE $\text{PerfClass} \gets \text{Upward Loss (UL)}$
    \ENDIF
\end{algorithmic}
} 
\end{algorithm}

\newpage
\label{Bibliography}
\setstretch{1}
\bibliographystyle{apalike}
\bibliography{reference_emc}

\newpage
\section*{Online Appendix}
\label{Online Appendix}
\appendix
\renewcommand{\thetable}{A\arabic{table}}
\setcounter{table}{0}

\renewcommand{\thefigure}{A\arabic{figure}}
\setcounter{figure}{0}

\renewcommand{\theequation}{A\arabic{equation}}
\setcounter{equation}{0}

\section{Institutional Background \& Data Appendix}
\subsection{Institutional Background Supplements}
\label{Institutional Background Supplements}

\begin{table}[H]
	\fontsize{9pt}{0.38cm}\selectfont
	\centering
	\caption{Market Reform Milestones}
	\label{tab:A1}
	\setlength{\tabcolsep}{7pt}
	\renewcommand{\arraystretch}{1.1}
	\begin{threeparttable}
		\begin{tabular}{p{0.8cm} p{12cm}}
			\hline
			Year &Milestone\\
			\hline
			\multicolumn{2}{l}{\textbf{Panel A. Corporatisation}} \\
			1995 & Electricity functions of the Public Utilities Board corporatised\\ &Singapore Power formed as a holding company \\
			1996 & SEP design process began \\
			\multicolumn{2}{l}{\textbf{Panel B. SEP}} \\
			1998 & SEP commenced\\
			&PowerGrid is SEP Administrator and PSO \\
			\multicolumn{2}{l}{\textbf{Panel C. NEMS}} \\ 
			2000 & Decision for further reform to obtain full benefits of competition\\
			&New market design process began \\
			2001 & Electricity industry legislation enacted\\
			&EMA established as industry regulator and PSO\\
			&EMC established as the NEMS wholesale market operator\\
			&First phase of retail contestability (retail contestability threshold gradually lowered in subsequent years) \\ 
			2003 & NEMS wholesale market trading began \\
			2004 & Vesting contract regime introduced\\
			&Interruptible loads (IL) began to participate in the reserves market \\
			2006 & First wholesale market trader joined the market and commenced trading as IL provider\\
			&First commercial generator since 2003 joined the market and started trading \\
			2008 & Sale of Tuas Power to China Huaneng Group in March, Senoko Power to Lion Consortium in September, and PowerSeraya to YTL Power in December\\
			&Embedded generators (EG) joined the market \\
			2009 & New EGs, small generators and incineration plants joined and started trading \\
			2010 & Vesting tender introduced to tender out a percentage of non-contestable electricity demand to generation companies for bidding \\
			2013 & Singapore's Liquefied Natural Gas (LNG) terminal started commercial operations\\
			&LNG vesting contract introduced \\
			2015 & Electricity futures trading commenced \\
			2016 & Demand Response programme introduced \\
			2018 & OEM launched and rolled out in stages \\ 
			2019 & Rollout of OEM across Singapore completed\\ 
			&Vesting contract regime rolled back to LNG vesting contract level \\
			2021 & First energy storage system (ESS) joined the market \\
			2022 & Electricity imports trial commenced \\
			2023 & NEMS completed 20 successful years of trading \\ 
			\hline
		\end{tabular}
		\begin{tablenotes}
			\fontsize{8pt}{0.35cm}\selectfont
			\item From \citet{emcreport2023market}
		\end{tablenotes}
	\end{threeparttable}
\end{table}

\subsection{Variable Definitions \& Summary Statistics}
\label{Variable Definitions and Summary Statistics}

\begin{table}[H]
	\fontsize{9pt}{0.38cm}\selectfont
	\centering
	\caption{Inputs Definitions}
	\label{tab:A2}
	\setlength{\tabcolsep}{7pt}
	\renewcommand{\arraystretch}{1.1}
	\begin{tabular}{p{2.8cm} p{11cm}}
		\hline
		Variables&Definition\\
		\hline
		
		\multicolumn{2}{l}{\textbf{\fontsize{10pt}{0.38cm}\selectfont Panel A.  In-market Variables}} \\
		\text{USEP\textsubscript{t-1}} & One-period time lag of USEP. \\
		
		\multicolumn{2}{l}{\textbf{Panel A.1  Demand}} \\
		\text{Demand\textsubscript{t}} & Forecasted half-hourly electricity demand for Singapore. \\
		\text{WEQ\textsubscript{t}} & Withdrawal Energy Quantity (WEQ) denotes the volume of electricity withdrawn by load facilities, as reported by the MSSL, SP Services. \\
		\text{OfferRatiobelow200\textsubscript{t}} \text{OfferRatiobelow300\textsubscript{t}} \text{OfferRatiobelow400\textsubscript{t}} \text{OfferRatioabove400\textsubscript{t}}&The offer ratios below \$200, \$300, and \$400 represent the cumulative volumes of energy offers priced at or below each respective threshold per MWh, while the offer ratio above \$400 captures the total volume of energy offers priced at or above \$400 per MWh.\\
		
		\multicolumn{2}{l}{\textbf{Panel A.2  Supply}} \\
		\text{Supplycushion\textsubscript{t}} & Supply Cushion represents the percentage of total supply available after matching off demand, measured every half hour. \\
		\text{CCGTGen\textsubscript{t}} & Combined-Cycle Gas Turbine (CCGT) Generation refers to the energy offered by CCGT units. \\
		\text{STGen\textsubscript{t}} & Steam Turbine (ST) Generation refers to the energy offered by ST units. \\
		\text{OTGen\textsubscript{t}} & Other Facilities (OT) Generation refers to the energy offered by other facilities. \\
		\text{OCGTGen\textsubscript{t}} & Open Cycle Gas Turbine (OCGT) Generation refers to the energy offered by OCGT units. \\
		\text{IEQ\textsubscript{t}} & Injection Energy Quantity (IEQ) pertains to the electricity amount injected by generation facilities, supplied by the MSSL, SP Services. \\
		\text{CCGTforcedoutage\textsubscript{t}} & CCGT forced outage refers to unanticipated outage volumes in CCGT units, measured on a half-hourly basis. \\
		\text{STforcedoutage\textsubscript{t}} & ST forced outage refers to unanticipated outage volumes in ST units, measured on a half-hourly basis. \\
		\text{OTforcedoutage\textsubscript{t}} & OT forced outage refers to unanticipated outage volumes in OT units, measured on a half-hourly basis. \\
		\text{OCGTforcedoutage\textsubscript{t}} & OCGT forced outage refers to unanticipated outage volumes in OCGT units, measured on a half-hourly basis. \\
		\text{Forcedoutage\textsubscript{t}} & Forced outage represents the total of forced outages from CCGT, ST, OT, OCGT, Import, Solar, and ESS units. \\
		\text{CCGTplannedoutage\textsubscript{t}} & CCGT planned outage represents the expected half-hourly outage volume for CCGT units. \\
		\text{STplannedoutage\textsubscript{t}} & ST planned outage represents the expected half-hourly outage volume for ST units. \\
		\text{OTplannedoutage\textsubscript{t}} & OT planned outage represents the expected half-hourly outage volume for OT units. \\
		\text{OCGTplannedoutage\textsubscript{t}} & OCGT planned outage represents the expected half-hourly outage volume for OCGT units. \\
		\text{Plannedoutage\textsubscript{t}} & Planned outage represents the total of planned outages from CCGT, ST, OT, OCGT, Import, Solar, and ESS units. \\
		
		\hline
	\end{tabular}
\end{table}

\begin{table}[H]
	\fontsize{9pt}{0.38cm}\selectfont
	\centering
	\caption*{Table A2 (Continued)}
	\setlength{\tabcolsep}{7pt}
	\renewcommand{\arraystretch}{1.1}
	\begin{tabular}{p{2.8cm} p{11cm}}
		\hline
		Variables&Definition\\
		\hline
		\multicolumn{2}{l}{\textbf{Panel A.3  Regulation}} \\
		\text{ConResReq\textsubscript{t}} & The system-level requirement from contingency reserve providers, who must be available for activation within a 10-minute response time and maintained for a minimum duration of 30 minutes. \\
		\text{ConResPrice\textsubscript{t}} & The market-determined price for the contingency reserve product. \\
		\text{Con\_Avail\textsubscript{t}} & The total capacity of contingency reserves available. \\
		\text{PriResReq\textsubscript{t}} & The system-level requirement from primary reserve providers, who must be available for activation within a 9-second response time and sustained for a minimum duration of 10 minutes. \\
		\text{PriResPrice\textsubscript{t}} & The market-determined price for the primary reserve product. \\
		\text{Pri\_Avail\textsubscript{t}} & The total capacity of primary reserves available. \\
		\text{RegReq\textsubscript{t}} & The system-level regulation requirement, where regulation refers to the standby generation tasked with fine-tuning or correcting frequency variations and imbalances between power demand and supply in the power system. \\
		\text{RegPrice\textsubscript{t}} & The market-determined price for the regulation product. \\
		\text{VestedQuantityMW\textsubscript{t}} \text{VestedQuantitykWh\textsubscript{t}} \text{VestingQuantity (\%)\textsubscript{t}}
		& The quantity (in MW, kWh or \% terms) for the purpose of hedging non-contestable consumer (NCC) load under the vesting contract. \\
		\text{VestingPrice\textsubscript{t}} & The vesting price is determined based on the long-run marginal cost of the most efficient technology that accounts for at least 25\% of the total electricity demand in Singapore. \\
		\text{ResCushion\textsubscript{t}} & Reserve Cushion refers to the excess reserve capacity available after dispatch, quantified on a half-hourly basis. \\
		
		\multicolumn{2}{l}{\textbf{\fontsize{10pt}{0.38cm}\selectfont Panel B. Domestic Macroeconomics Variables}} \\
		\text{Temperature\textsubscript{t}} & Daily average temperature in Singapore on day $t$ ($^\circ$C). \\
		\text{Total Rainfall\textsubscript{t}} & Daily average rainfall in Singapore on day $t$ (mm). \\
		\text{Wind Speed\textsubscript{t}} & Average wind speed in Singapore on day $t$ (km/h). \\
		\text{MPPI\textsubscript{m}} & Manufactured Producer Price Index in Singapore for month $m$, measuring average prices received by local manufacturers.\\
		\text{PMI\textsubscript{m}} &  Purchasing Managers’ Index in Singapore for month $m$, based on a weighted composite of five survey-based indicators reflecting manufacturing activity. \\
		\text{CPI \textsubscript{m}} & Consumer Price Index in Singapore for month $m$. It measures the average change over time in the prices paid by households for a fixed basket of consumer goods and services, including food, housing, transportation, healthcare, and education. It is a key indicator of inflation and cost of living. \\
		\text{IPI \textsubscript{m}} & Industrial Production Index in Singapore for month $m$. It measures the real output of the manufacturing, mining, and utilities sectors. The index tracks changes in the volume of production and serves as a key indicator of short-term economic activity and industrial performance.\\
		\hline
	\end{tabular}
\end{table}

\begin{table}[H]
	\fontsize{9pt}{0.38cm}\selectfont
	\centering
	\caption*{Table A2 (Continued)}
	\setlength{\tabcolsep}{7pt}
	\renewcommand{\arraystretch}{1.1}
	\begin{tabular}{p{2.8cm} p{11cm}}
		\hline
		Variables&Definition\\
		\hline
		
		\multicolumn{2}{l}{\textbf{\fontsize{10pt}{0.38cm}\selectfont Panel C. International Macroeconomics Variables}} \\
		\text{GPRD\textsubscript{t}} & Geopolitical Risk Daily serves as a news-based measure that quantifies adverse geopolitical events and their associated risks, encompassing both po-
		tential threats and realized events.\\
		\text{GPRD\_THREAT\textsubscript{t}} &  GPRD\_THREAT is a sub-index of GPRD that captures the potential threats of adverse geopolitical events.\\
		\text{GPRD\_ACT\textsubscript{t}} &  GPRD\_ACT is a sub-index of GPRD that focuses specifically on the realization of adverse geopolitical events.\\
		\text{Gas\_spot\textsubscript{t}} & Natural gas price on day $t$.\\
		\text{Gas\_futures\textsubscript{t}} & Futures price of natural gas on day $t$. \\
		\text{Fueloilpriceusmt\textsubscript{m}} &  Oil price in month $m$.\\
		\text{Oil\_futures\_price\textsubscript{t}} & Futures price of oil on day $t$. \\
		\text{ECPI\_mean\textsubscript{m}} & Energy Consumer Price Index in month $m$, measuring the average change over time in the retail prices of energy-related goods and services.\\
		\text{Exchange\_rate\textsubscript{t}} &  Exchange rate between SGD and USD on day $t$.\\
		
		\multicolumn{2}{l}{\textbf{Panel D. Seasonality Indicators}} \\
		\text{Weekday\textsubscript{i}} & For $i = 0$ to $6$, each value represents a day from Sunday to Saturday. The variable equals 1 if the date is \text{Weekday\textsubscript{i}}, and 0 otherwise.\\
		\hline
	\end{tabular}
\end{table}

\begin{table}[H]
	\fontsize{9pt}{0.25cm}\selectfont
	\makebox[\textwidth]{
		\rotatebox{90}{
			\begin{minipage}{\textheight}
				\centering
				\caption{Summary Statistics}
				\label{tab:A3}
				\setlength{\tabcolsep}{7pt}
				\renewcommand{\arraystretch}{1.1}
				\begin{tabular}{p{7.1cm} p{1.5cm} p{0.9cm} p{1.3cm} p{1cm} p{1cm} p{1cm} p{1cm} p{1cm} p{1cm}}
					\hline
					Variable & N & Mean & Std & 5th & 25th & 50th & 75th & 95th & 99th\\
					\hline 
					
					Uniform Singapore Energy Price (\$/MWh) & 7670 & 144.00 & 104.20 & 60.62 & 85.86 & 119.40 & 172.00 & 286.70 & 523.80\\
					System Demand (MW) & 7670 & 5226.00 & 802.30 & 3908.00 & 4572.00 & 5305.00 & 5898.00 & 6414.00 & 6614.00\\
					Primary Reserve Requirement (MW) & 7670 & 201.10 & 39.03 & 155.50 & 174.10 & 188.00 & 218.00 & 273.50 & 293.70\\
					Primary Reserve Price (\$/MWh) & 7670 & 3.68 & 19.49 & 0.01 & 0.04 & 0.23 & 1.34 & 15.82 & 57.12\\
					Primary Reserve Availability (MW) & 7670 & 414.70 & 94.78 & 295.10 & 344.30 & 406.50 & 475.80 & 545.50 & 606.60\\
					Contingency Reserve Requirement (MW) & 7670 & 542.60 & 68.85 & 378.90 & 522.60 & 558.90 & 593.60 & 613.90 & 623.10\\
					Contingency Reserve Price (\$/MWh) & 7670 & 11.93 & 24.33 & 0.08 & 1.46 & 4.73 & 12.52 & 47.16 & 104.70\\
					Contingency Reserve Availability (MW) & 7670 & 1128.00 & 219.70 & 803.80 & 988.00 & 1122.00 & 1252.00 & 1451.00 & 1596.00\\
					Regulation Requirement (MW) & 7670 & 104.20 & 14.71 & 81.73 & 94.17 & 100.00 & 117.20 & 124.80 & 126.50\\
					Regulation Price (\$/MWh) & 7670 & 45.39 & 69.62 & 4.42 & 13.07 & 28.12 & 58.43 & 117.70 & 245.00\\
					Combined-Cycle Gas Turbine (CCGT) Generation (MW) & 7670 & 4717.00 & 1259.00 & 2525.00 & 3698.00 & 4940.00 & 5797.00 & 6377.00 & 7079.00\\
					Steam Turbine (ST) Generation (MW) & 7670 & 438.20 & 501.30 & 2.56 & 3.00 & 132.70 & 846.70 & 1369.00 & 1722.00\\
					Other (OT) Facilities Generation (MW) & 7670 & 122.30 & 17.72 & 95.00 & 110.80 & 121.20 & 133.20 & 151.90 & 174.00\\
					Open-cycle Gas Turbine (OCGT) Generation (MW) & 7670 & 202.50 & 70.28 & 90.00 & 160.00 & 180.00 & 244.30 & 378.00 & 408.00\\
					Injection Energy Quantity (MW) & 7670 & 5114.00 & 826.50 & 3775.00 & 4436.00 & 5182.00 & 5801.00 & 6362.00 & 6601.00\\
					Withdrawal Energy Quantity (MW) & 7670 & 5099.00 & 831.00 & 3759.00 & 4412.00 & 5162.00 & 5791.00 & 6352.00 & 6588.00\\
					CCGT Forced Outage (MW) & 7670 & 31.53 & 86.42 & 0 & 0 & 0 & 7.40 & 222.00 & 379.50\\
					ST Forced Outage (MW) & 7670 & 3.33 & 23.50 & 0 & 0 & 0 & 0 & 3.38 & 94.33\\
					OT Forced Outage (MW) & 7670 & 0.39 & 2.85 & 0 & 0 & 0 & 0 & 0.58 & 17.00\\
					OCGT Forced Outage (MW) & 7670 & 0.65 & 7.48 & 0 & 0 & 0 & 0 & 0 & 4.79\\
					Forced Outage (MW) & 7670 & 35.91 & 90.30 & 0 & 0 & 0 & 17.00 & 239.40 & 389.00\\
					CCGT Planned Outage (MW) & 7670 & 644.50 & 555.00 & 0 & 212.50 & 475.00 & 942.50 & 1750.00 & 2213.00\\
					ST Planned Outage (MW) & 7670 & 436.60 & 377.80 & 0 & 233.00 & 250.00 & 702.00 & 1143.00 & 1583.00\\
					OT Planned Outage (MW) & 7670 & 29.43 & 49.80 & 0 & 0 & 0 & 47.80 & 132.00 & 187.00\\
					OCGT Planned Outage (MW) & 7670 & 15.73 & 37.08 & 0 & 0 & 0 & 0 & 105.00 & 109.00\\
					Planned Outage (MW) & 7670 & 1127.00 & 579.10 & 279.80 & 717.00 & 1047.00 & 1454.00 & 2243.00 & 2778.00\\
					Offer Ratio Below 100 & 7670 & 71.23 & 9.76 & 56.47 & 60.93 & 74.32 & 78.90 & 84.52 & 88.04\\
					Offer Ratio Below 200 & 7670 & 82.82 & 4.98 & 74.72 & 79.08 & 83.01 & 86.27 & 91.08 & 93.60\\
					Offer Ratio Below 300 & 7670 & 84.83 & 4.47 & 78.50 & 81.28 & 84.35 & 87.61 & 93.41 & 95.34\\
					Offer Ratio Below 400 & 7670 & 85.91 & 4.51 & 79.69 & 82.34 & 85.27 & 88.70 & 94.70 & 96.37\\
					Offer Ratio Above 400 & 7670 & 14.01 & 4.52 & 5.24 & 11.24 & 14.62 & 17.60 & 20.29 & 21.62\\
					Vested Quantity (MW) & 7305 & 1964.00 & 760.50 & 1007.00 & 1232.00 & 2057.00 & 2647.00 & 3017.00 & 3242.00\\
					Vested Quantity (kWh) & 7305 & 981757 & 380270 & 503650 & 615906 & 1029000 & 1323000 & 1509000 & 1621000\\
					Vesting Quantity (\%) & 7305 & 0.40 & 0.19 & 0.16 & 0.21 & 0.43 & 0.57 & 0.67 & 0.71\\
					Vesting Price (MWh) & 7305 & 160.70 & 34.08 & 101.30 & 137.30 & 159.90 & 190.00 & 216.70 & 238.60\\
					Reserve Cushion & 7670 & 48.43 & 9.47 & 31.62 & 42.32 & 49.72 & 54.83 & 61.07 & 65.84\\
					Supply Cushion & 7670 & 0.25 & 0.05 & 0.13 & 0.23 & 0.25 & 0.28 & 0.32 & 0.35\\	
					Industrial Production Index & 7670 & 83.25 & 22.09 & 49.93 & 62.86 & 84.31 & 99.90 & 123.60 & 131.00 \\
					Consumer Price Index & 7670 & 92.78 & 10.96 & 76.19 & 83.82 & 98.26 & 99.59 & 111.19 & 114.91 \\
					Manufactured Producer Price Index & 7670 & 106.33 & 10.21 & 90.06 & 98.07 & 106.56 & 113.31 & 123.48 & 131.17 \\
					Purchasing Manager Index & 7670 & 50.73 & 1.77 & 48.50 & 49.80 & 50.60 & 51.78 & 53.50 & 55.20 \\
					Total Rainfall & 7670 & 7.27 & 12.04 & 0.00 & 0.12 & 2.49 & 9.81 & 29.62 & 51.60 \\
					Temperature & 7670 & 27.74 & 1.12 & 25.88 & 27.00 & 27.76 & 28.56 & 29.49 & 30.01 \\
					Wind Speed & 7670 & 8.28 & 2.26 & 5.43 & 6.69 & 7.83 & 9.40 & 12.64 & 15.39 \\
					Fuel Oil Price & 7670 & 408.50 & 157.65 & 167.83 & 282.57 & 401.83 & 503.50 & 672.40 & 743.46 \\
					Oil Futures Price & 7670 & 93.77 & 26.18 & 54.73 & 72.85 & 94.83 & 111.29 & 135.25 & 163.84 \\
					Gas Spot Price & 7670 & 6.37 & 3.44 & 2.92 & 3.90 & 5.03 & 8.61 & 12.43 & 18.09 \\
					Gas Futures Price & 7670 & 6.65 & 3.58 & 3.17 & 4.02 & 5.19 & 9.06 & 12.95 & 19.78 \\
					Exchange Rate & 7670 & 1.41 & 0.13 & 1.25 & 1.33 & 1.37 & 1.46 & 1.70 & 1.75 \\
					ECPI & 7670 & 114.06 & 73.83 & 58.94 & 78.66 & 100.37 & 117.26 & 239.98 & 512.56 \\
					GPRD & 7670 & 101.38 & 50.74 & 42.06 & 69.01 & 92.17 & 121.95 & 186.12 & 282.84 \\
					GPRD ACT & 7670 & 98.54 & 64.86 & 23.81 & 57.21 & 87.03 & 125.57 & 209.46 & 309.37 \\
					GPRD THREAT & 7670 & 105.02 & 63.39 & 34.25 & 64.81 & 92.20 & 129.62 & 216.78 & 335.07 \\
					\hline
				\end{tabular}
			\end{minipage}
		}
	}
\end{table}

\subsection{Data Processing}
\label{Data Processing}
\subsubsection{Missing Values}
For random missing features, including 8 in-market features and 4 international macroeconomic features, we utilize time-series imputation techniques from the `imputTS` package in R. The in-market features include primary reserve availability, contingency reserve availability, OCGT generation, reserve cushion, and four offer-related variables. The international macroeconomic features comprise gas spot prices, gas futures prices, oil futures prices, and exchange rates. Importantly, each feature has less than 5\% missing values. Our imputation strategy adopts a comprehensive multi-method time-series approach, leveraging ten distinct techniques: interpolation methods (Linear, Spline, and Stineman interpolation), advanced statistical models (Structural Model with Kalman Smoothing and ARIMA State Space Representation with Kalman Smoothing), and simpler techniques (Last Observation Carried Forward, Next Observation Carried Backward, and imputation using Mean, Median, and Mode values). 

After applying each method individually, we compute the average of the imputed values from all techniques. This averaged value is used to fill the missing data, leveraging the combined strengths of the different methods for a more robust and reliable imputation. This approach ensures accurate handling of features with minimal missing data, maintaining the integrity and consistency of the dataset.

For systemically missing features, specifically the four vesting contract indicators, each with less than 5\% missing values, we apply advanced imputation techniques informed by \citet{du2023pypots}. These methods utilize five Transformer-based models designed for time series analysis: iTransformer, FreTS, SAITS, Transformer, and PatchTST. The missing values for these features are confined to the year 2003, as the vesting contract system was implemented in Singapore only from 2004 onwards. To address this gap, we generate imputed values using each of the five models, then compute their average to derive a robust final imputation. This procedure ensures systematic handling of missing data, preserving dataset consistency and alignment with the overall analytical framework.

For monthly variables, we extend them to a daily frequency by maintaining their values invariant within each month, thereby ensuring temporal consistency across variables. For the USEP\_lag variable, we address the missing value in the initial period by substituting it with the sample median value.

\subsubsection{Feature Normalization and Target Stationarity Adjustment}
Given our models update the parameters monthly, we perform a monthly (0,1) standardization by centering each feature through the subtraction of its monthly mean, then scaling through division by its monthly standard deviation. This process ensures that feature values are normalized within each month, facilitating cross-temporal comparisons and aligning with the requirements of our monthly out-of-sample forecasting framework. The importance of such standardization for effective forecasting has been emphasized in prior studies, including \citet{gu2020empirical} and \citet{bai2008forecasting}.

Additionally, we transform the USEP price series into log-returns, a step aimed at enhancing data stationarity by addressing unit root issues. Using the Augmented Dickey-Fuller (ADF) test \citep{dickey1979distribution} and the Kwiatkowski-Phillips-Schmidt-Shin (KPSS) test \citep{kwiatkowski1992testing}, we initially identified the presence of unit roots in the original price series \footnote{For ADF test, test statistics is -1.472 with \textit{p-value} 0.132, thus failing to reject the null that the process contains a unit root. For KPSS test, the test statistics is 1.163 with \textit{p-value} 0.001, which rejects the null that the process is weakly stationary.}—a common characteristic of high-volatility data such as electricity prices. After applying the log-return transformation, the tests confirmed that the unit root issue was resolved, achieving stationarity. This transformation is critical for time series forecasting \citep{campbell1998econometrics}, as it not only facilitates econometric analysis but also ensures mathematical consistency with price forecasting models \citep{cochrane2005bond}.

\section{Methodology Supplements}
\subsection{An illustration of Sample Splitting Scheme}
\label{An illustration of Sample Splitting Scheme}
\begin{figure}[H]
	\centering
	\caption{\centering Recursive Training, Validation, and Out-of-Sample Testing Process}
	\label{fig:A1}
	\includegraphics[width=1\textwidth]{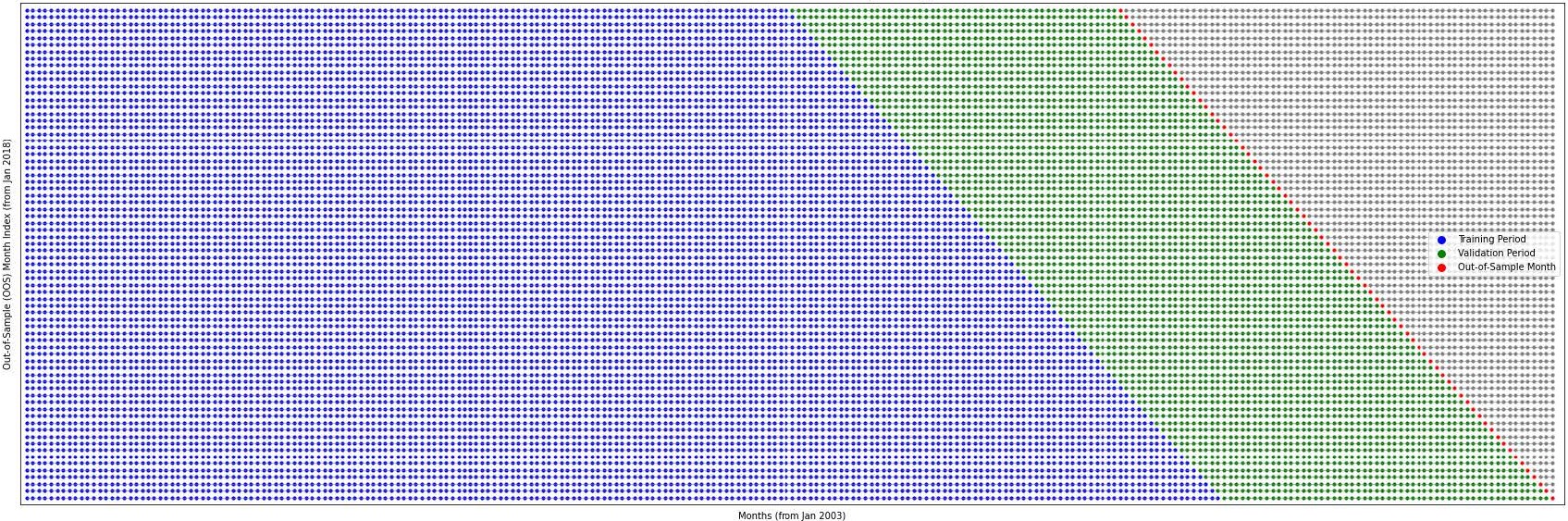}
\end{figure}

For instance, as depicted in Figure \ref{fig:A1}, the vertical axis represents the out-of-sample months, encompassing a total of 72 months, while the horizontal axis corresponds to the entire timeline of 252 months. For the first out-of-sample month (January 2018, the red dot shown in the top row of Figure \ref{fig:A1}), the training period spans January 2003 to June 2013 (126 months, the blue dots), and the validation period covers July 2013 to December 2017 (54 months, the green dots). For the subsequent out-of-sample month (February 2018), the training period is extended by one month, now covering January 2003 to July 2013 (127 months), while the validation period shifts forward by one month, spanning August 2013 to January 2018 (54 months). This process incrementally increases the training period while keeping the validation period length constant at 54 months. Such an iterative procedure ensures that each out-of-sample month is predicted using a model trained on progressively more extensive historical data, with the validation period continuously rolling forward to maintain temporal consistency. For any day $t$ in the first OOS month, it shares the same $t$-specific $\mathcal{T}_{t,1}=\{01Jan2003-31Jun2013\}$, $\mathcal{T}_{t,2}=\{01Jul2013-12Dec2017\}$ and $\mathcal{T}_{t,3}=\{01Jan2018-31Jan2018\}$, and analogous logic applied to other days in different OOS months.

\subsection{Models}
\label{appendix models}
This section describes models in detail, highlighting its two key characteristics. The first is the general framework for conditional mean forecasting, while the second focuses on the objective function used to estimate model parameters. All models aim to minimize the squared forecasting errors (henceforth, MSE). To address specific challenges, modifications such as the Huber loss are introduced, which employs a threshold-based step function to reduce the influence of outliers. Additionally, regularization is incorporated by adding parameterized penalties to the MSE objective, mitigating over-fitting and enhancing out-of-sample performance. As we present each model, we provide clear explanations of its foundational principles to help readers develop a comprehensive understanding of its strengths and limitations in out-of-sample forecasting.

\subsubsection{Simple Linear Models}
We use Ordinary Least Squares (OLS) regression with MSE as a complementary and interpretable benchmark for assessing the performance of advanced machine learning models in predicting USEP returns. Under linear framework, the model specification follows:

\begin{equation}
	\begin{aligned}
		\label{equ:A9}
		f(\mathcal{F}_t; \theta) = \mathcal{F}^{\prime}_t \theta
	\end{aligned}
\end{equation}
where \(\mathbf{z}_t\) represents the predictors at time \(t\), such as fuel prices, and supply cushion, and \(\theta\) is the parameter vector. The OLS objective function is defined as:
\begin{equation}
	\begin{aligned}
		\label{equ:A10}
		\mathcal{L}(\theta) = \frac{1}{T} \sum^T_{t=1} \left(r_{t+1} - f(\mathcal{F}_t; \theta)\right)^2.
	\end{aligned}
\end{equation}
where \(r_{t+1}\) is the realized USEP return at time \(t+1\), and \(T\) is the total number of periods. 

\subsubsection{Penalized Linear Models}
Penalized linear regression methods, such as Lasso, Ridge, and ENet, are particularly effective in addressing the limitations of OLS when applied to high-dimensional datasets. In such scenario, high dimensionality and multicollinearity among predictors can lead to unstable and over-fitted OLS estimates. The statistical model for our penalized linear regression mirrors the simple linear model described in Equation \ref{equ:A11}. Penalized regression extends this framework by incorporating a regularization term into the objective function, for instance, an Elastic Network (henceforth, ENet) penalty is formulated as:
\begin{equation}
	\begin{aligned}
		\label{equ:A11}
		\mathcal{L}(\theta, \lambda, \rho) = \frac{1}{T} \sum_{t=1}^{T} \left(r_{t+1} - f(\mathcal{F}_t; \theta) \right)^2 + \lambda \left[ \rho \sum_{j=1}^{P} |\theta_j| + (1-\rho) \sum_{j=1}^{P} \theta_j^2 \right],
	\end{aligned}
\end{equation}
where \(\lambda > 0\) controls the overall regularization strength,  \(\rho \in [0, 1]\) determines the balance between \(l_1\)- and \(l_2\)-regularization, and \( P \) is the number of predictors. These penalties serve distinct functions, making the choice of \(\rho\) critical for model performance. Lasso (\(\rho = 1\)) employs an \(l_1\)-regularization, which results in parsimonious model by setting some coefficients to exactly zero, thus exceling in selecting the variables. Ridge (\(\rho = 0\)), on the other hand, applies an \(l_2\)-regularization, shrinking all coefficients toward zero without reducing any to exactly zero. This approach is particularly effective in handling multicollinearity, ensuring stable and robust parameter estimates even when predictors are highly correlated. ENet (\(0 < \rho < 1\)) combines the strengths of Lasso and Ridge by balancing variable selection and coefficient shrinkage. 

\subsubsection{Dimension Reduction Models}
Dimension reduction methods, such as PCR and PLS, offer a compelling alternative to penalized linear models, particularly in high-dimensional datasets where predictors are highly correlated or exhibit multicollinearity. Penalized linear models achieve regularization through shrinkage and variable selection by penalizing the magnitude of regression coefficients. While effective in reducing over-fitting, penalized linear models can yield suboptimal results when predictors are strongly correlated, as the penalty term may overly constrain coefficients and lead to the exclusion of relevant predictors. For example, in cases where all predictors are equally informative but contain \textit{iid} noise, penalized models might fail to effectively capture the underlying signal, resulting in poor predictive performance. Dimension reduction methods address this limitation by aggregating predictors into a smaller set of components that encapsulate the most informative structure in the data, thereby mitigating the noise and correlation issues.

PCR achieves dimension reduction through a two-step process. First, Principal Component Analysis (PCA) identifies a small set of linear combinations of predictors that preserve the maximum variance and covariance structure among the predictors. Second, these components are used in a standard OLS regression model to predict the target variable. This approach regularizes the prediction problem by zeroing out coefficients on components with low variance, thereby reducing over-fitting. However, its emphasis on preserving the covariance structure of predictors often overlooks the direct relationship between predictors and the response variable, as the choice of components is made independently of the forecasting objective. Consequently, PCR may select components that retain shared variation among predictors but are less relevant for predicting the target.

PLS extends the dimension reduction framework by explicitly considering the relationship between predictors and the target variable. Unlike PCR, which selects components based solely on predictors' variance, PLS components are constructed to maximize their covariance with the response variable, ensuring stronger predictive associations. For each component, PLS assigns weights to predictors based on their univariate association with the response, thereby prioritizing variables with higher explanatory power while de-emphasizing weaker predictors. This targeted construction of components not only enhances predictive accuracy but also aligns more closely with the forecasting objective. In this sense, PLS addresses the primary limitation of PCR and represents a more refined approach to dimension reduction.

Mathematically, both methods can be expressed as reducing the dimensionality of predictors from \(P\) to \(K\) through a linear transformation:
\begin{equation}
	\begin{aligned}
		R = (\mathcal{F}\Omega_K)\theta_K + \tilde{E},
	\end{aligned}
\end{equation}
where \(R\) is the $T \times 1$ vector of $r_{i,t+1}$, \(\mathcal{F}\) is the $T \times P$ matrix of stacked predictors $\mathcal{F}_t$, \(\Omega_K\) is the transformation matrix that condenses predictors into \(K\) components, and \(\theta_K\) is the regression coefficient vector for the reduced components. In PCR, \(\Omega_K\) is derived through singular value decomposition to maximize predictor variance, while in PLS, \(\Omega_K\) is optimized to maximize the covariance between predictors and the response. Finally, given estimated $\Omega_K$, $\theta_K$ is estimated via OLS regression of $R$ on $Z \Omega_K$. For both models, \(K\) serves as a hyperparameter and is determined adaptively using the validation sample. In our analysis, we utilize the MSE objective function for both PCR and PLS. 

\subsubsection{Generalized Linear Model}
Linear models, while simple and interpretable, often fail to capture the complex relationships present in real-world data. These models assume a fixed linear relationship between predictors and response variables, which can lead to biased and overly simplistic estimates when the true relationship is nonlinear or involves intricate interactions.

To address these limitations, we utilize nonparametric methods, starting with GLM. This approach incorporates nonlinear transformations of the original predictors as additional additive terms within an otherwise linear framework. Our model achieves this enhancement through a \( K \)-term spline series expansion of the predictors. The generalized formulation is given by:
\begin{equation}
	f(\mathcal{F}_t; \theta, p(\cdot)) = \sum_{j=1}^P p({\mathcal{F}_t}_j) \theta_j,
\end{equation}
where \( p(\cdot) = (p_1(\cdot), p_2(\cdot), \dots, p_K(\cdot))^\prime \) represents a vector of basis functions, and the parameters \( \theta \) are organized as a \( K \times P \) matrix.  \( K \) is the number of spline terms per predictor. This formulation enables the model to flexibly adapt to nonlinear patterns in the predictors, thereby enhancing its ability to capture complex relationships.

In this analysis, we use splines of series order two, which extend the modeling framework to capture smooth nonlinear patterns in the data. For a predictor variable \( z \), the spline of series order two is expressed as:
\begin{equation}
	p(z) = \left( 1, z, (z - c_1)^2, (z - c_2)^2, \dots, (z - c_{K-2})^2 \right),
\end{equation}
where \( c_1, c_2, \dots, c_{K-2} \) are the nodes that partition the predictor space into regions. The linear term \( z \) captures global trends, while the quadratic terms \((z - c_k)^2\) capture local nonlinear variations around the nodes. This extension enables the model to approximate complex nonlinear relationships without imposing overly restrictive assumptions.

To ensure interpretability and enforce sparsity, we apply group lasso regularization to the spline-extended predictors. The penalty function is defined as:
\begin{equation}
	\label{equ:A15}
	\phi(\theta; \lambda, K) = \lambda \sum_{j=1}^P \left( \sum_{k=1}^K \theta_{j,k}^2 \right)^{1/2},
\end{equation}
where \( \lambda \) controls the strength of regularization, and \( \theta_{j,k} \) is the coefficient of the spline term. This penalty ensures that all spline terms associated with a predictor are either retained or excluded together, facilitating variable selection while preserving the ability to model nonlinear effects. We use MSE as the objective in the GLM framework, augmented by the penalty in Equation \ref{equ:A15}. The tuning parameters for this approach are \( \lambda \) and \( K \):
\begin{equation}
	\label{equ:A16}
	\mathcal{L}(\theta; \lambda, K) = \frac{1}{T} \sum^T_{t=1} \left(r_{t+1} - f(\mathcal{F}_t; \theta)\right)^2 + \lambda \sum_{j=1}^P \left( \sum_{k=1}^K \theta_{j,k}^2 \right)^{1/2}.
\end{equation}

\subsubsection{Tree-based Ensemble Models}
The GLM effectively captures the nonlinear impact of individual predictors on conditional mean of return. However, it does not account for interactions among predictors. Introducing multivariate functions to represent interactions significantly increases model complexity, as expanding univariate predictors with \( K \) basis functions multiplies the number of parameters by a factor of \( K \), while incorporating multiway interactions increases parameterization combinatorially. Without a priori assumptions about which interactions to include, such an approach becomes computationally infeasible.

Regression trees offer a nonparametric alternative that inherently captures interactions among predictors without explicitly predefining them. Unlike traditional regression models, regression trees divide the predictor space into partitions and approximate the outcome variable within each partition using simple averages. This property makes regression trees particularly effective for modeling complex relationships and interactions among predictors.

A regression tree grows by recursively splitting the data into bins based on predictor variables. Each split creates ``branches'' that segment the data into subsets, approximating the target function \( f(\cdot) \). This process continues until the data is divided into \( K \) partitions (or ``leaves''), resulting in a tree of depth \( L \). The prediction from a regression tree can be formally expressed as:
\begin{equation}
	\begin{aligned}
		f(\mathcal{F}_t; \theta, K, L) = \sum_{k=1}^{K} \theta_k \mathbf{1}_{\{\mathcal{F}_t \in C_k(L)\}},
	\end{aligned}
\end{equation}
where \( C_k(L) \) denotes the \( k \)-th partition at depth \( L \), \( \theta_k \) is the average value of the target variable within partition \( C_k(L) \), \( \mathbf{1}_{\{\mathcal{F}_t \in C_k(L)\}} \) is an indicator function that evaluates to 1 if \( \mathcal{F}_t \) belongs to partition \( C_k(L) \). 

The partitions \( C_k(L) \) are defined by the sequential splits of the tree, which are determined to minimize prediction error. Growing a tree involves identifying the predictor variable and split value that minimize forecast error, often measured by impurity. A common impurity metric is the \( \l_2 \) loss:
\begin{equation}
	\begin{aligned}
		H(\theta, C) = \frac{1}{|C|} \sum_{\mathcal{F}_t \in C} \left( r_{t+1} - \theta \right)^2,
	\end{aligned}
\end{equation}
where \( |C| \) is the number of observations in partition \( C \), \( r_{t+1} \) is the target variable. Given \(C \), it is clear the optimal value  \( \theta = \frac{1}{|C|} \sum_{\mathcal{F}_t \in C} r_t \) can minimize \( H \). Tree construction stops when the tree reaches a predefined depth \( L \), a maximum number of leaves \( K \), or when further splits do not significantly reduce impurity.

Regression trees present several advantages. First, they inherently model complex nonlinear relationships and interactions without requiring predefined assumptions. They are capable of handling both categorical and numerical predictors, making them suitable for diverse datasets. Additionally, their partitioning structure offers intuitive insights into decision-making rules, enhancing interpretability. Furthermore, predictions generated by regression trees are invariant to monotonic transformations of predictors, ensuring robustness across varying scales of input features.

Despite their advantages, regression trees exhibit several limitations. They are prone to overfitting, especially when grown too deep, resulting in excessive segmentation of the data and poor generalization to unseen samples. Additionally, single trees are highly sensitive to minor variations in the training data, leading to unstable predictions. To address these shortcomings, we incorporate three ensemble-based techniques: XGB, LGBM, and RF. These methods combine multiple regression trees to enhance stability, accuracy, and robustness.

XGB builds on the gradient boosting framework with innovations that enhance speed, accuracy, and robustness. It minimizes a Taylor-expanded loss function, with the default objective being MSE, and Huber optionally selected through validation to achieve the smallest validation loss. Regularization is applied through \( l_1 \)- and \( l_2 \)-regularization on leaf weights, helping to reduce overfitting and improve generalization. The learning rate (\( \eta \)) controls the step size in updating model parameters during optimization. A smaller learning rate allows for finer updates but requires more iterations, while a larger learning rate accelerates convergence but may lead to suboptimal solutions. To prevent overly complex trees, XGB introduces a minimum loss reduction threshold, \( \gamma \), which prunes splits that do not significantly decrease the loss. Key hyperparameters include \( n \), the number of trees; \( D \), the maximum depth of each tree; and \( \eta \), the learning rate. These parameters are optimized using validation data to ensure out-of-sample predictive performance.

In contrast, LGBM is designed with a focus on efficiency and scalability, making it particularly suitable for large datasets and high-dimensional features. It achieves computational efficiency through techniques such as histogram-based splitting and gradient-based one-side sampling, which significantly reduce memory usage and computational overhead while maintaining high predictive accuracy. Unlike the traditional level-wise growth used in many gradient boosting frameworks, LGBM adopts a leaf-wise growth strategy, where the leaf with the highest potential loss reduction is selected for splitting. This approach improves model accuracy with fewer trees but requires careful regularization to avoid overfitting. Furthermore, LGBM discretizes predictor variables into bins, simplifying the process of finding optimal split points. LGBM minimizes a differentiable loss function, allowing for the selection of MSE or a custom Huber objective based on validation performance. Similar to XGB, it employs regularization to iteratively manage model complexity and enhance generalization. The main hyperparameters include \( n \), the number of trees; \( D \), the maximum depth of each tree; and \( \eta \), the learning rate, all of which are tuned using validation data.

RF, in contrast to gradient boosting methods, is a bagging-based ensemble approach \citep{breiman2001random} that aggregates predictions from multiple independently constructed decision trees. Each tree is trained on a bootstrap sample of the data, introducing randomness and creating variation in the training dataset. At each split, RF considers a random subset of predictors, controlled by the hyperparameter \( F \), which specifies the maximum number of features used for splitting, thereby decorrelating the trees and improving ensemble stability. RF minimizes the MSE loss function, ensuring that predictions are optimized for the mean squared error criterion. Key hyperparameters include \( n \), the number of trees; \( D \), the maximum depth of each tree; and \( F \), the maximum number of split features. By averaging the predictions of these trees, RF effectively reduces variance, enhances diversity, and achieves robust generalization.

These three ensemble-based methods excel at capturing complex nonlinear relationships and interactions, particularly in high-dimensional datasets with multicollinearity. While XGB and LGBM focus on gradient boosting to iteratively minimize errors, RF leverages bagging to stabilize predictions. Each method has distinct strengths: XGB is robust and precise, making it ideal for applications demanding fine-tuned performance; LGBM is efficient and scalable, making it optimal for large-scale datasets and high-dimensional features; and RF is simple and interpretable, making it well-suited for preliminary analysis.

\subsubsection{Neural Network Models}
\label{appendix models nn}
ANN is the final nonlinear method explored in our analysis, known for its remarkable flexibility and capacity to model complex relationships.  As ``universal approximators", ANNs are theoretically capable of learning any smooth functional mapping under appropriate conditions. The strength of ANNs lies in their hierarchical structure, which facilitates the modeling of deep, nonlinear interactions among predictors—a concept commonly referred to as ``deep learning". However, this adaptability comes with some challenges, including increased computational demands and reduced interpretability, making ANNs some of the most parameter-intensive and least transparent modeling techniques available.

In our analysis, we use traditional ``feed-forward'' ANN networks (FFN), which consist of three fundamental components: an input layer, one or more hidden layers, and an output layer. The input layer directly represents the raw predictors, while the hidden layers capture complex relationships by applying nonlinear transformations to the input data. Finally, the output layer aggregates the transformed information to produce the final prediction. Each layer is composed of computational units called ``neurons", and the connections between neurons in consecutive layers, often referred to as ``synapses", transmit and transform information. 

To enable the modeling of nonlinear relationships, FFNs introduce hidden layers between the input and output layers. Each neuron in a hidden layer processes a weighted linear combination of its inputs and applies a nonlinear activation function \( f(\cdot) \), allowing the network to learn complex patterns in the data. The output of a neuron \( x_j^{(l)} \) in layer \( l \) can be expressed as:
\begin{equation}
	x_j^{(l)} = f\left( \sum_{i=1}^{n^{(l-1)}} w_{ij}^{(l)} x_i^{(l-1)} + b_j^{(l)} \right),
\end{equation}
where \( w_{ij}^{(l)} \) represents the weight connecting neuron \( i \) in layer \( l-1 \) to neuron \( j \) in layer \( l \), \( b_j^{(l)} \) is the bias term for neuron \( j \),  \( x_i^{(l-1)} \) is the output of neuron \( i \) in the previous layer, and \({n^{(l-1)}}\) is the number of neurons in layer \(l-1\). The final prediction of the FFN is obtained by propagating the transformed values through all layers, culminating in the output layer. For a network with \( L \) layers, the prediction \( \hat{r} \) can be expressed as:

\begin{equation}
	\hat{r} = f_{\text{out}}\left( \sum_{i=1}^{n^{(L-1)}} w_{i}^{(L)} x_i^{(L-1)} + b^{(L)} \right),
\end{equation}
where \( f_{\text{out}}(\cdot) \) is the activation function of the output layer, and \( n^{(L-1)} \) denotes the number of neurons in the final hidden layer. FFNs are powerful due to their ability to approximate complex nonlinear functions, with the hidden layers serving as feature extractors that learn representations from the raw input data. The general structure of FFN is shown as the following figure:

\begin{figure}[H]
	\caption{\centering Structure of NN3 (3 hidden layers) with 6, 6, 8 neurons in each layer}
	\label{fig:A2}
	\centering
	\includegraphics[width=0.8\textwidth]{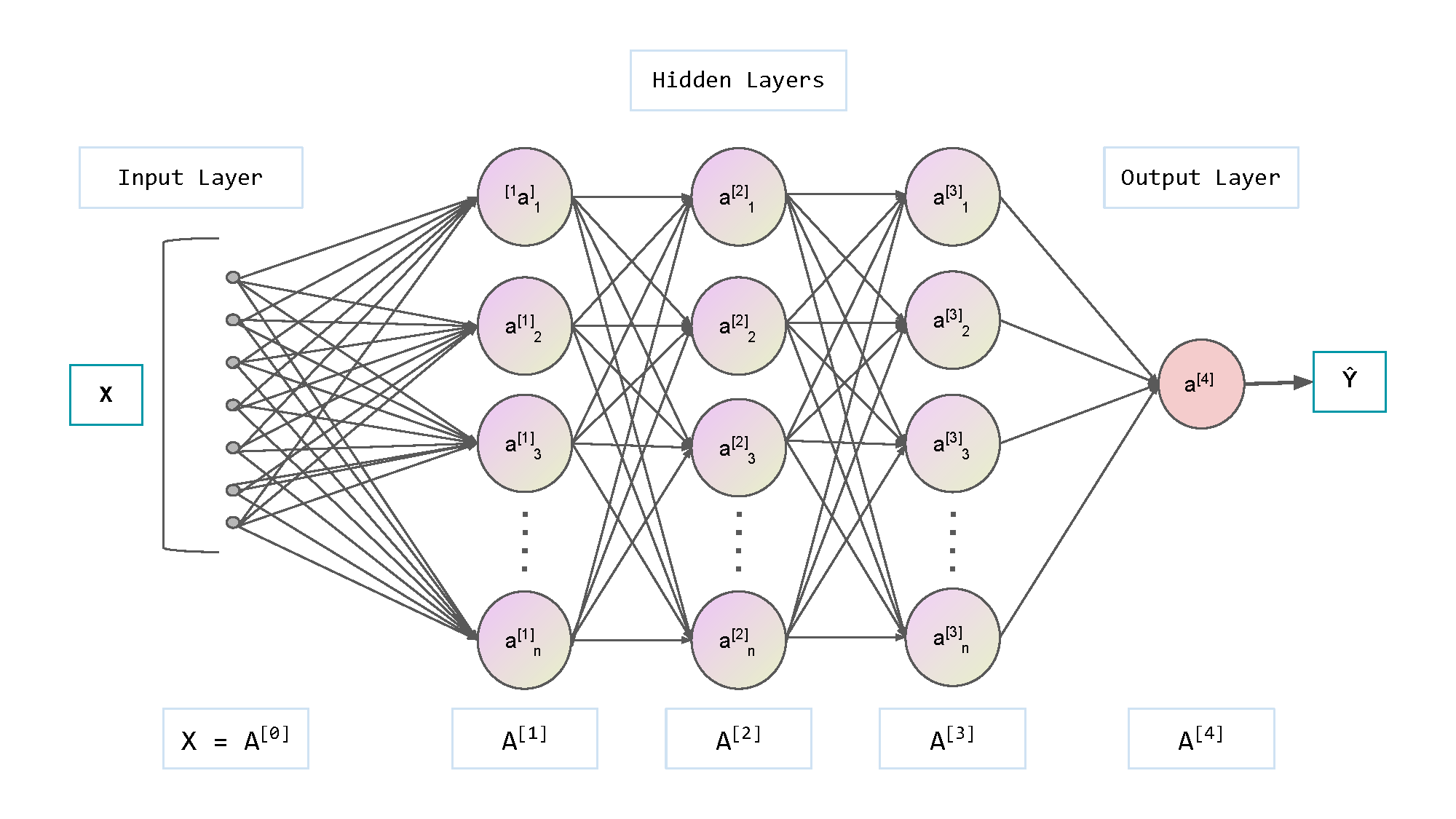}
\end{figure}

The design of FFNs requires careful consideration of key architectural elements, including the number of hidden layers, the number of neurons per layer, and the connectivity patterns between layers. FFNs are grounded in the theoretical framework of "universal approximators", capable of representing any smooth predictive association \citep{hornik1989multilayer, cybenko1989approximation}. Notably, deeper architectures often achieve superior performance with fewer parameters by leveraging hierarchical feature representations. However, increased network depth introduces several challenges, such as a higher risk of overfitting, greater optimization complexity, and issues related to numerical stability during training, including exploding or vanishing gradients.

Training deep neural networks is particularly challenging due to the large number of parameters and the inherently nonconvex nature of the optimization problem. This complexity makes it difficult to find globally optimal solutions, often requiring advanced optimization techniques to navigate the highly irregular loss landscape. Furthermore, the process of computing gradients and updating weights through backpropagation can encounter numerical instability. As the network depth increases, gradients may either grow excessively large (exploding gradients) or diminish to near zero (vanishing gradients), which hinders effective weight updates and slows convergence.

Given the computational infeasibility of exhaustively searching through all possible network configurations, we adopt a structured approach by predefining a range of architectures and evaluating their performance empirically. This strategy allows us to balance the trade-offs between model complexity and generalization, providing a systematic framework for optimizing neural network performance in practice.

To assess the impact of network complexity on performance, we construct architectures with up to five hidden layers, with the number of neurons per layer decreasing progressively based on a geometric pyramid rule. This design strategy balances the model's capacity for learning complex patterns with the risk of overfitting. The predefined architectures are outlined as follows:\\

\textbf{NN1}: One hidden layer with \(2^{10}\) neurons.\\

\textbf{NN2}: Two hidden layers with \(2^{10}\) and \(2^9\) neurons.\\

\textbf{NN3}: Three hidden layers with \(2^{10}\), \(2^9\), and \(2^8\) neurons.\\

\textbf{NN4}: Four hidden layers with \(2^{10}\), \(2^9\), \(2^8\), and \(2^7\) neurons.\\

\textbf{NN5}: Five hidden layers with \(2^{10}\), \(2^9\), \(2^8\), \(2^7\), and \(2^6\) neurons.\\

Each architecture is fully connected, and every neuron in a layer receives input from all neurons in the preceding layer. This comprehensive connectivity facilitates the network's ability to capture complex relationships within the data. By evaluating these architectures, we aim to quantify how increasing depth influences predictive performance and determine the optimal balance between model complexity and generalization. This structured approach offers a practical framework for designing neural networks tailored to datasets of varying sizes and complexities.

To introduce nonlinearity, we employ the rectified linear unit (ReLU) as the activation function for all neurons. ReLU is defined as:
\begin{equation}
	\text{ReLU}(x) = 
	\begin{cases} 
		0 & \text{if } x < 0, \\
		x & \text{if } x \geq 0,
	\end{cases}
\end{equation}
which encourages sparsity in activations and allows for efficient gradient computation.

As previously defined, the general structure of our neural network computes the output recursively through layers, with each neuron applying a non-linear activation function to the weighted sum of its inputs. The final output aggregates these computations across all layers. For completeness, we briefly summarize the key components. The input layer, represented as \( \mathbf{x}^{(0)} = [1, z_1, \dots, z_N]^\top \), includes the predictors augmented with a bias term. For each hidden layer \( l > 0 \), the output of neuron \( k \) is given by:
\(
x_k^{(l)} = \mathrm{ReLU}\left(\mathbf{x}^{(l-1)\prime} {\theta}_k^{(l-1)}\right),
\)
where \( {\theta}_k^{(l-1)} \) denotes the weights associated with neuron \( k \).
The final output is:
\(
g(\mathbf{z}_t; {\theta}) = \mathbf{x}^{(L-1)\prime} {\theta}^{(L-1)},
\)
The number of parameters in each hidden layer \( l \) is \( K^{(l)} \times \left(1 + K^{(l-1)}\right) \), with an additional \( 1 + K^{(L-1)} \) parameters for the output layer.

To estimate the parameters of the neural network, we minimize a \( l_1\) loss function on the prediction errors. Unlike tree-based models, which require step-by-step optimization of parameters, neural networks update all parameters simultaneously in each optimization step. This joint optimization process provides a significant computational advantage. However, the high complexity and nonconvexity of neural networks, combined with their extensive parameterization, make direct optimization computationally expensive and challenging. A practical solution to this issue is stochastic gradient descent (SGD), which approximates the gradient at each iteration by using small, randomly sampled subsets (batches) of the training data. SGD accelerates the optimization process by trading off some precision in gradient evaluation for substantial computational efficiency. This approach allows neural networks to be trained on large datasets while maintaining manageable computational costs. However, due to the inherent complexity of neural networks, regularization techniques are essential to prevent overfitting and ensure robust generalization.

In addition to \( l_1 \)-based weight penalization, we adopt four complementary regularization strategies, including adaptive learning rate scheduling, early stopping, batch normalization, and ensembling. These techniques collectively address the challenges of training deep neural networks, such as parameter overfitting and instability during optimization.

A critical factor in SGD optimization is the learning rate, which determines the step size for parameter updates. If the learning rate remains constant as the optimization progresses, it can lead to instability or premature convergence. To address this, we employ an adaptive learning rate scheduling approach \citep{kingma2014adam}, which dynamically reduces the step size as gradients approach zero, ensuring more stable convergence and mitigating the effects of gradient noise.

Early stopping is another key regularization technique. This approach begins with an initial parameter guess, such as weights initialized near zero. During optimization, parameters are iteratively updated to minimize errors on the training set, but predictions are also evaluated on a separate validation set. Optimization halts when validation errors start to increase, even if training errors continue to decrease. By stopping early, the model avoids overfitting to the training data and achieves better generalization.

Batch normalization \citep{ioffe2015batch} addresses a specific issue known as internal covariate shift, where the input distributions to hidden layers shift unpredictably across training iterations. For each mini-batch, the algorithm standardizes the inputs to each hidden layer by centering and scaling them to have a mean of zero and unit variance. This stabilization improves the training process, accelerates convergence, and enhances the generalization capability of the network, particularly for deep architectures.

Lastly, we employ an ensemble strategy \citep{hansen1990neural, dietterich2000ensemble} to reduce prediction variance and improve robustness. Multiple neural networks are trained using different random initialization seeds, and their predictions are averaged to produce the final output. This ensemble approach mitigates the stochastic variability introduced by SGD and ensures more stable and reliable predictions.

\subsubsection{Grid of Hyperparameter}
\begin{table}[H]
	\scriptsize
	\centering
	\caption{Model Frameworks, Hyperparameters, Mnemonics, and Values}
	\label{tab:A4}
	\setlength{\tabcolsep}{2pt}
	\renewcommand{\arraystretch}{1.1}
	\begin{threeparttable}
		\begin{tabular}{p{1.5cm} p{2cm} p{5cm} p{0.2cm} p{9cm}}
			\hline
			\textbf{Model} & \textbf{Framework} & \textbf{Hyperparameters} && \textbf{Values} \\
			\hline
			LASSO & sklearn & $\alpha$: L1 penalty && \texttt{\{'alpha': np.linspace(1e-4, 1e-1, 50)\}} \\
			Ridge & sklearn & $\alpha$: L2 penalty && \texttt{\{'alpha': np.linspace(1e-4, 1e-1, 50)\}} \\
			ENet & sklearn & (1) $\alpha$: L1 penalty; (2) l1\_ratio: L1 to L2 penalty ratio && \texttt{\{'alpha': np.linspace(1e-4, 1e-1, 50), 'l1\_ratio': [0.1, 0.3, 0.5, 0.7, 0.9]\}} \\
			PLS & sklearn & n\_components: number of components && \texttt{\{'n\_components': [i for i in range(1, 201)]\}} \\
			PCR & sklearn & n\_PCs: number of components && \texttt{\{'n\_PCs': [i for i in range(1, 201)]\}} \\
			GLM & group\_lasso & (1) knots: number of knots to tick the groups label; (2) lmd: the group sparsity penalty; (3) l1\_reg: the coefficient sparsity penalty && \texttt{\{'knots': [3], 'lmd': np.linspace(1e-4, 1e-1, 50), 'l1\_reg': [1e-4, 1e-3, 1e-2, 1e-1]\}} \\
			RF & sklearn & (1) n\_estimators: number of trees; (2) max\_depth: max depth of each tree; (3) max\_features: max number of split features && \texttt{\{'n\_estimators': [100, 200, 300, 400], 'max\_depth': [1, 2, 3, 4, 5, 6], 'max\_features': [5, 10, 20, 30, 50, 70, 100], 'random\_state': [20000524]\}} \\
			LGBM(+H) & lightgbm & (1) objective: default is MSE, self-defined is $huber\_obj$; (2) max\_depth: max depth of each tree; (3) n\_estimators: number of trees; (4) learning\_rate: optimization rate && \texttt{\{'objective': [None, huber\_obj], 'max\_depth': [1, 2, 3, 4, 5, 6], 'n\_estimators': [100, 200, 300, 400], 'random\_state': [20000524], 'learning\_rate': [0.001, 0.01, 0.1]\}} \\
			XGB(+H) & xgboost & (1) n\_estimators: \# of trees; (2) max\_depth: max depth of each tree; (3) learning\_rate: optimization rate; (4) objective: default is MSE; (5) huber\_slope: $\xi$ in the Huber loss, the threshold && \texttt{\{'n\_estimators': [100, 200, 300, 400], 'max\_depth': [1, 2, 3, 4, 5, 6], 'random\_state': [20000524], 'learning\_rate': [0.001, 0.01, 0.1], 'objective': [None, 'reg:pseudohubererror'], 'huber\_slope': [max(np.quantile(np.abs(residuals), q), 1) for q in quantiles]\}} \\
			NN1-NN5 & TensorFlow & (1) n\_layers: number of dense layers; (2) l1: L1 penalty; (3) epochs: number of training iterations; (4) learning\_rate: optimization rate; (5) base\_neurons: $\#$ of neurons in each layer; (6) monitor='val\_loss': for early stropping, the criteria related to loss; (7) ensemble=10: less random-driven; (8) activation='relu', activation function; (9) batch\_size; (10) BatchNormalization; (11) optimizers.Adam; (12) patience: for early stropping, the criteria related to $\#$ of consecutive rounds  && \texttt{\{'n\_layers': [1 , 2, 3, 4 , 5], 'loss': ['mse'], 'l1': [1e-5, 1e-4, 1e-3], 'learning\_rate': [.001, .01, .1], 'batch\_size': [int(X\_train.shape[0]/50)], 'epochs': [100], 'random\_state': [20000524], 'BatchNormalization': [True], 'patience':[5], 'verbose': [0], 'monitor':['val\_loss'],  'dropout\_rate': [0.2], 'base\_neurons': [5, 10]\}} \\
			\hline
		\end{tabular}
		\begin{tablenotes}
			\fontsize{8pt}{0.35cm}\selectfont
			\item \textit{Notes}: The table summarizes key model frameworks, hyperparameters, and value ranges for machine learning algorithms.
		\end{tablenotes}
	\end{threeparttable}
\end{table}

\subsection{Statistical Tests}
\label{appendix Statistical Tests}
\subsubsection{Diebold-Mariano Test} 
\label{appendix DM}
We adapt the \citet{francis1995comparing} test to our context by comparing the time-series average of prediction errors from each model. Specifically, to evaluate the forecasting performance of method (1) relative to method (2), we define the test statistic as \(DM_{1,2} = \bar{d}_{1,2} / \widehat{\sigma}_{\bar{d}_{1,2}}\), where  
\begin{equation}
	\begin{aligned}
		d_{1,2; t+1} = \frac{1}{\mathcal{T}_{3, t+1}} \sum_{t=1}^{\mathcal{T}_{3, t+1}}\left[\left(\widehat{e}_{t+1}^{(1)}\right)^2 - \left(\widehat{e}_{t+1}^{(2)}\right)^2\right],
	\end{aligned}
\end{equation}
\(\widehat{e}_{t+1}^{(1)}\) and \(\widehat{e}_{t+1}^{(2)}\) denote the prediction errors for \(r_{t+1}\) at time \(t\) from each method, and \(\mathcal{T}_{3, t+1}\) is the number of days in the testing sample (out-of-sample month \(t+1\)). The terms \(\bar{d}_{1,2}\) and \(\widehat{\sigma}_{\bar{d}_{1,2}}\) represent the mean and the Newey-West standard error of \(d_{1,2; t+1}\) over the testing sample, respectively.  

This modified DM test statistic, which is now based on a single time series \(d_{1,2; t+1}\) of error differences with minimal autocorrelation, is more likely to satisfy the regularity conditions required for asymptotic normality, thereby providing reliable \(p\)-values for model comparison.

The DM statistic can also be interpreted as the \(t\)-statistic associated with the intercept \(\alpha\) in the following regression:  
\begin{equation}
	\begin{aligned}
		(r_{t+1} - \widehat{r}_{t+1}^{(1)})^2 - (r_{t+1} - \widehat{r}_{t+1}^{(2)})^2 = \alpha + \varepsilon_{t+1},
	\end{aligned}
\end{equation}
where \(\varepsilon_{t+1}\) is the error term. The null hypothesis asserts that the two models exhibit equivalent predictive performance (\(\alpha = 0\)), while the alternative hypothesis posits a difference in their predictive accuracy (\(\alpha \neq 0\)). 

\subsubsection{Equal Predictive Ability Test}
\label{Equal Predictive Ability Test}
However, as emphasized by \citet{diebold2015comparing}, the model-free DM test is intended for comparing forecasts, not intended for comparing models. For robustness check, we also apply other widely-used equal predictive ability (EPA) tests \citep{bianchi2021bond, goulet2022machine}, such as the MSFE-adjusted CW \citep{clark2007approximately} test, Model Confidence Sets  (MCS) \citep{hansen2011model} test, and Giacomini and White (GW) \citep{giacomini2006tests} test, to comapre the relative performance between sophisitcated models and simple OLS.

\paragraph{Clark-West Test}
The \citet{clark2007approximately} test is an adjusted version of the DM test designed to address specific challenges in comparing forecasting models, particularly in the presence of bias or nested models. The CW test introduces a bias adjustment term to the test statistic. This adjustment is essential when comparing forecasts that may be biased, as the standard DM test could incorrectly favor one forecast over another due to inherent bias. By incorporating this adjustment, the CW test provides a more accurate assessment of predictive accuracy. The test is particularly well-suited for comparing nested models, where one model is a restricted version of another. In such scenarios, the DM test may underestimate the variance of forecast error differences, leading to unreliable results. The CW test corrects for this by modifying the test statistic to account for the nesting structure, ensuring more robust comparisons.

The CW statistic can be interpreted as the \(t\)-statistic corresponding to the intercept \(\alpha\) in the following regression equation:  
\begin{equation}
	\begin{aligned}
		(r_t - \widehat{r}^{(1)}_{t+1})^2 - (r_t - \widehat{r}^{(2)}_{t+1})^2 + \underbrace{(\widehat{r}^{(1)}_{t+1} - \widehat{r}^{(2)}_{t+1})^2}_{\text{bias adjustment}} = \alpha + \varepsilon_t,
	\end{aligned}
\end{equation}
where \(\varepsilon_t\) represents the error term. The term \((\widehat{r}^{(1)}_{t+1} - \widehat{r}^{(2)}_{t+1})^2\) serves as a bias adjustment, ensuring a fair and robust comparison between models. Under the null hypothesis, the two models have equivalent predictive performance (\(\alpha = 0\)). The alternative hypothesis posits differing predictive accuracy between the models (\(\alpha \neq 0\)).

\paragraph{Model Confidence Sets Test}
The MCS test, introduced by \citet{hansen2011model}, provides a robust framework for comparing multiple forecasting models by identifying a set of models that cannot be statistically rejected as having the best predictive performance. Unlike pairwise tests such as the DM or CW tests, the MCS procedure allows for simultaneous comparison of multiple models, thus mitigating the risk of multiple testing biases.

The MCS test proceeds iteratively by evaluating a loss differential measure, such as the mean squared forecast error (MSFE), across all models. At each iteration, the model with the worst performance (based on a statistical test, such as a \(t\)-test or \(F\)-test) is removed from the set. The procedure continues until no model can be rejected at the chosen confidence level.

Mathematically, for \(k\) competing models with losses \(\ell_{t,j}\) for model \(j\) at time \(t\), the loss differential between models \(j\) and \(l\) is defined as:
\[
d_{j,l;t} = \ell_{t,j} - \ell_{t,l}.
\]

The MCS statistic is based on these loss differentials, and the null hypothesis assumes that all models in the current set have equal predictive ability. The test outputs the final "confidence set" of models, which contains those not rejected as inferior at the specified significance level. The MCS procedure is particularly useful in our context, where we aim to evaluate multiple sophisticated models against simpler benchmarks such as OLS, to identify a subset of models that demonstrate superior performance.

\paragraph{Giacomini-White Test}
The GW test, developed by \citet{giacomini2006tests}, is designed to evaluate conditional predictive ability, making it particularly suitable for comparing models when their performance may vary over time or under different economic conditions. Unlike the DM and CW tests, which focus on unconditional predictive ability, the GW test considers whether one model consistently outperforms another across varying states of the data-generating process.

The GW test is formulated as a regression of the loss differential between two models on a set of state-dependent variables that capture the prevailing economic or market conditions. For two competing models with losses \(\ell_{t,j}\) and \(\ell_{t,l}\), the GW regression is:
\[
d_{j,l;t} = \ell_{t,j} - \ell_{t,l} = \alpha + \boldsymbol{z}_t^\top \boldsymbol{\beta} + \varepsilon_t,
\]
where \(\boldsymbol{z}_t\) represents a vector of state variables (e.g., lagged returns, volatility measures, or macroeconomic indicators); \(\boldsymbol{\beta}\) measures the influence of these state variables on the loss differential.

Under the null hypothesis, the models have equal conditional predictive ability, i.e., \(\boldsymbol{\beta} = 0\). The GW test evaluates whether \(\boldsymbol{\beta}\) is statistically different from zero, indicating that the performance differential between the models is influenced by the state variables.

In our application, the GW test is particularly insightful for assessing whether sophisticated models outperform simpler benchmarks like OLS under specific market conditions, such as periods of high volatility or market stress. By incorporating state variables, the GW test complements unconditional comparisons and offers a nuanced view of model performance dynamics.

\subsubsection{Huber Loss}
The smooth and differentiable nature of MSE facilitates efficient optimization. However, MSE's  excessively high sensitivity to outliers presents a drawback, as large residuals are squared, disproportionately influencing the objective function and distorting parameter estimates.

In contrast, MAE provides robustness against outliers by applying a linear penalty to residuals, thereby limiting the influence of extreme values on parameter estimates. Nevertheless, MAE lacks sensitivity to small deviations and is non-smooth, posing challenges in optimization due to the presence of non-differentiable points.

To address these limitations, we adopt the Huber loss function, which combines the strengths of both MSE and MAE while mitigating their respective drawbacks. The Huber loss achieves this balance through a piecewise formulation that transitions between quadratic and linear behaviors. Specifically:
\begin{equation}
	\begin{aligned}
		H(x; \xi) =
		\begin{cases} 
			\frac{1}{2}x^2 & \text{if } |x| \leq \xi, \\
			\xi|x| - \frac{1}{2}\xi^2 & \text{if } |x| > \xi,
		\end{cases}
	\end{aligned}
\end{equation}
where \(\xi\) is a threshold parameter. For residuals \(|x| \leq \xi\), the Huber loss behaves like MSE, penalizing small deviations to preserve sensitivity. For residuals \(|x| > \xi\), the Huber loss transitions to MAE, reducing the influence of outliers. The hybrid nature of the Huber loss makes it particularly well-suited for environments like the electricity market, which are characterized by high volatility and frequent outliers. 

\subsubsection{Bai-Perron Multiple Breakpoints Test}

The Bai-Perron Multiple Breakpoints Test \citep{bai1998estimating} is a statistical framework for identifying multiple structural changes in linear models for time series data, allowing for efficient segmentation of time series into homogeneous sub-periods. When implemented with the Pruned Exact Linear Time (Pelt) algorithm and the Radial Basis Function (RBF) model, this method becomes efficient and flexible for detecting both linear and non-linear breakpoints.

The objective of the test is to partition the time series \( \{y_t\}_{t=1}^T \) into \(m+1\) segments, where each segment exhibits distinct statistical properties. The detection process minimizes a global cost function, defined as:
\begin{equation}
	\begin{aligned}
		\text{Cost} = \sum_{i=1}^{m+1} C(y_{t_{i-1}+1:t_i}) + \beta m,
	\end{aligned}
\end{equation}
where \( C(y_{t_{i-1}+1:t_i}) \) is the cost function for segment \(i\), capturing the intra-segment dissimilarity (e.g., RBF kernel distance for non-linear models), and \(\beta\) is a penalty term that regulates the number of breakpoints \(m\), balancing model complexity and goodness-of-fit. The data is divided into \(m+1\) contiguous segments \( \{[1, t_1], [t_1+1, t_2], \dots, [t_m+1, T]\} \), where \( t_i \) represents the position of the \(i\)-th breakpoint.

For the RBF model, the cost function measures the similarity between data points using radial basis functions:
\begin{equation}
	\begin{aligned}
		C(y_{t_{i-1}+1:t_i}) = \sum_{j, k \in [t_{i-1}+1:t_i]} \exp\left(-\gamma \|y_j - y_k\|^2\right),
	\end{aligned}
\end{equation}
where \(\gamma > 0\) controls the sensitivity of the kernel to the distance between points. The penalty term \(\beta\) adjusts the model's granularity. A higher \(\beta\) favors fewer breakpoints by penalizing additional segments.

Building on the Bai-Perron approach, the Pelt algorithm optimizes the cost function using a dynamic programming approach with pruning, ensuring computational efficiency. Its complexity is \(O(T \log T)\), making it suitable for large-scale datasets \citep{killick2012optimal}.

\subsubsection{Pettitt Test}
The Pettitt Test, proposed by \citet{pettitt1979non}, is a non-parametric statistical method designed to detect a single changepoint in a time series by evaluating whether the distribution before and after a candidate changepoint is significantly different. Given a time series \( \{y_t\}_{t=1}^T \), the test computes a statistic \( K \), which is the maximum absolute value of \( U_k \), defined as:
\begin{equation}
	\begin{aligned}
		U_k = \sum_{i=1}^k \sum_{j=k+1}^T \text{sign}(y_i - y_j),
	\end{aligned}
\end{equation}
where the sign function is given by:
\begin{equation}
	\begin{aligned}
		\text{sign}(z) = 
		\begin{cases} 
			1 & \text{if } z > 0, \\
			0 & \text{if } z = 0, \\
			-1 & \text{if } z < 0.
		\end{cases}
	\end{aligned}
\end{equation}
The Pettitt Test statistic \( K \) is calculated as:
\begin{equation}
	\begin{aligned}
		K = \max_k |U_k|,
	\end{aligned}
\end{equation}
indicating the most likely location of a changepoint in the time series. The test evaluates the null hypothesis \( H_0 \), which assumes that the data comes from a single homogeneous distribution (no changepoint). Rejection of \( H_0 \) implies the existence of a structural change.

The significance of the test is determined by the p-value, computed as:
\begin{equation}
	\begin{aligned}
		p = 2 \exp\left( -\frac{6K^2}{T^3 + T^2} \right),
	\end{aligned}
\end{equation}
where \( T \) is the length of the time series.

\subsection{Decomposing OOS Performance: An Illustration of Trend-based Probability Framework}
\label{Trend-based Probability Framework}
\begin{figure}[H]
	\centering
	\caption{Tolerance Interval of Performance Gain Relative to the Lag Benchmark}
	\label{fig:A3}
	\includegraphics[width=0.9\textwidth]{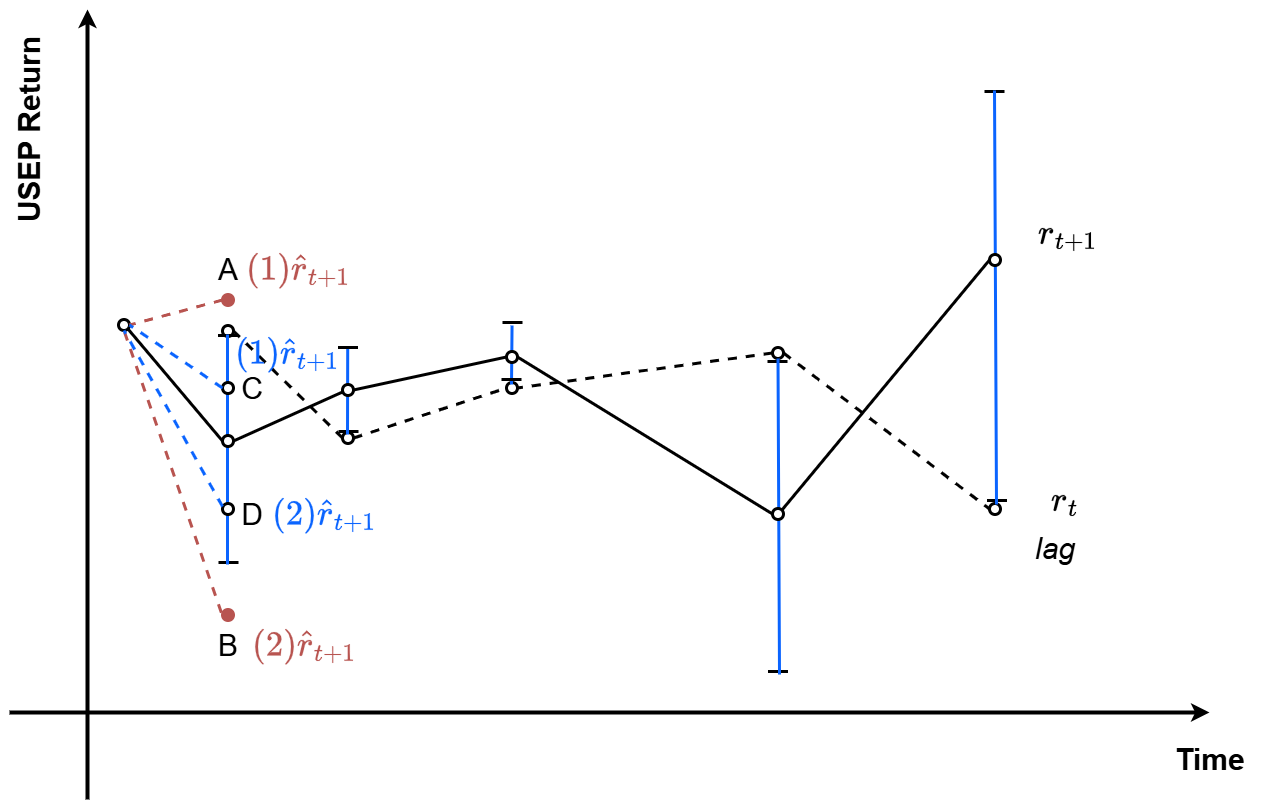}
\end{figure}

We use \(r_t\) as the benchmark for illustration. As shown in Figure \ref{fig:A3}, our predictions can either fall within the prediction interval or outside of it. Predictions outside the interval, indicating excessive mispricing or performance loss, can be categorized into two types: false trend (e.g., point A) and excessive right trend (e.g., point B). Conversely, predictions within the interval, where our model outperforms the benchmark, indicate appropriate trends or performance gains. These can also be classified into two types: appropriate weak trend (e.g., point C) and appropriate strong trend (e.g., point D).

In summary, all predictions can be categorized into four types: false trend, excessive right trend, appropriate weak trend, and appropriate strong trend, providing a comprehensive framework to trace the source of performance gain. Building on these classifications, we derive specific criteria under two scenarios, \(r_{t+1} < r_t\) and \(r_{t+1} > r_t\), as outlined in the following evaluation standards:

\noindent \textit{If \( r_{t+1} < r_t \) (Downward trend):}
\[
\begin{cases}
	\text{False Trend} & \hat{r}_{t+1} > r_t \\
	\text{Excessive Right Trend} & \hat{r}_{t+1} < 2r_{t+1} - r_t \\
	\text{Appropriate Weak Trend} & r_{t+1} < \hat{r}_{t+1} \leq r_t \\
	\text{Appropriate Strong Trend} & 2r_{t+1} - r_t \leq \hat{r}_{t+1} \leq r_{t+1}
\end{cases}
\]

\medskip

\noindent \textit{If \( r_{t+1} \geq r_t \) (Upward trend):}
\[
\begin{cases}
	\text{False Trend} & \hat{r}_{t+1} < r_t \\
	\text{Excessive Right Trend} & \hat{r}_{t+1} > 2r_{t+1} - r_t \\
	\text{Appropriate Weak Trend} & r_t \leq \hat{r}_{t+1} < r_{t+1} \\
	\text{Appropriate Strong Trend} & r_{t+1} \leq \hat{r}_{t+1} \leq 2r_{t+1} - r_t
\end{cases}
\]

\subsection{Feature Importance}
\label{appendix Feature Importance}
\subsubsection{Sum of Squared Deviation}
The SSD, introduced by \citet{dimopoulos1995use}, is a sensitivity-based measure designed to evaluate the importance of variables in models that leverage differentiability. As such, SSD is not applicable to non-differentiable models, such as tree-based algorithms, dimensionality reduction methods like PCR and PLS, or GLM. By quantifying the sensitivity of the model output with respect to changes in a specific input variable, SSD provides insights into the overall contribution of each variable to the model's behavior. This approach has been successfully applied in various research contexts, including asset pricing models \citep{gu2020empirical} and bond return predictability \citep{bianchi2021bond}.

Formally, SSD for the \(j\)-th variable is defined as:
\begin{equation}
	\begin{aligned}
		SSD_j = \frac{1}{|\mathcal{T}_1|} \sum_{t \in \mathcal{T}_1} \left( \frac{\partial f(z, \theta)}{\partial z_j} \Bigg|_{z = z_{t}} \right)^2,
	\end{aligned}
\end{equation}
where \(z\) denotes the vector of input variables and \(z_j\) is the \(j\)-th element of \(z\). The function \(f(z, \theta)\) refers to the model output as a function of input variables \(z\) and parameters \(\theta\). The partial derivative \(\frac{\partial f(z, \theta)}{\partial z_j}\) measures the sensitivity of the model output to changes in the \(j\)-th input variable at each training sample \(z_t\). By averaging the squared sensitivities across all training samples, SSD provides a comprehensive metric of the variable's influence on the model's output.

\subsubsection{Mean Decrease Gini}
MDG introduced by \citet{breiman2002manual} is a widely recognized metric for evaluating variable importance in tree-based models. It quantifies the total reduction in node impurity attributed to a specific feature across all splits in the decision trees. The importance of a feature \(j\) is calculated as the cumulative impurity reduction for all nodes where the feature is used for splitting:
\begin{equation}
	\begin{aligned}
		\text{MDG}_j = \sum_{\text{nodes where } j \text{ is used}} \Delta I,
	\end{aligned}
\end{equation}
where \(\Delta I\) represents the reduction in impurity at a given split, calculated as:
\begin{equation}
	\begin{aligned}
		\Delta I = I_{\text{parent}} - \left( w_{\text{left}} \cdot I_{\text{left}} + w_{\text{right}} \cdot I_{\text{right}} \right),
	\end{aligned}
\end{equation}
with \(I_{\text{parent}}, I_{\text{left}},\) and \(I_{\text{right}}\) denoting the impurity of the parent node and the left and right child nodes, respectively. The weights \(w_{\text{left}}\) and \(w_{\text{right}}\) correspond to the proportions of samples allocated to the left and right child nodes.

\subsection{Evaluating Conditional Mean Misspecification via Conditional Variance}
\label{Evaluating Conditional Mean Misspecification via Conditional Variance}
Given that misspecification of the conditional mean model typically induces misspecification in the conditional variance model, with access to high-frequency intraday electricity price series, we further evaluate the robustness of identified best-performing conditional mean models from the perspective of conditional variance. Specifically, we model their predictive errors using a GJR-GARCH($p$,$q$) and evaluate how well the resulting conditional variance estimates align with high-frequency realized variance proxies. In our framework, we utilize two types of realized variances as proxies for the true daily variance:

\noindent \textbf{Intra-day Realized Variance.} This proxy uses the sum of intra-day squared returns across 48 intervals within a day, computed as:
\begin{equation}
	\begin{aligned}
		RV_t^{(intra)} = \sum_{i=1}^{48} r_{i,t}^2,
	\end{aligned}
\end{equation}
where \(r_{i,t,h}\) represents the return in the \(i\)-th interval of day \(t\), with 48 denoting the total number of 30-minute intervals within a single day.

\noindent \textbf{Whole(Intra-day + Overnight Realized Variance).}  This method extends the first proxy by incorporating the squared overnight return, ensuring the overnight information is captured in addition to intra-day returns.

\begin{equation}
	\begin{aligned}
		RV_t^{(\text{whole})} = \left( \ln P_{1,t} - \ln P_{48,t-1} \right)^2 + \sum_{i=1}^{48} r_{i,t}^2,
	\end{aligned}
\end{equation}	
where \(P_{1,t}\) denotes the opening price of day \(t\), and \(P_{48,t-1}\) represents the closing price of the previous day \(t-1\). The term \(\ln P_{1,t} - \ln P_{48,t-1}\) captures the log return from the closing price of day \(t-1\) to the opening price of day \(t\), reflecting the overnight volatility.

The conditional mean model is defined as \(r_{t+1} = f(X_t, \theta) + \epsilon_{t+1}\). If \(f(X_t, \theta)\) is misspecified, \(\epsilon_{t+1}\) becomes large. The GJR-GARCH(1,1) model for conditional variance is expressed as:
\begin{equation}
	\begin{aligned}
		h_{t+1} = \omega + \alpha \epsilon_t^2 + \beta h_t + \gamma \epsilon_t^2 I_{\epsilon_t < 0},
	\end{aligned}
\end{equation}	
where \(\alpha \geq 0\) ensures positive variance. If \(\epsilon_t\) is large, \(h_{t+1}\) will also be large. Thus, a misspecified conditional mean model typically leads to larger conditional variance estimates.

\begin{figure}[H]
	\caption{Realized Volatility vs Conditional Volatility Forecast}
	\includegraphics[width=1\linewidth]{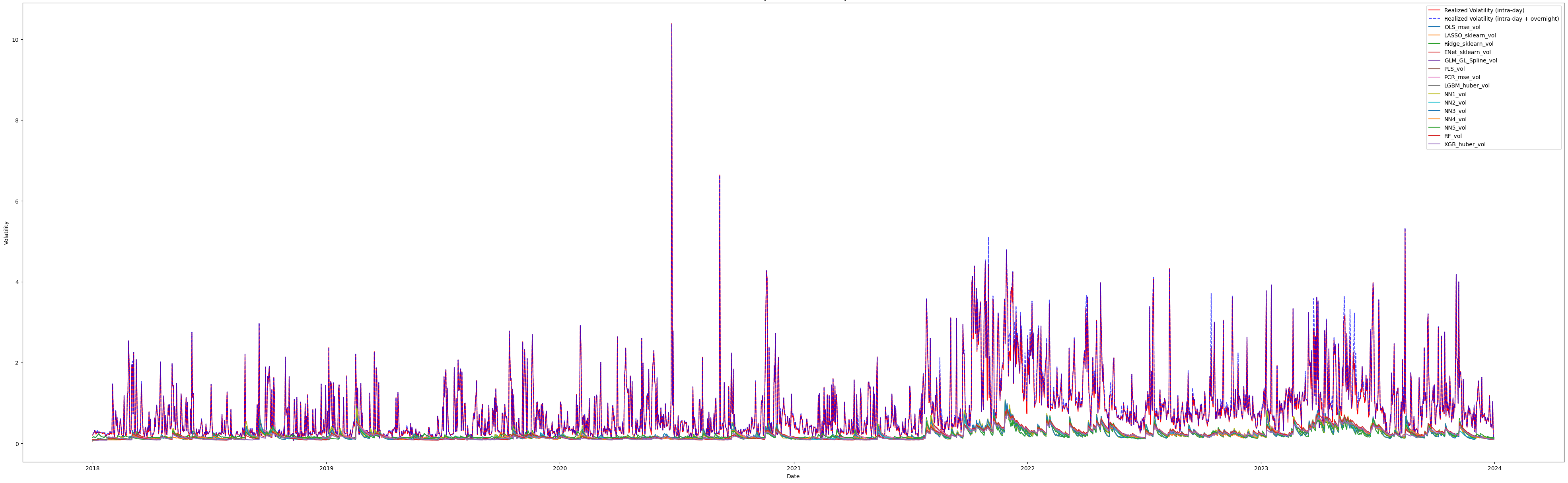}
	\label{fig:a4}
\end{figure}

By definition, the MSE is given by:
\begin{equation}
	\begin{aligned}
		MSE = \frac{1}{T} \sum (h_t - \hat{h}_t)^2,
	\end{aligned}
\end{equation}
where \(\hat{h}_t\) represents the estimated conditional variance. In our setting, \(h_t > \hat{h}_t\) (as illustrated in Figure \ref{fig:a4}), a larger \(\hat{h}_t\) (stemming from misspecified conditional mean models) results in a smaller MSE. Thus, the best-performing models naturally exhibit the largest MSE.

We also utilize the Quasi Likelihood (QLIKE) loss function to evaluate model performance:		
\begin{equation}
	\begin{aligned}
		\tilde{L}(h_t, \hat{h}_t) = \frac{h_t}{\hat{h}_t} - \log\left(\frac{h_t}{\hat{h}_t}\right) - 1
	\end{aligned}
\end{equation}

\section{Result Supplements}
\label{Result Supplements}
\subsection{Additional Tables}
\label{Additional Tables}

\begin{table}[H]
	\fontsize{9pt}{0.38cm}\selectfont
	\centering
	\caption{\centering CW Test for Nested Models}
	\label{tab:A5}
	\setlength{\tabcolsep}{7pt}
	\renewcommand{\arraystretch}{1.1}
	
	\textbf{Panel A: Penalized Linear Models}
	\vspace{0.2cm}
	
	\begin{tabular}{p{2.5cm} p{1.5cm} p{1.5cm} p{1.5cm}}
		\hline
		& (1) & (2) & (3) \\
		& LASSO & Ridge & ENet \\
		\hline
		LASSO & 0 & 3.09*** & 3.18*** \\
		Ridge & 6.91*** & 0 & 7.24*** \\
		ENet  & -1.87* & 2.44** & 0 \\
		\hline
	\end{tabular}
	
	\vspace{0.5cm}
	
	\textbf{Panel B: Neural Network Models}
	\vspace{0.2cm}
	
	\begin{tabular}{p{2cm} p{1.5cm} p{1.5cm} p{1.5cm} p{1.5cm} p{1.5cm}}
		\hline
		& (1) & (2) & (3) & (4) & (5) \\
		& NN1  & NN2   & NN3   & NN4   & NN5 \\
		\hline
		NN1 & 0      & 4.83*** & 4.73*** & 6.00*** & 6.98*** \\
		NN2 & 3.35***  & 0     & 5.24*** & 5.11*** & 6.04*** \\
		NN3 & 2.76***  & 3.85*** & 0     & 4.59*** & 5.84*** \\
		NN4 & 2.92***  & 3.21*** & 3.32*** & 0     & 4.86*** \\
		NN5 & 2.72***  & 3.58*** & 4.09*** & 4.48*** & 0 \\
		\hline
	\end{tabular}
\end{table}

By Table \ref{tab:A5}, the performance of each pair of models is assessed by comparing their diagonal scores. For example, Ridge (column) relative to Lasso (row) scores 3.09, whereas Lasso (column) relative to Ridge (row) scores 6.91, indicating that Lasso outperforms Ridge. The results reveal a consistent ranking: ENet \(>\) Lasso \(>\) Ridge, and NN5 \(>\) NN4 \(>\) NN3 \(>\) NN2 \(>\) NN1.

\begin{table}[H]
	\fontsize{9pt}{0.38cm}\selectfont
	\centering
	\caption{\centering Equal Predictive Ability (EPA) Tests}
	\label{tab:A6}
	\setlength{\tabcolsep}{7pt}
	\renewcommand{\arraystretch}{1.1}
	\begin{tabular}{p{3cm} p{3cm} p{4cm} p{3cm}}
		\hline
		&(1)&(2)&(3)\\
		& $p-values_{CW}$ & $p-values_{MCS}$ & $p-values_{GW}$ \\ 
		\hline
		LASSO              & 0.0019***     & 1.000***$i$ (0.000$e$)     & 0.0000*** \\
		Ridge              & 0.7876        & 1.000***$i$ (0.7504***$i$) & 0.6870    \\
		Enet               & 0.0137**      & 1.000***$i$ (0.000$e$)     & 0.0000*** \\
		PLS                & 0.1781        & 1.000***$i$ (0.000$e$)     & 0.0000*** \\
		PCR                & 0.0004***     & 1.000***$i$ (0.000$e$)     & 0.0000*** \\
		GLM                & 0.3579        & 1.000***$i$ (0.000$e$)     & 0.0000*** \\
		RF                 & 0.0000***     & 1.000***$i$ (0.1182*$i$)   & 0.0002*** \\
		XGB(+H)            & 0.0137***     & 1.000***$i$ (0.000$e$)     & 0.0000*** \\
		LGBM(+H)           & 0.0077***     & 1.000***$i$ (0.000$e$)     & 0.0000*** \\
		NN1                & 0.0001***     & 1.000***$i$ (0.000$e$)     & 0.0000*** \\
		NN2                & 0.0002***     & 1.000***$i$ (0.000$e$)     & 0.0000*** \\
		NN3                & 0.0009***     & 1.000***$i$ (0.0004$e$)    & 0.0000*** \\
		NN4                & 0.0016***     & 1.000***$i$ (0.000$e$)     & 0.0000*** \\
		NN5                & 0.0003***     & 1.000***$i$ (0.000$e$)     & 0.0000*** \\
		$Ensemble_{avg}$   & 0.0092***     & 1.000***$i$ (0.000$e$)     & 0.0000*** \\
		$Ensemble_{op}$    & 0.7480        & 1.000***$i$ (0.000$e$)     & 0.0000*** \\
		$Ensemble_{wp}$    & 0.0000***     & 1.000***$i$ (0.000$e$)     & 0.0000*** \\
		$Ensemble_{\theta=0.9}$ & 0.4180   & 1.000***$i$ (0.000$e$)     & 0.0000*** \\
		$Ensemble_{\theta=1}$ & 0.0111**   & 1.000***$i$ (0.000$e$)     & 0.0000*** \\
		\hline
	\end{tabular}
	\begin{tablenotes}
		\fontsize{8pt}{0.35cm}\selectfont
		\item \textit{Notes}: For CW and GW tests, significance levels are denoted by "*", "**", and "***", corresponding to 10\%, 5\%, and 1\%, respectively. For MCS test, "*", "**", and "***" indicate significance at the 90\%, 75\% and 50\% levels respectively. The suffix $i$ represents the model is included into the confidence set, while $e$ denotes the exclusion. The values in parentheses represent the \textit{p-values} for the OLS model, while the values outside the parentheses represent the \textit{p-values} for the competing model.
	\end{tablenotes}
\end{table}

As shown in Table \ref{tab:A6}, in the CW test, Ridge, PLS, GLM, \(Ensemble_{op}\), and \(Ensemble_{\theta=0.9}\) show no significant difference from OLS, while all other sophisticated models outperform OLS significantly. In the MCS test, only Ridge and RF are comparable to OLS, with all other models performing significantly better. In the GW test, Ridge is the only model not significantly different from OLS, while all other models show superior performance. Ridge is the only model that fails to achieve statistically significant improvement over OLS across all three tests, likely due to its underperformance in strong-factor environments. Overall, these results highlight that, apart from Ridge, our sophisticated machine learning models outperform OLS at least in one of the three tests, demonstrating their robustness and predictive power.

\newpage
\begin{table}[H]
	\fontsize{7pt}{0.35cm}\selectfont
	\makebox[\textwidth]{
		\rotatebox{90}{
			\begin{minipage}{\textheight}
				\centering
				\caption{Decomposing OOS Performance (Lag as Benchmark): A Trend-based Probability Framework}
				\label{tab:A7}
				\setlength{\tabcolsep}{7pt}
				\renewcommand{\arraystretch}{1.1}
				\begin{threeparttable}
					\begin{tabular}{p{1.9cm} p{0.001cm} p{0.4cm} p{0.7cm} p{1cm} p{1cm} p{0.001cm} p{0.4cm} p{0.7cm} p{1cm} p{1cm} p{0.001cm} p{0.4cm} p{0.7cm} p{1cm} p{1cm} p{1cm} p{0.5cm}  p{0.5cm}}
						\hline
						&&\multicolumn{4}{c}{\textbf{Down}}&&\multicolumn{4}{c}{\textbf{Up}}&&\multicolumn{7}{c}{\textbf{All}}\\
						&&(1)&(2)&(3)&(4)&&(5)&(6)&(7)&(8)&&(9)&(10)&(11)&(12)&(13)&(14)&(15)\\
						Model&&False trend&Excessive trend&Appropriate weak trend&Appropriate strong trend&&False trend&Excessive trend&Appropriate weak trend&Appropriate strong trend&&False trend&Excessive trend&Appropriate weak trend&Appropriate strong trend&Performance Gain&Over Estimate&Under Estimate\\
						\cline{1-1} \cline{3-6} \cline{8-11} \cline{13-19}
						OLS&&0.1434&0.164&0.4607&0.2318&&0.1594&0.1509&0.4288&0.2609&&0.152&0.157&0.4436&0.2474&0.691&0.4044&0.5956\\
						LASSO&&0.1336&0.1356&0.4715&0.2593&&0.1279&0.1245&0.4518&0.2958&&0.1305&0.1296&0.461&0.2789&0.7398&0.4085&0.5915\\
						Ridge&&0.1454&0.164&0.4578&0.2328&&0.162&0.1535&0.4297&0.2549&&0.1543&0.1584&0.4427&0.2446&0.6874&0.403&0.597\\
						ENet&&0.1326&0.1405&0.4656&0.2613&&0.1321&0.1194&0.451&0.2975&&0.1324&0.1292&0.4578&0.2807&0.7385&0.4099&0.5901\\
						PCR&&0.1365&0.1493&0.4558&0.2583&&0.1355&0.1304&0.4305&0.3035&&0.136&0.1392&0.4423&0.2825&0.7248&0.4217&0.5783\\
						PLS&&0.1346&0.1346&0.4578&0.2731&&0.133&0.1262&0.4356&0.3052&&0.1337&0.1301&0.4459&0.2903&0.7362&0.4204&0.5796\\
						GLM&&0.1454&0.1257&0.4921&0.2367&&0.1185&0.1219&0.4604&0.2992&&0.131&0.1237&0.4751&0.2702&0.7453&0.3939&0.6061\\
						XGB(+H)&&0.1523&0.1041&0.5088&0.2348&&0.1074&0.1211&0.4876&0.2839&&0.1283&0.1132&0.4975&0.2611&0.7586&0.3743&0.6257\\
						LGBM(+H)&&0.1523&0.112&0.5206&0.2151&&0.1091&0.1228&0.4766&0.2916&&0.1292&0.1178&0.497&0.256&0.7531&0.3738&0.6262\\
						RF&&0.164&0.0825&0.5737&0.1798&&0.1262&0.1125&0.5362&0.2251&&0.1438&0.0986&0.5536&0.204&0.7576&0.3026&0.6974\\
						NN1&&0.165&0.1041&0.5147&0.2161&&0.1364&0.1125&0.4672&0.2839&&0.1497&0.1086&0.4893&0.2524&0.7417&0.361&0.639\\
						NN2&&0.164&0.1159&0.5039&0.2161&&0.1228&0.1168&0.4885&0.272&&0.1419&0.1164&0.4957&0.246&0.7417&0.3624&0.6376\\
						NN3&&0.1582&0.1051&0.5275&0.2092&&0.1142&0.1134&0.4996&0.2728&&0.1346&0.1095&0.5126&0.2433&0.7558&0.3528&0.6472\\
						NN4&&0.1611&0.1041&0.5305&0.2043&&0.1245&0.1057&0.4842&0.2856&&0.1415&0.105&0.5057&0.2478&0.7535&0.3528&0.6472\\
						NN5&&0.1827&0.0845&0.5206&0.2122&&0.11&0.11&0.4868&0.2933&&0.1438&0.0981&0.5025&0.2556&0.7581&0.3537&0.6463\\
						$Ensemble_avg$&&0.1336&0.1051&0.5079&0.2534&&0.1125&0.1006&0.48&0.3069&&0.1223&0.1027&0.4929&0.2821&0.775&0.3848&0.6152\\
						$Ensemble_{op}$&&0.1395&0.1031&0.5285&0.2289&&0.1091&0.1108&0.4936&0.2864&&0.1232&0.1073&0.5098&0.2597&0.7695&0.367&0.633\\
						$Ensemble_{wp}$&&0.1326&0.0825&0.5589&0.2259&&0.104&0.0963&0.5098&0.2899&&0.1173&0.0899&0.5326&0.2602&0.7928&0.3501&0.6499\\
						$Ensemble_{wp}$&&0.1306&0.1081&0.5138&0.2475&&0.1194&0.1083&0.4842&0.2882&&0.1246&0.1082&0.4979&0.2693&0.7672&0.3775&0.6225\\
						$Ensemble_{theta=1}$&&0.1316&0.1051&0.5147&0.2485&&0.1142&0.1006&0.4731&0.312&&0.1223&0.1027&0.4925&0.2825&0.775&0.3852&0.6148\\				
						\hline
					\end{tabular}
					\begin{tablenotes}
						\fontsize{6pt}{0.35cm}\selectfont
						\item \textit{Notes}: All values are presented as ratios. Columns (1) to (4) correspond to down trends, defined as instances where \( r_{t+1} < r_t \). The numerator represents the number of occurrences of the decomposing trend within the down trend, while the denominator represents the total number of down trends. For example, the first column is computed as \( \frac{N_{\text{False Trend in Down}}}{N_{\text{Down}}} \). Columns (5) to (8) pertain to up trends, where \( r_{t+1} \geq r_t \), with the numerator representing the number of occurrences of the decomposing trend within the up trend, and the denominator representing the total number of up trends. Similarly, columns (9) to (15) consider both down and up trends. The denominator for these columns represents the total number of observations. The numerators for columns (9) to (12) represent the respective decomposing trends. For column (13), the numerator is the sum of appropriate weak trends and appropriate strong trends. Column (14) considers the sum of excessive trends and appropriate strong trends in the numerator. Finally, for column (15), the numerator includes false trends combined with appropriate weak trends.
					\end{tablenotes}
				\end{threeparttable}
			\end{minipage}
		}
	}
\end{table}

\begin{table}[H]
	\fontsize{9pt}{0.38cm}\selectfont
	\centering
	\caption{\centering Model Performance Based on Common Metrics}
	\label{tab:A8}
	\setlength{\tabcolsep}{7pt}
	\renewcommand{\arraystretch}{1.1}
	\begin{tabular}{p{4.5cm} p{2.5cm} p{2.5cm}}
		\hline
		Model & MSE & MAE\\
		\hline
		OLS & 0.059328 & 0.148757 \\
		LASSO & 0.057869 & 0.137307\\
		Ridge & 0.059329 & 0.148776 \\
		ENet & 0.057626 & 0.137268  \\
		PCR & 0.057799 & 0.138726  \\
		PLS & 0.056989 & 0.136861 \\
		GLM & 0.050256 & 0.129359\\
		XGB(+H) & 0.050292 & 0.129559  \\
		LGBM(+H) & 0.050647 & 0.129237 \\
		RF & 0.058318 & 0.140729 \\
		NN1 & 0.055691 & 0.136932 \\
		NN2 & 0.055481 & 0.136998 \\
		NN3 & 0.054400 & 0.135859 \\
		NN4 & 0.054488 & 0.135149 \\
		NN5 & 0.053103 & 0.134397 \\
		Ensemble\_avg & 0.052825 & 0.128600 \\
		Ensemble\_weight & 0.049257 & 0.126359 \\
		\hline
	\end{tabular}
\end{table}

\begin{table}[H]
	\fontsize{9pt}{0.38cm}\selectfont
	\centering
	\caption{\centering Mean-Variance Portfolio Results ($\gamma$=2)}
	\label{tab:A9}
	\setlength{\tabcolsep}{7pt}
	\renewcommand{\arraystretch}{1.1}
	\begin{tabular}{p{1.5cm} p{2.5cm} p{1.5cm} p{2cm} p{2.7cm} p{1.5cm} p{1.8cm}}
		\hline
		&Average Optimal Weight&Average Rf Weight&Std Dev Optimal Weight&Number of Positive Optimal Weight&Average Utility&Total Utility\\
		\hline
		OLS&0.10638&0.89362&2.35852&1077&0.15099&330.81730\\
		LASSO&0.09462&0.90538&2.92926&1090&0.14140&309.81327\\
		Ridge&0.10623&0.89377&2.36008&1083&0.15104&330.92132\\
		ENet&0.08640&0.91360&2.96408&1086&0.14456&316.72770\\
		PCR&0.08414&0.91586&3.04852&1077&0.15432&338.11657\\
		PLS&0.08422&0.91578&3.11817&1071&0.15876&347.84654\\
		GLM&0.15690&0.84310&4.21015&1081&0.22146&485.21984\\
		XGB+H&0.23099&0.76901&4.68172&1085&0.27721&607.35906\\
		LGBM+H&0.25483&0.74517&4.28186&1100&0.24890&545.33038\\
		RF&0.20965&0.79035&3.94577&999&0.18248&399.80651\\
		NN1&0.30028&0.69972&3.25467&1067&0.17451&382.34243\\
		NN2&0.36126&0.63874&3.44518&1075&0.19086&418.17854\\
		NN3&0.40388&0.59612&3.51366&1100&0.20305&444.89230\\
		NN4&0.36929&0.63071&3.56410&1110&0.19169&419.99582\\
		NN5&0.41118&0.58882&3.59367&1123&0.20878&457.44784\\
		
		\hline
	\end{tabular}
\end{table}

\begin{table}[H]
	\fontsize{9pt}{0.38cm}\selectfont
	\centering
	\caption{\centering Mean-Variance Portfolio Results ($\gamma$=4)}
	\label{tab:A10}
	\setlength{\tabcolsep}{7pt}
	\renewcommand{\arraystretch}{1.1}
	\begin{tabular}{p{1.5cm} p{2.5cm} p{1.5cm} p{2cm} p{2.7cm} p{1.5cm} p{1.8cm}}
		\hline
		&Average Optimal Weight&Average Rf Weight&Std Dev Optimal Weight&Number of Positive Optimal Weight&Average Utility&Total Utility\\
		\hline
		OLS&0.05319&0.94681&1.17926&1077&0.07549&165.40865\\
		LASSO&0.04731&0.95269&1.46463&1090&0.07070&154.90664\\
		Ridge&0.05312&0.94688&1.18004&1083&0.07552&165.46066\\
		ENet&0.04320&0.95680&1.48204&1086&0.07228&158.36385\\
		PCR&0.04207&0.95793&1.52426&1077&0.07716&169.05828\\
		PLS&0.04211&0.95789&1.55909&1071&0.07938&173.92327\\
		GLM&0.07845&0.92155&2.10507&1081&0.11073&242.60992\\
		XGB+H&0.11549&0.88451&2.34086&1085&0.13860&303.67953\\
		LGBM+H&0.12741&0.87259&2.14093&1100&0.12445&272.66519\\
		RF&0.10483&0.89517&1.97289&999&0.09124&199.90325\\
		NN1&0.15014&0.84986&1.62733&1067&0.08725&191.17121\\
		NN2&0.18063&0.81937&1.72259&1075&0.09543&209.08927\\
		NN3&0.20194&0.79806&1.75683&1100&0.10153&222.44615\\
		NN4&0.18464&0.81536&1.78205&1110&0.09585&209.99791\\
		NN5&0.20559&0.79441&1.79683&1123&0.10439&228.72392\\	
		\hline
	\end{tabular}
\end{table}

\begin{table}[H]
	\fontsize{9pt}{0.38cm}\selectfont
	\centering
	\caption{\centering Mean-Variance Portfolio Results ($\gamma$=5)}
	\label{tab:A11}
	\setlength{\tabcolsep}{7pt}
	\renewcommand{\arraystretch}{1.1}
	\begin{tabular}{p{1.5cm} p{2.5cm} p{1.5cm} p{2cm} p{2.7cm} p{1.5cm} p{1.8cm}}
		\hline
		&Average Optimal Weight&Average Rf Weight&Std Dev Optimal Weight&Number of Positive Optimal Weight&Average Utility&Total Utility\\
		\hline
		OLS&0.04255&0.95745&0.94341&1077&0.06040&132.32692\\
		LASSO&0.03785&0.96215&1.17170&1090&0.05656&123.92531\\
		Ridge&0.04249&0.95751&0.94403&1083&0.06041&132.36853\\
		ENet&0.03456&0.96544&1.18563&1086&0.05782&126.69108\\
		PCR&0.03366&0.96634&1.21941&1077&0.06173&135.24663\\
		PLS&0.03369&0.96631&1.24727&1071&0.06350&139.13862\\
		GLM&0.06276&0.93724&1.68406&1081&0.08858&194.08794\\
		XGB+H&0.09239&0.90761&1.87269&1085&0.11088&242.94362\\
		LGBM+H&0.10193&0.89807&1.71274&1100&0.09956&218.13215\\
		RF&0.08386&0.91614&1.57831&999&0.07299&159.92260\\
		NN1&0.12011&0.87989&1.30187&1067&0.06980&152.93697\\
		NN2&0.14450&0.85550&1.37807&1075&0.07634&167.27142\\
		NN3&0.16155&0.83845&1.40547&1100&0.08122&177.95692\\
		NN4&0.14772&0.85228&1.42564&1110&0.07668&167.99833\\
		NN5&0.16447&0.83553&1.43747&1123&0.08351&182.97914\\
		\hline
	\end{tabular}
\end{table}

\begin{table}[H]
	\fontsize{9pt}{0.38cm}\selectfont
	\centering
	\caption{\centering MSE for Intra-day and Whole Realized Variance}
	\label{tab:A12}
	\setlength{\tabcolsep}{7pt}
	\renewcommand{\arraystretch}{1.1}
	\begin{tabular}{p{5cm} p{4cm} p{4cm}}
		\hline
		&$MSE_{intra}$&$MSE_{whole}$\\
		\hline
		OLS&0.979084&1.037952\\
		LASSO&0.969818&1.027295\\
		Ridge&0.978901&1.037761\\
		ENet&0.971069&1.028569\\
		PCR&0.971679&1.029235\\
		PLS&0.974008&1.031658\\
		GLM&0.98825&1.046194\\
		XGB(+H)&0.985014&1.04282\\
		LGBM(+H)&0.984255&1.042083\\
		RF&0.955949&1.012179\\
		NN1&0.973866&1.031579\\
		NN2&0.972085&1.029677\\
		NN3&0.974669&1.032354\\
		NN4&0.977363&1.034921\\
		NN5&0.979931&1.037811\\
		realized\_volatility\_intra&0&0.006536\\
		realized\_volatility\_whole&0.006536&0\\
		\hline
	\end{tabular}
\end{table}

Table \ref{tab:A12} highlights that our top three models, GLM, XGB(+H), and LGBM(+H), consistently report the highest MSE. This finding underscores the robustness of our top three models in capturing volatility dynamics while maintaining the lowest predicted volatility.

\begin{table}[H]
	\fontsize{9pt}{0.38cm}\selectfont
	\centering
	\caption{\centering QLIKE for Intra-day and Whole Realized Variance}
	\label{tab:A13}
	\setlength{\tabcolsep}{7pt}
	\renewcommand{\arraystretch}{1.1}
	\begin{tabular}{p{5cm} p{4cm} p{4cm}}
		\hline
		Model&$QLIKE_{intra}$&$QLIKE_{whole}$\\
		\hline
		OLS&1.833401&1.904022\\
		LASSO&1.953975&2.023013\\
		Ridge&1.833984&1.904631\\
		ENet&1.960155&2.029138\\
		PCR&1.952731&2.0218\\
		PLS&1.969248&2.038498\\
		GLM&2.264765&2.340503\\
		XGB(+H)&2.205248&2.280392\\
		LGBM(+H)&2.213538&2.28936\\
		RF&2.077745&2.147426\\
		NN1&2.121816&2.197077\\
		NN2&2.074851&2.147887\\
		NN3&2.155568&2.231179\\
		NN4&2.102093&2.174075\\
		NN5&2.107397&2.181434\\
		realized\_volatility\_intra&0&0.003124\\
		realized\_volatility\_whole&0.00259&0\\
		\hline
	\end{tabular}
\end{table}

The interpretation of QLIKE is consistent with that of MSE, where higher QLIKE values indicate better performance. As shown in Table \ref{tab:A13}, we reach the same conclusion: our top three models—GLM, XGB(+H), and LGBM(+H)—achieve the highest QLIKE scores, further reinforcing their superior performance.

\subsection{Additional Figures}
\label{Additional Figures}

\begin{figure}[H]
	\centering
	\caption{\centering Model Complexity, Number of Features \& $R^2_{OOS}$}
	\vspace{-0.4in}
	\caption*{\fontsize{10pt}{0.35cm}\selectfont }
	Panel A: Model Complexity \& $R^2_{OOS}$ \\
	\resizebox{\textwidth}{!}{
		\begin{minipage}{\textwidth}
			\begin{multicols}{2}
				\includegraphics[width=0.9\linewidth]{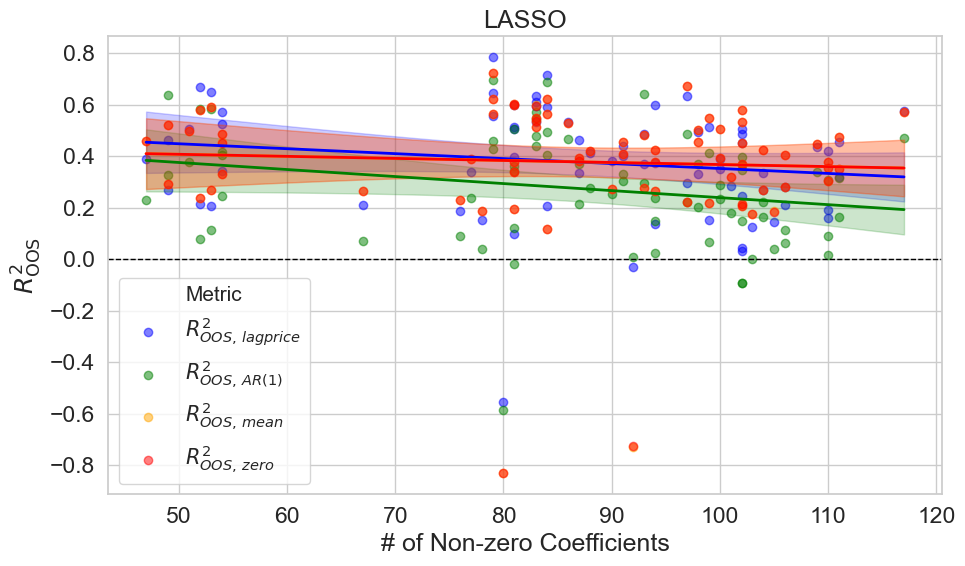}
				\includegraphics[width=0.9\linewidth]{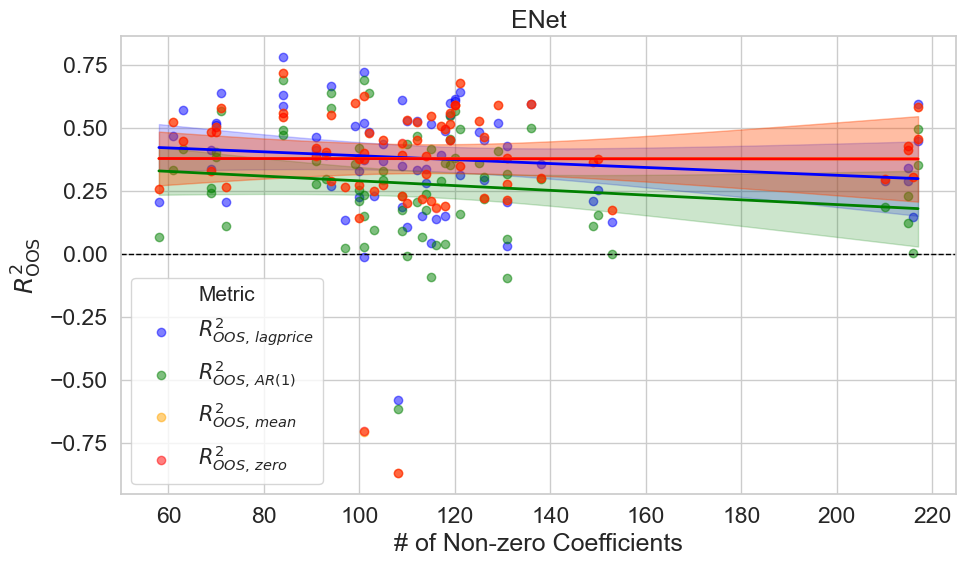}
				\includegraphics[width=0.9\linewidth]{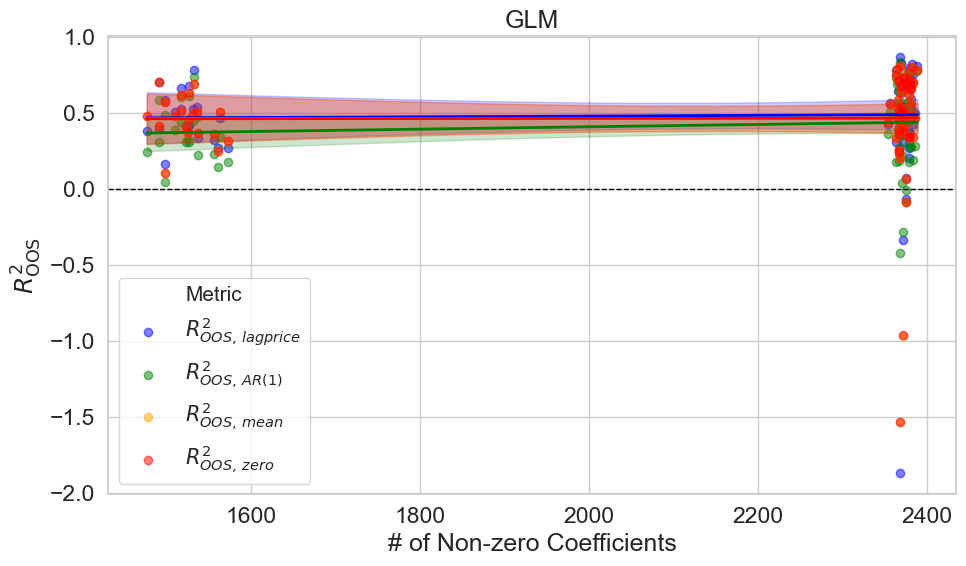}
				\includegraphics[width=0.9\linewidth]{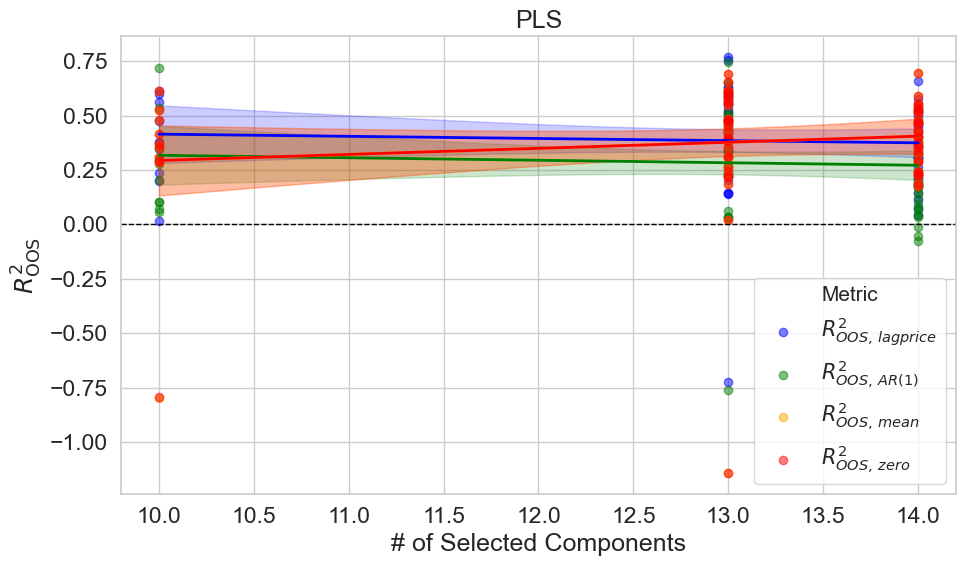}
				
				\includegraphics[width=0.9\linewidth]{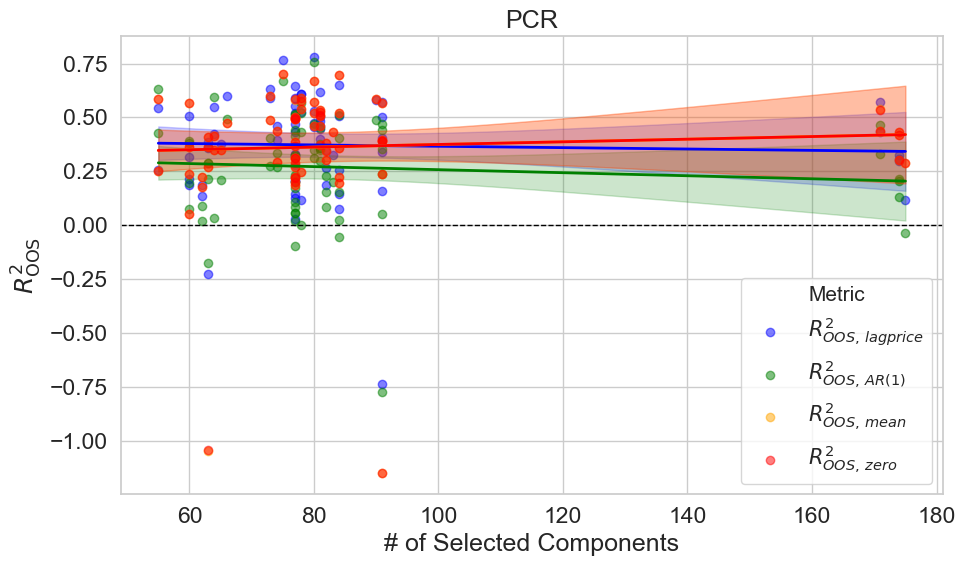}
				\includegraphics[width=0.9\linewidth]{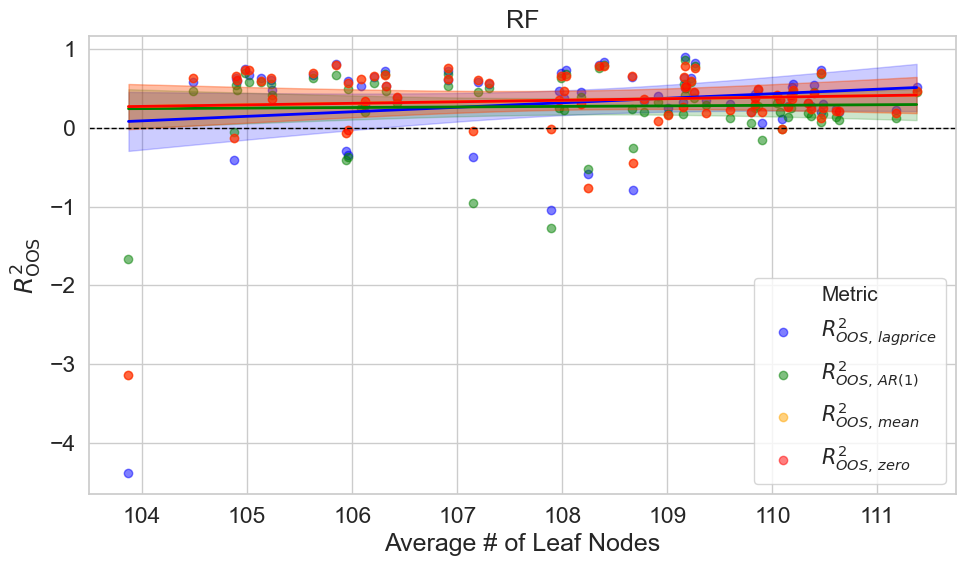}
				\includegraphics[width=0.9\linewidth]{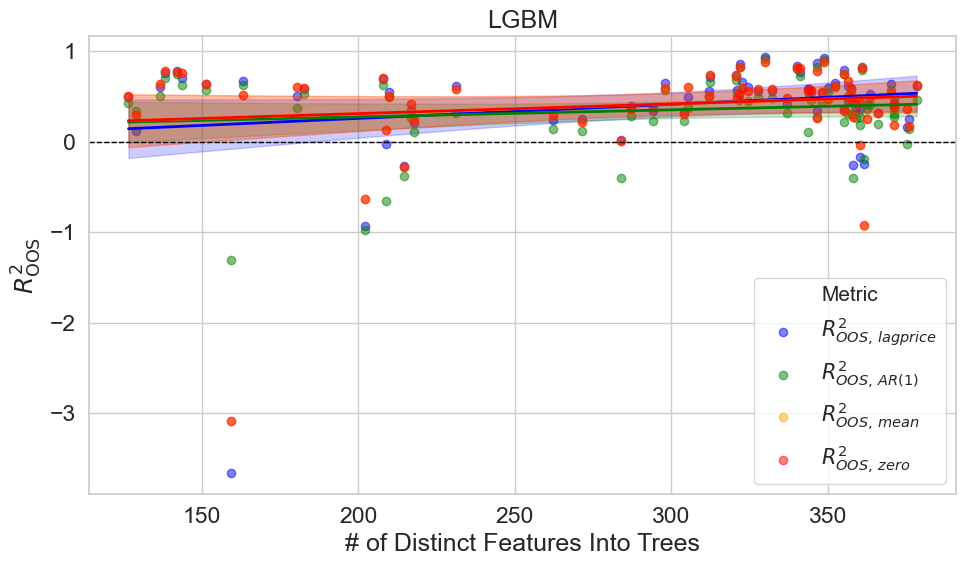}
				\includegraphics[width=0.9\linewidth]{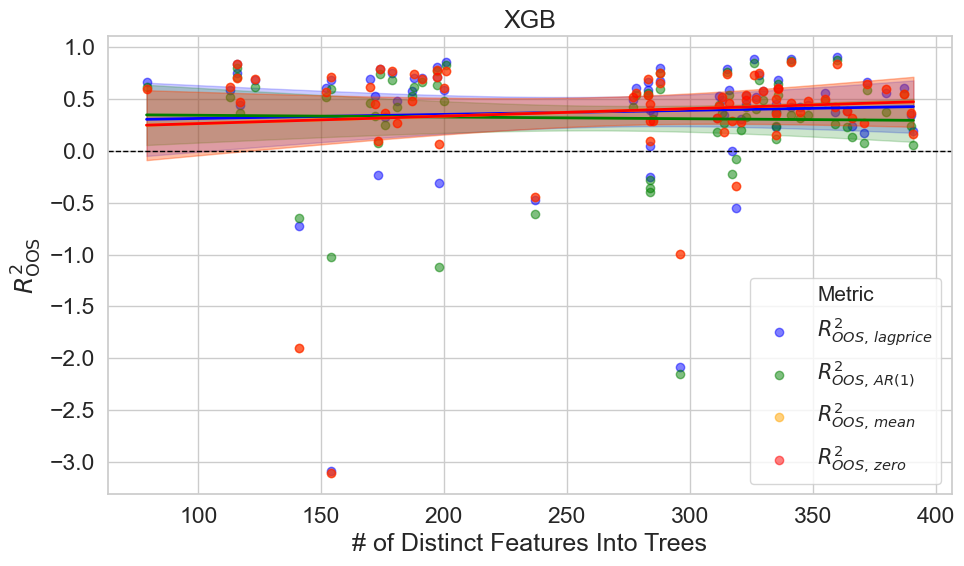}
			\end{multicols}
		\end{minipage}
	}
	\vspace{0.5em}
	Panel B: Number of Features \& $R^2_{OOS}$\\
	\begin{minipage}{0.48\linewidth}
		\centering
		\includegraphics[width=0.9\linewidth]{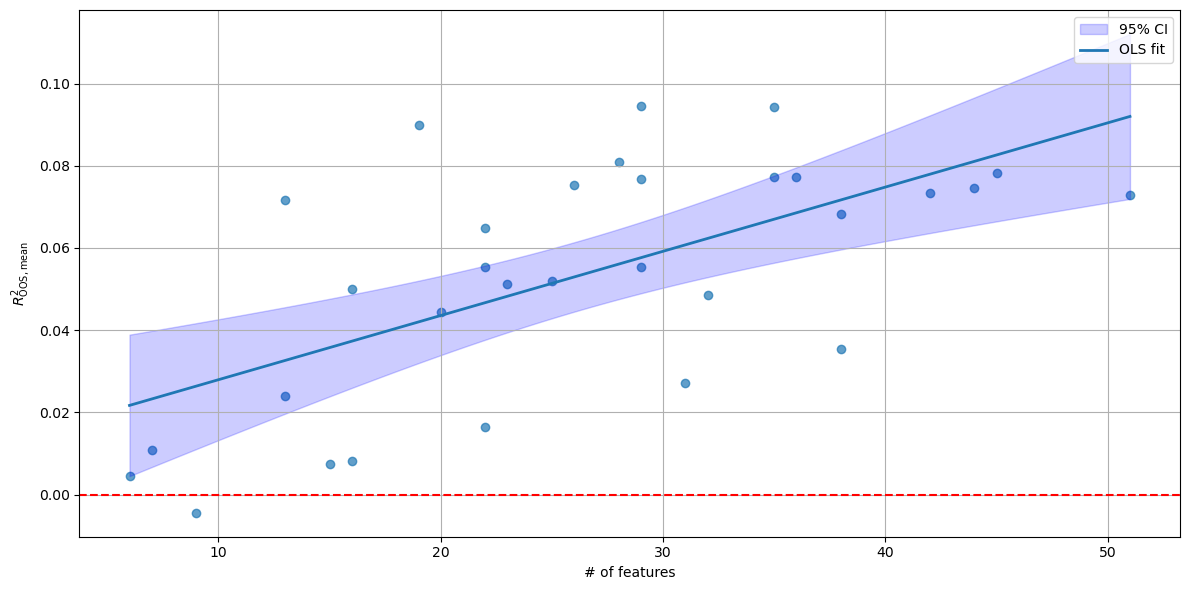}
	\end{minipage}
	\hfill
	\begin{minipage}{0.48\linewidth}
		\centering
		\includegraphics[width=0.9\linewidth]{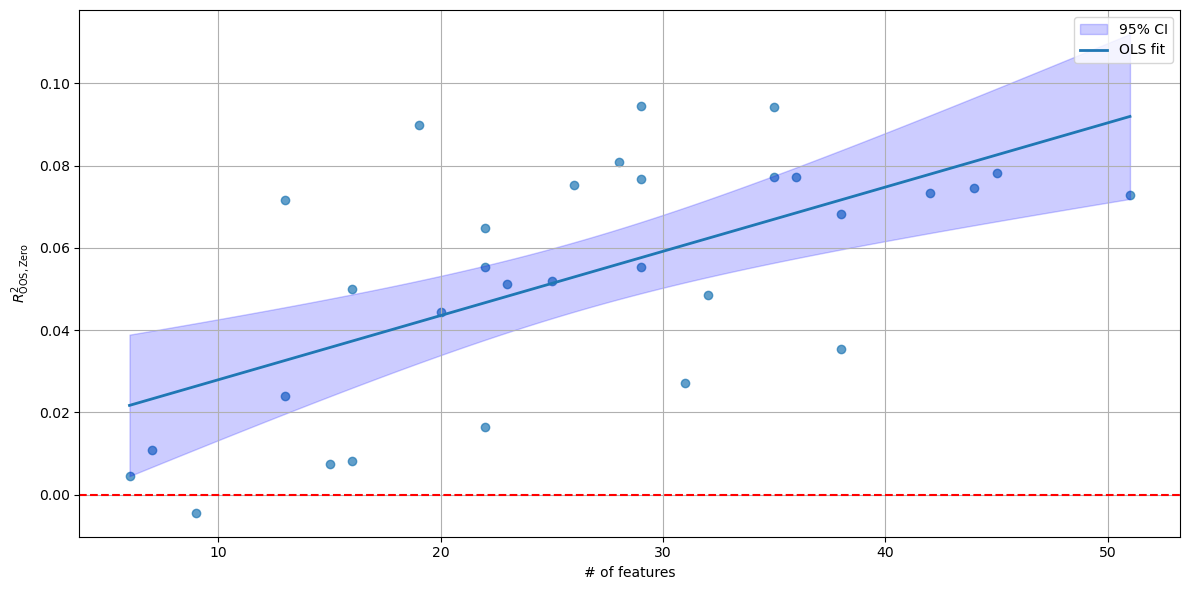}
	\end{minipage}
	\label{fig:a5}
\end{figure}

\begin{figure}[H]
	\caption{\centering Feature Importance Based on \(R^2\) Reduction Across All Models}
	\caption*{\fontsize{10pt}{0.35cm}\selectfont Rankings of top-50 most influential variables in terms of overall model contribution. Characteristics are ordered based on the sum of their ranks over all models, with the most influential characteristics on the top and the least influential on the bottom. Columns correspond to the individual models,
		and the color gradients within each column indicate the most influential (dark blue) to the least influential (white) variables.}
	\includegraphics[width=1\linewidth]{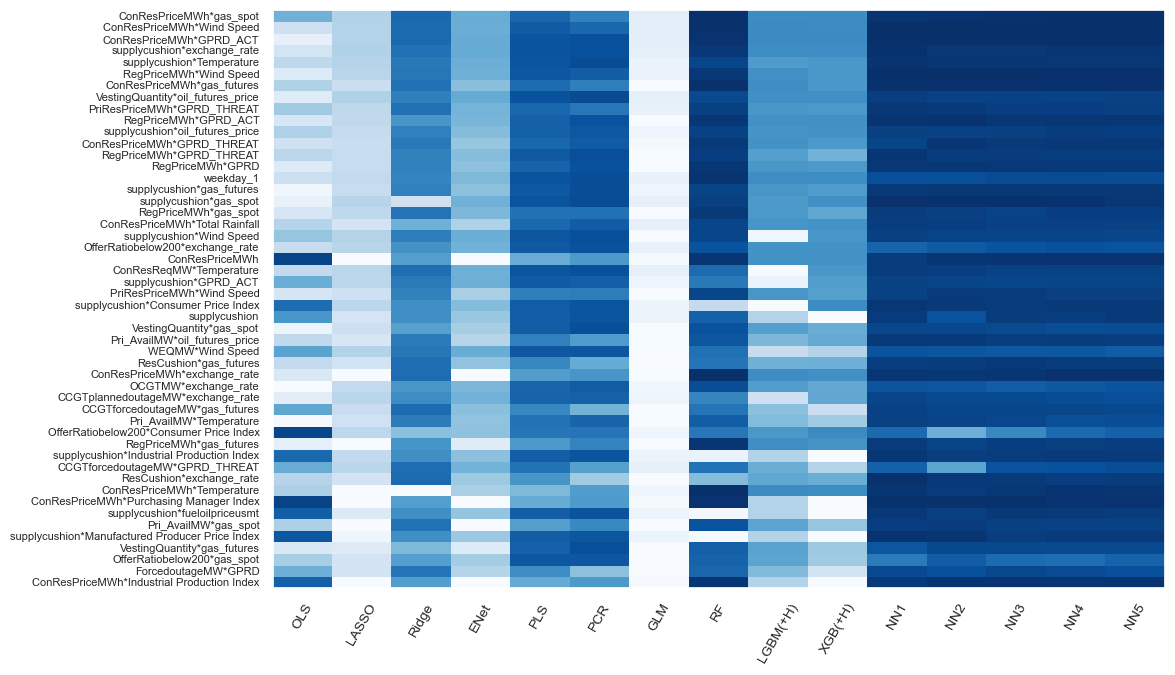}
	\label{fig:a6}
\end{figure}

To provide a comprehensive picture of feature importance using reductions in $R^2$, we examine the average rankings of features aggregated across all models. Specifically, we combine the feature importance rankings from each model to obtain an overall ranking of importance. \footnote{Figure \ref{fig:a6} presents the ordering of all characteristics based on their cumulative importance across models. The overall ranks are computed by first determining the relative contribution of each variable within a given model, ranking them accordingly, and then summing these ranks across all models. Variables with the highest total ranks are positioned at the top, while those with the lowest ranks are at the bottom. The color gradients within each column provide a visual representation of the model-specific rankings, with darker shades indicating higher importance and lighter shades representing lower importance.}. For clarity, only the top 50 features are presented in Figure \ref{fig:a6}. The results are generally consistent with the above frequency-based analysis. Notably, the model disagreement of USEP\_lag excludes it from the top 50 average rankings \footnote{We compute the standard deviation of feature ranks across models and OOS months, defining it as a measure of model disagreement for each feature within each OOS month. Based on this metric, we identify the top and bottom 10\% of features exhibiting the highest and lowest levels of model disagreement, respectively. The features that consistently appear in the top 10\% across all 72 OOS months include USEP\_lag, weekday 0 (Sunday), and three temperature interaction features: Offer Ratio below 200*Temperature, Offer Ratio below 300*Temperature, and WEQMW*Temperature. These features are predominantly favored by linear models, whereas non-linear models assign them relatively lower importance. In contrast, the features most frequently ranked within the bottom 10\% of model disagreement, appearing in at least 65 out of 72 OOS months, comprise contingency reserve price interaction features, raw material price interaction features, and GPRD interaction features. Specifically, these features include ConResPriceMWh*GPRD\_ACT, ConResPriceMWh*Wind Speed, VestingQuantity*oil\_\\futures\_price, supplycushion*gas\_futures, and PriResPriceMWh*GPRD\_THREAT.}.

The distribution of colors reveals the model similarity. Among the penalized linear models, LASSO and ENet share a similar pattern, while Ridge deviates from this, which aligns with our performance results that ENet performs similarly to LASSO rather than Ridge. Consistent patterns are clustered among the same model category, i.e., PLS and PCR behave similarly, LGBM(+H) and XGB(+H) display high agreement, as do NNs, while GLM, RF, and OLS each display independent color distributions. Notably, consistent with Figure \ref{fig:7}, GLM relies heavily on the lag term (more specifically, the higher polynomial of the lag terms) rather than the interactions between in-market and macro features, which explains why the GLM column is nearly white. Based on the behavior of our GLM, in our data setting, a higher polynomial of the same variable within a category is possibly more important than the interaction of many variables across categories, as in the setting of \citet{bianchi2021bond}.

\begin{figure}[H]
	\caption{\centering Feature Importance Based on Shapley Value Across Each Model}
	\caption*{\fontsize{10pt}{0.35cm}\selectfont Variable importance for the top-20 most influential variables in each model. Variable importance is an average over all training samples. Variable importance within each model is normalized to sum to one.}
	\begin{multicols}{2}
		\begin{figure}[H]
			\includegraphics[width=1\linewidth]{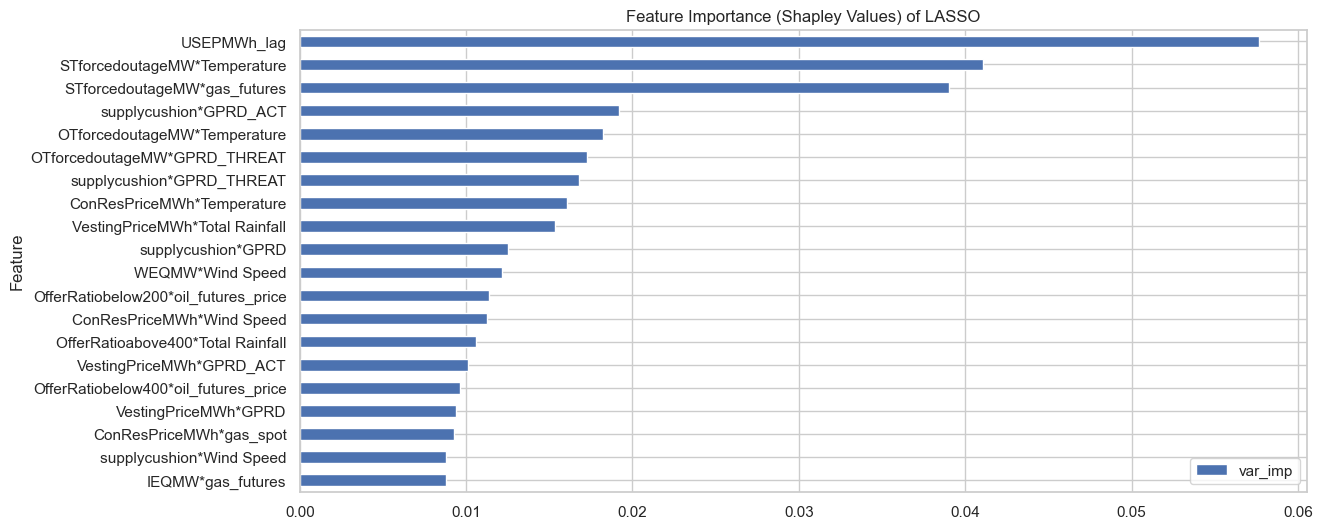}
			\includegraphics[width=1\linewidth]{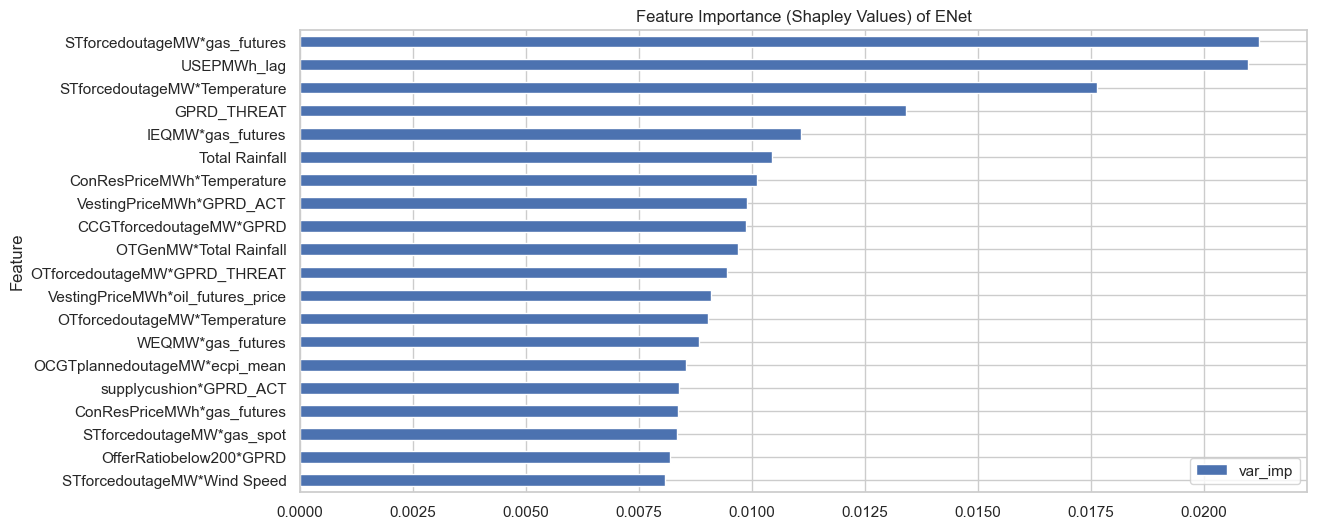}
			\includegraphics[width=1\linewidth]{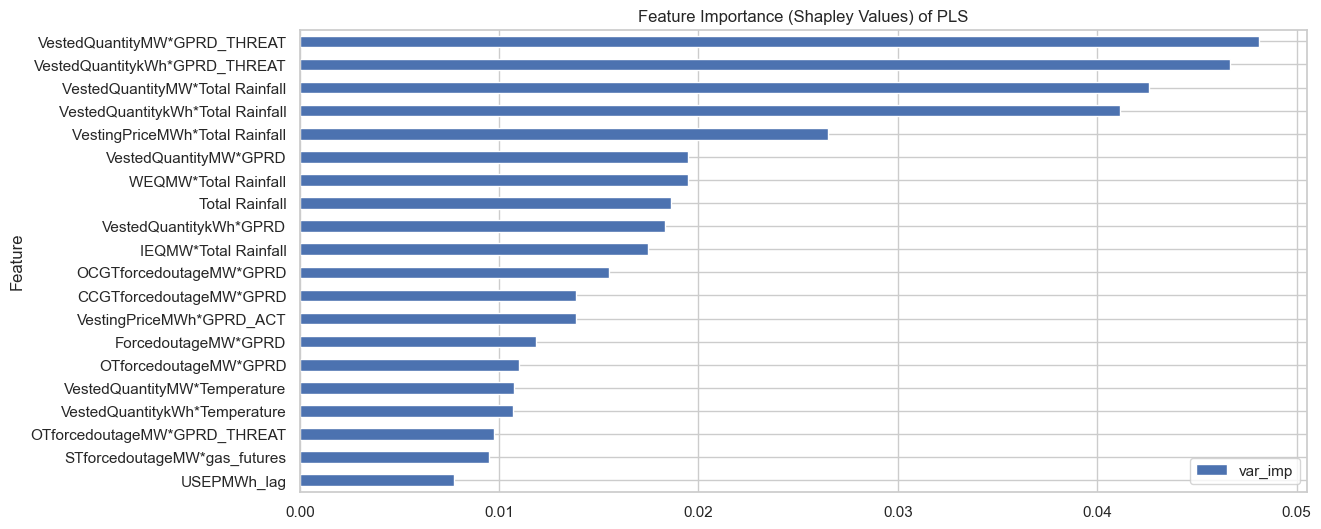}
			\includegraphics[width=1\linewidth]{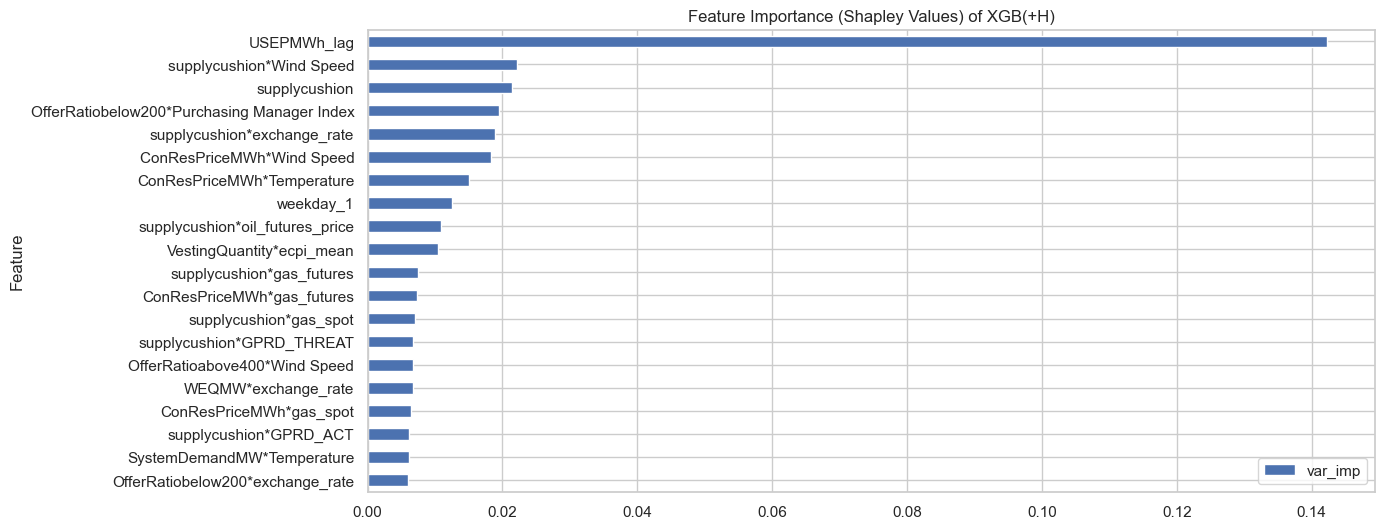}
			\includegraphics[width=1\linewidth]{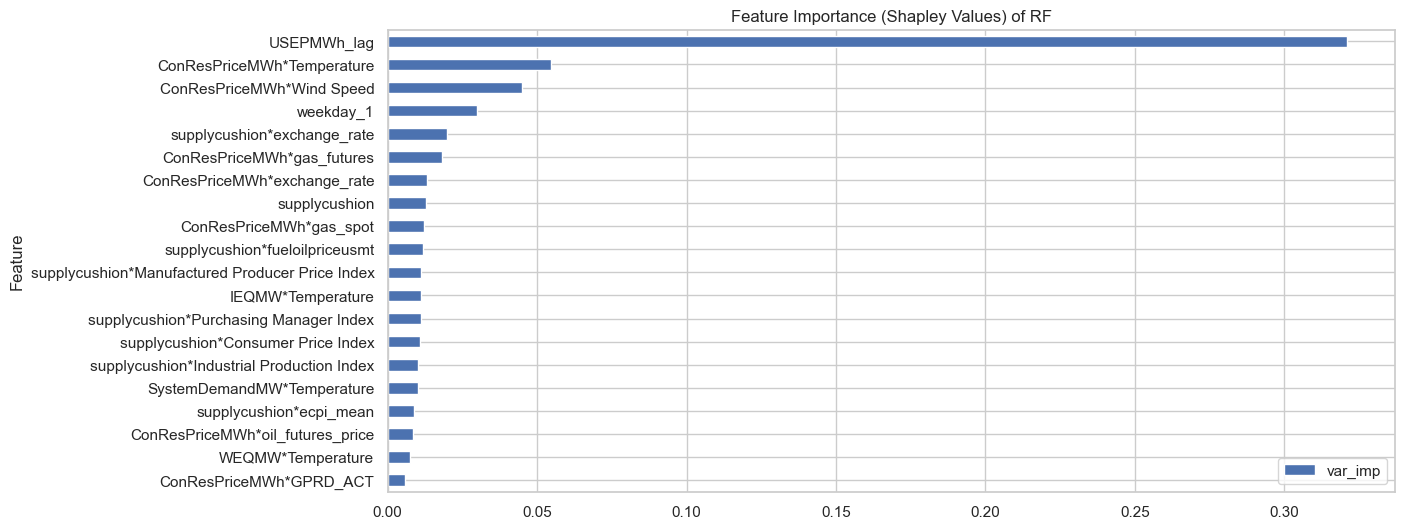}
		\end{figure}
		\begin{figure}[H]
			\includegraphics[width=1\linewidth]{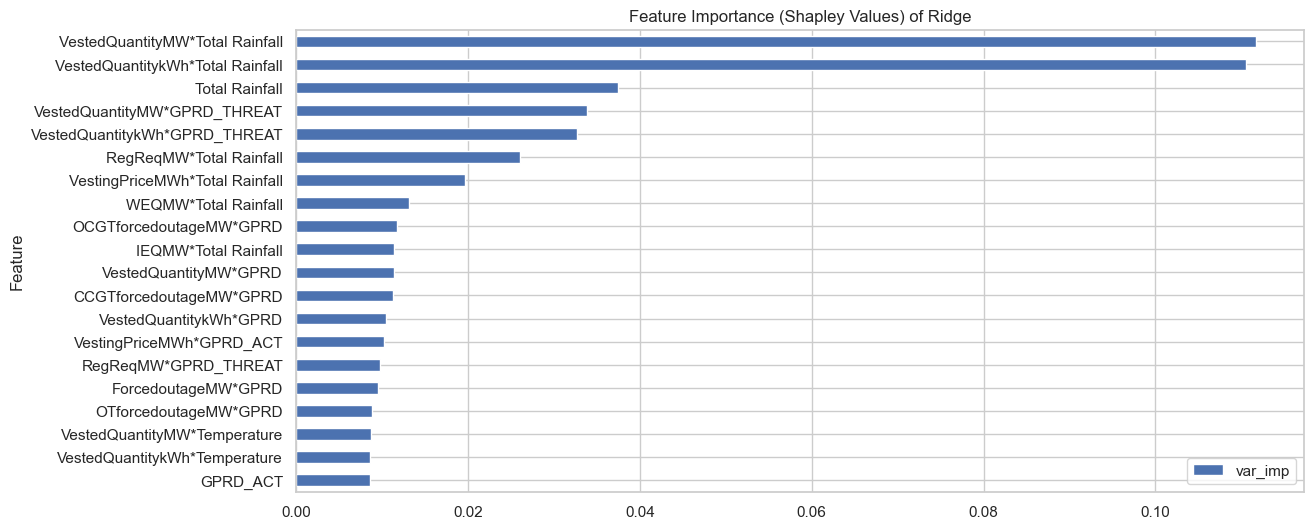}
			\includegraphics[width=1\linewidth]{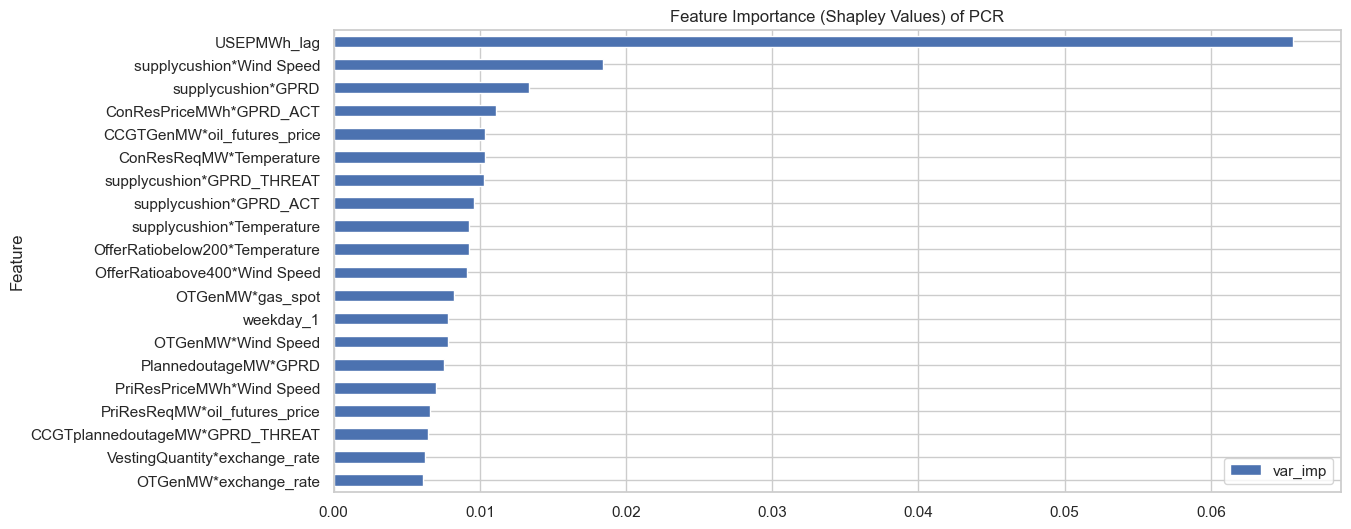}
			\includegraphics[width=1\linewidth]{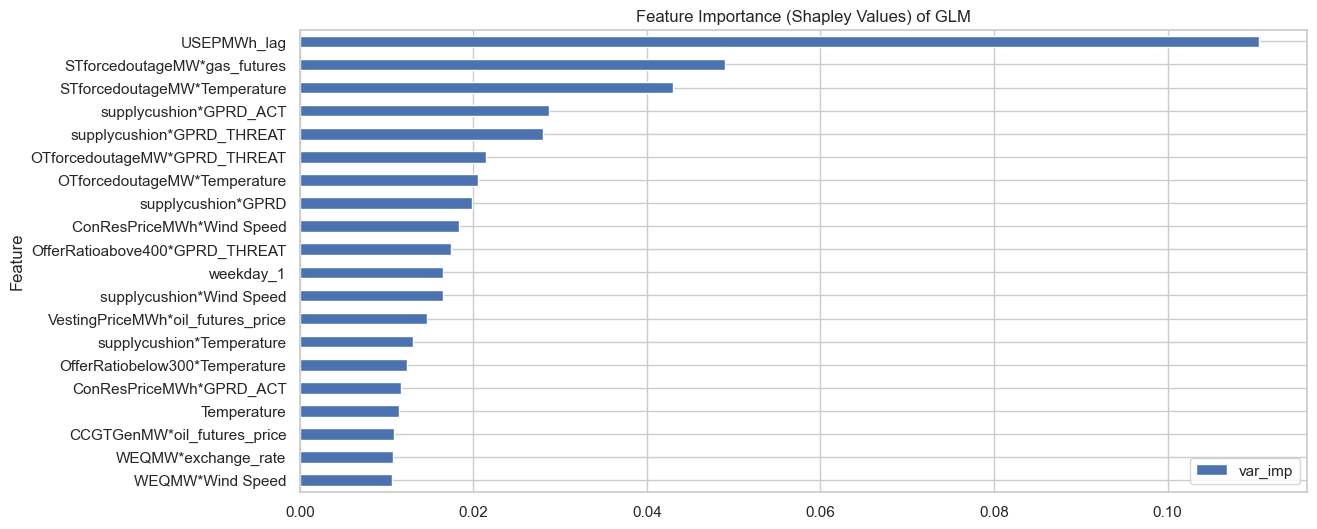}
			\includegraphics[width=1\linewidth]{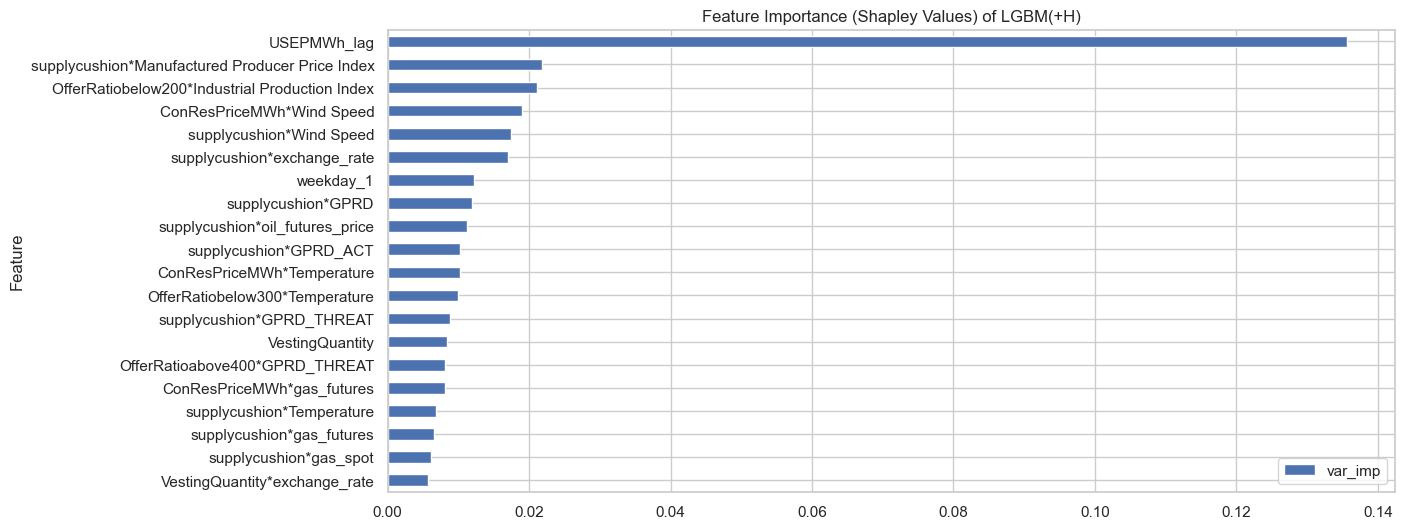}
			\includegraphics[width=1\linewidth]{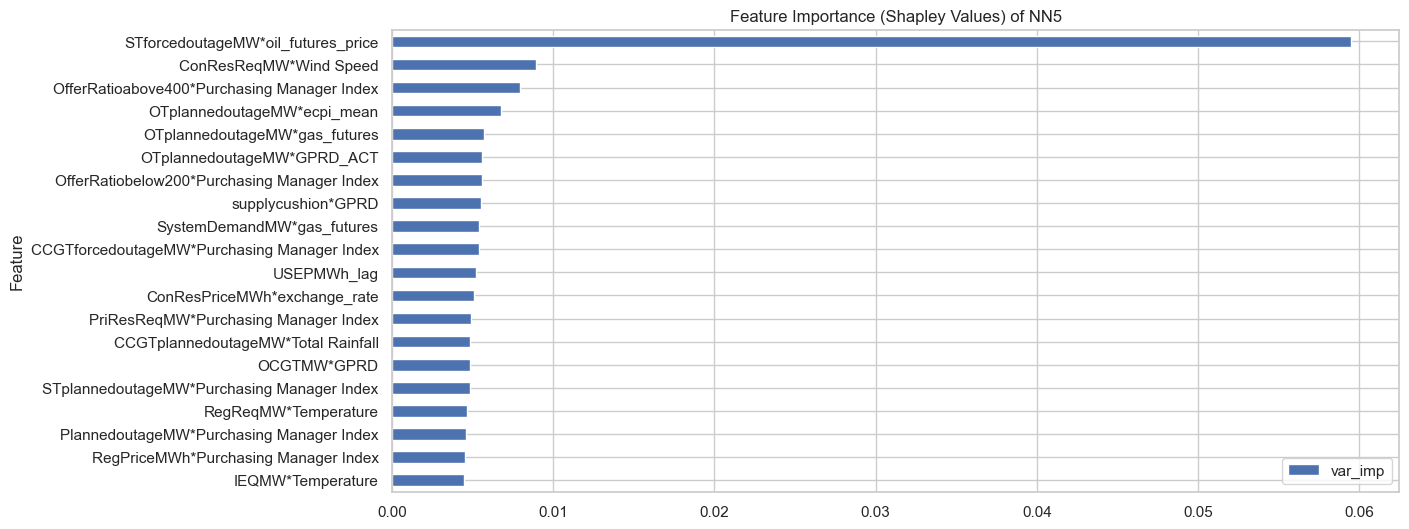}
		\end{figure}
	\end{multicols}
	\label{fig:a7}
\end{figure}

For feature importance based on Shapley value across each model, as shown in Figure \ref{fig:a7}, similar to the $R^2$ reduction results, interactions matter more than individual features. Not surprisingly, compared with the distribution of magnitudes based on $R^2$ reduction, the distribution based on Shapley value does not display sparsity in penalized linear models, as Shapley value cannot reflect the statistical properties of the models. USEP\_lag consistently emerges as the most frequently top-ranked feature across models, showing less disagreement across models compared to the $R^2$ reduction results. Other features that frequently appear in the top five rankings are consistent with the $R^2$ reduction results. The primary distinction is that Shapley value emphasize the importance of outage-related variables.

\begin{figure}[H]
	\caption{Feature Importance Based on SSD Across Each Model}
	\begin{multicols}{2}
		\begin{figure}[H]
			\includegraphics[width=1\linewidth]{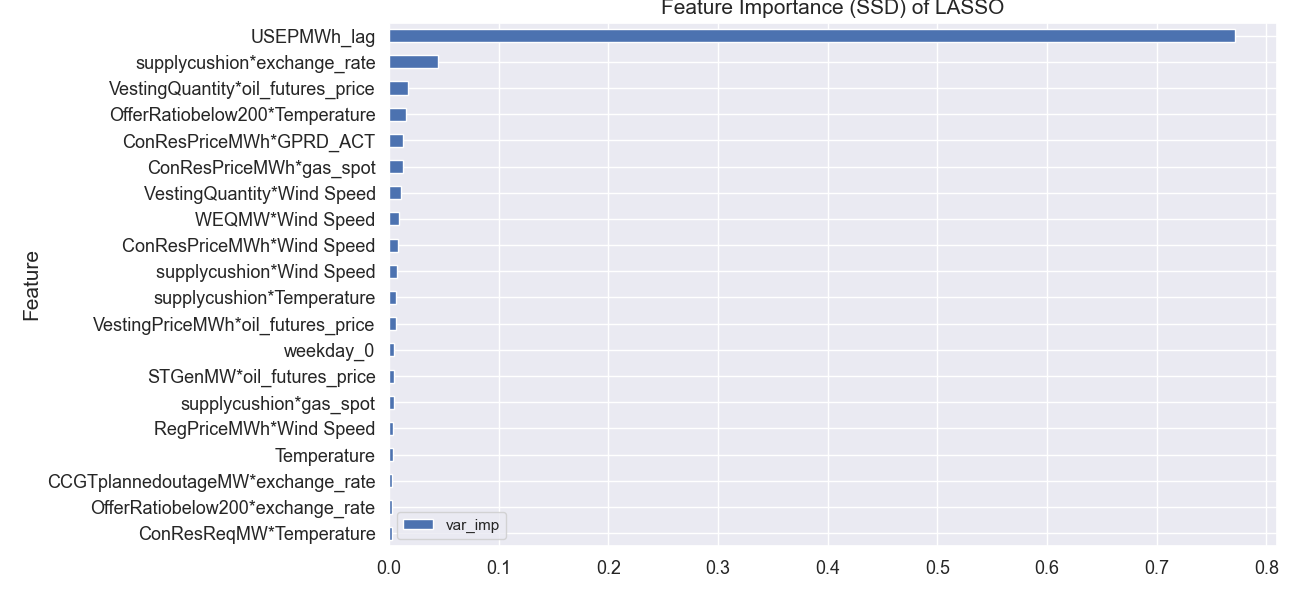}
			\includegraphics[width=1\linewidth]{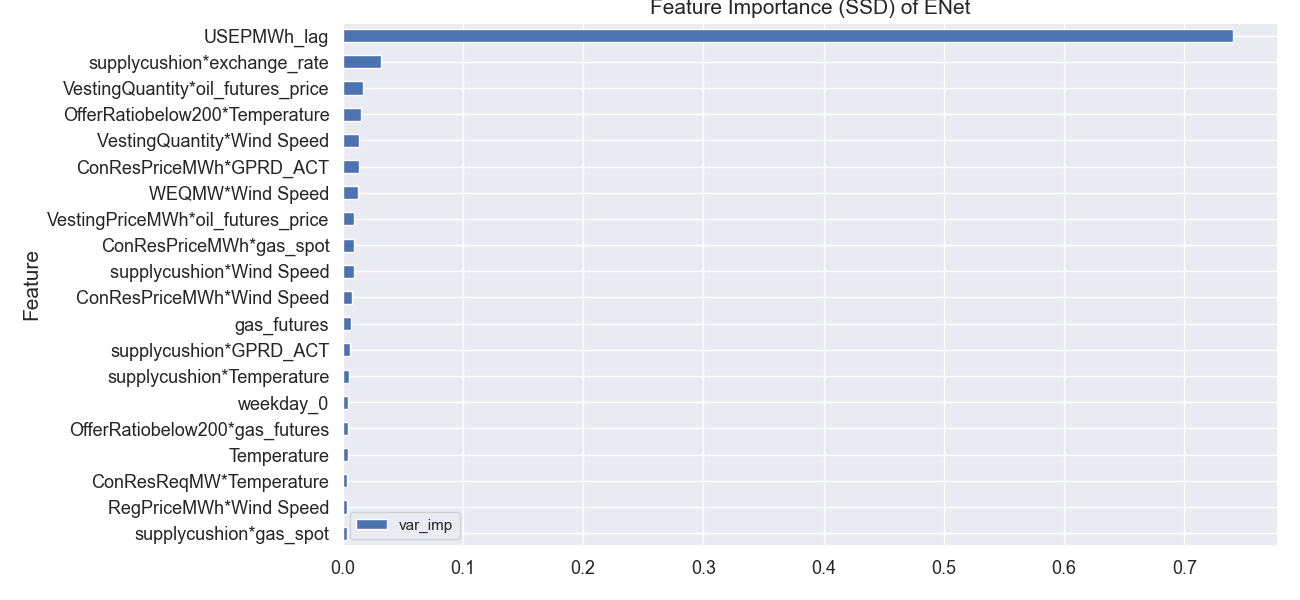}
		\end{figure}
		\begin{figure}[H]
			\includegraphics[width=1\linewidth]{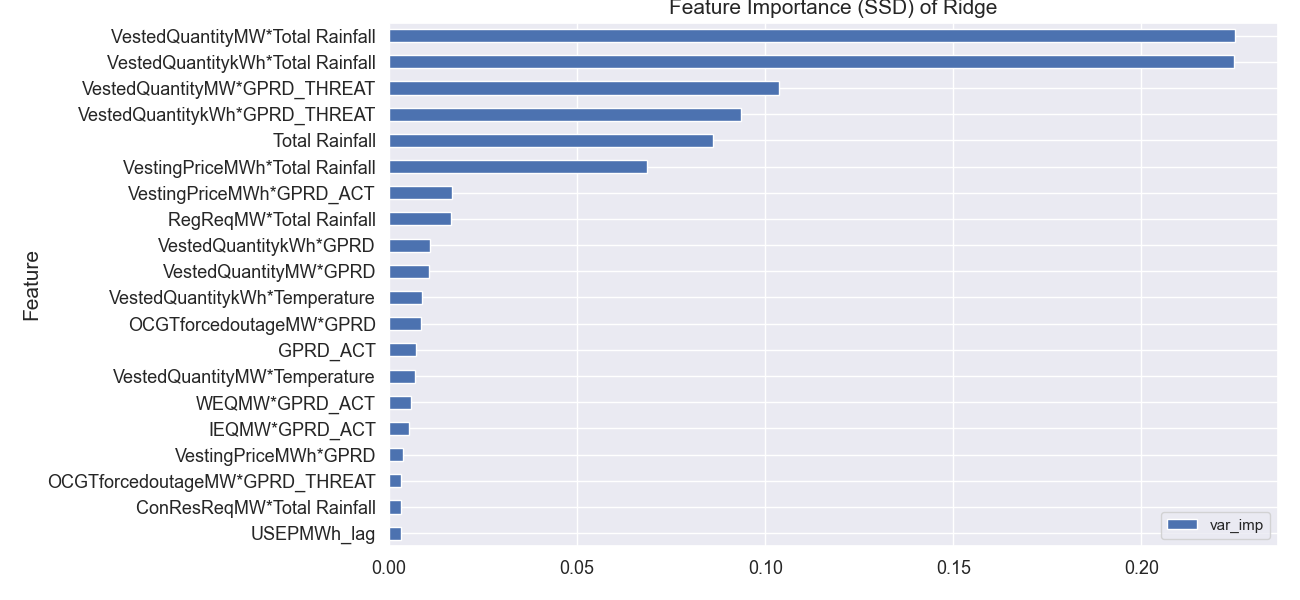}
			\includegraphics[width=1\linewidth]{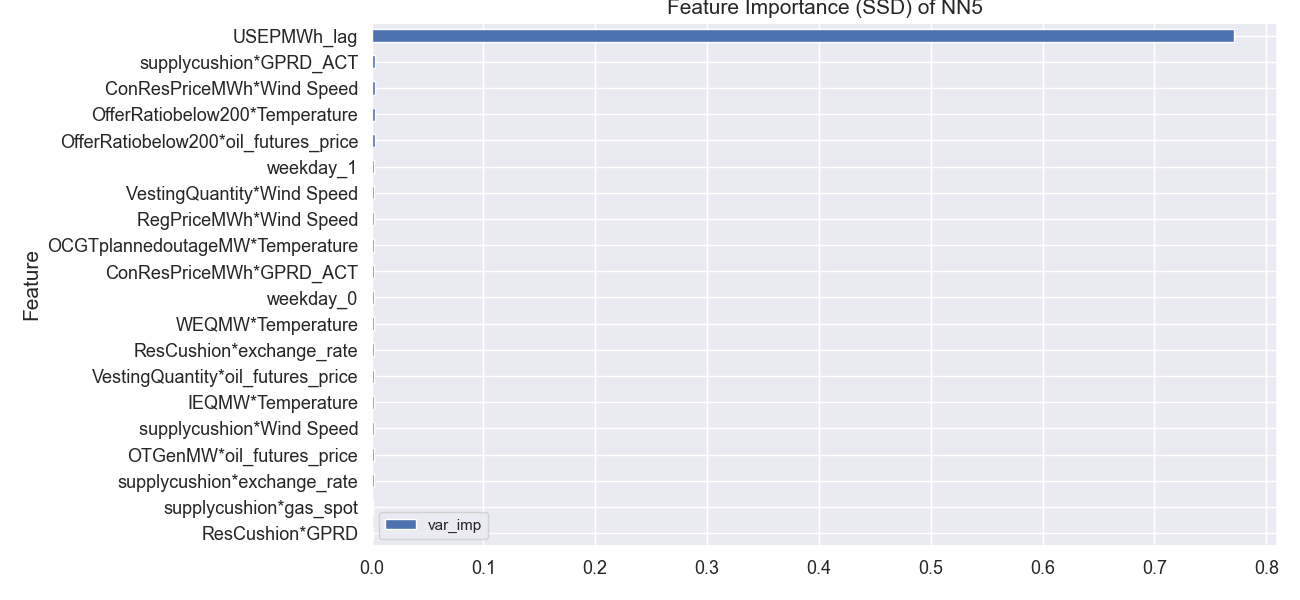}
		\end{figure}
	\end{multicols}
	\label{fig:a8}
\end{figure}

\begin{figure}[H]
	\caption{Feature Importance Based on SSD Across All Models}
	\centering
	\includegraphics[width=1\textwidth]{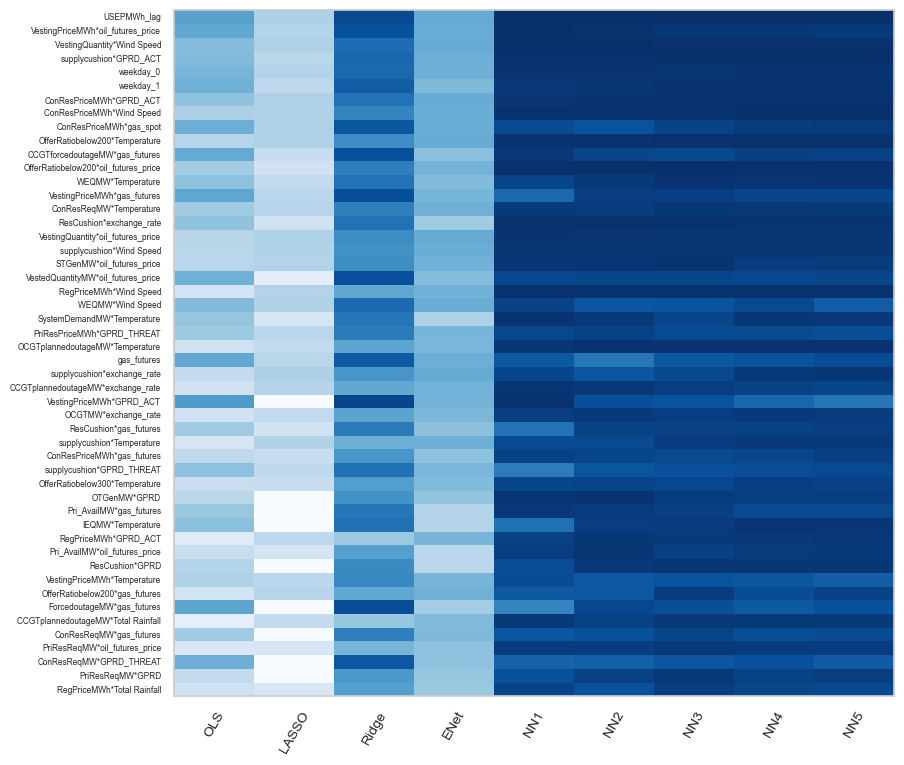}
	\label{fig:a9}
\end{figure}

\newpage
Figure \ref{fig:a8} presents the SSD results for each model, which closely align with those obtained from the \(R^2\) reduction and Shapley Value analyses. Consistent with the \(R^2\) reduction and Shapley Value results, USEP\_lag emerges as the most frequently ranked top feature across models. Other features frequently appearing in the top five rankings also mirror the findings from the \(R^2\) reduction and Shapley Value, highlighting the robustness of our analysis. Specifically, these features include raw material prices, supply cushion, vested quantity, weather-related variables, GPRD-related features, and the offer ratio below 200. Figure \ref{fig:a9} presents the aggregated SSD results across all models, which are consistent with the individual model results, further confirming the reliability of these findings.

\begin{figure}[H]
	\caption{Feature Importance Based on MDG Across Each Model}
	\begin{multicols}{2}
		\begin{figure}[H]
			\includegraphics[width=1\linewidth]{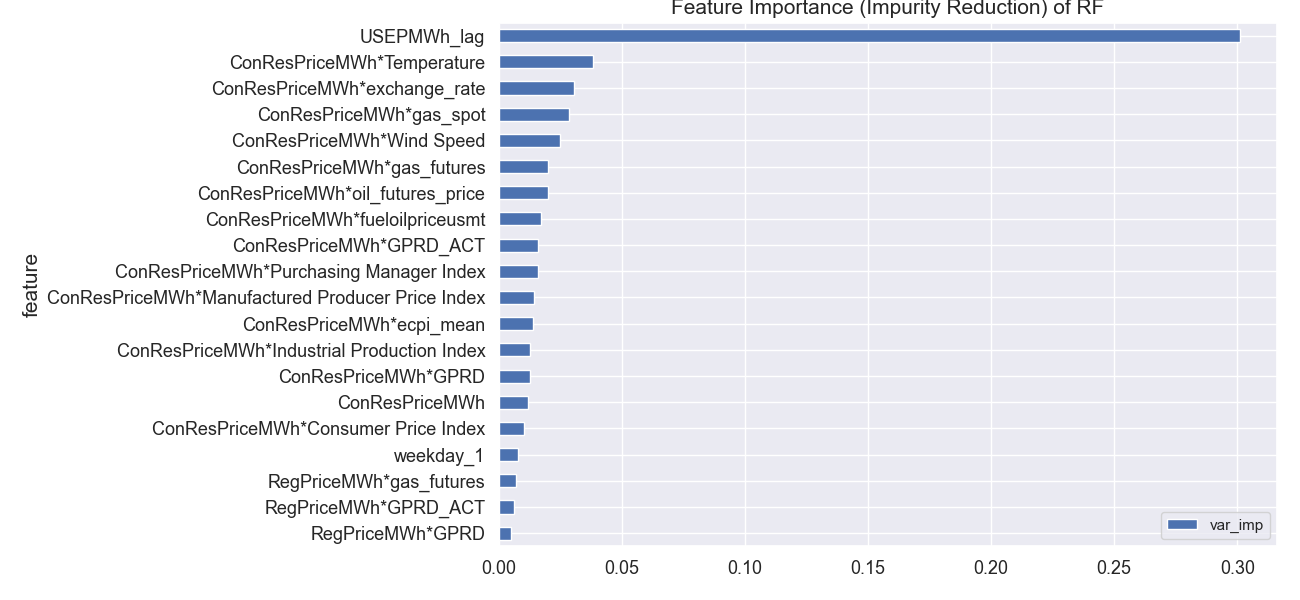}
			\includegraphics[width=1\linewidth]{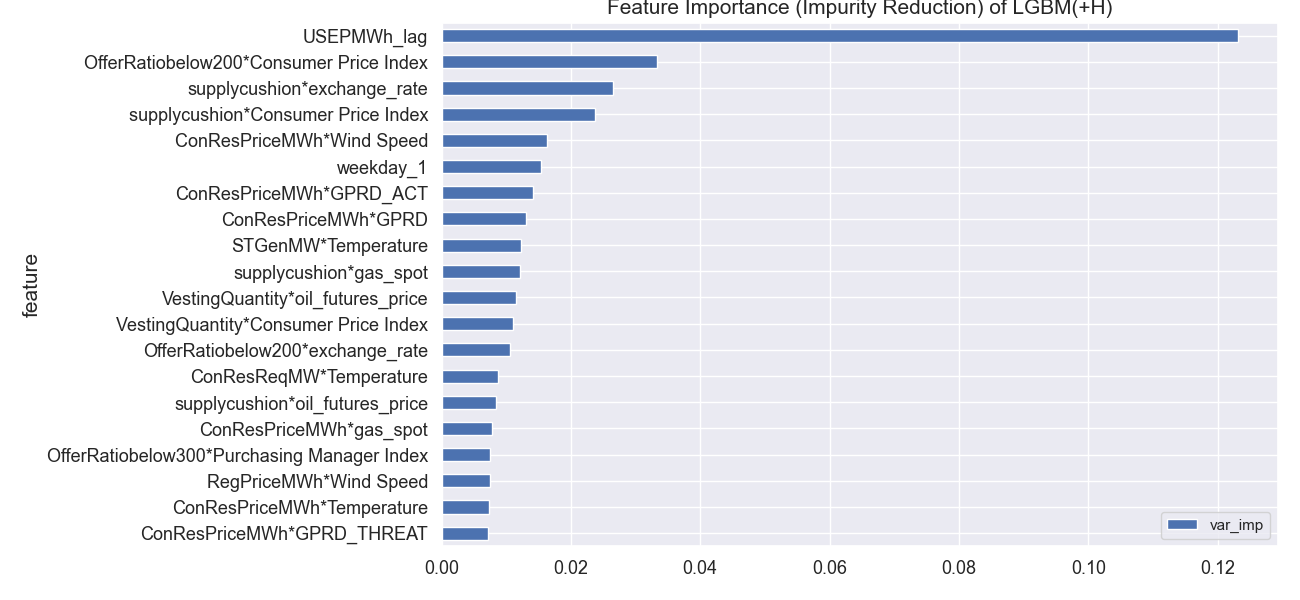}
		\end{figure}
		\begin{figure}[H]
			\includegraphics[width=1\linewidth]{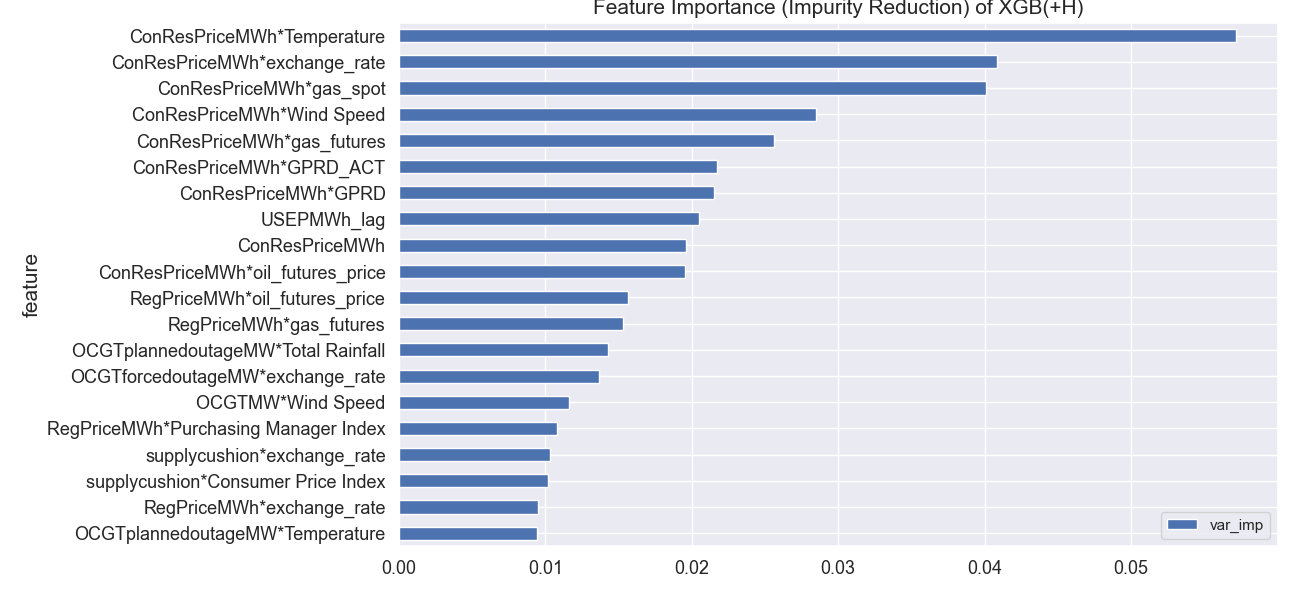}
		\end{figure}
	\end{multicols}
	\label{fig:a10}
\end{figure}

Figure \ref{fig:a10} reports the MDG results for each model, which show strong alignment with the findings from the \(R^2\) reduction, Shapley Value, and SSD methods. As with the other methods, USEP\_lag consistently ranks as the most influential feature across models. Other features frequently appearing among the top five include contingency reserve price, raw material prices, weather-related variables, supply cushion, macroeconomic indicators such as exchange rate and CPI, and the offer ratio below 200. These results further emphasize the robustness of our analysis.

\begin{figure}[H]
	\caption{Feature Importance Based on MDG Across All Models}
	\centering
	\includegraphics[width=1\textwidth]{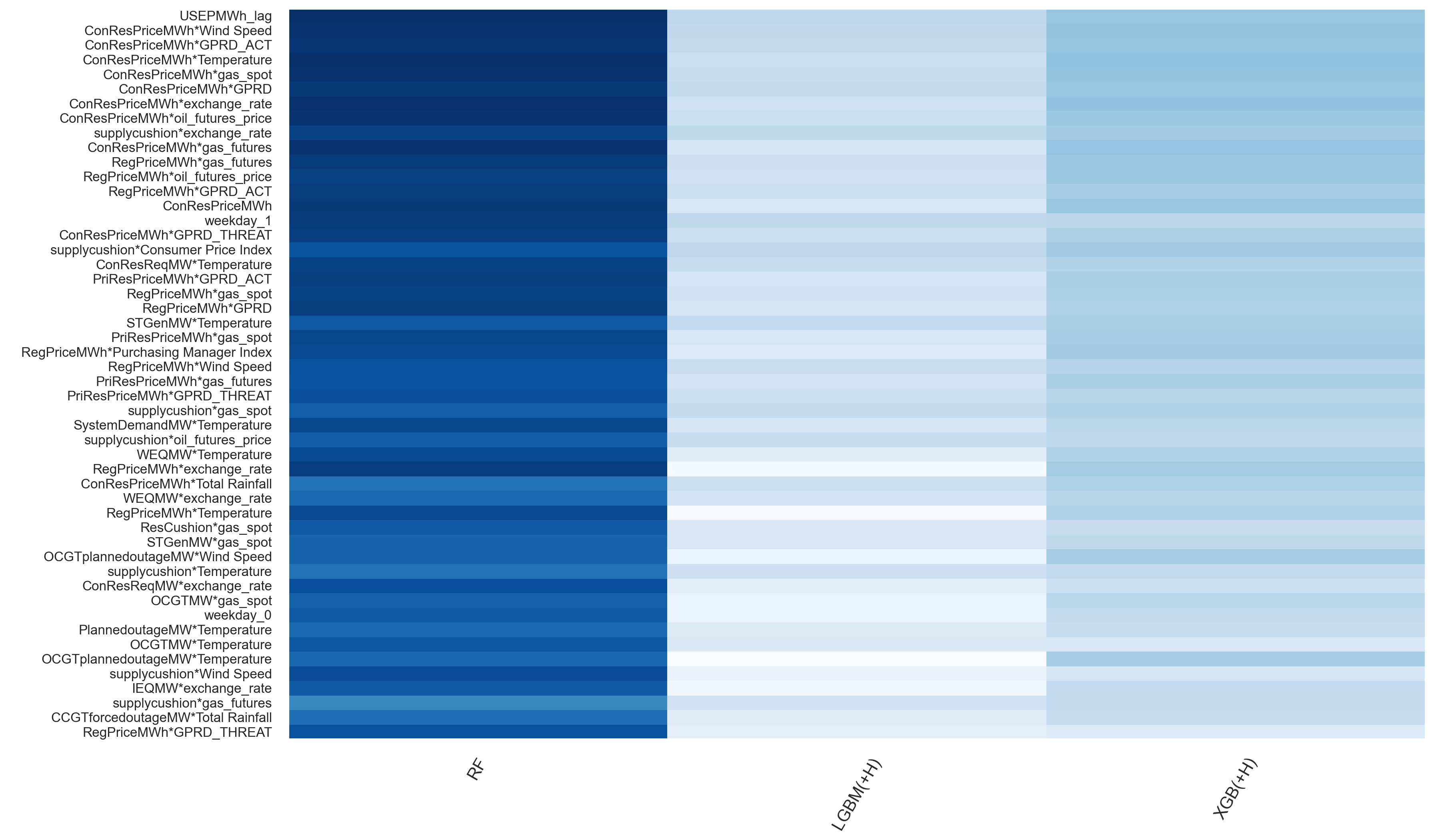}
	\label{fig:a11}
\end{figure}

Figure \ref{fig:a11} provides the aggregated MDG results across all models, which reaffirm the patterns observed in individual model analyses, demonstrating the reliability and consistency of these findings.

\begin{figure}[H]
	\caption{Mean-Variance Optimal Portfolio ($\gamma$=2)}
	\includegraphics[width=1\linewidth]{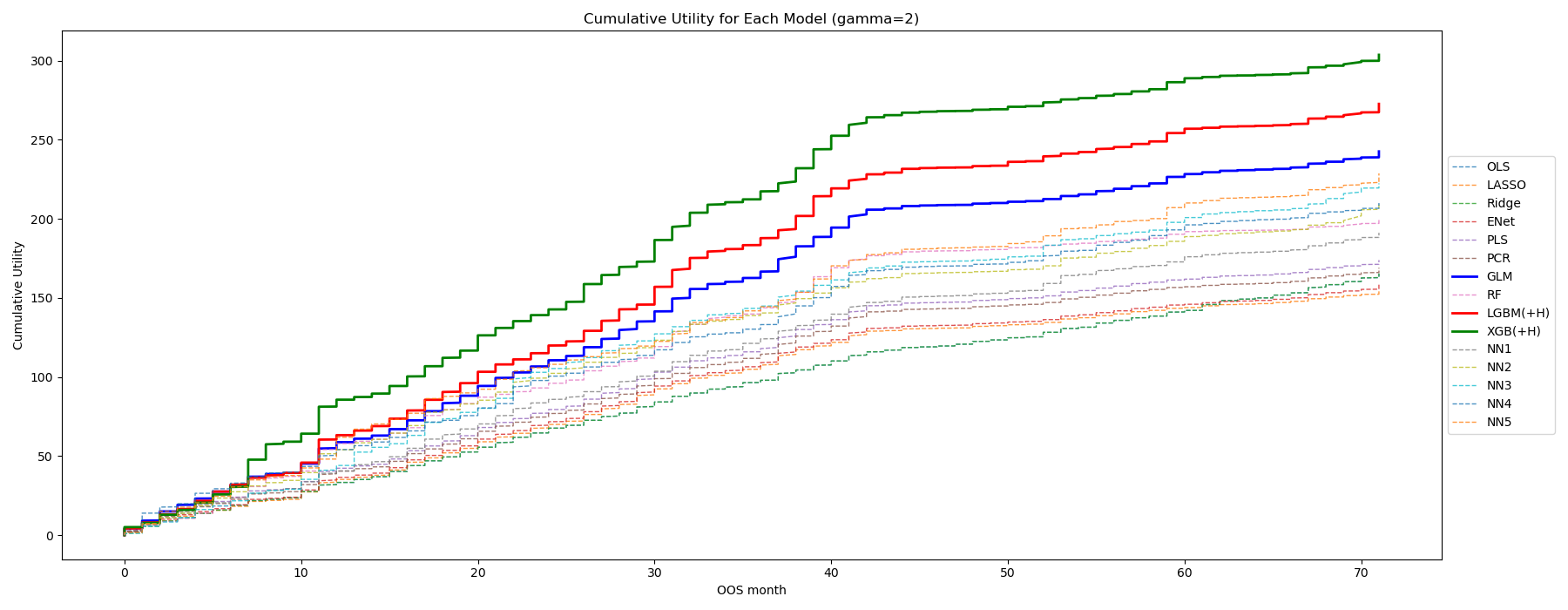}
	\includegraphics[width=1\linewidth]{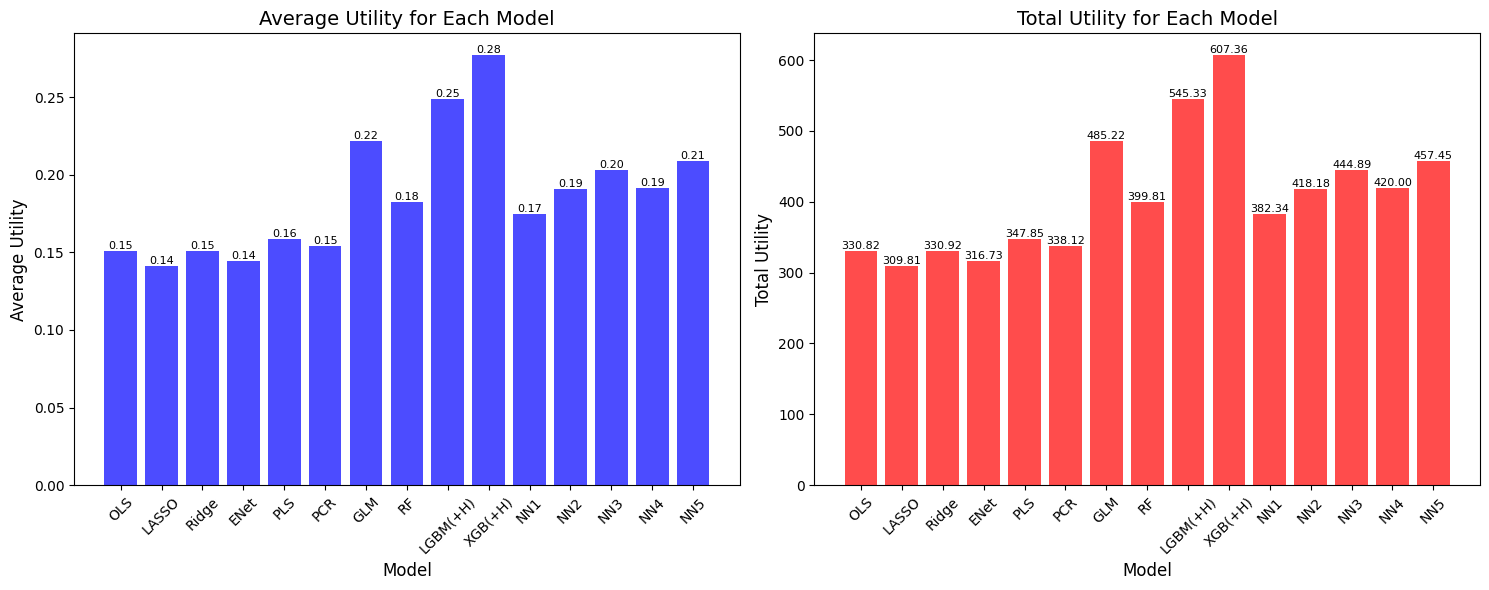}
	\label{fig:a12}
\end{figure}

\begin{figure}[H]
	\caption{Mean-Variance Optimal Portfolio ($\gamma$=4)}
	\includegraphics[width=1\linewidth]{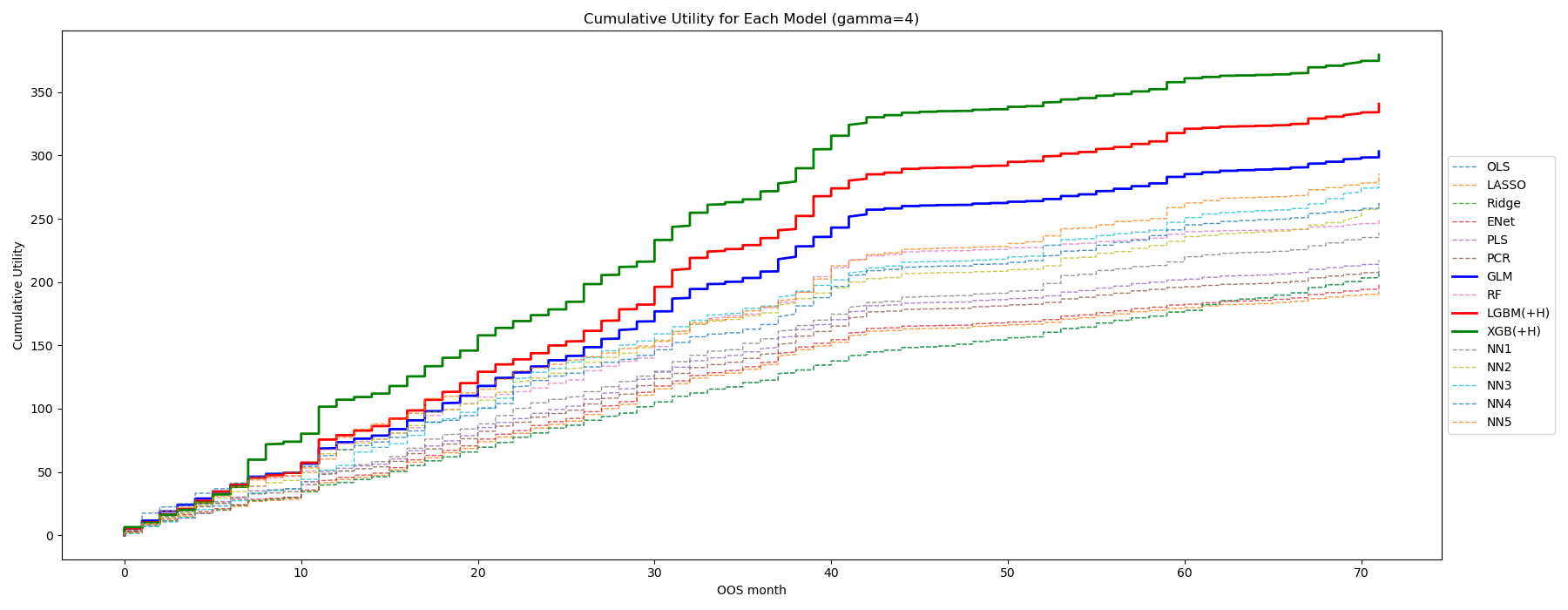}
	\includegraphics[width=1\linewidth]{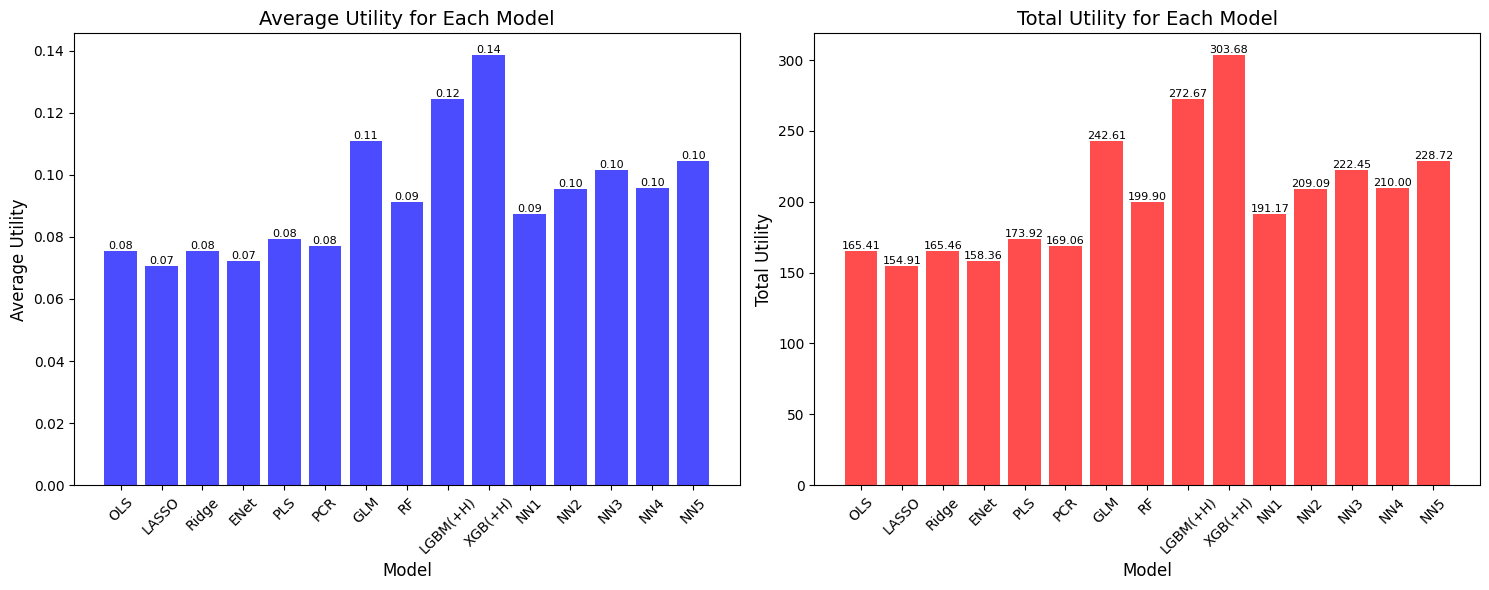}
	\label{fig:a13}
\end{figure}

\begin{figure}[H]
	\caption{Mean-Variance Optimal Portfolio ($\gamma$=5)}
	\includegraphics[width=1\linewidth]{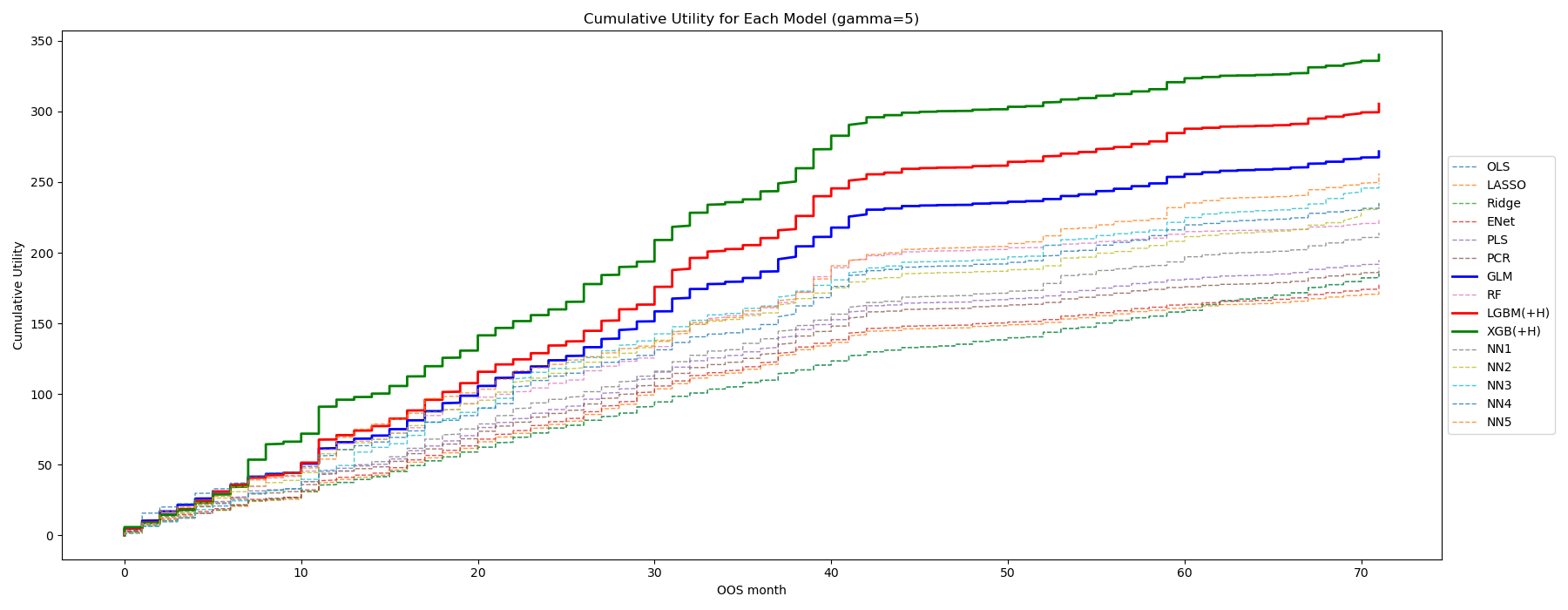}
	\includegraphics[width=1\linewidth]{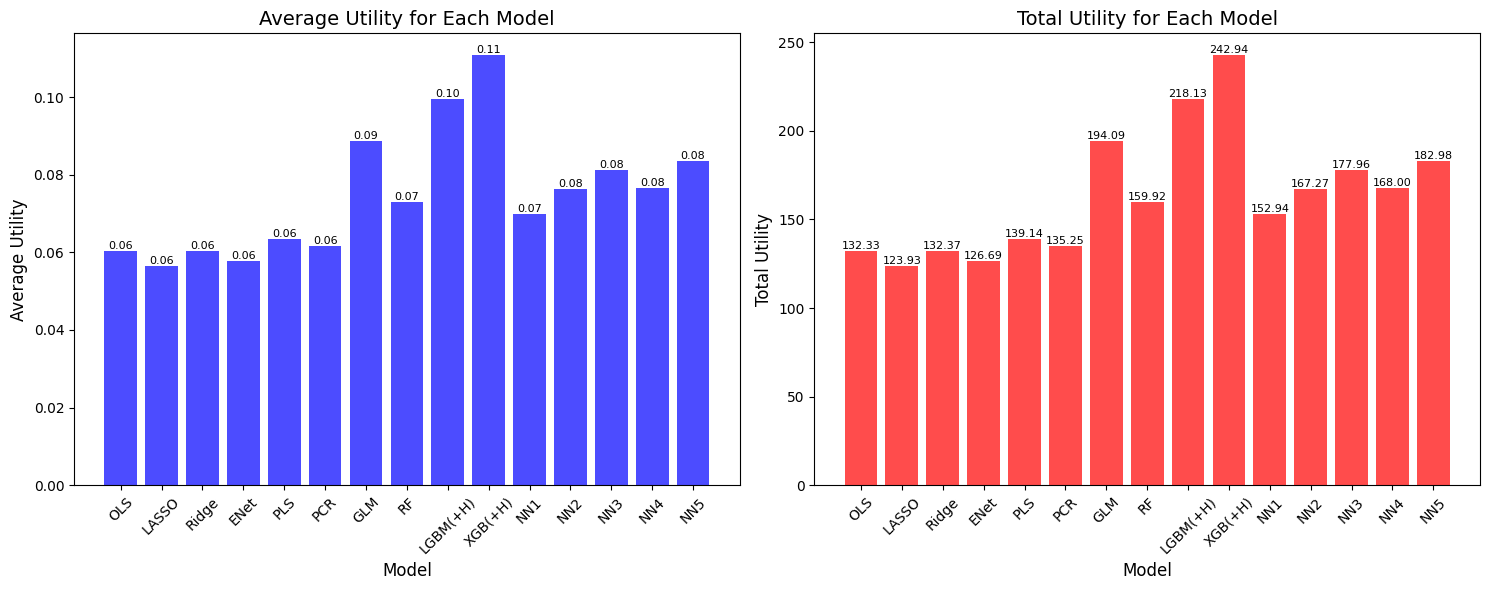}
	\label{fig:a14}
\end{figure}

\newpage
\subsection{Monte Carlo Simulations}
\label{simulation}
To demonstrate the finite sample performance of all machine learning models, we simulate a (latent) 3-factor model for the conditional mean of returns $r_{t+1}$ for $t = 1, 2, \dots, T$:

\begin{align}
	& r_{t+1} = \mu_{t+1} + e_{t+1} = f(Z_t) + e_{t+1}, \\
	& e_{t+1} = \lambda_t v_{t+1} + \epsilon_{t+1}, \\
	& Z_t = (p_t, s_t, x_t, (1, x_t) \otimes C_t), \\
	& \lambda_t = (C_{1,t}, C_{2,t}, C_{3,t}) \otimes x_t, \\
	& v_{t+1} \sim N(0, \sigma_v^2 \times I_3), \\
	& \epsilon_{t+1} \sim t_n(0, \sigma_{\epsilon}^2),
\end{align}
where $p_t$ is the price at time $t$; $s_t$ captures the weekly seasonality; $x_t$ is a univariate time series representing a macro feature; $C_t$ is a $T \times P_C$ matrix of in-market features ($P_C \geq 3$). The $j$-th in-market feature is denoted as $C_{j,t}$, where $j \in \{1, \dots, P_C\}$. The three (latent) factors are the interactions of the first three in-market features with the macro feature, which is based on our empirical findings of feature importance (note that $\epsilon_{t+1}$ also captures $p_t$ as a (latent) factor). The variances $\sigma_v^2$ and $\sigma_{\epsilon}^2$ are calibrated to $0.00037873^2$ and $3.5702^2$, so that the average $R^2$ of latent factors is 0.043777\% \footnote{Specifically, we match the average $R^2$ of latent factors with the average $R^2$ reductions of the three most important interaction covariates ($ConResPriceMWh*GPRD\_ACT$, $ConResPriceMWh*Wind\_Speed$, and $ConResPriceMWh*gas\_spot$) across all models and OOS months.} and the annualized variance of simulated $r_{t+1}$ is 13.3297, which is the sample variance of return in the real data.

Given $r_{t+1} = \log\left(\frac{p_{t+1}}{p_t}\right)$, we initialize $p_1$ and generate the rest of the time series recursively. For initialization, we employ the sample mean of price 143.9712. 

Following \citet{keles2012comparison}, we simulate the weekly seasonality as:
\begin{equation}
	s_t = \alpha + \beta |\sin(\frac{\pi t}{7} - \phi)|,
\end{equation}
where $\alpha$, $\beta$ and $\phi$ are estimated by regressing polynomial de-trended returns $r^{\text{de-trend}}_{t+1}$ on $|\sin(\frac{\pi t}{7} - \phi)|$, which are calibrated as -0.21883, 0.88886 and 0.20005 respectively. 

Then, We simulate the univariate time series $x_t$ as an AR(1) process:
\begin{equation}
	x_t = \rho x_{t-1} + \mu_t,
\end{equation}
where $\mu_t \sim N(0, 1 - \rho^2)$ and $\rho = 0.95$ to ensure high persistence and stationarity.

In addition, for any $j \in \{1, \dots, P_C\}$, we simulate the in-market feature $C_{j,t}$ as:
\begin{equation}
	C_{jt} = \text{MonStand}(\bar{C}_{j,t}),
\end{equation}
where $\bar{C}_{j,t} = \rho_j \bar{C}_{j,t-1} + \varepsilon_{j,t}$, $\rho_j \sim U[0.9, 1]$, and $\varepsilon_{j,t} \sim N(0, 1 - \rho_j^2)$. $\text{MonStand}(\cdot)$ is the monthly $(0,1)$-standardization function.

Two types of $f(\cdot)$ are considered:
\begin{align}
	& \text{Model 1:} \quad f(Z_t) = (p_t, s_t, C_{1,t}, C_{2,t}, C_{3,t} \times x_t) \theta_1, \\
	& \text{Model 2:} \quad f(Z_t) = (p_t, s_t, C_{1,t}^2, C_{1,t} \times C_{2,t}, \text{sgn}(C_{3,t} \times x_t)) \theta_2.
\end{align}
In both cases, $f(\cdot)$ only depends on 5 covariates, so there are only 5 non-zero entries in $\theta$, denoted as $\theta_1$ and $\theta_2$ respectively for the two models.  Eq (10) is simple and sparse linear model, while Eq (11) involves a nonlinear covariate $C_{1,t}^2$, a nonlinear interaction term $C_{1,t} \times C_{2,t}$, and a discrete variable $\text{sgn}(C_{3,t} \times x_t)$. $\theta_1$ and $\theta_2$ (both $5 \times 1$ vectors) are calibrated to achieve predictive $R^2$ values of 35\%. Specifically, 
\begin{equation}
	\begin{aligned}
		& \theta_1 = (-0.0013, -0.0032, 0.0059, 0.0037, 0.0053)^{\prime} \\
		& \theta_2 = (-0.0026, 0.0082, 0.0074, 0.0010, 0.0052)^{\prime}.
	\end{aligned}
\end{equation}

Throughout, we fix $T=3600$, and $P_x = 2$, while comparing the cases of $P_C = 100$ and $P_C = 50$, to demonstrate the effect of increasing dimensionality.

For each Monte Carlo sample, we divide the whole time series into 3 consecutive subsamples of 7:2:1 for training, validation and testing. Specifically, we estimate each of the two models in the training sample, using OLS, LASSO, Ridge, ENet, PLS, PCR, GLM, RF, LightGBM, XGBoost, and NNs that we adopt for the empirical part, then choose the tuning parameters for each method in the validation sample, and calculate the prediction errors in the test sample. For benchmark, we also compare the pooled OLS with all covariates and that using the oracle model.

First, we report the average $R^2$s both in-sample (IS) and out-of-sample (OOS) for each model and each method over 100 Monte Carlo repetitions in Table \ref{tab:A14}. Similar to empirical part, we select historical mean and zero as benchmarks. For model 1, LASSO, ENet, GLM and tree-based ensembles deliver the best and almost identical (if we keep 2 decimals) OOS $R^2$ with historical mean or zero benchmarks. This is not surprising given that the true model is sparse
and linear in the input covariates. The advanced neural network methods tend to overfit, so their performance is quite good in in-sample but worse in out-of-sample. By contrast, for model 2, these NNs clearly dominate Lasso and ENet, because the latter cannot capture the nonlinearity in the model. PLS outperforms PCR in both the linear model 1 and the nonlinear model 2. When $P_C$ increases, the IS $R^2$ tends to increase while the OOS $R^2$ decreases. Hence, the performance of all methods deteriorates as overfitting exacerbates.

Next, we report the average variable selection frequencies of 6 particular covariates and the average of the remaining $P_C - 6$ covariates for model 1 and 2 in \ref{tab:A15}, using Lasso and ENet. We focus on these methods because they all impose the $l1$ penalty and hence encourage variable selection. As expected, for the model 1, the true covariates are ($p_t$, $s_t$, $C_{1,t}$, $C_{2,t}$ and $C_{3,t} \times x_t$) are selected more often than the correlated yet redundant
covariate ($x_t$). By contrast, the remaining covariates (noise) are rarely selected. For model 2, the true covariates are $(p_t, s_t, C_{1,t}^2, C_{1,t} \times C_{2,t}, \text{sgn}(C_{3,t} \times x_t))$, and we observe that $p_t, s_t$ are selected more often in model 2 than the model 1.

Overall, the simulation results suggest that the machine learning methods are successful in (1) capturing non-linear relationship between features; (2) singling out informative variables, even though highly correlated covariates are difficult to distinguish.

\begin{table}[H]
	\fontsize{7pt}{0.35cm}\selectfont
	\makebox[\textwidth]{
		\rotatebox{90}{
			\begin{minipage}{\textheight}
				\centering
				\caption{Comparison of Predictive $R^2$s for Machine Learning Algorithms in Simulations}
				\label{tab:A14}
				\setlength{\tabcolsep}{7pt}
				\renewcommand{\arraystretch}{1.1}
				\begin{threeparttable}
					\begin{tabular}{p{0.9cm} p{0.9cm} p{0.9cm} p{0.9cm} p{0.9cm} p{0.1cm} p{0.9cm} p{0.9cm} p{0.9cm} p{0.9cm} p{0.3cm} p{0.9cm} p{0.9cm} p{0.9cm} p{0.9cm} p{0.1cm} p{0.9cm} p{0.9cm} p{0.9cm} p{0.9cm}}
						\hline
						&\multicolumn{9}{c}{Model 1}&&\multicolumn{9}{c}{Model 2}\\
						\cline{2-10} \cline{12-20}
						&\multicolumn{4}{c}{$P_C=50$}&&\multicolumn{4}{c}{$P_C=100$}&&\multicolumn{4}{c}{$P_C=50$}&&\multicolumn{4}{c}{$P_C=100$}\\
						\cline{2-5} \cline{7-10} \cline{12-15} \cline{17-20}
						&\multicolumn{2}{c}{IS}& \multicolumn{2}{c}{OOS} &&\multicolumn{2}{c}{IS}& \multicolumn{2}{c}{OOS}&&\multicolumn{2}{c}{IS}& \multicolumn{2}{c}{OOS}&&\multicolumn{2}{c}{IS}& \multicolumn{2}{c}{OOS}\\
						\cline{2-5} \cline{7-10} \cline{12-15} \cline{17-20}
						Model &$R^{2,mean}_{OOS}$& $R^{2,zero}_{OOS}$& $R^{2,mean}_{OOS}$& $R^{2,zero}_{OOS}$ &&$R^{2,mean}_{OOS}$& $R^{2,zero}_{OOS}$& $R^{2,mean}_{OOS}$& $R^{2,zero}_{OOS}$ &&$R^{2,mean}_{OOS}$& $R^{2,zero}_{OOS}$& $R^{2,mean}_{OOS}$& $R^{2,zero}_{OOS}$ && $R^{2,mean}_{OOS}$& $R^{2,zero}_{OOS}$& $R^{2,mean}_{OOS}$& $R^{2,zero}_{OOS}$\\
						\cline{2-5} \cline{7-10} \cline{12-15} \cline{17-20}
						
						OLS & 0.1051 & 0.1056 & 0.0198 & 0.0231 &  & 0.1397 & 0.1402 & -0.0356 & -0.0322 &  & 0.1072 & 0.1074 & 0.0217 & 0.0246 &  & 0.1413 & 0.1415 & -0.0348 & -0.0317 \\
						LASSO & 0.0744 & 0.0749 & 0.0646 & 0.0678 &  & 0.0758 & 0.0763 & 0.0646 & 0.0677 &  & 0.0767 & 0.0769 & 0.0673 & 0.0701 &  & 0.0777 & 0.0779 & 0.0670 & 0.0698 \\
						Ridge & 0.1151 & 0.1156 & 0.0020 & 0.0054 &  & 0.1592 & 0.1597 & -0.0731 & -0.0695 &  & 0.1171 & 0.1173 & 0.0029 & 0.0059 &  & 0.1607 & 0.1609 & -0.0743 & -0.0711 \\
						ENet & 0.0745 & 0.0750 & 0.0647 & 0.0678 &  & 0.0766 & 0.0771 & 0.0641 & 0.0672 &  & 0.0771 & 0.0773 & 0.0672 & 0.0700 &  & 0.0784 & 0.0786 & 0.0666 & 0.0694 \\
						PLS & 0.0999 & 0.1004 & 0.0302 & 0.0335 &  & 0.1179 & 0.1184 & 0.0100 & 0.0133 &  & 0.1016 & 0.1018 & 0.0319 & 0.0347 &  & 0.1191 & 0.1193 & 0.0111 & 0.0140 \\
						PCR & 0.0210 & 0.0215 & 0.0068 & 0.0101 &  & 0.0355 & 0.0360 & 0.0015 & 0.0048 &  & 0.0221 & 0.0223 & 0.0077 & 0.0106 &  & 0.0362 & 0.0364 & 0.0018 & 0.0048 \\
						GLM & 0.0797 & 0.0802 & 0.0679 & 0.0710 &  & 0.0817 & 0.0822 & 0.0672 & 0.0703 &  & 0.0816 & 0.0819 & 0.0709 & 0.0736 &  & 0.0836 & 0.0839 & 0.0700 & 0.0728 \\
						RF & 0.1906 & 0.1910 & 0.0743 & 0.0774 &  & 0.1962 & 0.1967 & 0.0741 & 0.0772 &  & 0.1840 & 0.1842 & 0.0775 & 0.0802 &  & 0.2087 & 0.2090 & 0.0760 & 0.0787 \\
						LGBM(+H) & 0.1602 & 0.1606 & 0.0725 & 0.0756 &  & 0.1694 & 0.1699 & 0.0712 & 0.0743 &  & 0.1600 & 0.1602 & 0.0751 & 0.0778 &  & 0.1678 & 0.1680 & 0.0749 & 0.0776 \\
						XGB(+H) & 0.1611 & 0.1616 & 0.0726 & 0.0757 &  & 0.1631 & 0.1636 & 0.0717 & 0.0748 &  & 0.1619 & 0.1621 & 0.0762 & 0.0789 &  & 0.1676 & 0.1678 & 0.0744 & 0.0771 \\
						NN1 & 0.1676 & 0.1681 & 0.0029 & 0.0061 & & 0.1816 & 0.1821 & 0.0049 & 0.0082 &  & 0.1664 & 0.1666 & 0.1073 & 0.1001 &  & 0.1743 & 0.1745 & 0.1084 & 0.1013 \\
						NN2 & 0.1504 & 0.1509 & 0.0162 & 0.0194 &  & 0.2319 & 0.2323 & 0.0117 & 0.0150 &  & 0.1576 & 0.1578 & 0.1208 & 0.1237 &  & 0.2307 & 0.2309 & 0.1105 & 0.1134 \\
						NN3 & 0.1800 & 0.1804 & 0.0174 & 0.0207 &  & 0.3461 & 0.3464 & 0.0125 & 0.0158 & & 0.1957 & 0.1959 & 0.1203 & 0.1232 &  & 0.3449 & 0.3450 & 0.1138 & 0.1167 \\
						NN4 & 0.2407 & 0.2411 & 0.0161 & 0.0194 & & 0.3435 & 0.3439 & 0.0143 & 0.0176 & & 0.2105 & 0.2107 & 0.1191 & 0.1220 & & 0.3541 & 0.3543 & 0.1151 & 0.1180 \\
						NN5 & 0.2210 & 0.2215 & 0.0123 & 0.0156 & & 0.3635 & 0.3638 & 0.0123 & 0.0156 & & 0.1967 & 0.1969 & 0.1160 & 0.1189 & & 0.3435 & 0.3436 & 0.1124 & 0.1153 \\
						Oracle & 0.0717 & 0.0722 & 0.0663 & 0.0695 & & 0.0718 & 0.0724 & 0.0663 & 0.0695 & & 0.0740 & 0.0742 & 0.0687 & 0.0715 & & 0.0739 & 0.0742 & 0.0687 & 0.0715 \\
						\hline
					\end{tabular}
					\begin{tablenotes}
						\fontsize{6pt}{0.35cm}\selectfont
						\item \textit{Notes}: In this table, we report the average in-sample (IS) and out-of-sample (OOS) $R^2$ for model 1 and 2 using Ridge, Lasso, Elastic Net (ENet), generalized linear model with group lasso (GLM), random forest (RF), gradient boosted regression trees (LGBM(+H) and XGB(+H)), and five architectures of neural networks (NN1, $\dots$, NN5), respectively. "Oracle" stands for using the true covariates in a pooled-OLS regression. We fix $T=3600$, comparing $P_C = 100$ with $P_C=50$. The number of Monte Carlo repetitions is 100.
					\end{tablenotes}
				\end{threeparttable}
			\end{minipage}
		}
	}
\end{table}

\begin{table}[H]
	\fontsize{9pt}{0.38cm}\selectfont
	\centering
	\caption{\centering Comparison of Average Variable Selection Frequencies in Simulations}
	\label{tab:A15}
	\setlength{\tabcolsep}{7pt}
	\renewcommand{\arraystretch}{1.1}
	\begin{tabular}{p{3cm} p{1.5cm} p{1cm} p{1cm} p{1cm} p{1cm} p{1cm} p{2cm} p{1cm}}
		\hline
		&Model&$p_t$&$s_t$&$x_t$&$C_{1,t}$&$C_{2,t}$&$C_{3,t} \times x_t$ &Noise \\
		\hline
		Panel A: Model 1 & & & & & & & & \\
		\hline
		$P_C = 50$ & LASSO & 1 & 0.12 & 0.1 & 0.15 & 0.13 & 0.14 & 0.10 \\
		& ENet & 1 & 0.12 & 0.11 & 0.16 & 0.13 & 0.15 & 0.10 \\
		$P_C = 100$ & LASSO & 1 & 0.07 & 0.07 & 0.09 & 0.07 & 0.08 & 0.06 \\
		& ENet & 1 & 0.08 & 0.08 & 0.1 & 0.09 & 0.11 & 0.07 \\
		\hline
		Panel B: Model 2 & & & & & & & & \\
		\hline
		$P_C = 50$ & LASSO & 1 & 0.12 & 0.13 & 0.15 & 0.1 & 0.13 & 0.10 \\
		& ENet & 1 & 0.15 & 0.14 & 0.19 & 0.11 & 0.15 & 0.11 \\
		$P_C = 100$ & LASSO & 1 & 0.08 & 0.07 & 0.08 & 0.04 & 0.07 & 0.06 \\
		& ENet & 1 & 0.1 & 0.08 & 0.1 & 0.04 & 0.08 & 0.06 \\
		\hline
	\end{tabular}
\end{table}

\end{document}